\documentclass[12pt]{nuthesis}	
\usepackage{graphicx}

\usepackage{mathtools}
\usepackage{amsmath, amsfonts, amssymb}
\usepackage{array}
\usepackage{ptdr-definitions}
\usepackage{subfig}
\usepackage{xspace}

\usepackage[symbol*]{footmisc}
\DefineFNsymbolsTM{myfnsymbols}{
  \textdagger    \dagger
  \textdaggerdbl \ddagger
}%
\setfnsymbol{myfnsymbols}

\usepackage{hyperref}

\newcommand{\ns}{\unit{ns}}

\graphicspath{{figs/}}

\author{Andrey Pozdnyakov} 

\title{Search for the Higgs Boson Decays to  a Photon and Two Leptons\\ with Low Dilepton
  Invariant Mass.}

\field{Physics and Astronomy}
\graduationmonth{December}

\graduationyear{2015}

\begin{document}
\frontmatter		
\maketitle		
\copyrightpage		


\abstract		

A search for a Higgs boson decay $\PH\to\gamma^*\gamma\to\ell\ell\gamma$ is presented.
The analysis is performed using proton-proton collision data recorded by the CMS detector
at the CERN LHC at a centre-of-mass energy of 8\TeV, corresponding to an integrated
luminosity of 19.7\fbinv.  The selected events have an opposite-sign muon or electron pair
and a high transverse momentum photon.  No excess above background has been found in the
three-body invariant mass range $120<m_{\ell\ell\gamma}<150\,\GeV$, and limits have been
derived for the Higgs boson production cross section times branching fraction for the
$\PH\to\gamma^*\gamma\to\ell\ell\gamma$ decay, where the dilepton invariant mass is less
than 20\GeV.  For a Higgs boson with $m_\PH=125\GeV$, a 95\% confidence level (CL)
exclusion observed (expected) limit is 6.7 ($5.9^{+.2.8}_{-1.8}$) times the standard model
prediction.

Additionally, a search for $\PH\to(\JPsi)\gamma\to\mu\mu\gamma$ process is presented, and
an upper limit at 95\% CL on the branching fraction of the $\PH\to(\JPsi)\gamma$ decay for
the 125\GeV Higgs boson is set at $1.5\times10^{-3}$.

\acknowledgements I would like to thank my mentor and advisor Mayda Velasco for all the
support.  I also wish to thank Stoyan Stoynev, Michael Schmitt, Brian Pollack, Nathaniel
Odell, Chia-Ming Kuo and Chiu-Ping Chang for the help and feedback provided for the
analysis.

\preface		

As it is clear from the title of this dissertation, you will learn something new about the
Higgs boson from it. In Chapter \ref{sec:intro} I will give a brief introduction to the
Standard Model (SM) of Particle Physics and describe the Higgs mechanism.  There I will
also motivate the search for the particular decay of the Higgs boson into two leptons and
a photon, which is the main subject of the dissertation.

Then, in Chapter \ref{sec:cms} I will give a short description of the CMS detector
including its main subsystems relevant for the analysis.  There are more than 2000
scientists working on the CMS experiment and everyone contributes to the detector support
in order to ensure its smooth operation. Personally, I was responsible for the beam timing
detector at CMS, during 2012 data-taking.  I implemented the online software for
monitoring of the beam arrival times.  This work is described in Section~\ref{sec:bptx}.

In Chapter~\ref{sec:ana} I come back to the main topic of the dissertation and describe
all the details of the search analysis for $\PH\to\gamma^*\gamma\to\ell\ell\gamma$ and
$\PH\to(\JPsi)\gamma\to\mu\mu\gamma$ decays. And in Chapter~\ref{sec:results} I present the
results and conclude.  I hope you will enjoy the reading.

%
%
%
%
%
%
%
%

\clearpage\phantomsection 
\tableofcontents	

\clearpage\phantomsection 
\listoftables		

\clearpage\phantomsection 
\listoffigures		

\mainmatter             


\chapter{Introduction}
\label{sec:intro}

\section{Standard Model theory and Higgs Mechanism}
\label{sec:SM}

The Standard Model (SM) theory of Particle Physics is one of the greatest achievements of
human mind.  Based on the principles of symmetry it provides a framework for describing
the fundamental interactions between elementary particles.  One of the successes of the
theory was the prediction of $W^{\pm}$ and $\Z^0$ bosons, the carriers of the
\textit{weak} interactions, which were discovered afterwards.  Naively one expected those
bosons to be mass-less, just as a photon is a mass-less carrier of the
\textit{electromagnetic} force, but they are not. In fact, quite the opposite, the mass of
the $W$ boson is now measured to be $m_W = 80.4\,\GeV$, and the mass of the $\Z$ is $m_\Z
= 91.2\,\GeV$.  In order to explain the heavy weak bosons, a neat trick was invented by
introducing a set of new scalar fields through what is now called the
Englert-Brout-Higgs-Guralnik-Hagen-Kibble mechanism, or Higgs mechanism for short.  The
Higgs mechanism is of great relevance to the topic of this dissertation, hence I will
describe it in more detail.

The SM is a Quantum Field Theory presented in the Lagrangian formalism.  The Lorentz
invariant Lagrangian density function of the SM (further in the text I will simply say,
the Lagrangian) can be broken down into two parts:
\begin{equation}\label{eq:l}
  \mathcal{L}_{SM} =  \mathcal{L}_{QCD} + \mathcal{L}_{EW}.
\end{equation}
The first term in (\ref{eq:l}) describes the interactions between quarks and gluons, and
this theory is called Quantum Chromodynamics (QCD).  The details of QCD is not discussed
here, instead, one could refer to e.g.~\cite{Peskin,PDG} for this description.  The second
term in (\ref{eq:l}) represents the Electro-Weak theory (EW) and it is the term we are
interested in.

The EW theory is based on the gauge group $SU(2)_L\times U(1)_Y$, with four gauge vector
fields $A^1, A^2, A^3, B$ and two coupling constants $g$ and $g'$.  The left-handed
fermion fields transform as doublets under $SU(2)_L$ group, while the right-handed fields
transform as singlets under this group, that is:
\begin{equation}
\psi_i =
\left(\!
    \begin{array}{c}
      \nu_i \\
      \ell_i
    \end{array}
  \!\right)_L
\quad
\left(\!
    \begin{array}{c}
      u_i \\
      d_i
    \end{array}
  \!\right)_L
\quad
(u_i)_R \quad (d_i)_R \quad (\ell_i)_R,
\end{equation}
where $i=1,2,3$ for the three families of fermions.  In addition, a complex doublet field
$\Phi$ is introduced in order to generate the masses of weak bosons and fermions:
\begin{equation}
\Phi \equiv
\left(\!
    \begin{array}{c}
      \phi_1 \\
      \phi_2
    \end{array}
    \!\right)
\end{equation}

The EW Lagrangian can be written as:
\begin{equation}
  \mathcal{L}_{EW} =   \mathcal{L}_{g} + \mathcal{L}_{\Phi} + \mathcal{L}_{f} + \mathcal{L}_{Y},
\end{equation}
where the $\mathcal{L}_g$ term describes the interactions of the $A^i$ and $B$ fields,
$\mathcal{L}_{\Phi}$ is a component for the scalar field, $\mathcal{L}_f$ is the fermionic
kinetic term, and $\mathcal{L}_Y$ gives the Yukawa interaction between fermions and field
$\Phi$.

In order to explain the Higgs mechanism, let's describe the bosonic plus scalar part of
the theory in more detail.  Its Lagrangian is given by:
\begin{equation}
\label{eq:lagr0}
  \mathcal{L}_{g} + \mathcal{L}_{\Phi} =
  -\frac14 \Fmn^a\Fmn^a - \frac14 \Bmn\Bmn + (D_\mu\Phi)^\dagger D_\mu\Phi
  - \lambda\left(\Phi^\dagger\Phi - \frac {\upsilon^2}{2}\right)^2,
\end{equation}
where:
\begin{equation}
  \Fmn = \partial_\mu A_\nu^a - \partial_\nu A_\mu^a + g\varepsilon^{abc}A_\mu^bA_\nu^c,
  \quad
  \Bmn = \partial_\mu B_\nu - \partial_\nu B_\mu.
\end{equation}
The co-variant derivative is defined as:
\begin{equation}
  D_\mu\Phi = \partial_\mu\Phi - i \frac{g}{2}\tau^aA_\mu^a\Phi - i  \frac{g'}{2}B_\mu\Phi,
\end{equation}
where the Pauli matrices, $\boldsymbol{\tau}^a$, act on the two-component field $\Phi$.

As stated above, the Lagrangian in (\ref{eq:lagr0}) is invariant under $SU(2)\times U(1)$
group, with the generators $\boldsymbol{T}^a = \frac12\boldsymbol{\tau}^a$ and $\mathbf{Y}
= \frac12\boldsymbol{1}$. This Lagrangian describes the interactions of the massless
fields at high energies, $E \gtrsim 1\TeV$.  In order to describe the theory at low
energies we need to determine the state of the system with minimal energy -- the ground
state, and rewrite the Lagrangian in terms of the excitations above the ground state.  The
fluctuations of the fields above that ground state correspond to particles.

Because the potential term of the scalar field in $\mathcal{L}_{\Phi}$ is written in such
a specific way (known as the Mexican hat potential), it produces degenerate ground states
of the field.  Following the conventions in~\cite{Rubakov}, let's pick the ground state of
the $\Phi$ field as:
\begin{equation}
  \Phi^{vac} =
  \left(\!
    \begin{array}{c}
      0 \\ \frac{\upsilon}{\sqrt2}
    \end{array}
    \!\right),
\end{equation}
where $\upsilon$ is a constant called Higgs vacuum expectation value, which has a value of
246\GeV.

Once the ground state of $\Phi$ is chosen, the Lagrangian is no longer symmetric under
$SU(2)\times U(1)$, but it remains symmetric under a new generator, $\boldsymbol{Q}$:

\begin{equation}
  \boldsymbol{Q} =
  \left(
    \begin{matrix}
      1 & 0  \\
      0 & 0
    \end{matrix}
  \right),
\end{equation}
which can also be expressed as:

\begin{equation}
  \boldsymbol{Q = T^3 - Y}.
\end{equation}
Here $\boldsymbol{Q}$ is an generator of electric charge, $\boldsymbol{T}^3$ is the
generator of isospin and $\boldsymbol{Y}$ is the hypercharge.  Hence, the new Lagrangian
is invariant under the new, $U(1)_{EM}$ group, which is a sub-group of $SU(2)\times
U(1)_Y$.

In order to write down the Lagrangian at low energies, we define the excitation of the
field $\Phi$ near its vacuum as:
\begin{equation}
  \Phi =
  \left(\!
    \begin{array}{c}
      0 \\ \frac{\upsilon}{\sqrt2} + \frac{\chi}{\sqrt2}
    \end{array}
    \!\right),
\end{equation}
where $\chi(x)$ is a real scalar field.  Substituting this in eq.~(\ref{eq:lagr0}) and
carrying on the calculation (omitted here, see e.g.~\cite{Rubakov}), we can write the
quadratic part of the Lagrangian as:

\begin{align}
\label{eq:lagr1}
  \mathcal{L}^{(2)} = & -\frac12\Wmn^+\Wmn^- + m_W^2W_\mu^+W_\mu^- - \frac14\Fmn\Fmn - \\
  ~             & -\frac14\Zmn\Zmn+\frac{m_Z^2}{2}Z_\mu Z_\mu +\frac12(\partial_\mu\chi)^2 -\frac{m_\chi^2}{2}\chi^2, \nonumber
\end{align}
where:
\begin{equation*}
  \Wmn^{\pm} = \partial_\mu W_\nu^{\pm} - \partial_\nu W_\mu^{\pm}, \quad
  \Fmn = \partial_\mu A_\nu - \partial_\nu A_\mu, \quad
  \Zmn = \partial_\mu Z_\nu - \partial_\nu Z_\mu
\end{equation*}
and the fields $A^1,A^2,A^3,B$ are transformed into $W^{\pm}, Z$ and $A$ according to:
\begin{align}
\label{eq:WZA}
  W^{\pm}  &= \frac{1}{\sqrt2}(A^1 \mp iA^2), \nonumber\\
  Z~~      &= -B\sin\theta_W + A^3\cos\theta_W, \\
  A~~      &= ~~B\cos\theta_W + A^3\sin\theta_W. \nonumber
\end{align}
Here the $\theta_W$ is called weak mixing angle and defined as:
\begin{equation*}
  \tan\theta_W = \frac{g'}{g}.
\end{equation*}

The masses in (\ref{eq:lagr1}) are composed from parameters, $g, g',
\lambda$ from the original Lagrangian~(\ref{eq:lagr0}) and the Higgs
vacuum expectation value, $\upsilon$, as follows:
\begin{equation}
  \label{eg:Hm1}
  m_W = \frac{g\upsilon}{2}, \quad m_Z = \frac{\sqrt{g^2+g'^2}\upsilon}{2}, \quad m_\chi = \sqrt{2\lambda}\upsilon.
\end{equation}
The fields in (\ref{eq:WZA}) now correspond to the well known, massive $W^{\pm}$ and
$\Z^0$ bosons, and the photon, $\gamma$.  Hence, the constructed theory with Lagrangian
(\ref{eq:lagr1}) is a combined theory of the electromagnetic and weak interactions.  This,
in essence, describes the Electroweak symmetry breaking and the Higgs mechanism.  One of
the consequences and the prediction of the theory is the existence of a new particle,
corresponding to the scalar field $\chi$, called the Higgs boson, with the mass, $m_\chi
\equiv m_\PH = \sqrt{2\lambda}\upsilon$.

As recent as 2012, a new particle with the mass of 125\GeV was discovered at the Large
Hadron Collider (LHC) by the ATLAS\footnote{A Toroidal LHC Apparatus} and
CMS\footnote{Compact Muon Solenoid} experiments~(\cite{atlas_h,cms_h}). This particle is
now widely accepted as \textit{the} Higgs boson predicted by the Standard Model, which is
another success of the theory.  Experimentally, the Higgs boson is studied at the LHC by
colliding protons at very high energies, which enables its production, and detecting the
decay products of the boson.  In Section~\ref{sec:prod} I describe the physics mechanisms
by which the Higgs boson is produced at the LHC and in Section~\ref{sec:decay} I give an
overview of its decays and experimental sensitivity of detecting them.

Since the SM is fully constrained (there are no free parameters), the properties of the
Higgs boson, including its decay branching fractions, are predicted by the theory.  Hence,
any deviations from these predictions, observed experimentally, would point out to the New
Physics, also called the Physics Beyond the Standard Model (BSM).  It is understood that
the SM theory, in general, is not complete and deviations from it are expected.
Therefore, a search for such deviations is now a priority of the LHC experiments.  There
are no evidence for any deviations found so far.  In that respect the rare decays of the
Higgs boson are interesting, because, while they are rare in the SM, they could be
enhanced one way or another within the BSM models.  The main topic of my dissertation is a
search for the $\PH\to\gamma^*\gamma\to\ell\ell\gamma$ rare decay, described in
Section~\ref{sec:Htollg}.

\section{Higgs boson production at the LHC}
\label{sec:prod}
The LHC \cite{LHC} is a proton-proton synchrotron collider build in the underground
circular tunnel at the European Organization for Nuclear Research (CERN) on the border
between France and Switzerland.  It has a circumference of 27 km and was designed to
accelerate the proton beams to the energies of 7\TeV per beam.  That energy however has
not been reached yet and the main collision data, taken in 2011 and 2012, are with 3.5 and
4\TeV per beam, i.e. 7 and 8\TeV center-of-mass energy.

At high energies of the LHC, a proton is no longer a composite of just the $uud$ quarks,
but it consists of a spectrum of gluons and quarks of all flavors -- commonly called
\textit{partons}.  The relative composition of those partons and their dependence of the
proton energy are described by the Parton Distribution Functions (PDF) \cite{pdf4lhc0}.
It turns out that at the LHC the dominant component of the proton is a gluon, thus the
hard collision processes are dominated by the gluon-gluon interactions.

In order to understand the production of the Higgs boson, we need to know its coupling to
other particles of the SM.  In the Lagrangian of the eq.~(\ref{eq:lagr1}) only quadratic
terms are kept.  If we were to expand it, the three-particle interactions with weak bosons
emerge in the full Lagrangian, and the coupling of these interactions is given by:
\begin{equation}
  \label{eq:HVV}
  g_{HVV} = \frac{2m_V^2}{\upsilon}.
\end{equation}
Therefore, the interactions of the $W$ and $\Z$ bosons with the Higgs boson is
proportional to their mass squared.  Furthermore, the fermionic part of the
$\mathcal{L}_{EM}$ was omitted in (\ref{eq:lagr0}).  Once included, after the symmetry
breaking described in Sec.~\ref{sec:SM}, the interaction term of the fermions with the
Higgs field is given by:
\begin{equation}
  \mathcal{L}_{F} = \sum_f \frac{gm_f\chi}{2M_W} \bar{\psi_f}\psi_f,
\end{equation}
which means that the coupling of the Higgs boson to the fermions is proportional to the
mass of a fermion:
\begin{equation}
  \label{eq:Hff}
  g_{Hf\bar{f}} = \frac{gm_f}{2M_W} =  \frac{m_f}{\upsilon}.
\end{equation}

\begin{figure}[b]
  \centering
  \includegraphics[width=0.6\textwidth]{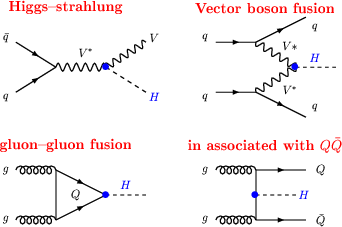}
  \caption[Diagrams of the Higgs boson production processes at hadron colliders.]
  {Diagrams of the Higgs boson production processes at hadron colliders, image from \cite{higgs-theory}.}
  \label{fig:Higgs-prod}
\end{figure}

\begin{figure}[t]
  \centering
  \includegraphics[width=0.6\textwidth]{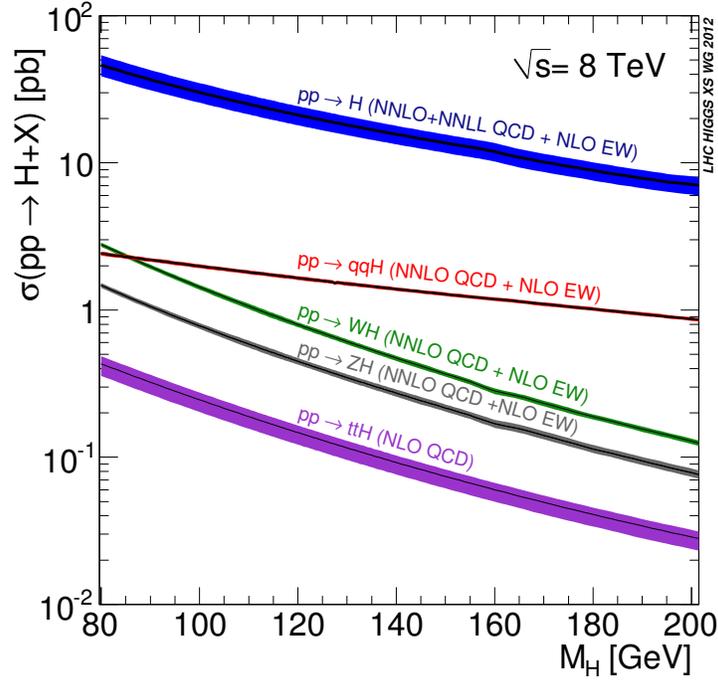}
  \caption{Predicted Higgs boson production cross sections vs $m_\PH$ at the LHC for
    $\sqrt{s}=8\TeV$.}
  \label{fig:Higgs-prod-cs}
\end{figure}

The leading Higgs boson production processes at hadron colliders are shown in
Fig.~\ref{fig:Higgs-prod}.  The relative production rates for a SM Higgs boson with
$m_\PH=125\GeV$ at $\sqrt{s} = 8\TeV$ at the LHC are: gluon-gluon fusion (ggF) -- about
88\%; Vector Boson Fusion (VBF) -- 7\%; associated production with a Z or W boson (VH) --
5\%; and $t\bar{t}$ fusion (ttH) -- 0.4\%. Figure~\ref{fig:Higgs-prod-cs} shows the
production cross sections versus a Higgs boson mass.

Even though the gluon fusion process dominates, there are experimental advantages of the
VBF and VH modes: tagging events with extra particles and reducing the backgrounds.  In
the VH production, the tag is based on the leptonic decays of the Z/W bosons: missing
transverse energy due to neutrinos ($E_T^{miss}$), and/or the presence of charged leptons
-- W($\ell\nu$)H, Z($\ell\ell$)H and Z($\nu\nu$)H.  Let me note the branching ratios of
those decays: $\mathcal{B}(W\to\ell\nu) \approx 10\%$, $\mathcal{B}(Z\to\ell\ell) \approx
3.4\%$ per lepton, $\mathcal{B}(Z\to\nu\nu) \approx 20\%$.  The typical VBF tag requires
an event with two jets with large invariant mass, $m_{jj} > 500$\,GeV, and large angular
separation between the jets.

\subsection{Background processes}
\begin{figure}[h]
  \centering
  \includegraphics[width=0.8\textwidth]{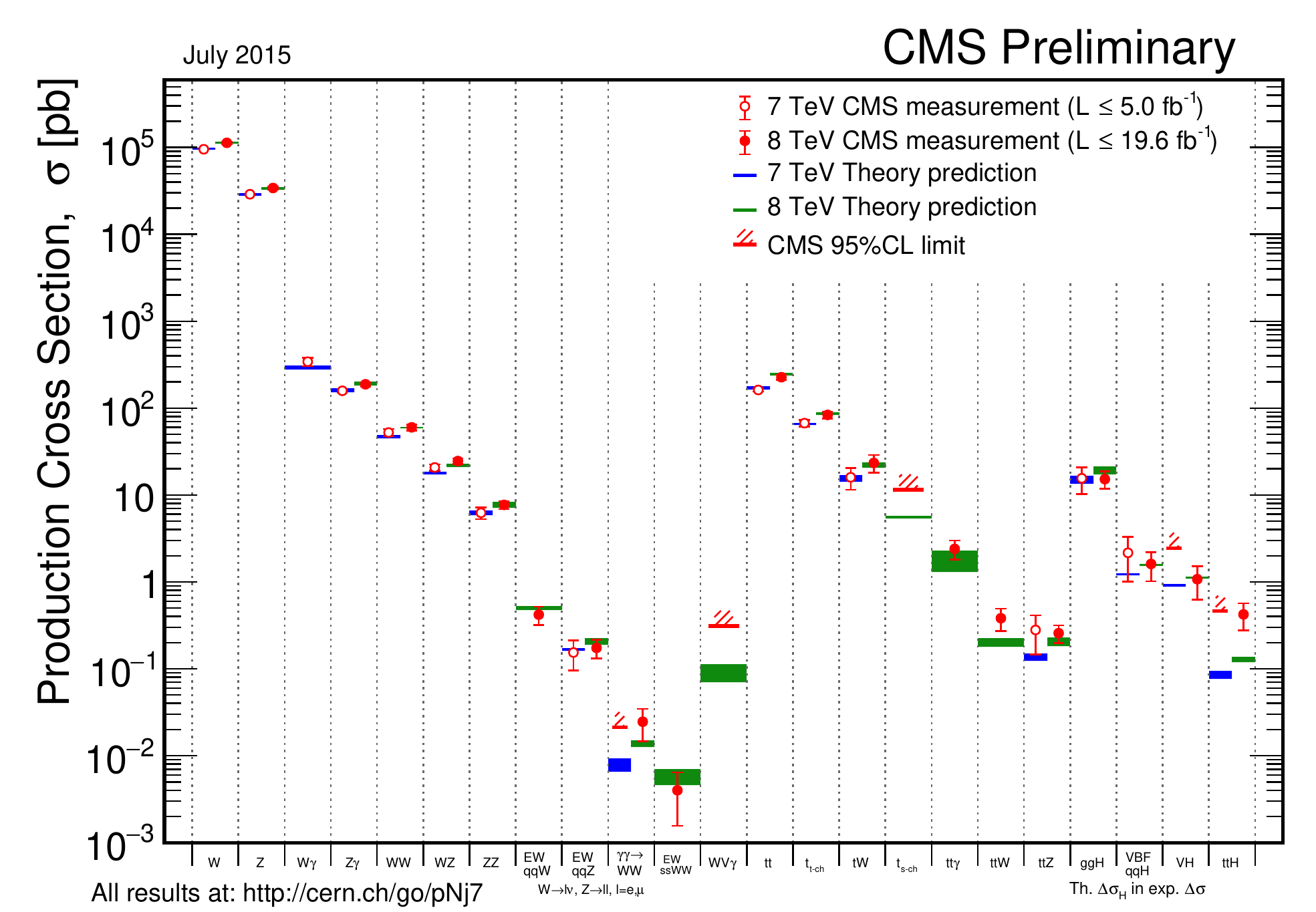}
  \caption{Cross sections of the SM processes at LHC.}
  \label{fig:cs-SM}
\end{figure}

Since the protons are collided at the LHC, the QCD part of the SM presented in
eq.~(\ref{eq:l}) becomes quite relevant in experimental observation of the Higgs boson and
studying its properties.  The total inelastic cross section at $\sqrt{s}=7\TeV$ is
measured to be around 60\unit{mb} \cite{cs-ATL,cs-CMS} and the QCD processes contribute a
large part to this cross section. Therefore, it is the dominant background, which we have
to deal with when searching for new processes and particles. The Electroweak SM processes
also have large cross sections, compared to the Higgs boson production. For example, the
cross sections for the $W$ or $\Z$ bosons production are on the order of
$10^{4}-10^{5}\unit{pb}$, while the total Higgs boson production cross section is about
20\unit{pb}.  Figure~\ref{fig:cs-SM} shows a summary of the cross sections of the EW
processes measured by CMS~\cite{CMS-cs-summary}.

\newpage
\section{Decays of the Higgs boson}
\label{sec:decay}

\begin{figure}[b]
  \begin{center}
    \subfloat[]{\includegraphics[width=0.2\textwidth]{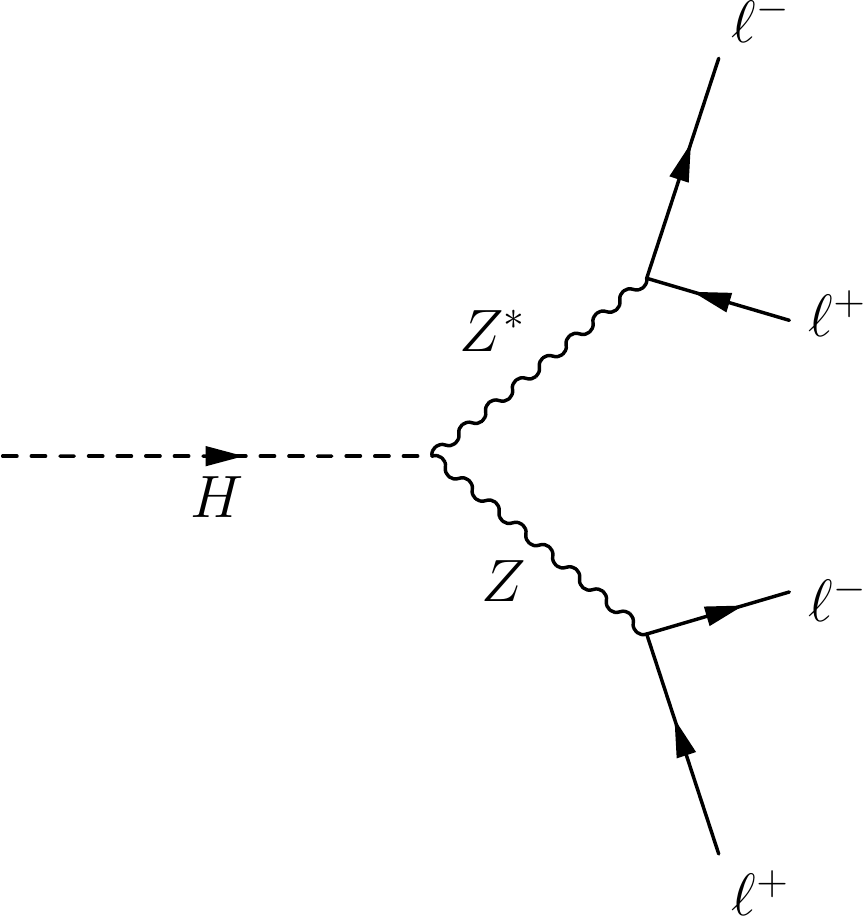}\label{fig:dia-H4l}}
    \qquad
    \subfloat[]
    {
      \includegraphics[width=0.2\textwidth]{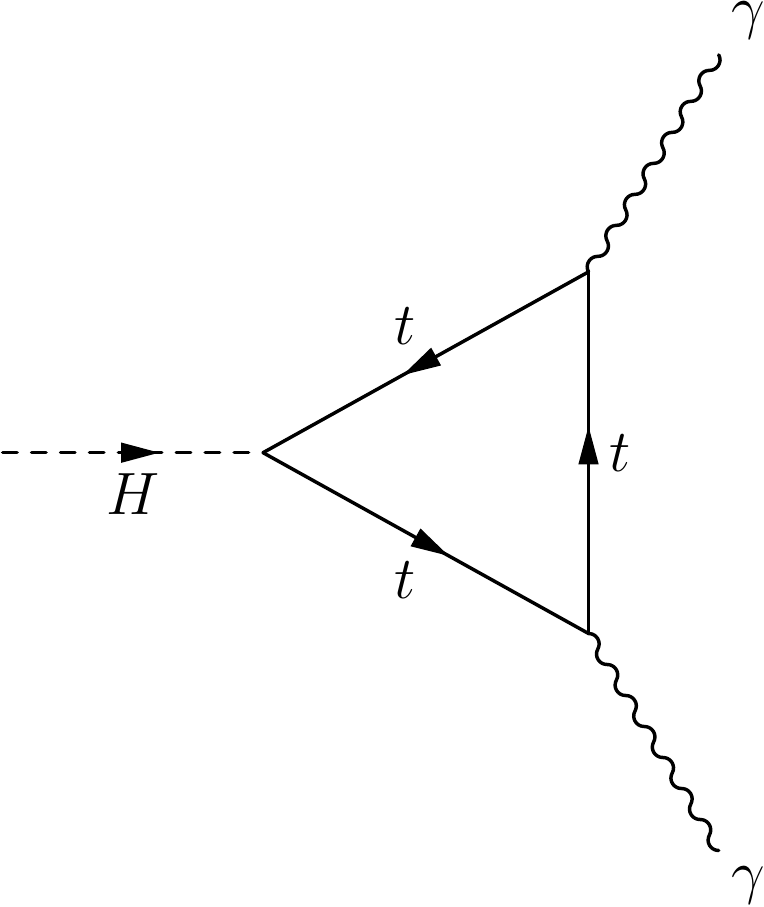}~~~
      \includegraphics[width=0.2\textwidth]{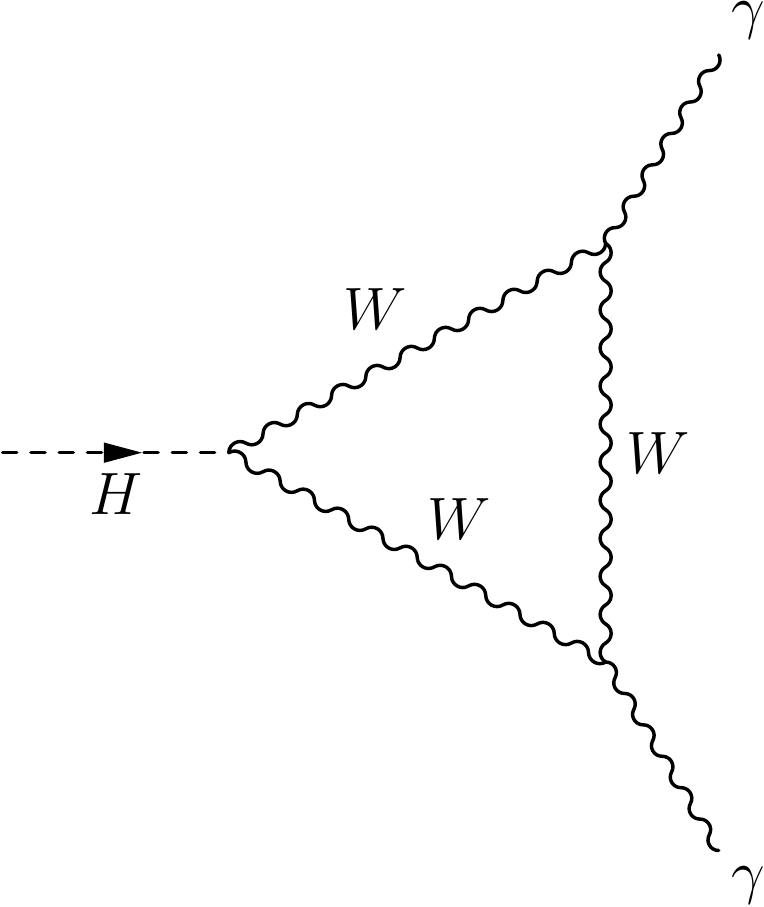}\label{fig:dia-Hgg}
    }
    \caption{Dominant diagrams for golden decay channels of the SM Higgs boson:
      a) $\PH\to ZZ^*\to4\ell$, and b) $\PH\to\gamma\gamma$ processes.}
    \label{fig:dia-golden}
  \end{center}
\end{figure}

\begin{figure}[t]
  \centering
  \includegraphics[width=0.35\textwidth]{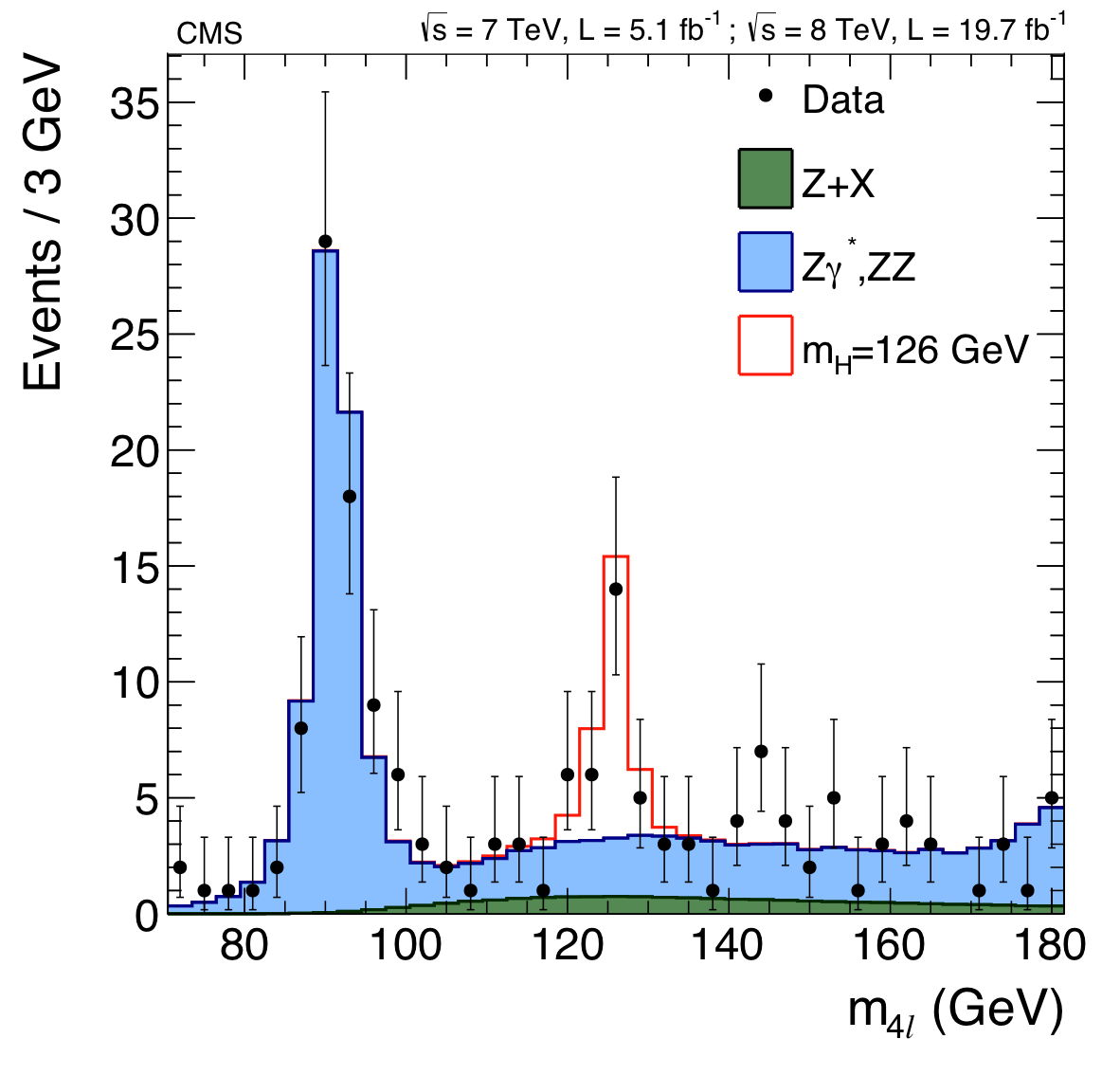}~
  \includegraphics[width=0.35\textwidth]{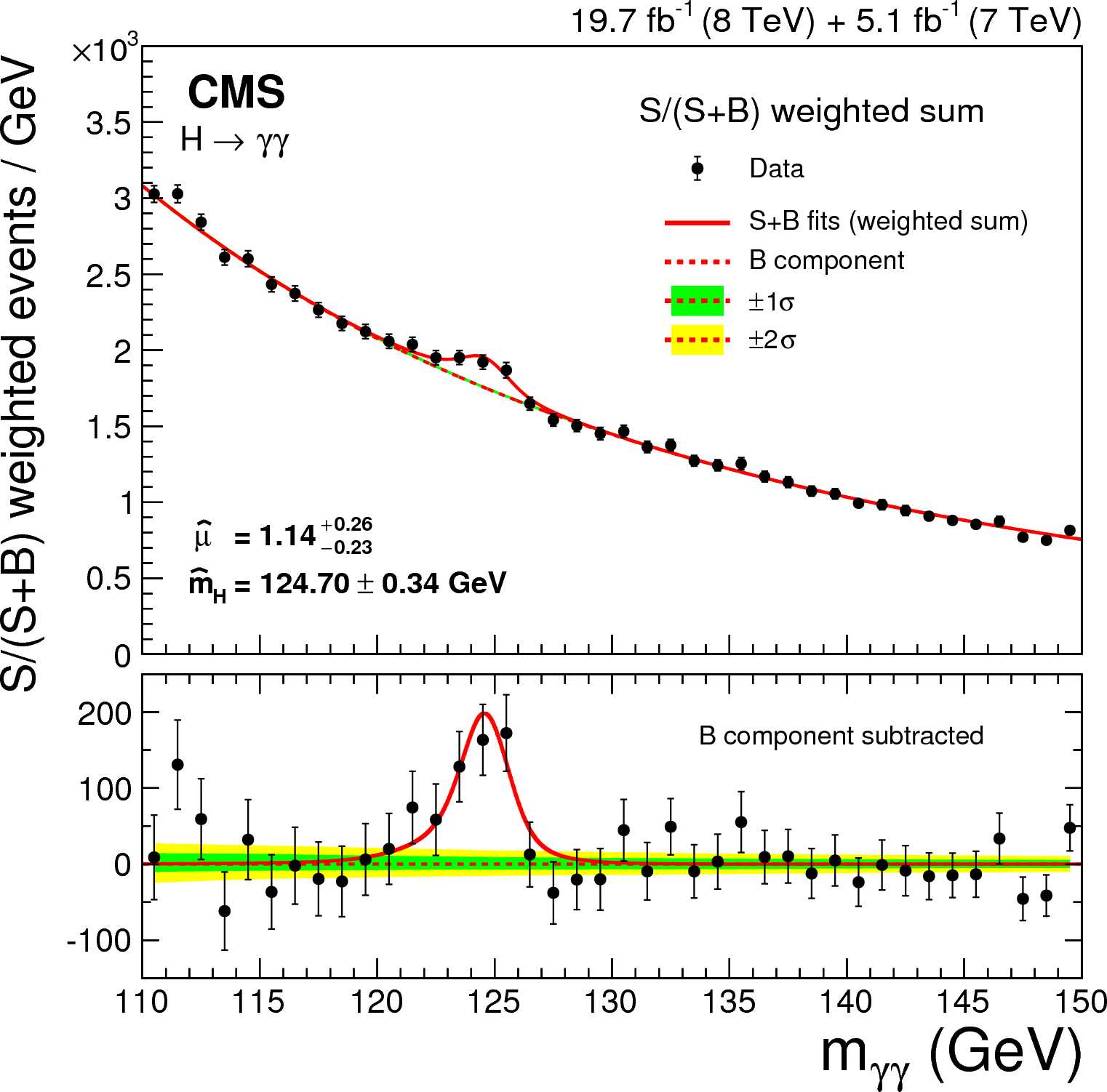}~
  \caption{Invariant mass distributions of the Higgs boson candidates
    from $ \PH\to ZZ^*\to4\ell$ and $\PH\to\gamma\gamma$ analysis at CMS.}
  \label{fig:Hvv_m125}
\end{figure}

Based on eqs.~(\ref{eq:HVV}) and (\ref{eq:Hff}), and taking into account phase space
constraint, the largest decay rates of the Higgs boson for $m_\PH=125\GeV$ arise from
$\PH\to b\bar{b}$ and $\PH\to WW$.  These two processes, however, are difficult from the
experimental point of view. In the first one, the final state involves hadronic jets,
hence it is overwhelmed by background processes, like $\Pp\Pp\to b\bar{b}$. The second
process involves jets as well, from $W\to qq$ decay, or missing energy from neutrinos in
$W\to\ell\nu$ mode, which makes it impossible to reconstruct accurately the invariant mass
of the Higgs boson candidate.  The two most sensitive decay channels at the LHC are
$\PH\to ZZ\to 4\ell$ and $\PH\to\gamma\gamma$, the so called \textit{golden channels} of
the Higgs boson decays. The leading diagrams for these processes are shown in
Figure~\ref{fig:dia-golden}. The Higgs boson couples to the $\Z$ boson directly but it
does not couple to the photon, therefore the $\PH\to\gamma\gamma$ process occurs via loops
as shown in Fig.~\ref{fig:dia-Hgg}.  The dominant contributions in the loops come from the
heaviest candidates: top quark and $W$ boson (which contribute to the total amplitude with
opposite signs).  Figure~\ref{fig:Hvv_m125} shows the key plots from the two
\textit{golden channel} analyses at CMS -- the invariant mass distributions of the Higgs
boson candidates.  Clear resonant peaks at the same mass, $m_\PH=125\GeV$, manifest the
existence of the particle.  Surely, many other modes are searched for by both ATLAS and
CMS.  Particularly, a search for the direct decays of the Higgs boson to
fermions~\cite{pz-fermions} shows the evidence for $\PH\to\tau\tau$ decay, with the same
$m_\PH$.

Overall, the observed decay rates of the Higgs boson are in-line with the predictions of
the SM.  Now it is the time to look for the rare processes, such as $\PH\to
V\gamma\to\ell\ell\gamma$, for example, where $V=\gamma^*, \Z, \JPsi$ or $\Upsilon$.
While these processes are rare in the SM, they could be enhanced by the presence of the
New Physics.  In the next section I describe the $\PH\to\ell\ell\gamma$ process in more
detail.

\section{Higgs boson decays into \texorpdfstring{$\ell^+\ell^-\gamma$}{l+l-gamma} final state}
\label{sec:Htollg}

\begin{figure}[t]
  \begin{center}
    \subfloat[]{\includegraphics[width=0.25\textwidth]{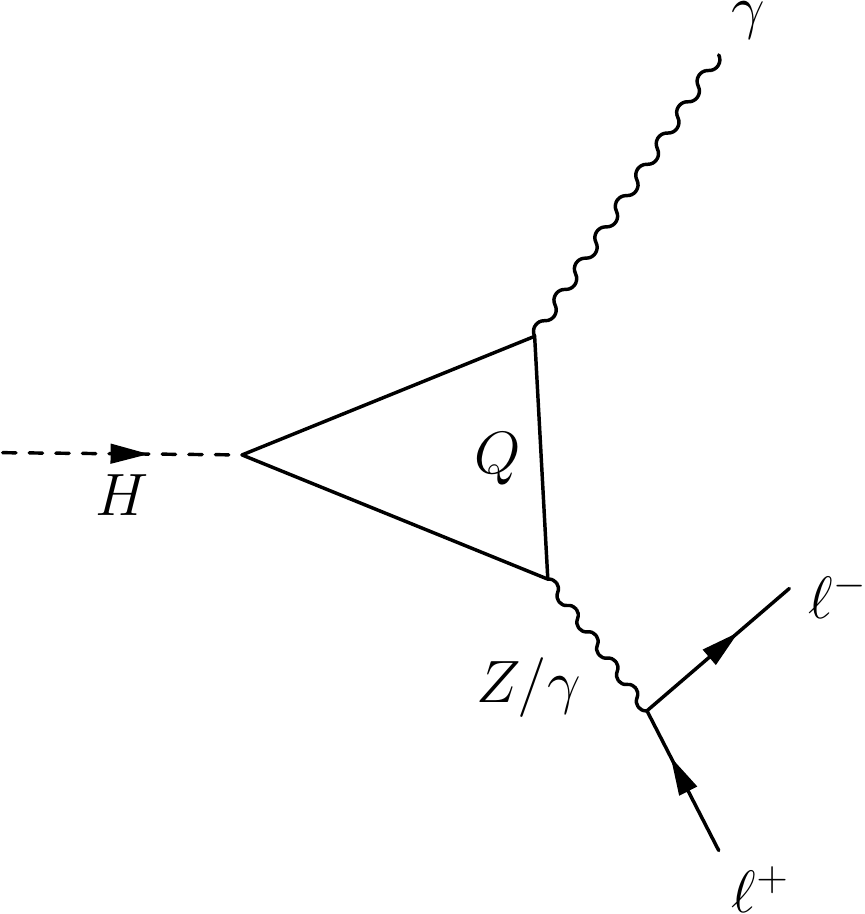}} \hfill
    \subfloat[]{\includegraphics[width=0.25\textwidth]{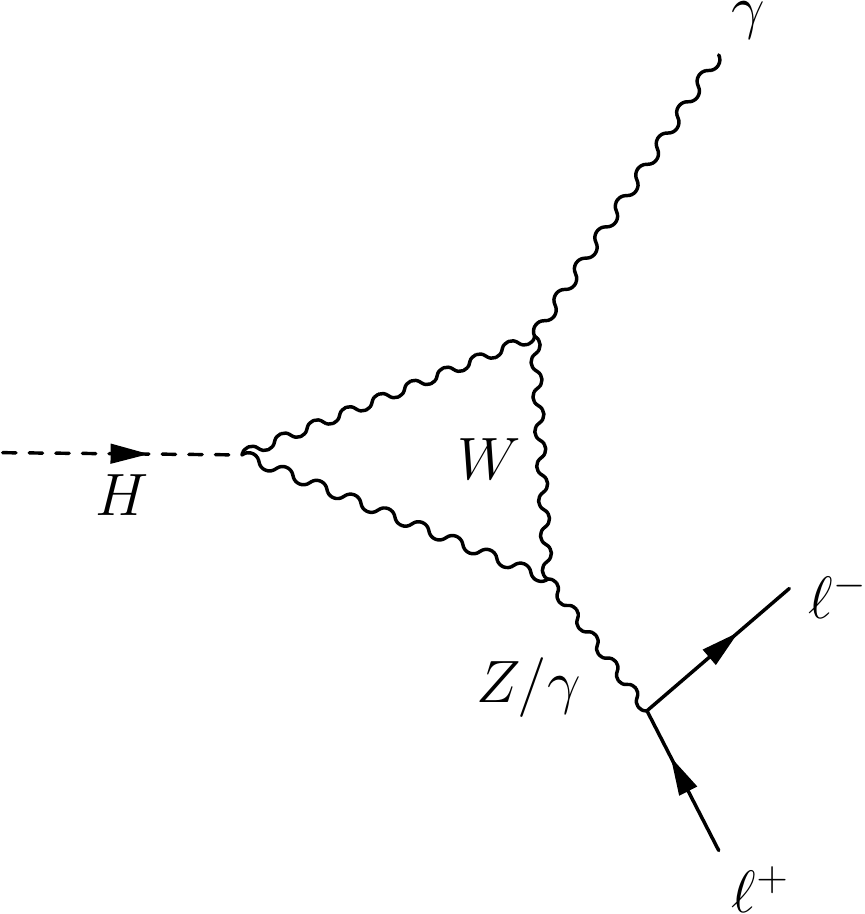}} \hfill
    \subfloat[]{\includegraphics[width=0.25\textwidth]{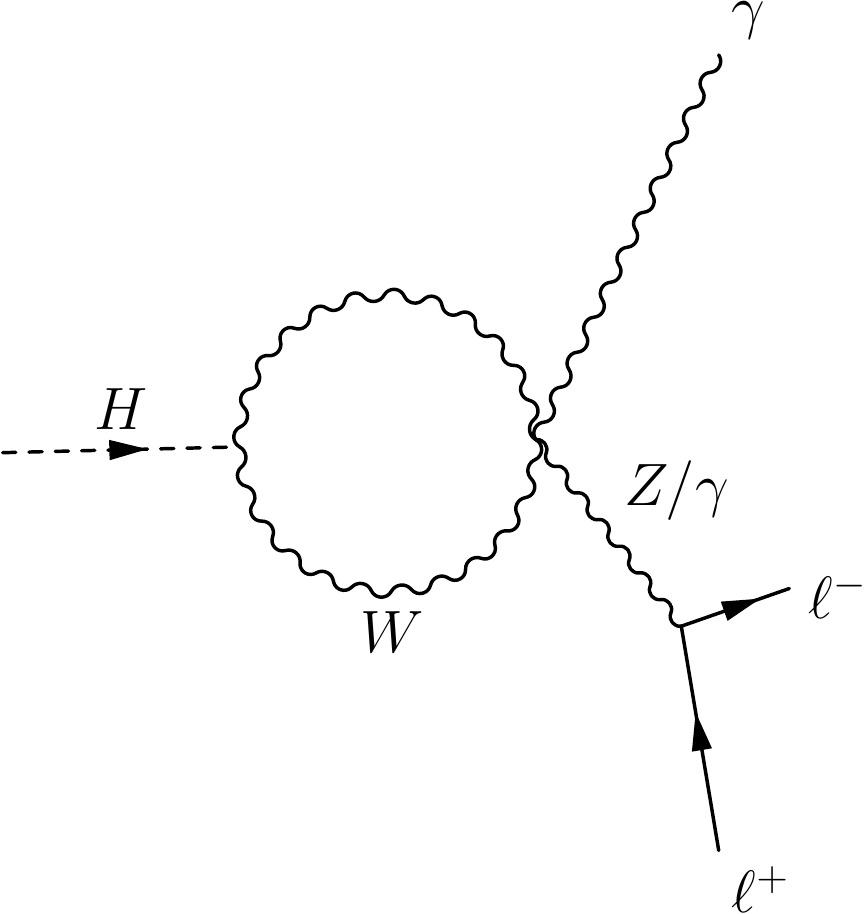}}\\
    \subfloat[]{\includegraphics[width=0.2\textwidth]{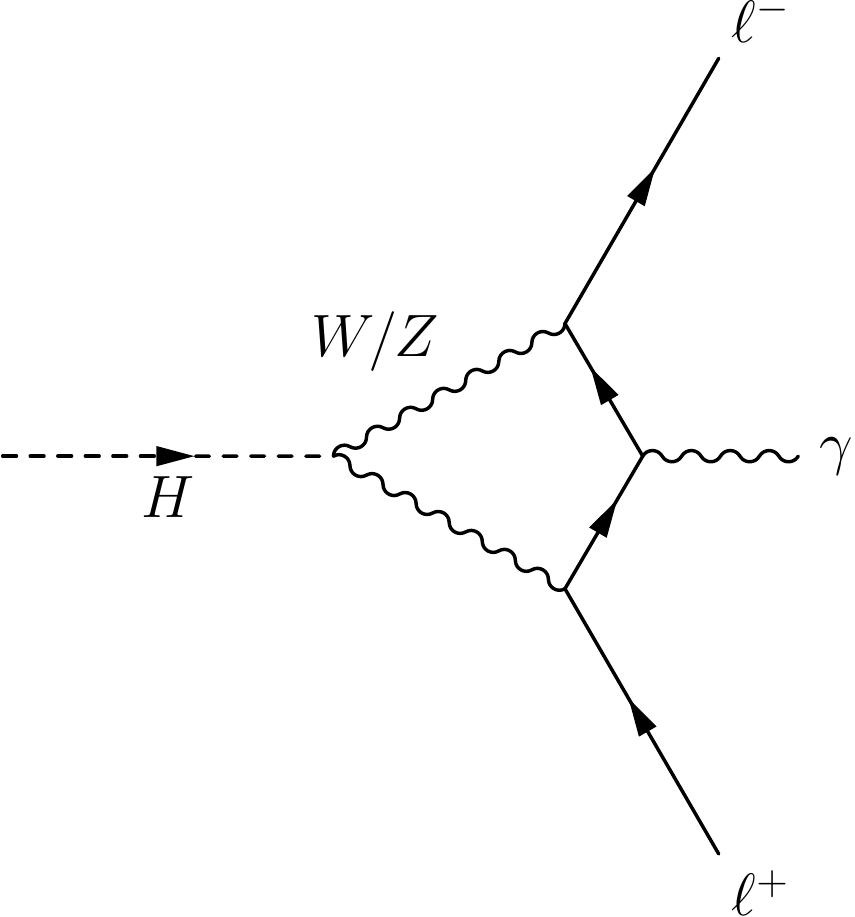}} \hfill
    \subfloat[]{\includegraphics[width=0.2\textwidth]{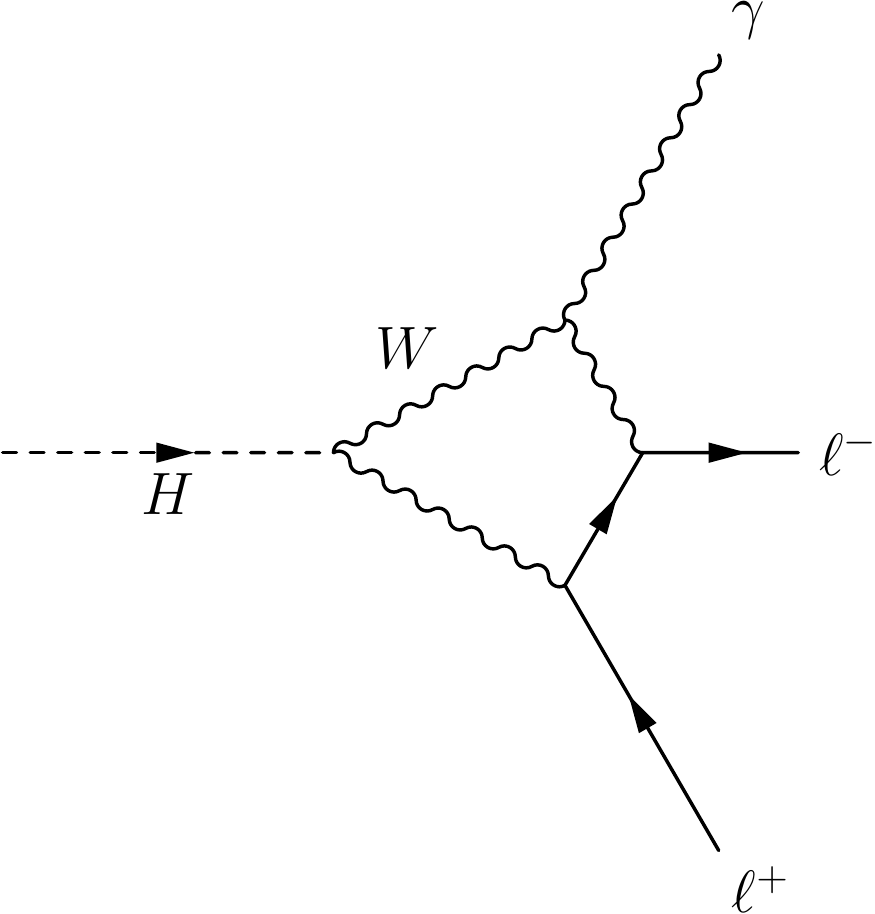}} \hfill
    \subfloat[]{\includegraphics[width=0.2\textwidth]{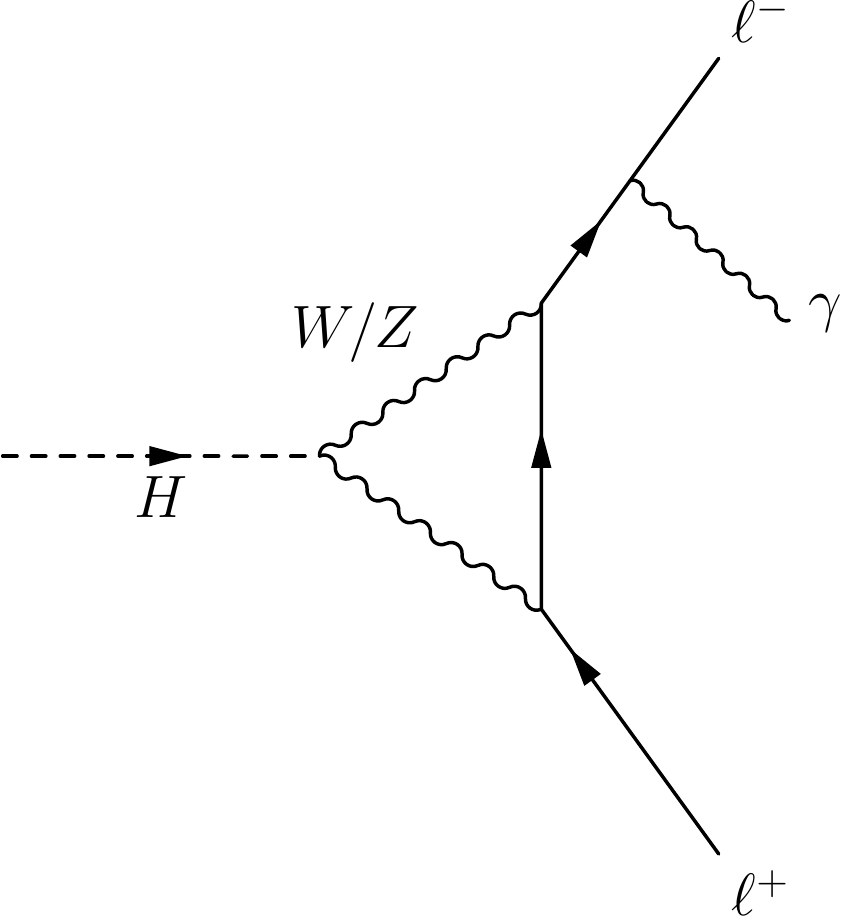}} \qquad
    \subfloat[]{\includegraphics[width=0.2\textwidth]{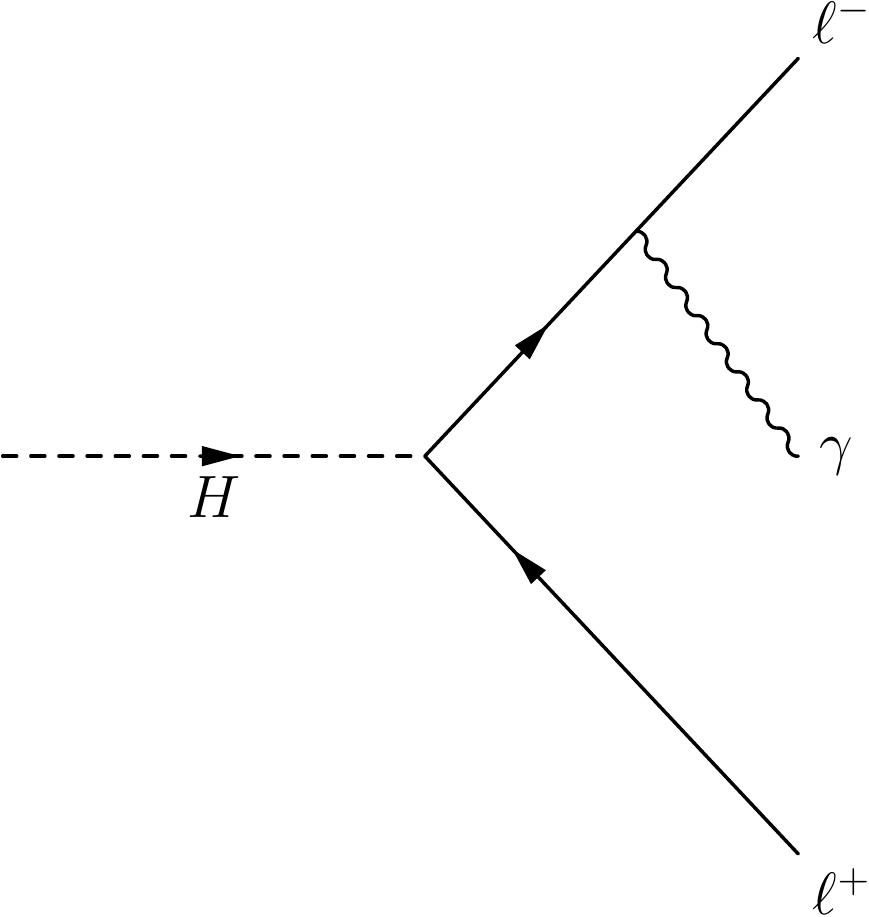}}
    \caption[Diagrams contributing to $\PH \to \ell\ell\gamma$ process.]
    {Diagrams contributing to $\PH \to \ell\ell\gamma$ process. The contributions
      from diagrams (a), (b), and (c) dominate.  Higher order contributions from diagrams
      (d), (e) and (f) are negligible.  The final-state radiation of $\PH\to\ell\ell$
      decay (g) is important at high dilepton invariant mass.}
    \label{fig:dia-dalitz}
  \end{center}
\end{figure}

The decay of the Higgs boson into $\ell\ell\gamma$ final state ($\ell$ = $\mu$ or \Pe),
although rare, provides valuable information to enhance our understanding of the
properties of the newly discovered boson.  The dominant contributions to this decay come
from the loop-induced processes, $\PH\to\gamma^*\gamma$ and $\PH \to \Z\gamma$, where one
of the photons or a $\Z$ boson converts internally into two leptons, as illustrated in
diagrams (a),(b),(c) of Fig.~\ref{fig:dia-dalitz}. These are the so-called \textit{loop}
or \textit{pole} diagrams, where the \textit{pole} refers to the $\gamma^*$ and $\Z^*$
poles.  There are also contributions from the processes represented by box-diagrams, which
do not have the $\Z^*/\gamma^*$ poles (d,e,f), and the final-state radiation (FSR) in the
$\PH\to\ell\ell$ process (g).  Other contributions include $\PH \to V(q\bar{q})\gamma
\to\ell\ell\gamma$ processes, where $V$ denotes a vector meson (like $\JPsi$ and
$\Upsilon$) that decays to $\ell\ell$ pair. This process is discussed in
Sec.\ref{sec:hjp}.

\newcommand{\mffSq}{m_{\bar{f}f}^2}

The calculation of the various contributions to the $\PH\to f\bar{f}\gamma$ decay (here
$f$ denotes all kinematically accessible charged fermions, i.e. $f$ = $\Pe$, $\mu$,
$\tau$, u, d, s, c, b) were demonstrated in Refs.~\cite{Abba96,Dicus13,Chen12,Passarino}.
The phase space of the $f\bar{f}\gamma$ final state can be naturally parametrized by the
invariant mass, $m_{f\bar{f}}$, of the outgoing fermions. Hence,the decay rate of this
process can be expressed as a function of $m_{f\bar{f}}$ as:

\begin{equation}
  \label{eq:int-mff}
\frac{d\Gamma(\PH\to \bar{f}f\gamma)}{d\mffSq} =
\frac{1}{256\pi^3 m_{\PH}^3}\int\limits_{(m_{\bar{f}\gamma}^2)_{min}}^{(m_{\bar{f}\gamma}^2)_{max}}dm_{\bar{f}\gamma}^2\sum_{spin}|\mathcal{M}|^2,
\end{equation}
where the limits of integration are given by
\begin{eqnarray}
  \label{eq:int-lim}
  (m_{\bar{f}\gamma}^2)_{min} &= m_f^2 + \frac 12(m_\PH^2-\mffSq)\left(1 - \sqrt{1 - \frac{4m_f^2}{\mffSq}}\right),\\
  (m_{\bar{f}\gamma}^2)_{max} &= m_f^2 + \frac 12(m_\PH^2-\mffSq)\left(1 + \sqrt{1 - \frac{4m_f^2}{\mffSq}}\right).
\end{eqnarray}

The full matrix-element, $\mathcal{M}$, can be expressed as:
\begin{equation}
  \label{eq:M}
  \sum_{spin}|\mathcal{M}|^2 = C (\mathcal{A}_\gamma, \mathcal{A}_\Z, \mathcal{B}_\Z, \mathcal{B}_W),
\end{equation}
where $\mathcal{A}_\gamma$, $\mathcal{A}_\Z$ and $\mathcal{B}_\Z$, $\mathcal{B}_W$ are the
amplitudes for the pole and box diagrams of Fig.~\ref{fig:dia-dalitz}, respectively.  The
full expressions for these amplitudes, as well as the matrix elements of eq.~(\ref{eq:M}),
are given in Ref.~\cite{Abba96}.  It was found that the contribution from the box diagrams
is quite small. If only the \textit{pole} diagrams are considered, the expression for
$\frac{d\Gamma}{d\mffSq}$ (\ie $m_{\bar{f}f}$ distribution) can be written as:

\begin{equation}
  \label{eq:mff-pole}
\begin{split}
  \frac{d\Gamma}{d\mffSq} = &
  \frac{\alpha^4 m_W^2}{(8\pi)^3 \sin^6{\theta_W} m_{\PH}^3}
  \bigg[
  \sin^4{\theta_W} \frac{|\mathcal{A}_\gamma(\mffSq)|^2}{\mffSq} + 2\sin^2{\theta_W}\upsilon_f
  \Re\left(\frac{\mathcal{A}_\gamma(\mffSq)\mathcal{A}_\Z^*(\mffSq)}{ (\mffSq - m_\Z^2) - im_\Z\Gamma_\Z} \right) +\\
  & + \frac{(1 + \upsilon_f^2) \mffSq |\mathcal{A}_\Z(\mffSq)|^2 }{ (\mffSq - m_\Z^2)^2 + m_\Z^2\Gamma_\Z^2}
  \bigg](m_\PH^2 - \mffSq)\sqrt{1 - \frac{4m_{f}^2}{\mffSq}}\bigg[(m_\PH^2 + 2m_f^2 - \mffSq)^2 +\\
  & + \frac13(m_\PH^2 - \mffSq)^2 \left( 1 - \frac{4m_{f}^2}{\mffSq}\right)
  \bigg]
\end{split}
\end{equation}

\begin{figure}[h]
  \centering
  \includegraphics[width=0.75\textwidth]{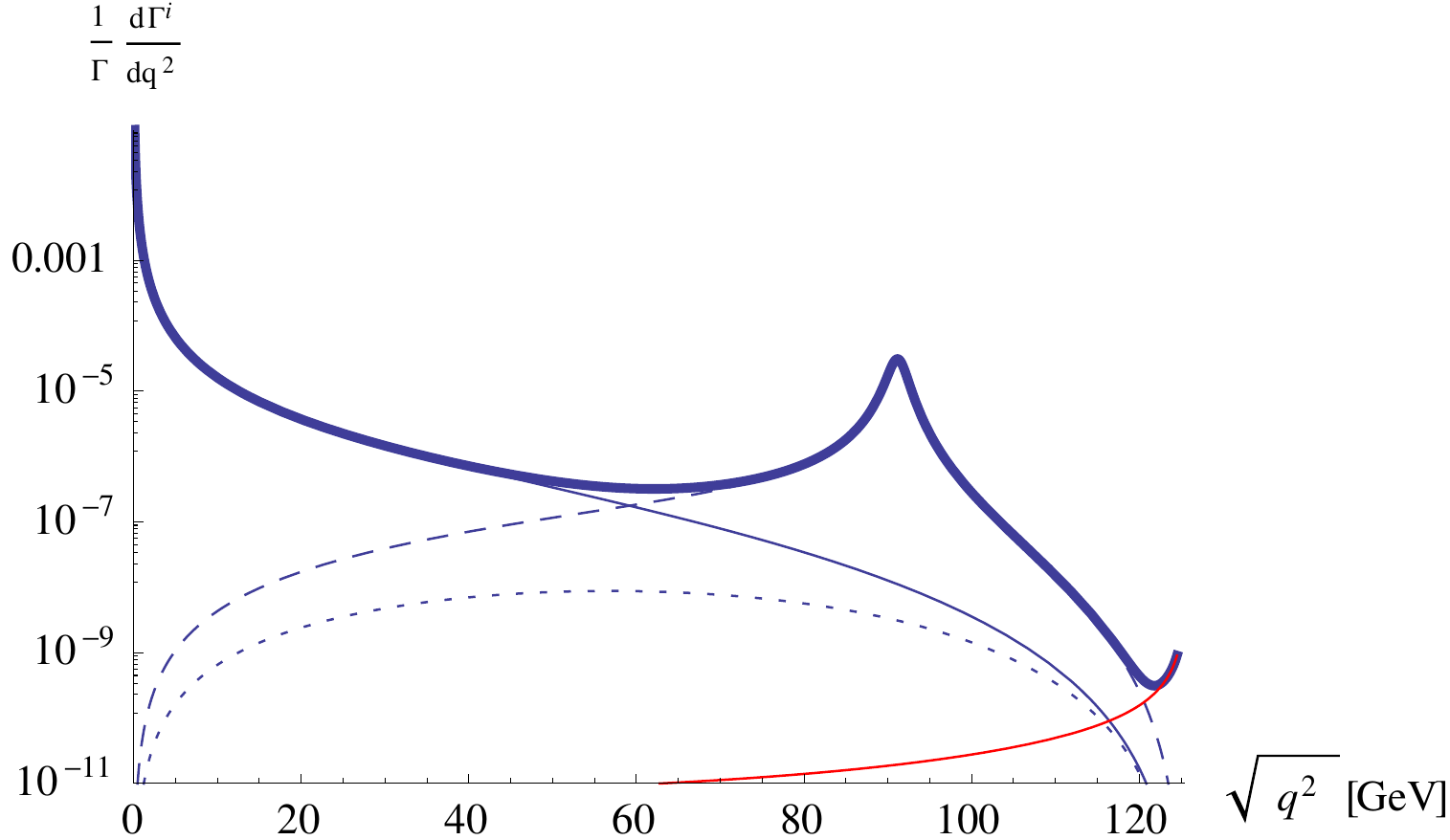}\\
  \vspace{25pt}
  \includegraphics[width=0.75\textwidth]{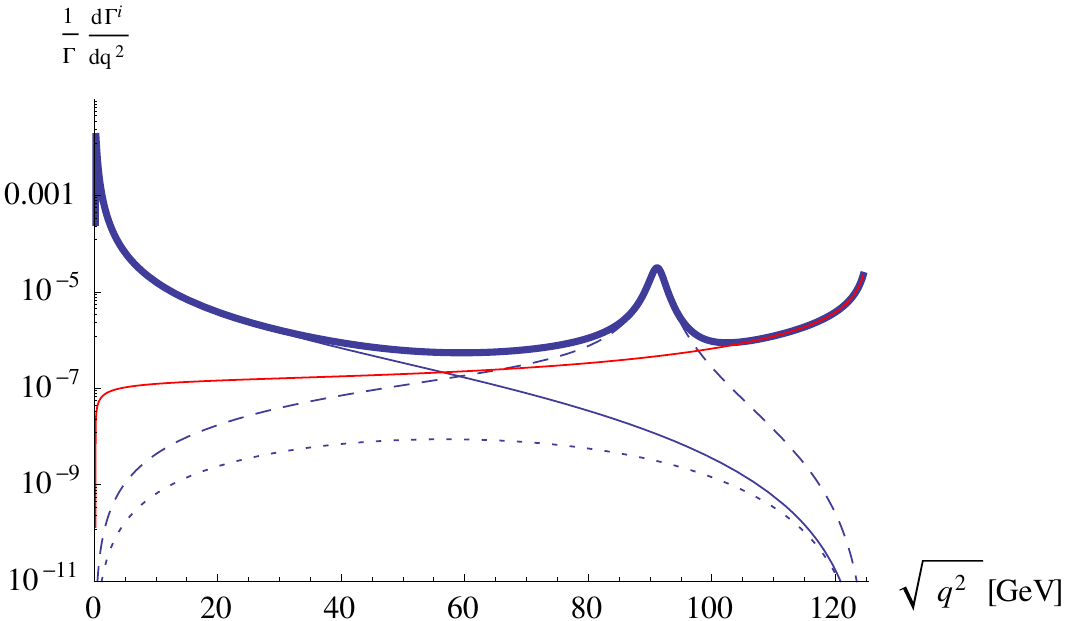}
  \caption[The invariant mass distribution of the two leptons from $\PH\to\ell\ell\gamma$
  decay normalized by $\Gamma(\PH\to\gamma\gamma)$ in the electron and muon
  channels for $m_\PH=125\GeV$.]
  {The invariant mass distribution of the two leptons from $\PH\to\ell\ell\gamma$
    decay normalized by $\Gamma(\PH\to\gamma\gamma)$ in the electron (top) and muon
    (bottom) channels for $m_\PH=125\GeV$.  The thin solid line denotes the contribution
    from the $\gamma^*$ pole diagrams, the dashed line shows the contribution from the
    $\Z^*$ pole diagrams, the red line denotes the contribution of the tree diagrams of
    $\PH\to\mu\mu$ with FSR photon, and the dotted line is the contribution from the
    four-point box diagrams.  The thick blue line gives the total sum. Figures from
    Ref.~\cite{Htollg-FB-Sun}.}
  \label{fig:qll-theo}
\end{figure}

The results of the calculations are illustrated in Fig.~\ref{fig:qll-theo} for the muon
and electron channels (\ie $f=\mu$ or \Pe).  These distribution reveal a few features of
the $\PH\to\ell\ell\gamma$ process.  First of all, there is an obvious peak at the \Z
mass, which arises from the $\Z^*\to\ell\ell$ pole contribution.  Both ATLAS and CMS
collaborations have performed a search for $\PH \to \Z\gamma\to\ell\ell\gamma$ decay, with
$m_{\ell\ell} > 50\GeV$ selection~\cite{atl-HZG, cms-HZG}. The results are consistent with
the SM predictions and the upper limits on the $\sigma/\sigma_{SM}$ are set at
${\sim}10\times$SM.  Secondly, there is a peak at small $m_{\ell\ell}$, which is due to
the photon pole $\gamma^*\to\ell\ell$.  It is important to point out that there is no
singularity at low mass. This can be seen from the integration limits in
eq.~(\ref{eq:int-lim}), which become equal at $m_{\ell\bar{\ell}}^2 = 4 m_\ell^2$, hence
the integral in eq.~(\ref{eq:int-mff}) vanishes. The effect due to the photon pole is
significant, and compatible with the pole at \Z mass, hence one expects to obtain a
similar search sensitivity. In fact, as a result of this dissertation, the sensitivity of
the $\PH\to\gamma^*\gamma$ channel for the SM Higgs boson search turns out to be higher
than of the $\PH\to\Z^*\gamma$ channel. The small contributions of the box diagrams are
also illustrated in the Fig.~\ref{fig:qll-theo}. They are usually neglected in the
simulation by the Monte Carlo (MC) programs.  Finally, at high invariant mass,
$m_{\ell\ell}>100\GeV$, one observes a rise of the curve. It comes from the FSR process,
$\PH\to\ell\ell\to\ell\ell\gamma$, which strength is proportional to the lepton mass
($g_{Hf\bar{f}} = \frac{m_f}{\upsilon}$). Thus, it is more pronounced in the muon channel
than in the electron channel. The $\PH\to\mu\mu$ process by itself is crucial for
understanding the SM, and, of course, it has been searched for by the ATLAS and
CMS~\cite{atl-Hmm,cms-Hmm}. The upper limits on the $\sigma/\sigma_{SM}$ are set at
${\sim}$8 times the SM prediction, consistent with the expected sensitivity.

In the above description I am differentiating the individual processes
$\PH\to\gamma^*\gamma$ and $\PH\to\Z\gamma$, with the final state of
$\ell\ell\gamma$. Strictly speaking this is not correct, since these processes are
ill-defined from the gauge invariance point of view. They interfere and have contributions
from the non-pole diagrams, which are also mentioned. Therefore one should refer to the
total $\PH\to\ell\ell\gamma$ process instead. Nevertheless, I will continue using this
notation, but one has to keep that subtlety in mind (see also discussion in
Ref.~\cite{Passarino}).  Experimentally, the separation of the two processes is achieved
by selecting on the dilepton invariant mass.  For the main subject of this dissertation
only the $m_{\ell\ell}<20\GeV$ part of the $\PH\to\gamma^*\gamma\to\ell\ell\gamma$
spectrum is considered.  However, sometimes, a looser requirement of 50\GeV is used, as
explicitly mentioned.  This decay process is often referred to as \textit{Higgs Dalitz
  decay} in analogy to the $\pi_0 \to \Pe^+\Pe^-\gamma$ decay, induced by an internal
conversion of one of the photons, and named after the physicist Richard Dalitz.

\subsection{Details on the \texorpdfstring{$\PH\to\gamma^*\gamma\to\ell\ell\gamma$}{H to g*g to llg} process}
The expected rate of the $\PH\to\gamma^*\gamma\to ff\gamma$ decay for a Higgs boson mass
of 125\GeV is about 7--10\% of the rate of $\PH\to\gamma\gamma$
decay~\cite{Firan07,Dicus14}, while it is 54\% for $\PH\to \Z\gamma$ process~\cite{YR3}.
If only leptonic decay channels are considered, the corresponding fractions become: 
\small
\begin{equation*}
  \frac{\Gamma(\PH\to\gamma^*\gamma \to \Pe \Pe\gamma)}{\Gamma(\PH\to\gamma\gamma)} \sim 3.5\%,\quad
  \frac{\Gamma(\PH\to\gamma^*\gamma \to \mu\mu\gamma)}{\Gamma(\PH\to\gamma\gamma)} \sim 1.7\% \quad \text{and} \quad
  \frac{\Gamma(\PH\to\Z\gamma\to \ell\ell\gamma)}{\Gamma(\PH\to\gamma\gamma)} \sim 2.3\%.
\end{equation*}
\normalsize

\begin{table}[b]
  \begin{center}
    \begin{tabular}{ c|cccc|cc|cc}
      \multicolumn{5}{c|}{} &\multicolumn{4}{c}{$\mathcal{B}(\PH\to\ell\ell\gamma) \times 10^{-5}$}\\
      & \multicolumn{4}{c|}{$\sigma(\Pp\Pp\to\PH + X)$, fb}  & \multicolumn{2}{c|}{$m_{\ell\ell}<50\,\GeV$} & \multicolumn{2}{c}{$m_{\ell\ell}<20\,\GeV$}\\
      \hline
      $m_\PH$  &gg &VBF & ZH & WH & $\mu\mu\gamma$ & $\Pe\Pe\gamma$ & $\mu\mu\gamma$ & $\Pe\Pe\gamma$\\
      \hline
      120 & 20.9 & 1.65 & 0.47 & 0.81 & 3.73 & 7.75 & 3.21 & 7.25 \\
      125 & 19.3 & 1.58 & 0.42 & 0.70 & 3.83 & 8.07 & 3.33 & 7.45 \\
      130 & 17.9 & 1.51 & 0.37 & 0.62 & 3.83 & 8.01 & 3.28 & 7.37 \\
      135 & 16.6 & 1.45 & 0.33 & 0.54 & 3.64 & 7.62 & 3.09 & 6.97 \\
      140 & 15.4 & 1.39 & 0.29 & 0.48 & 3.32 & 6.87 & 2.82 & 6.42  \\
      145 & 14.5 & 1.33 & 0.26 & 0.42 & 2.89 & 5.96 & 2.48 & 5.62  \\
      150 & 13.6 & 1.28 & 0.23 & 0.37 & 2.35 & 4.87 & 2.00 & 4.49 \\
    \end{tabular}
  \end{center}

  \caption[Cross sections of the SM Higgs boson production  and branching fraction of the $\PH\to\ell\ell\gamma$ decay process.]
  {Cross sections of the SM Higgs boson   production at $\sqrt{s}=8\TeV$ for each production channel;
    and the branching fraction of the $\PH\to\ell\ell\gamma$ decay process in the muon and electron channels,
    for $m_{\ell\ell} < 50\,\GeV$ and $m_{\ell\ell} < 20\,\GeV$.}
  \label{tab:BR}
\end{table}

The full information of the Higgs boson production cross section~\cite{YR3}, and the
branching fraction of the Dalitz decay mode into leptonic final states is presented in
Table~\ref{tab:BR}.  The branching fractions depend on the upper cut on the $m_{\ell\ell}$
and on the mass of the lepton, as emphasized previously.  The numbers reported in
Table~\ref{tab:BR} are given for two selections: $m_{\ell\ell}<20$ and $<50\GeV$.  The
branching fractions are obtained using MCFM\,6.6 program~\cite{MCFM}, where only the
\textit{poll} diagrams are included. MCFM reports the values of $\sigma\cdot\mathcal{B}$,
the cross section times the branching fraction, without the NNLO corrections to the Higgs
boson production. This correction is about ${\sim}$1.16, almost independent of the
$m_\PH$, and taken into account in the values given in Table~\ref{tab:BR}. Muons in the
MCFM calculations are assumed to be massless. Once their mass is taken into account, it
changes the low $m_{\ell\ell}$ part of the spectra and results in the reduction of the
branching fraction by ${\sim}$3--4\%, which is also taken into account in
Table~\ref{tab:BR}.

\begin{figure}[b]
  \centering
  \includegraphics[width=0.7\textwidth]{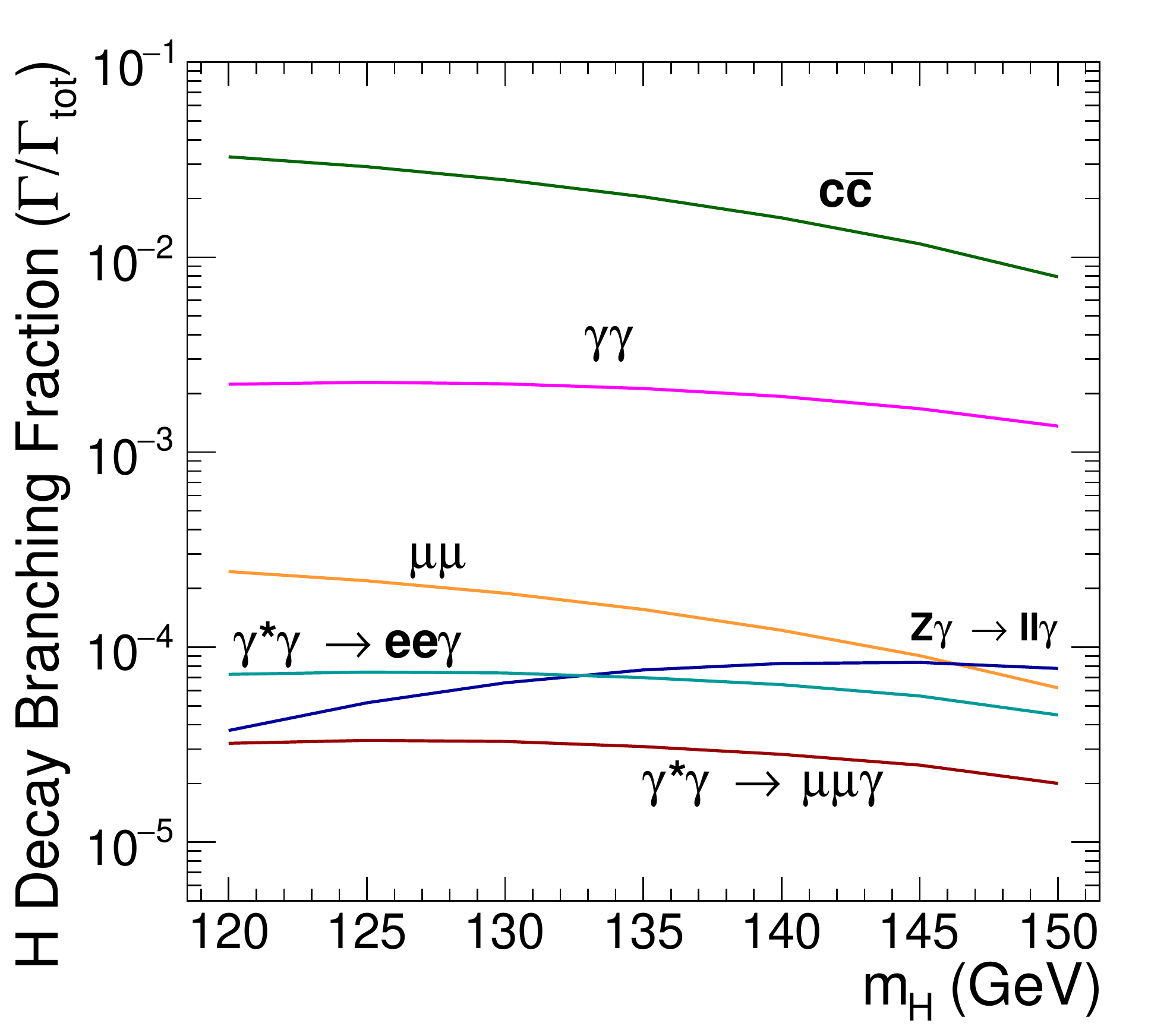}
  \caption[Predicted Higgs boson decay branching fractions in SM.]
  {Predicted Higgs boson decay branching fractions of selected processes.  Here
    the $\mathcal{B}(\PH\to\ell\ell\gamma)$ is shown with $m_{\ell\ell}<20\GeV$
    requirement.}
  \label{fig:BR}
\end{figure}

For comparison with other processes, Fig.~\ref{fig:BR} shows the branching fractions of
the Higgs boson decays into various relevant final states. The
$\mathcal{B}(\PH\to\ell\ell\gamma)$ for the Dalitz mode is shown for
$m_{\ell\ell}<20\GeV$.  Even though the Dalitz decay rate is the smallest, there are
certain advantages of searching for this process:

\begin{itemize}
\item It is sensitive to new physics (NP) beyond the standard model (BSM) via loops.  That
  is, the existence of a new particle could enhance the rate of this decay.  In the
  absence of the NP, it gives an extra handle on the measurement of the Higgs boson
  couplings.  Particularly, it offers a reliable determination of the primary vertex,
  which becomes useful in \textit{high pile-up} environment (see Sec.~\ref{sec:pileup}),
  while, \eg in the $\PH\to\gamma\gamma$ decay the vertex would become harder to
  reconstruct.

\item The $\PH\to\gamma^*\gamma\to \Pe\Pe\gamma$ decay is distinct from the
  $\PH\to\gamma\gamma$ followed by a conversion of a photon to an $\Pe^+\Pe^-$ pair in the
  detector. Experimentally, however, these two processes may be difficult to distinguish
  if the photon conversions are not properly identified.  Hence, one process can become a
  background for another, and it is important to understand their relative contributions
  for the crucial measurement of the $\PH\to\gamma\gamma$ decay.

\item It consists of non-trivial angular correlations that could result in a
  forward-backward asymmetry in the presence of the NP, manifested itself through CP
  violation in $\PH_{\bar{f}f}$ effective coupling~\cite{Htollg-FB-Sun,Htollg-FB-Kor}.
  This feature provides an additional test for the SM and the properties of the Higgs
  boson.
\end{itemize}

It may be also of an interest to mention that the same effective coupling, with inverted
diagrams, is involved in $\Pe^+\Pe^-\to\PH\gamma$ process, which is possibly accessible at
the future linear electron-positron colliders.  The calculation of this process was
performed in Ref.~\cite{eeHg}.


In this dissertation the search for $\PH\to\gamma^*\gamma\to\ell\ell\gamma$ is
presented. The search is performed in muon and electron channels, for the Higgs boson mass
range between 120 and 150\GeV, and it is described in Section~\ref{sec:ana}.  All results
are based on proton-proton collision data recorded in 2012 with the CMS detector at the
LHC at $\sqrt{s}=8\TeV$, corresponding to the integrated luminosity of 19.7\fbinv.

\subsection{\texorpdfstring{$\PH\to V\gamma\to\ell\ell\gamma$}{H to V+gamma}}
\label{sec:hjp}
\begin{figure}[b]
  \centering
  \includegraphics[width=0.35\textwidth]{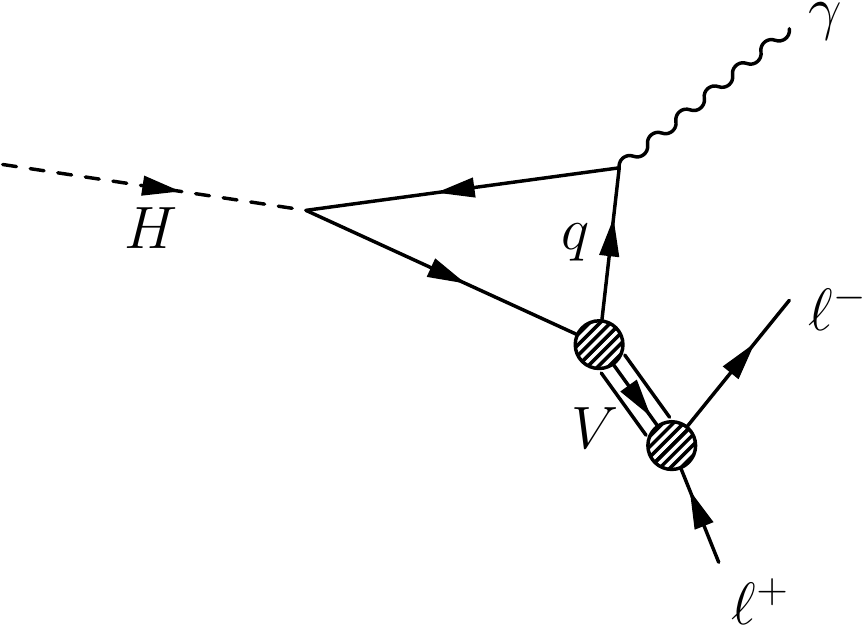}~\qquad
  \includegraphics[width=0.35\textwidth]{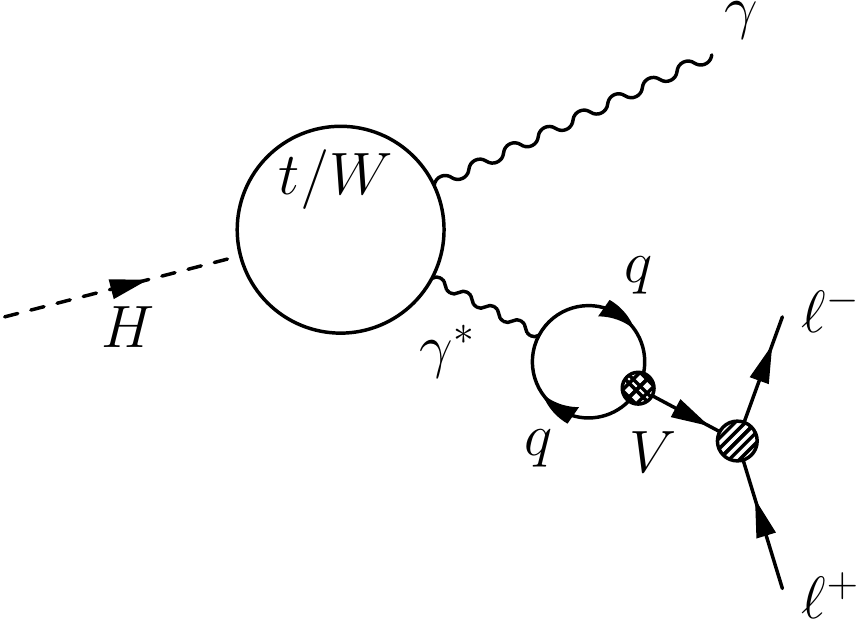}~
  \caption{Main diagrams contributing to the Higgs boson decay through a vector resonance,
    $\PH\to V\gamma\to(\ell\ell)\gamma$.}
  \label{fig:dia-Hjp}
\end{figure}

In addition to the Higgs Dalitz decay, the result for the $\PH \to
(\JPsi)\gamma\to\mu\mu\gamma$ search, at $m_\PH=125\GeV$, will be presented in
Section~\ref{sec:res-Hjp}.  This decay, allows us to test the Higgs boson couplings to the
charm quark, as suggested in Refs.~\cite{hToJPsiGamma-2013,hToJPsiGamma-2014}. It is a
promising way to measure this coupling at the LHC.  There are two mechanisms through
which the $\PH\to V\gamma\to\ell\ell\gamma$ decay occurs (see the diagrams in
Fig.~\ref{fig:dia-Hjp}):

\begin{itemize}
\item A direct process, where the Higgs boson couples to a ${Q}\bar{Q}$ pair (${Q}=c,b$),
  with an FSR radiation of a photon. In this process the $q\bar{q}$ pair hadronizes into a
  vector meson ($V = \JPsi, \Upsilon$), which decays to a pair of leptons.
\item An indirect process, where the Higgs boson decays through a usual t/W loop to a
  $\gamma\gamma^*$ pair with a subsequent decay of the $\gamma^*$ to the vector meson,
  through the ${Q}$-loop.
\end{itemize}

It was calculated in Ref.~\cite{hToJPsiGamma-2014} that the contribution from the indirect
process is in fact larger, and the interference between the two processes is
destructive. In the case of $\Upsilon$, this interference is nearly complete. The widths
of these decays are expected to be:
\begin{eqnarray}
  \label{eq:hVg-width}
  \Gamma_{\JPsi\gamma}    &= 1.42\times [(1.0 \pm 0.017)\kappa_\gamma - (0.087 \pm 0.012)\kappa_c]^2 \times 10^{-8} \GeV,\\
  \Gamma_{\Upsilon\gamma} &= 0.11\times [(1.0 \pm 0.009)\kappa_\gamma - (1.058 \pm 0.045)\kappa_b]^2 \times 10^{-8} \GeV,
\end{eqnarray}
where $\kappa_\gamma$, $\kappa_c$ and $\kappa_b$ parametrize the strength of the
$\PH\gamma\gamma$ and $\PH{QQ}$ couplings.  Taking the total width of the Higgs boson at
$\Gamma_\PH = 4.20\,\MeV$ and $\kappa_\gamma = \kappa_c = \kappa_b= 1$, they obtain:

\begin{eqnarray}
  \label{eq:hV-br}
  \mathcal{B}_{SM}(\PH \to \JPsi + \gamma)    =& 2.79_{+0.16}^{-0.15} \times 10^{-6},\\
  \mathcal{B}_{SM}(\PH \to \Upsilon + \gamma) =& 8.4_{+19.3}^{-8.2} \times 10^{-10}.
\end{eqnarray}
ATLAS has performed the search for these decays, see Ref.~\cite{atl-hjp} for their
results.  First estimates of the bounds on the $\PH{QQ}$ couplings are discussed in
Ref.~\cite{Perez15}, based on the ATLAS and CMS results.


\chapter{Experimental Apparatus}
\label{sec:cms}
\section{Compact Muon Solenoid}
\begin{figure}[t]
  \centering
  \includegraphics[width=0.85\textwidth]{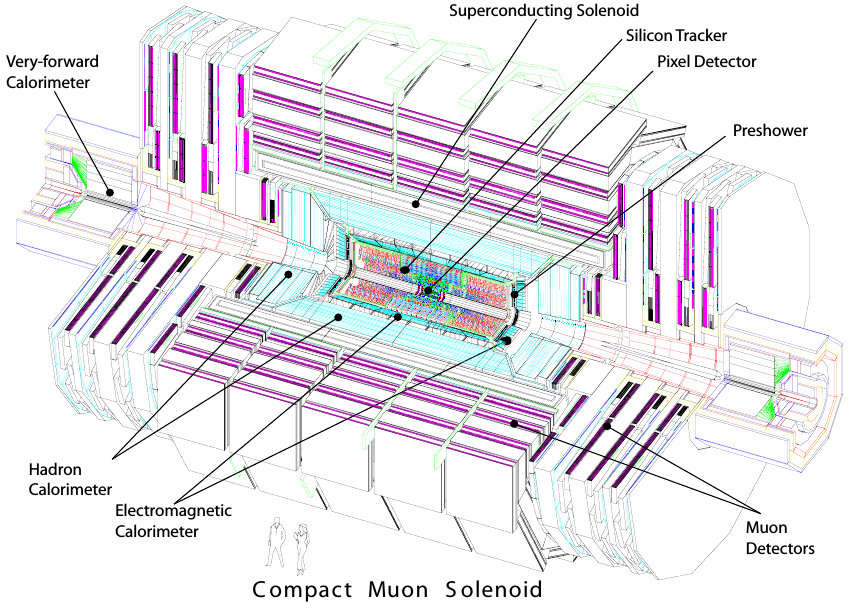}
  \caption{The view of the CMS detector.}
  \label{fig:cms}
\end{figure}

Compact Muon Solenoid (CMS)~\cite{cms-jinst} is a general purpose detector located at one
of the four points of the LHC ring, where the beams collide. It was designed to
reconstruct most of the outgoing particles of the collision: charged leptons and hadrons,
neutral hadrons, and photons.  A detailed description of the CMS detector can be found in
Ref.~\cite{cms-jinst}, below I provide only a short overview of the detector.  The CMS
coordinate system is oriented such that the $x$-axis points to the center of the LHC ring,
the $y$-axis points vertically upward and the $z$-axis along the anticlockwise-beam
direction.  The azimuthal angle $\phi$ is measured in the $xy$ plane, with $\phi=0$ along
the positive $x$ axis and $\phi = \pi/2$ along the positive $y$ axis and the radial
coordinate in this plane is denoted by $r$.  The polar angle $\theta$ is defined in the
$rz$ plane and pseudorapidity variable is defined as $\eta = -\ln[\tan (\theta/2)]$. The
momentum component transverse to the beam direction, denoted by $\PT$, is computed from
the $x$- and $y$-components, and the transverse energy is defined as $E_T =
E\sin{\theta}$.  The central feature of the CMS apparatus is a superconducting solenoid of
6\unit{m} internal diameter, providing a magnetic field of 3.8\unit{T}.  The magnet
largely determines the geometry of the detector, see Fig.~\ref{fig:cms}.  Within the
superconducting solenoid volume are a silicon pixel and strip tracker, a lead tungstate
(PbWO$_4$) crystal electromagnetic calorimeter (ECAL), and a brass and scintillator hadron
calorimeter (HCAL), each composed of a barrel and two endcap sections. The muon system is
composed of the gas-ionization detectors embedded in the steel flux-return yoke outside
the solenoid. Extensive forward calorimetry complements the coverage provided by the
barrel and endcap detectors.  The detector is nearly hermetic, \ie covers almost full
4$\pi$ open angle.  This allows to perform energy balance measurements in the plane
transverse to the beam direction, thus provide a measurement of missing transverse energy,
$\ETm$, associated to neutrinos or other weakly interacting particles.

\begin{figure}[t]
  \centering
  \includegraphics[width=0.8\textwidth]{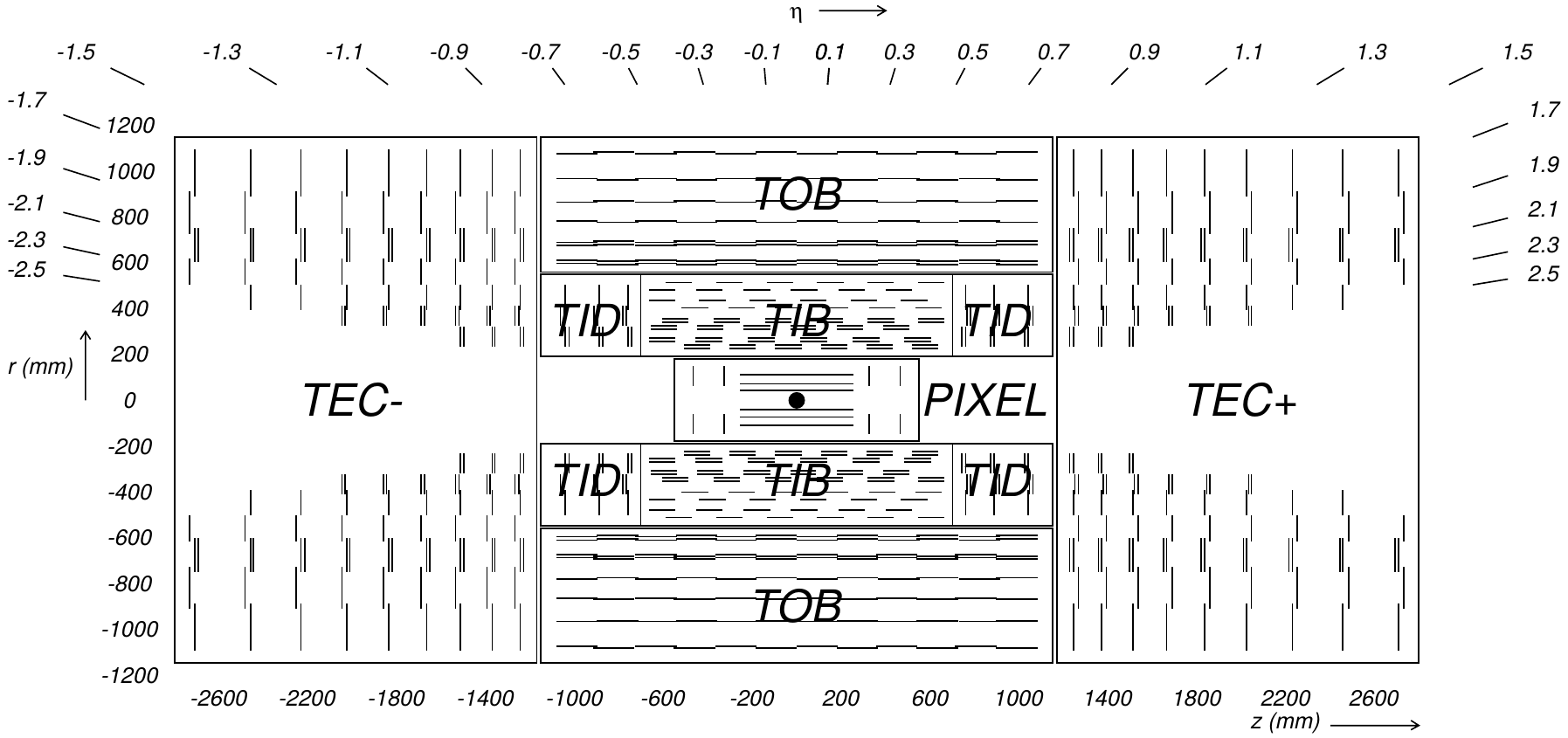}
  \caption[Schematic cross section through the CMS tracker.]
  {Schematic cross section through the CMS tracker. Each line represents a detector module. 
    Double lines indicate back-to-back modules which deliver stereo hits.}
  \label{fig:tracker}
\end{figure}

The silicon tracker measures charged particles within the pseudorapidity range
$\abs{\eta}<2.5$.  It consists of 1440 silicon pixel and 15\,148 silicon strip detector
modules.  The inner pixel detector is housed in a cylindrical volume of 1\unit{m} long and
30\unit{cm} in diameter. It consists of about 66\unit{M} pixels of size
$100\times150\micron$, distributed over three barrel layers and two endcap disks.  With
analogue signal interpolation, a hit resolution of $10\times20\micron$ is achieved.  The
silicon strip tracker is divided into four sub-detectors: outer barrel, inner barrel,
inner disk, and endcap. All active components are housed in a cylindrical volume of length
5.4\unit{m} and diameter of 2.4\unit{m}. Modules laying within $r<60\cm$ have a strip
pitch between 80 and 120\micron, which is increased to ${\sim}$120 to 200\micron for
$r>60\cm$. A schematic view of the tracker system is shown in Fig.~\ref{fig:tracker}.

\begin{figure}[t]
  \centering
  \includegraphics[width=0.4\textwidth]{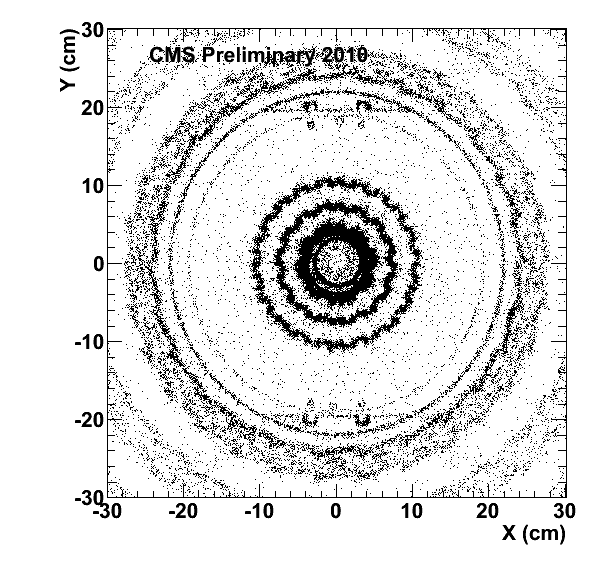}~
  \includegraphics[width=0.4\textwidth]{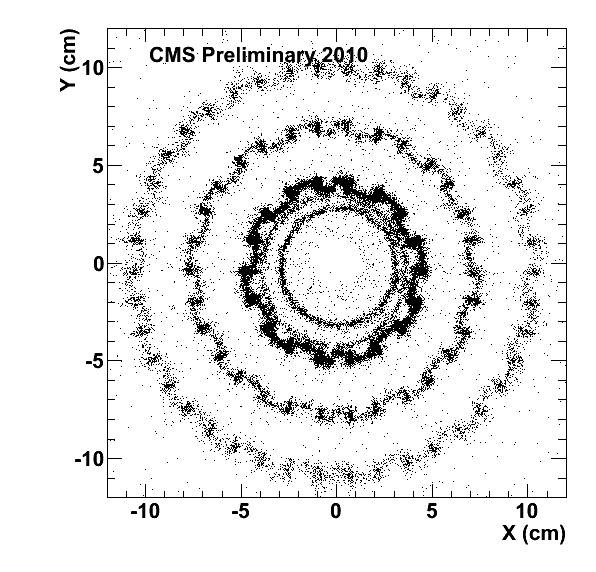}
  \caption{CMS tracker cross section view obtained by reconstructing the verteces of the
    photon conversions, $\gamma\to\Pe^+\Pe^-$.}
  \label{fig:tracker-conv}
\end{figure}

Material of the tracking volume itself affects the overall event topology and
reconstruction through electron bremsstrahlung, photon conversions and nuclear
interactions. It also affects the trajectories of charged tracks because of multiple
scattering and energy loss. Photon conversions are used by CMS as ``radiographie''
measurements of the tracker material~\cite{CMS-PAS-TRK-10-003} by reconstructing the
vertex position of the two electrons from $\gamma\to\Pe^+\Pe^-$, as illustrated in
Fig.~\ref{fig:tracker-conv}. On this figure the LHC beam pipe is also clearly visible. In
fact, the beam pipe is a physical boundary between the $\gamma^*\to\Pe\Pe$ process, where
the photon converts internally, and the $\gamma\to\Pe\Pe$, where the conversion occurs on
the beam pipe or in the detector. This information is used in the analysis to separate the
two processes.

Large magnetic field allows for a precise measurement of the tracks momenta. For
non-isolated particles of $1 < \pt < 10\,\GeV$ and $\abs{\eta} < 1.4$, the track
resolutions are typically 1.5\% in \pt, and 25--90 (45--150)\mum in the transverse
(longitudinal) impact parameter \cite{TRK-11-001}.  For a muon with \PT=100\GeV the
resolution on the \PT measured in the tracker alone is about 2\% in the barrel region.
The pixel and tracker systems also play an important role in the reconstruction of the
primary interaction vertices, and identification of the converted photons by recovering the
electron track, see Sec.~\ref{sec:reco}.

The ECAL is distributed in a barrel region $\abs{\eta} < 1.48$ and two endcaps that extend
up to $\abs{\eta} = 3$.  A lead and silicon-strip preshower detector is located in front
of the ECAL endcaps in order to improve the identification of $\pi^0 \to\gamma\gamma$
events.  Initial calibration of the calorimeter was done with the test beam and the
achieved resolution can be parametrized as follows:

\begin{equation}
  \left(\frac{\sigma}{E}\right)^2 = \left(\frac{0.028}{\sqrt{E}}\right)^2 + \left(\frac{0.12}{E}\right)^2 +(0.003)^2.
\end{equation}
Further calibration of the calorimeter is performed with collision data, using
$Z\to\Pe\Pe$ events, where electrons are reconstructed as photons, and $Z\to\mu\mu\gamma$
events, where the photon is radiated off the muon in the final state.

The HCAL surrounds the ECAL volume and covers the region $\abs{\eta} < 3$.  Iron forward
calorimeters with quartz fibers, read out by photomultipliers, extend the detector
coverage up to $\abs{\eta} = 5$.  The resolution of HCAL obtained after the calibration
with the test beam is:
\begin{align}
  \left(\frac{\sigma}{E}\right)^2 &= \left(\frac{0.9}{\sqrt{E}}\right)^2 +(0.045)^2 \quad \text{in Barrel and Endcap};\nonumber\\
  \left(\frac{\sigma}{E}\right)^2 &= \left(\frac{1.72}{\sqrt{E}}\right)^2 +(0.09)^2 \quad \text{in Forward}.
\end{align}
With collision data the calibration of HCAL is performed using isolated charged tracks,
with momenta between 40 and 50\GeV.  The momentum measurement of the tracks is obtained in
the tracker with high accuracy.  When a (hadron) track riches the calorimeters, it
deposits all of its energy, thus allowing to calibrate HCAL using the energy measured in
the tracker~\cite{HCAL-calib}. For this purpose only the tracks with small energy deposits
in ECAL (minimum ionizing particles) are selected.

\begin{figure}[t]
  \centering
  \includegraphics[width=0.7\textwidth]{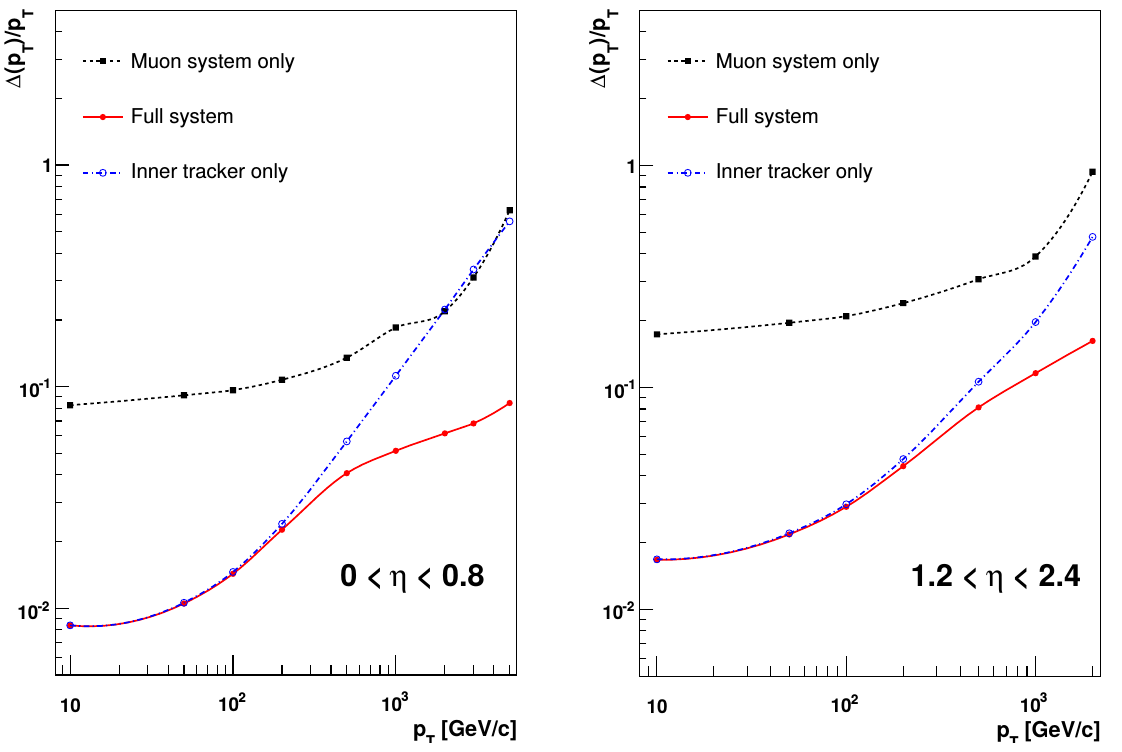}
  \caption[The muon transverse momentum resolution as a function of the transverse momentum.]
  {The muon transverse momentum resolution as a function of the transverse momentum ($\PT$)
    using the muon system only, the inner tracking only, and both. Left panel: $|\eta| < 0.8$, right
    panel: $1.2 < |\eta| < 2.4$}
  \label{fig:mu}
\end{figure}

Muons penetrate the whole detector with minimal interaction and are identified in
gas-ionization detectors.  Figure~\ref{fig:mu} shows the improvement of the momentum
resolution of muons on top of the tracker system.  Before the start of the LHC, the
alignment and calibration of the muon sub-detectors was performed with data, using
atmospheric muons reaching the detector. Then, with collision data the momentum of the
muons is calibrated using $\JPsi\to\mu\mu$ and $\Z\to\mu\mu$ events.

\subsection{Trigger}
Two-tier online trigger system is implemented in CMS in order to reduce the rate of the
events collected on tape.  The first tier, called the Level-1 (L1) trigger, composed of
custom hardware processors, uses the basic information from the calorimeters and muon
detectors to select the most interesting events in a fixed time interval of less than
4\mus.  If the event satisfies the L1 selection criteria, it is processed further; if not,
then it is discarded. The L1 trigger reduces the event rate to ${\sim}100$\unit{kHz}.  At
the second tier, called the high level trigger (HLT), more sophisticated selection is
performed. It combines the kinematic information of multiple trigger objects (particle
candidates), in order to keep the most interesting events for the offline analysis.  Total
HLT rate is about 100\unit{Hz}, i.e. about ${\sim}$100 events per second is saved on tape
for further analysis.


\subsection{Pile-up}
\label{sec:pileup}

\begin{figure}[bh]
  \centering
  \includegraphics[width=0.55\textwidth]{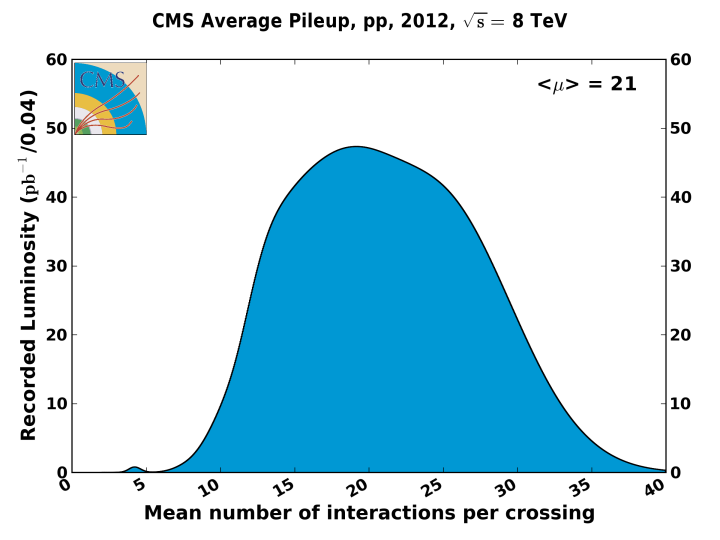}
  \caption{Average pileup distribution in $\Pp\Pp$ data of 2012.}
  \label{fig:pileup}
\end{figure}

Each bunch of the LHC beam contains more than $10^{10}$ protons in it, hence there is a
large probability for multiple p-p interactions per bunch.  In 8\TeV collisions, there was
on average 21 interactions per bunch crossing, shown in Fig.~\ref{fig:pileup}. This
phenomena is called \textit{pile-up} and it results in the reconstructing of multiple
primary interaction vertices. Luckily, the signal processes that are interesting for
physics analysis are so rare that they never happen twice in the same bunch
crossing. Thus, only one primary vertex is chosen per event, which is most likely to
correspond to the hardest interaction, see Sec.~\ref{sec:reco}.  However, the extra
interactions spoil the purity of the event reconstruction and this needs to be taken into
account at the analysis level.

\section{Beam timing measurement at CMS}
\label{sec:bptx}
For measuring the position of the beam in the beam pipe, there are 1032 beam position
monitors (BPM) installed around the LHC.  The majority of the BPMs (860 of the 1032) are
built out of four electrostatic button pick-up electrodes that are installed symmetrically
around the beam pipe.  A more detailed description of the BPMs can be found in
Refs.~\cite{LHC-BEAM,BPM,LHC-design}.  When a proton bunch travels around the pipe, it
induces a mirror current of free-moving electrons on the surface of the pipe. Traveling
over the electrode surface of the button pick-up, this current gives rise to a signal on
the button surface. This signals from the BPM provides an opportunity for a precise
measurement of the timing and structure of the incoming beams, as well as the
characteristics of individual bunches.

\begin{figure}[t]
  \centering
  \includegraphics[width=.5\textwidth]{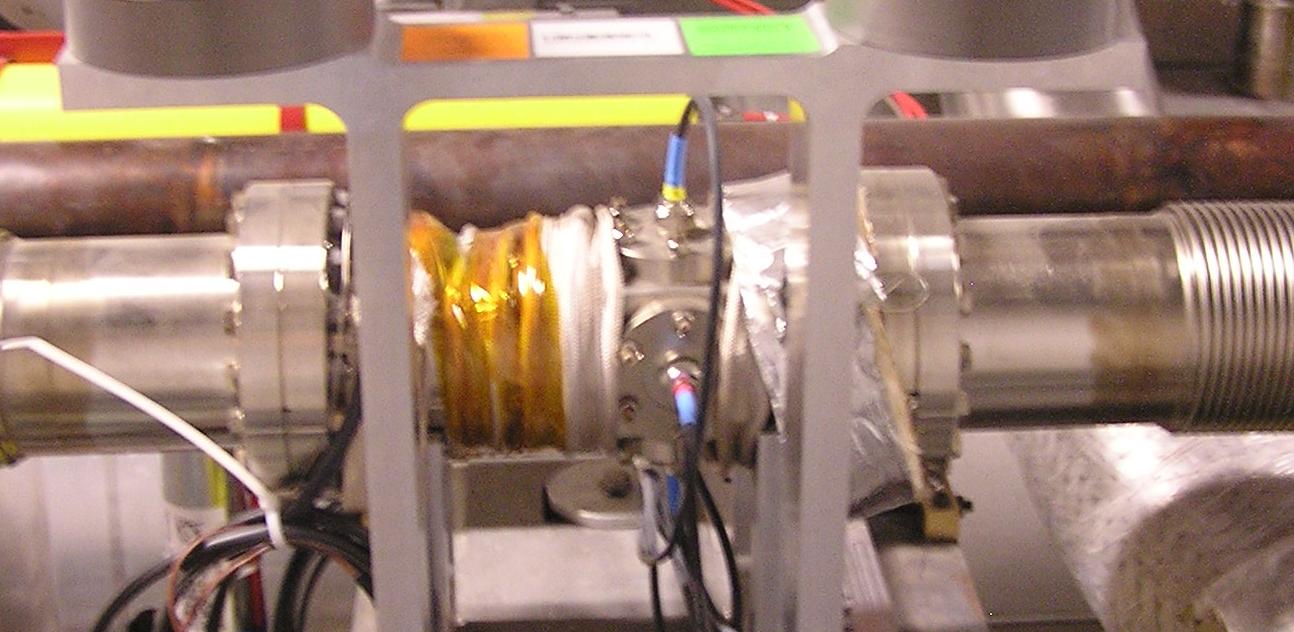}
  \caption[One of the CMS BPTX elements in the LHC tunnel.]
  {One of the CMS BPTX elements in the LHC tunnel. Three of
    the four button connectors (blue) can be seen.}
  \label{fig:bptx_element}
\end{figure}

The two BPMs closest to the interaction point of each LHC experiment are reserved for the
timing measurements and are called the Beam Pick-up Timing eXperiment (BPTX) detectors,
which for CMS are located approximately 175\unit{m} on either side of the interaction
point.  One BPM element contains four pickup buttons located in a single vertical plane
and orientated $90\de$ with respect to each other, see Fig.~\ref{fig:bptx_element}.  For
the BPTX use at CMS all four buttons are connected in parallel to provide the maximum
signal.  Each pickup only sees a single beam.

\begin{figure}[b]
  \centering
  \includegraphics[width=0.55\textwidth]{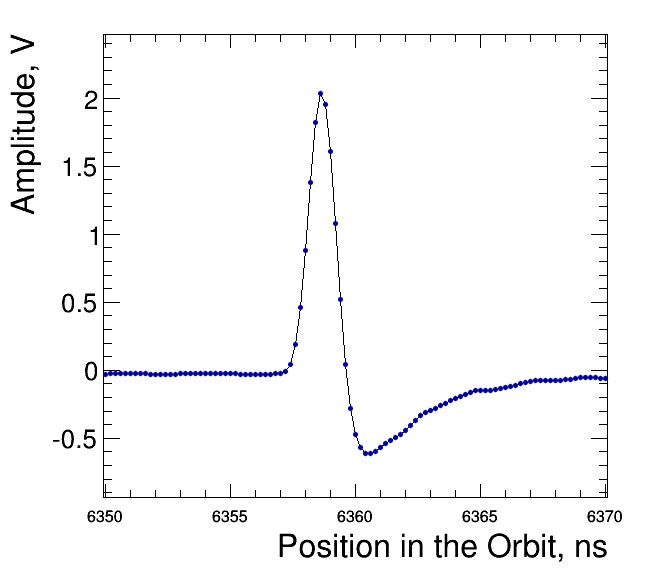}
  \caption[A typical BPTX pulse.]
  {A typical BPTX pulse as seen on the oscilloscope. The markers represent the
    measured points.}
  \label{fig:bptx_pulse}
\end{figure}

When the signal arrives at the counting room at CMS, it is split into four equal copies.
One copy is dedicated to the trigger and enters the BPTX logic crate.  The other signals
are available for monitoring with high sampling oscilloscopes.  The signals serve a dual
purpose; they are used both for monitoring of timing related beam conditions and for the
L1 trigger.  Figure~\ref{fig:bptx_pulse} shows a typical BPTX pulse signal. Its
characteristics are a steep leading flank, followed by a shallow trough after which the
signal slowly returns to the baseline.

The BPTX trigger electronics is implemented in NIM modules.  The key module of the system
is the Ortec's constant fraction discriminator (CFD).  This module is designed to operate
at 200 MHz frequency~\cite{CFD935}.  For BPTX purposes it experiences 40~MHz frequency
during nominal 25\ns bunch spacing beams, or 20\unit{MHz} with 50\ns bunch spacing.  The
threshold of the discriminator is manually adjustable on the module.  As the intensity of
the beam changes one may need to adjust the threshold accordingly in order to maintain
100\% efficiency.  The plain NIM BPTX(1,2) trigger signals are taken to various logic
units in order to provide an AND, an OR and exclusive AND signals.  Then the signals are
sent to the L1 Trigger.  The logical AND of the two BPTX signals is used to gate other
triggers (both at the L1 and HLT) with collidable beam crossings.

\begin{figure}[b]
  \centering
  \includegraphics[width=.40\textwidth]{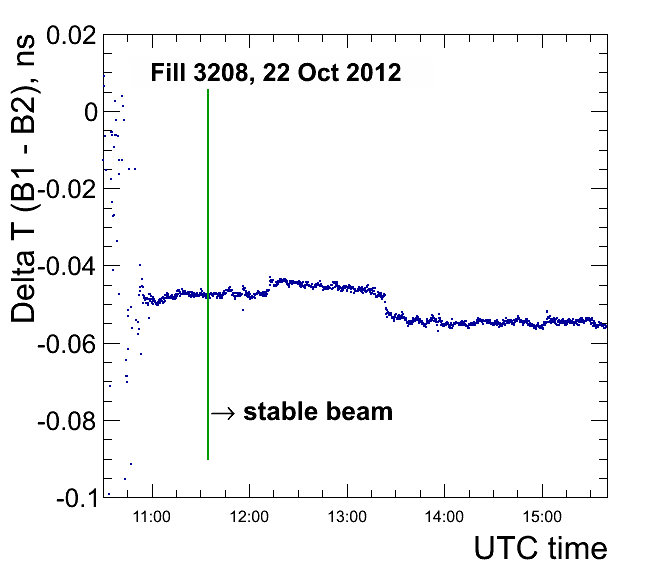}~
  \includegraphics[width=.55\textwidth]{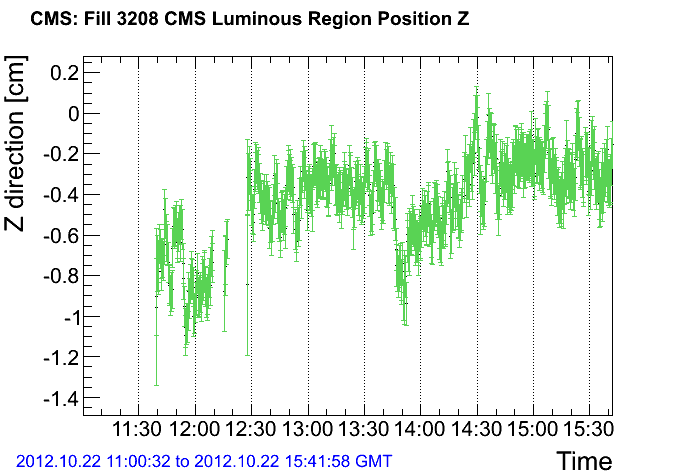}~
  \caption[Cogging measurement of BPTX versus time for a particular LHC fill. 
  Average $z$ position of the beamspot as reconstructed by the tracking system.]
  {Cogging measurement of BPTX versus time for a particular LHC fill (left);
    Average $z$ position of the beamspot as reconstructed by the tracking system (right).
    The correlations between the two indicate the sensitivity of the beamspot to the
    moving beams.}
  \label{fig:cogging}
\end{figure}

Upon injection of the beam into LHC an estimate of the beam crossing position near the IPs
can be obtained from a BPTX based timing measurement. As mentioned above the copies of the
analog BPTX signals are also fed into the oscilloscopes (LeCroy WR
104MXi-A~\cite{WaveRunner}).  The timing measurements is then performed on those
oscilloscopes.  The time resolution of the BPTX-based timing is better than 0.1\ns, which
is sufficient to distinguish between adjacent RF buckets (2.5\ns).  The result of this
measurement, the relative time difference between the two BPTX signals in nanoseconds, is
published through the LHC data interchange protocol (DIP)~\cite{dip} and picked up for
publication on the LHC Vistar web page~\cite{vistar} as \textit{BPTX: deltaT of IP (B1 -
  B2)} and displayed for the world to see.  This is commonly referred to as the
\textit{cogging measurement} (Fig.~\ref{fig:cogging}), and provides a first estimate of
the beam interaction position along $z$ direction.

At the time of writing this dissertation the Run-2 of the LHC operation is started. The
BPTX system of CMS is one of the first to see the beams from the LHC. Normal operation of
the BPTX detector is now re-established with beams and shows excellent performance.  New
developments of the electronics and software for the system are also ongoing. First of
all, the NIM based logic is to be replaced with a programmable VME board, the V1495 module
by CAEN, which would do the analog logic of the two beams and provide signals to the L1
trigger system. Secondly, the oscilloscope based measurement is to be be replaced with the
hardware, using the newly developed ADC uTCA board.

\chapter{Physics Analysis: Search for \texorpdfstring{$\PH\to\ell\ell\gamma$}{H to llg}}
\label{sec:ana}

\section{Features of the decay}
\label{sec:features}

Before going into details of the event reconstruction and selection, the basic features of
the $\PH\to\gamma^*\gamma\to\ell\ell\gamma$ final state are described.

\begin{figure}[t]
  \begin{center}
    \includegraphics[width=0.44\textwidth]{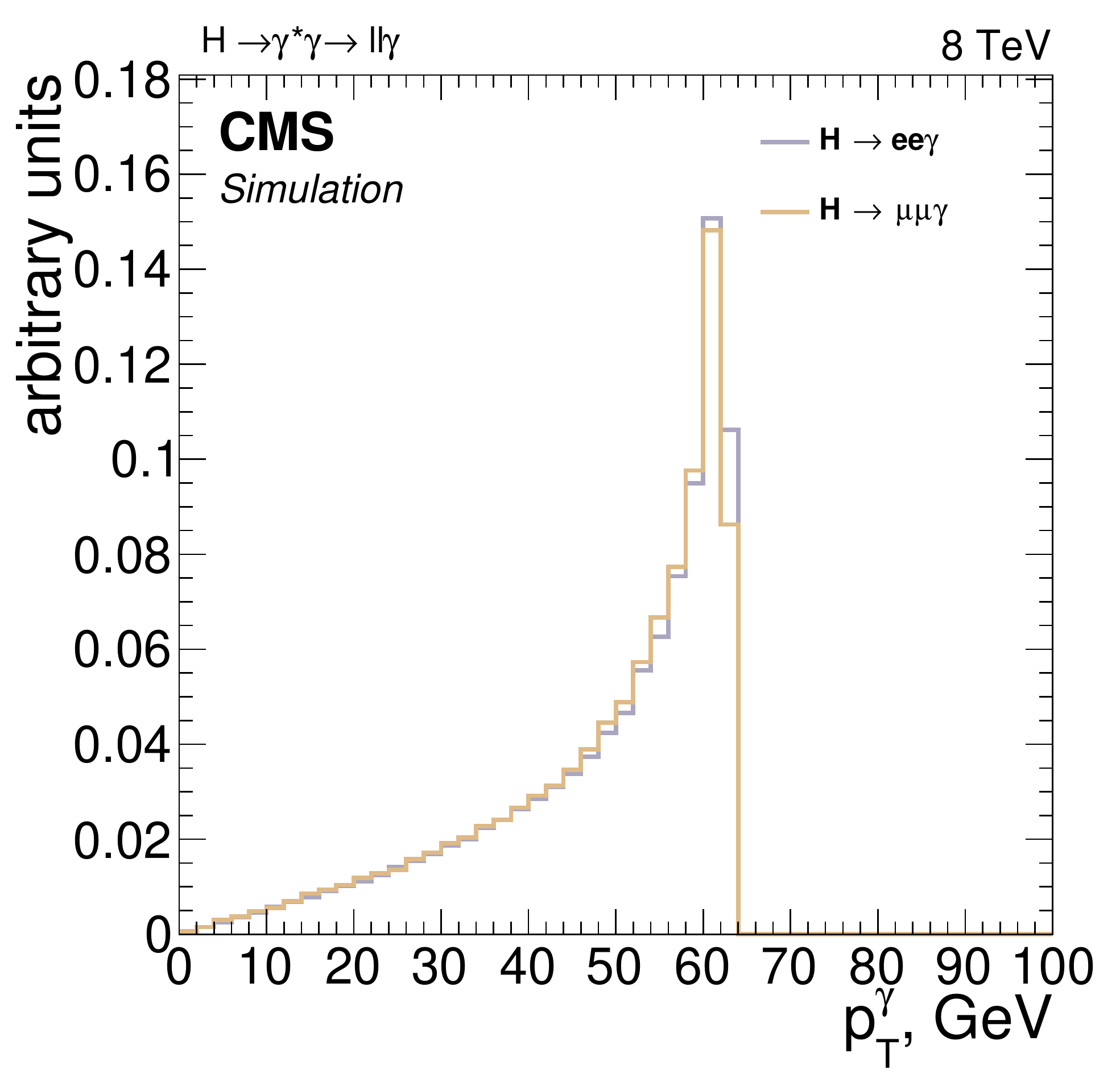}
    \includegraphics[width=0.44\textwidth]{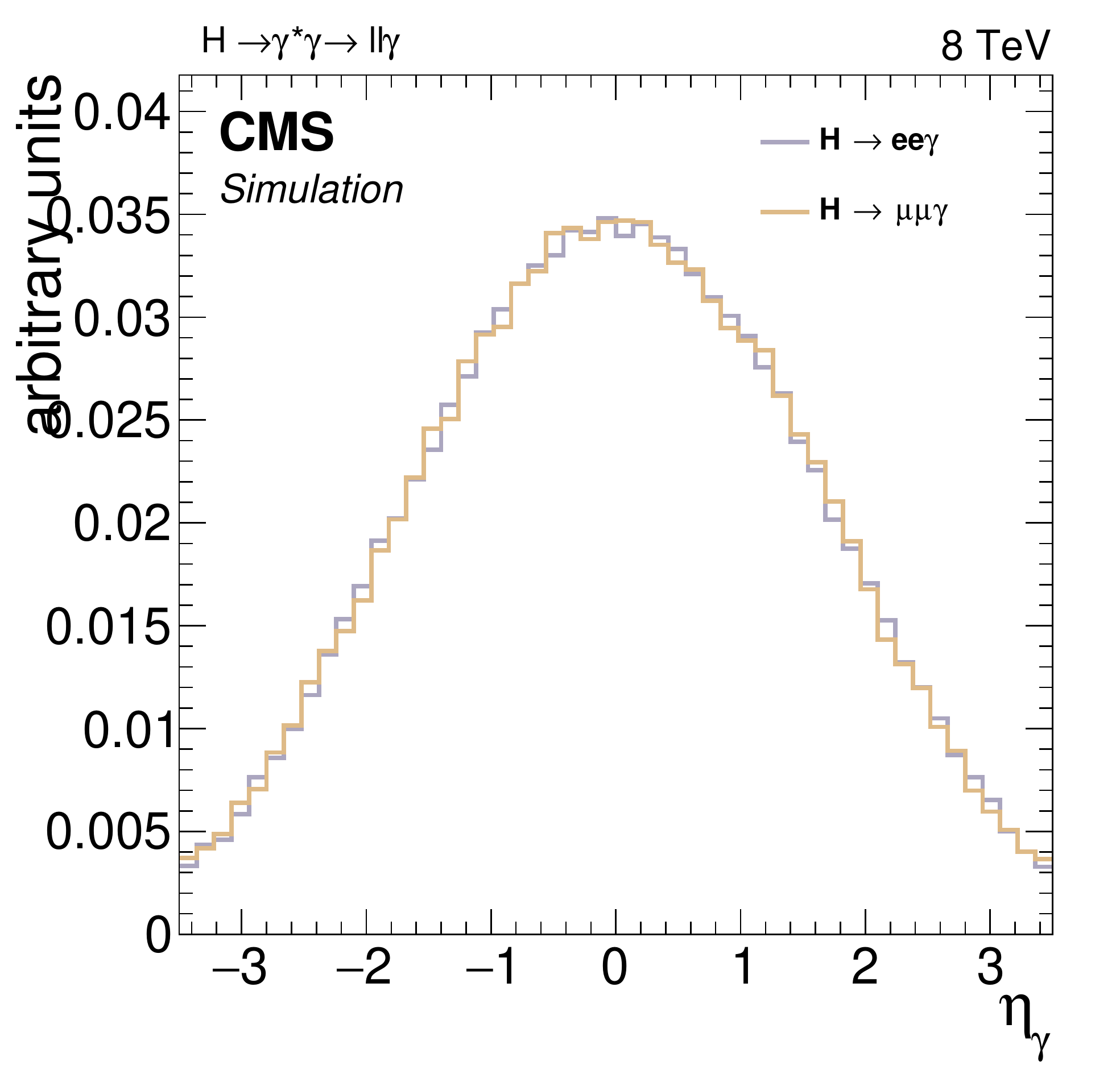}
    \caption[Photon transverse momentum and $\eta$ distributions at generator level for $m_\PH=125\GeV$.]
    {Photon transverse momentum and $\eta$ distributions at generator level for $m_\PH=125\GeV$.
      The $p_T$ distribution illustrates the fact that the photon is energetic (peaked at 60 GeV).
      From the $\eta$ distribution we can see that the signal events are produced predominantly
      centrally in $\eta_{\gamma}$.}
    \label{fig:LHE-gamma}
  \end{center}
\end{figure}

Due to a heavy Higgs boson, the $\gamma$ and $\gamma^*$ from its decay are highly
energetic (boosted), and predominantly central, see Fig.~\ref{fig:LHE-gamma}.  Therefore,
a stringent selection on their $\PT$, as well as $\Delta R$, is possible, and those
requirements reject a large part of the backgrounds.  Because the $\gamma^*$ is boosted,
the two leptons from its decay are anti-correlated in their transverse momenta.  Also
because of the boost and low dilepton invariant mass of the $\gamma^*\to\ell\ell$ decay,
the leptons in the final state are very close to each other in $\Delta R_{\eta,\phi}$.  In
the case of electron channel, this feature prevents us from reconstructing two electrons:
they are merged into a single shower in the ECAL and can not be resolved.  In order to
overcome this problem a dedicated identification criteria was developed, based on the
multivariate analysis (MVA) technique, described in Section~\ref{sec:reco-el}.  In the
muon channel the situation is better: both muons can be well reconstructed. However, a
\textit{loose} identification (ID) criteria has to be used in order to increase the
reconstruction efficiency.  The invariant mass, $m_{\ell\ell}$, of the two leptons is
close to the photon pole mass, $m_{\gamma^*} = 2m_{\ell}$, for the majority of events, see
Fig.~\ref{fig:LHE-diLep-mass}.  Hence, in order to isolate the contribution from the
Dalitz decay, the main analysis is limited to the phase space with $m_{\ell\ell} <
20\GeV$.

\section{Simulated samples}
\label{sec:sim}

The first challenge of the analysis was to produce a proper simulation of the signal
samples, in order to obtain the description of the Higgs boson signal events to be used in
the search.  The samples for Dalitz signal are produced at the leading-order of QCD, using
the \MADGRAPH~5 matrix-element generator with anomalous Higgs Effective coupling
model~\cite{MAD5,ANO-HEFT}.  The output events of \MADGRAPH are further showered with
{\PYTHIA6.426}~\cite{pythia6} and undergo the full CMS detector simulation with \GEANT4.
The samples are generated for the gluon-gluon and vector bosons fusion, and associated
production with a vector boson production processes.  Associated Higgs boson production
with a \ttbar pair is ignored due to its small contribution.  The kinematic distributions
of the \MADGRAPH samples were also cross-checked with the output of \MCFM program and
found to be consistent.  The parton distribution function (PDF) set used to produce these
samples is CTEQ6L1~\cite{CTEQ6L}.  The pile-up events are also introduced in the
simulation using a sample of the minimum bias events.  The simulated samples are often
referred to as Monte Carlo (MC) samples and I will use those terms interchangeably in the
later text.

\begin{figure*}[t]
  \begin{center}
    \includegraphics[width=0.3\textwidth]{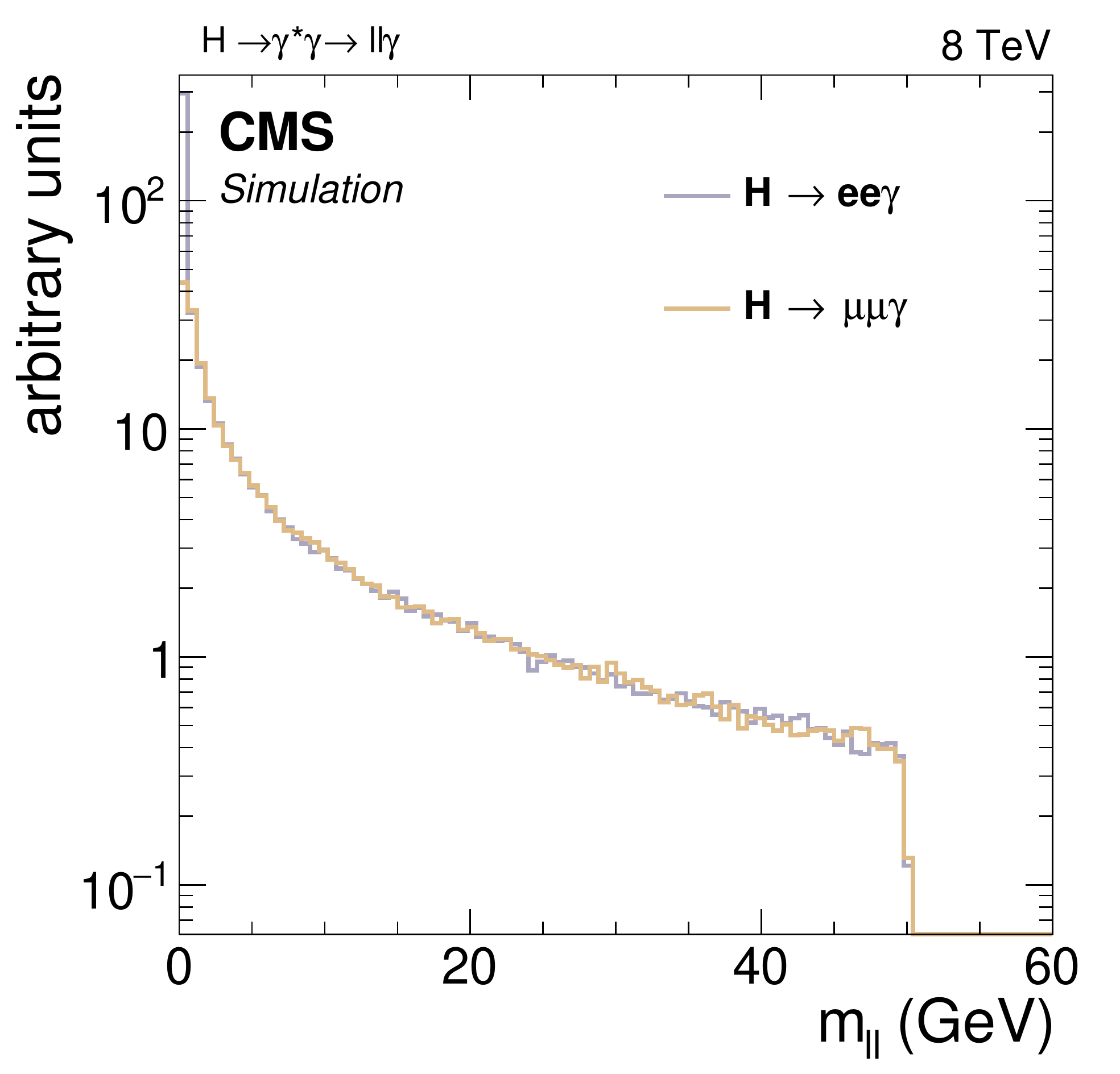}~
    \includegraphics[width=0.3\textwidth]{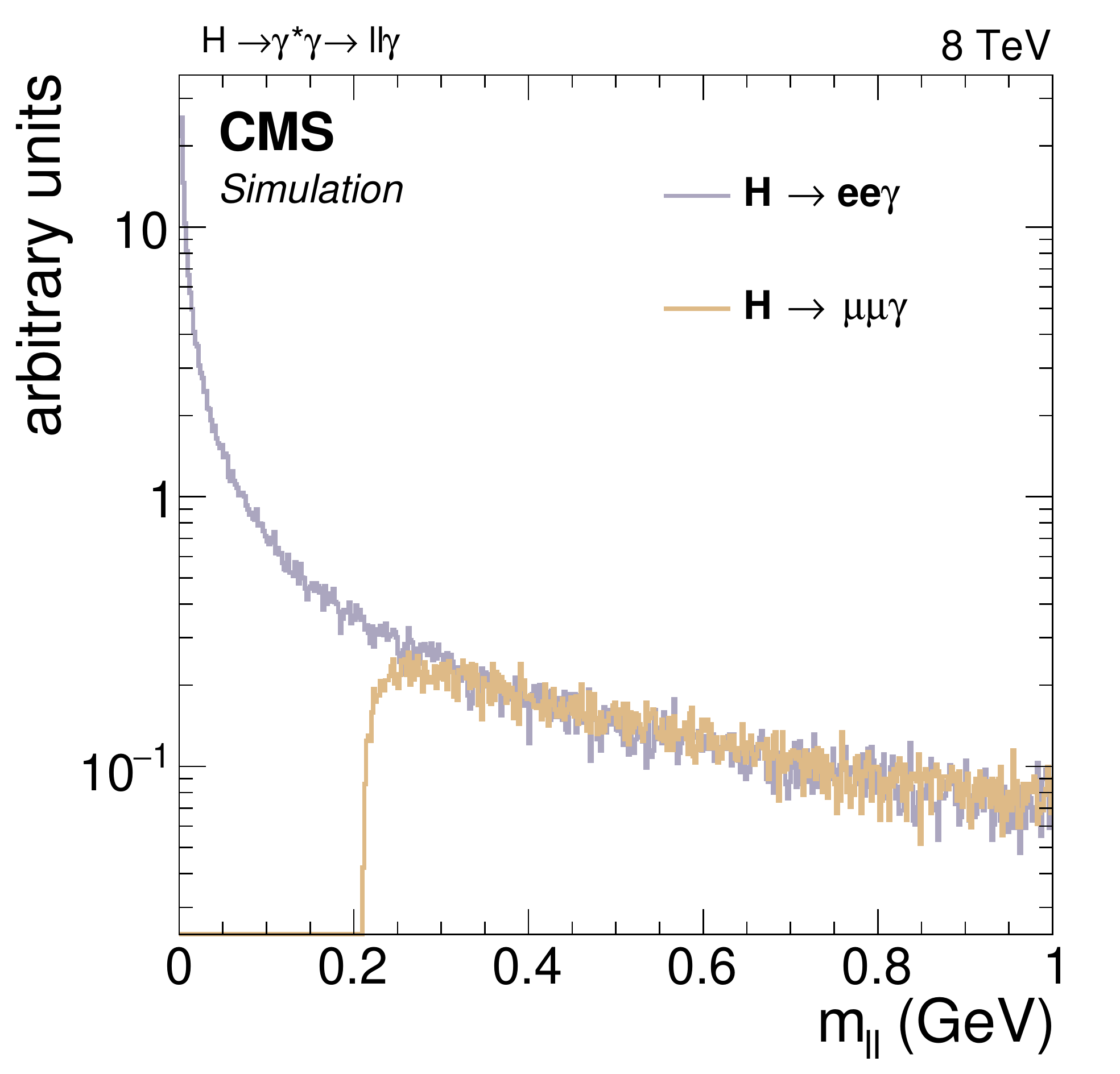}~
    \includegraphics[width=0.3\textwidth]{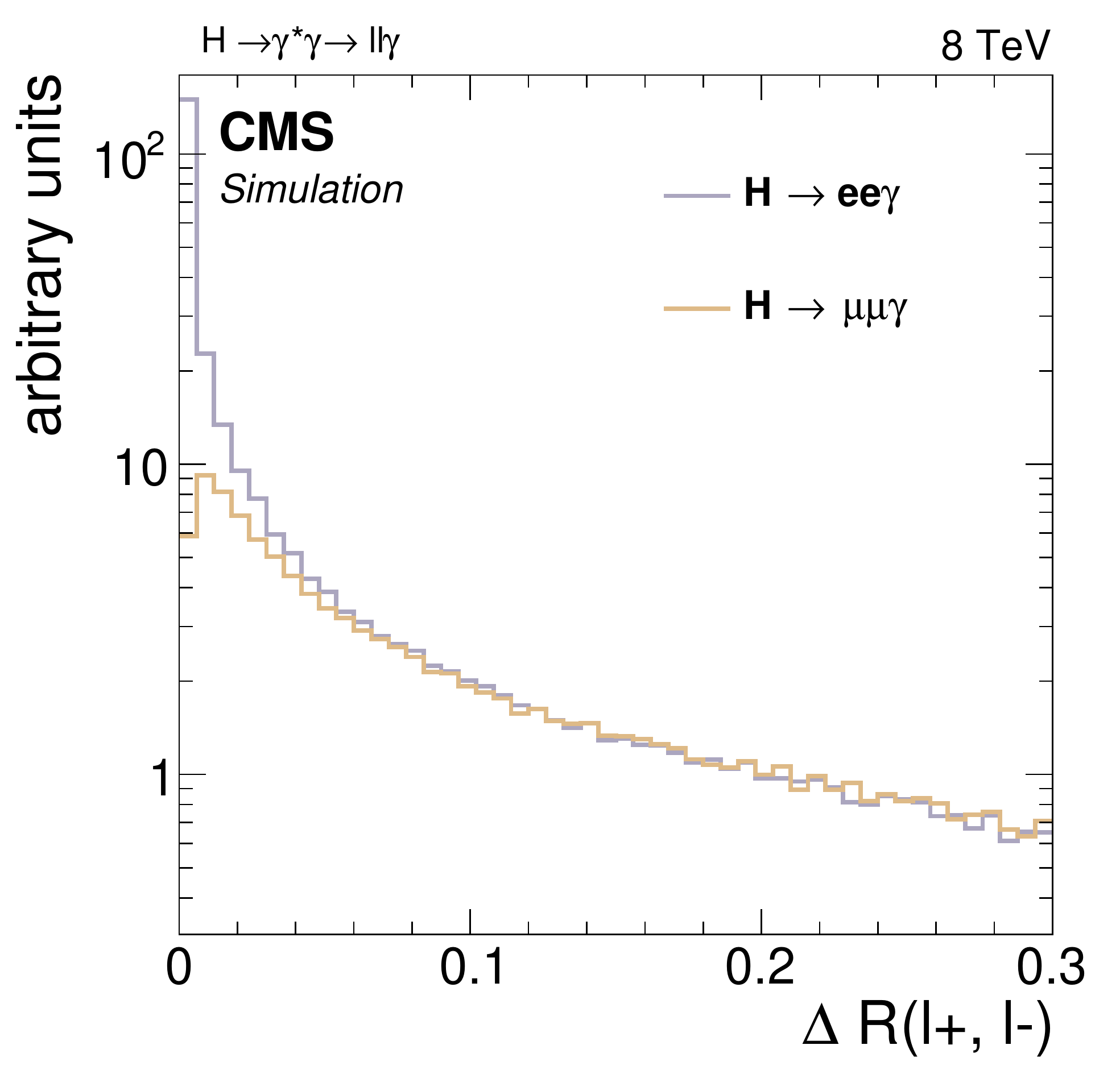}~
    \caption[Dilepton invariant mass distributions from $H \to \ell\ell\gamma$ Dalitz
    signal.]
    {The dilepton invariant mass distributions from $H \to \ell\ell\gamma$ Dalitz signal
      (Left: full range; Middle: zoomed into the low mass region, which shows the effect
      of non-zero lepton masses).  Right: $\Delta R(\ell_1, \ell_2)$ distribution. All
      plots are made at generator level for $m_\PH\,=\,125\GeV$, before FSR.}
    \label{fig:LHE-diLep-mass}
  \end{center}
\end{figure*}

In order to obtain the correct results for the Dalitz signal process one has to take into
account the mass of the leptons: $m_{e}=0.000511\GeV$, $m_{\mu}=0.1057\GeV$. The masses
make a difference to the natural cut-off from $\gamma^*\to\ell\ell$ process at
$m_{\ell\ell} > 2m_{\ell}$, which can be seen in Fig.~\ref{fig:LHE-diLep-mass} of
$m_{\ell\ell}$ and $\Delta R(\ell\ell)$ distributions..

A sample for $\PH\to(\JPsi)\gamma$ signal is produced using \PYTHIA8
generator~\cite{pythia8}. The polarization of the \JPsi is not correctly taken into
account by the generator. Therefore, this sample is additionally reweighted to simulate
100\% polarization of the \JPsi, see Appendix~\ref{sec:app-jpsi}.

The SM Higgs boson production cross sections are taken from Ref.~\cite{YR3}.  The
branching fractions for the $\PH\to\gamma^*\gamma\to\ell\ell\gamma$ signal processes are
estimated at the next-to-leading order (NLO) in QCD using \MCFM, as described in
Sec.~\ref{sec:Htollg}.  Using those numbers one can estimated the total number of signal
events produced with 19.7\fbinv of integrated luminosity, which are given in
Table~\ref{tab:sig-yield-before-acc} for the gluon fusion process.  Only part of those
events could be reconstructed, when all of the final state particles have large enough
momenta to reach the detector volume.  This fraction, called the signal event acceptance,
$a = \frac{N^{sel}}{N^{tot}}$, depends on the selection.  With the basic
selection\footnote{Typical selection consists of a photon with $\PT>30\,\GeV$ in
  $|\eta^{\gamma}| < 2.5$ and two leptons with $\pt^{\ell_1} > 23\GeV$ and $\pt^{\ell_2} >
  4\GeV$, both in $|\eta^{\mu}| < 2.4$, see Section~\ref{sec:sel} for the motivation of
  these choices.} applied to the generator level particles, an acceptance of ${\sim}55\%$
is obtained, i.e. about half of the events from the Table~\ref{tab:sig-yield-before-acc}
could be reconstructed in the detector.  Efficiencies of the reconstruction will be
discussed in Section~\ref{sec:reco} after the event reconstruction is described.

\begin{table}[t]
  \caption[Number of expected signal events at $\sqrt{s} = 8\,\TeV$ with $19.7\,\fbinv$.]
  {Number of signal events within $m_{\ell\ell} < 50\,\GeV$, and $m_{\ell\ell} < 20\,\GeV$
    expected to be produced at $\sqrt{s} = 8\,\TeV$ with $19.7\,\fbinv$,
    before acceptance and reconstruction effects, in gluon fusion process.}
  \label{tab:sig-yield-before-acc}
\begin{center}
  \begin{tabular}{ c|c|c|c|c}
& \multicolumn{2}{c|}{$m_{\ell\ell}<50\,\GeV$}& \multicolumn{2}{c}{$m_{\ell\ell}<20\,\GeV$}\\
    $m_{H}$ & $\PH\to\mu\mu\gamma$ & $\PH\to \Pe\Pe\gamma$ & $\PH\to\mu\mu\gamma$ & $\PH\to \Pe\Pe\gamma$\\
    \hline
    120 & 15.0 & 30.7 & 12.9 & 23.9 \\
    125 & 14.4 & 29.4 & 12.2 & 22.6 \\
    130 & 13.3 & 27.2 & 11.3 & 20.9 \\
    135 & 11.7 & 23.9 & 10.0 & 18.5 \\
    140 &  9.9 & 20.3 & 8.4  & 15.5 \\
    145 &  8.2 & 16.6 & 6.9  & 12.7 \\
    150 &  6.3 & 12.8 & 5.4  & 9.9  \\
\end{tabular}
\end{center}
\end{table}

For the $\PH\to(\JPsi)\gamma$ decay the branching fraction is taken from
Ref.~\cite{hToJPsiGamma-2014}, where for a SM Higgs boson the prediction is
$\mathcal{B}(\PH\to(\JPsi)\gamma) = (2.8\pm0.2)\times10^{-6}$.

The pile-up event simulation does not match exactly the pile-up conditions in data. Hence,
in order to achieve a better agreement, the simulated samples are reweighted based on the
number of simulated primary vertices.  Furthermore, the reconstruction efficiencies of the
physics object do not match exactly between the data and simulation. In some cases this is
also accounted for by reweighting of the sample, in other cases a systematic uncertainty
is applied, see Section~\ref{sec:syst} for details.  Overall, the effect of the
reweightings is smaller than 2\% on the predicted yield of the signal.  The energy and
momentum resolution of the photon and muons in simulated events are also corrected to
match the resolution in data.

\section{Background Estimation and Analysis Strategy}
The background estimation is data-driven: it is determined from a fit to the data
distribution of the reconstructed three-body invariant mass, $m_{\ell\ell\gamma}$.  The
strategy of the analysis is to use that fit as the background model and search for the
signal peaks on top of it.

\subsection{Background Composition: Muon Channel}
The main irreducible background is the Drell--Yan (DY) initial state radiation (ISR)
process: $\Pp\Pp \to \gamma^* + \gamma \to \mu\mu\gamma$, with low dilepton invariant
mass. There is also a contribution from FSR events off the Z-peak:
$\Pp\Pp\to\gamma^*/Z\to\mu\mu\gamma$. However the contribution from the second process has
to be small in the signal region, which is far from the Z-mass peak, $m_{\mu\mu\gamma} >
110\,\GeV$.

Major reducible background is a DY+jet process, $\Pp\Pp \to \gamma^* + jet \to \mu\mu\ +
jet$, where a $jet$ in the final state is mis-identified as a photon.

Even though the background estimation in the analysis is data-driven, I have also made an
attempt to describe the backgrounds with MC simulation, see Appendix~\ref{sec:app-mu-bkg}.
It is however not used in the analysis because the agreement between data and simulation
was found not satisfactory.  The reason for the disagreement is due to the difficulty to
implement the jet matching between the NLO process generated by \MADGRAPH and its
showering by \PYTHIA~\cite{MG-jets}. Nevertheless from this study, I can conclude that
DY+$\gamma$ consists of approximately 40\% of the total background, while DY+jet is the
rest, about 60\%.

\subsection{Background Composition: Electron Channel}
Backgrounds in the electron channel are like in the muon channel, but in addition, there
is a large contribution from QCD events due to the topology of two very close electrons,
and from $\gamma\gamma$ events, where one of the photons converts in the detector material
or on the beam pipe. See Section~\ref{sec:reco-el} for details. The fit to the three-body
mass distribution in data is used as the background model.

\clearpage

\section{Event reconstruction}
\label{sec:reco}

\subsection{Primary vertex}
\label{sec:PV}

As mentioned in Section~\ref{sec:pileup} multiple $\Pp\Pp$-interactions occur
per-collision. A deterministic annealing algorithm~\cite{TRK-11-001} is used to identify
all vertices from those interactions.  A vertex with the highest scalar sum of the
${\pt^2}$ of its associated tracks is chosen as \textit{the} primary vertex (PV).  The PV
must have the reconstructed longitudinal position ($z$) within 24\unit{cm} of the
geometric center of the detector and the transverse position ($x$-$y$) within 2\unit{cm}
of the beam interaction region.

\subsection{Particle-Flow algorithm}
\label{sec:pf}
The global event reconstruction (also called particle-flow (PF) event
reconstruction~\cite{PF09,PF10}) is based on reconstructing and identifying each single
particle with an optimized combination of the information from all subdetectors. In this
process, the identification of the particle type (photon, electron, muon, charged hadron,
neutral hadron) plays an important role in the determination of the particle direction and
energy. Photons are identified as ECAL energy clusters not linked to the extrapolation of
any charged particle trajectory to the ECAL. Electrons are identified as a primary charged
particle track and potentially many ECAL energy clusters corresponding to this track
extrapolation to the ECAL and to possible bremsstrahlung photons emitted along the way
through the tracker material. Muons are identified as a track in the central tracker
consistent with either a track or several hits in the muon system, associated with an
energy deficit in the calorimeters. Charged hadrons are identified as charged particle
tracks neither identified as electrons, nor as muons. Finally, neutral hadrons are
identified as HCAL energy clusters not linked to any charged hadron trajectory, or as ECAL
and HCAL energy excesses with respect to the expected charged hadron energy deposit.

The energy of photons is directly obtained from the ECAL measurement, corrected for
zero-suppression effects. The energy of electrons is determined from a combination of the
track momentum at the main interaction vertex, the corresponding ECAL cluster energy, and
the energy sum of all bremsstrahlung photons attached to the track. The energy of muons is
obtained from the corresponding track momentum. The energy of charged hadrons is
determined from a combination of the track momentum and the corresponding ECAL and HCAL
energy, corrected for zero-suppression effects and for the response function of the
calorimeters to hadronic showers. Finally, the energy of neutral hadrons is obtained from
the corresponding corrected ECAL and HCAL energy.

Based on the PF candidates it is useful to construct isolation variables as follows.  For
an object reconstructed with transverse momentum, ${\pt}_0$, in a given direction,
($\eta_0,\phi_0$), one defines a cone in $\eta-\phi$ plane, with radius $R_0$, such that
$\DR_{\eta\phi} \equiv \sqrt{ (\eta-\eta_0)^2 + (\phi-\phi_0)^2} < R_0$.  One then
calculates the transverse energy of all particle candidates within this cone, relative to
${\pt}_0$, separating them by type: $I_{ch} =
\sum\limits_{pf}^{\DR<R_0}{\pt}^{charged}/{\pt}_0$ for charged hadrons, $I_{neu} =
\sum{\pt}^{neu}/{\pt}_0$ for neutral hadrons, and $I_{pho} = \sum{\pt}^{pho}/{\pt}_0$ for
the photons.  Of course, the original object, for which the isolation variable is
constructed, is excluded from the sums.  The energy of the particles associated to the
pile-up interaction vertex is also measured, $I_{PU} = \sum{\pt}^{PU}/{\pt}_0$.
Furthermore, the average energy associated to pile-up particles, $\rho$, is calculated.
All these isolation variables are used in the identification criteria (ID) of the photons
and muons, as discussed below.

\subsection{Photons}
\label{sec:pho}

The photons are reconstructed using the electromagnetic calorimeter and their energy is
obtained from a sum of ECAL crystals.  A set of crystals with energy deposition are
combined into clusters. The arrays of clusters, which contain all of the energy of a
photon are called superclusters.  In the barrel section of the ECAL, an energy resolution
of about 1\% is achieved for unconverted or late-converting photons in the tens of \GeV
energy range. The remaining barrel photons have a resolution of about 1.3\% up to a
pseudorapidity of $|\eta| = 1$, rising to about 2.5\% at $|\eta| = 1.4$, see
Ref.~\cite{EGM-14-001}.

After the basic reconstruction in ECAL, the identification criteria are applied in order
to better separate photons from jets and electrons.  The observables used in the photon ID
are: the ratio of the energy in the hadron calorimeter towers behind the supercluster to
the electromagnetic energy in the supercluster; the transverse width in $\eta$ of the
electromagnetic shower; the PF isolation variables, $I_{ch}, I_{neu}, I_{pho}, \rho$,
calculated in the cone $R_0=0.3$.  Specific selection based on these variables was
initially optimized on simulated samples of $W+\gamma$ and $W+jet$ events to maintain
approximately 80\% identification efficiency for a photon.  Furthermore, a veto on the
hits in the innermost layer of the pixel detector is applied to avoid misidentifying an
electron as a photon. Such veto however, allows for the electrons produced from the
conversion of the photon on the material of the pixel detector.  The efficiency of the
photon identification is measured with the ``tag-and-probe'' method using $\Z \to
\Pe^+\Pe^-$ events in data and MC, where the electrons are reconstructed as photon
showers.  The efficiency of the pixel veto is estimated with $\Z\to\mu\mu\gamma$ events,
where the photon is produced via FSR.  The total efficiency is found to be 80\% (88\%) for
a photon with $\ET>30\,(50)$\GeV and $|\eta^{\gamma}|<1.44$.

The photon energy resolution is further improved by using a multivariate regression
technique developed for $\PH \to \gamma\gamma$ analysis. See Ref.~\cite{cms-Hgg-Legacy}
for an extended description of the technique.  The energy scale corrections are applied to
the reconstructed photons in data and the smearing corrections to the photons in MC
events. These corrections are necessary due to imperfect knowledge of the detector and its
simulation.  The underlying causes are known to be from: a) tracker material simulation,
b) underestimation of uncertainty in the individual crystal calibration and c) residual
differences between the actual ECAL geometry and its simulation.  These corrections are
also derived using $\Z\to \Pe\Pe$ events, where the electrons are reconstructed as
photons.

\subsection{Converted photons}
\label{sec:conversions}
A photon interacting in the material of the detector often converts to a pair of
electrons, $\Pe^+\Pe^-$. The probability of such interaction to occur before the last
three layers of the tracker is 20--50\% in the barrel, and up to 60\% in the
endcap~\cite{cms-Hgg-Legacy}.  A method of reconstructing the tracks from the conversion
electrons was developed in Ref.~\cite{Marinelli} and used in $\PH\to\gamma\gamma$
analysis, as described in Ref.~\cite{cms-Hgg-Legacy}.  Fully reconstructed conversions are
used in the PF reconstruction algorithm: the association of electron-track pairs with
energy deposits in the ECAL avoids the photons being misidentified as charged hadrons,
thus improving the determination of the photon isolation, as already discussed.


\subsection{Muons}
\label{sec:reco-mu}

Muons are measured in the pseudorapidity range $\abs{\eta}<2.4$, with detection planes
made using three technologies: drift tubes, cathode strip chambers, and resistive plate
chambers. Matching muons to tracks measured in the silicon tracker results in a relative
transverse momentum resolution for muons with $20 <\pt < 100\,\GeV$ of 1.3--2.0\% in the
barrel and better than 6\% in the endcaps. The \pt resolution in the barrel is better than
10\% for muons with \pt up to 1\TeV~\cite{cms-mu-7TeV}.  For low-\PT muons used in the
analysis the resolution is between 0.8\% and 3\% depending on $\eta$.

\begin{figure}[ht]
  \centering
  \includegraphics[width=0.75\textwidth]{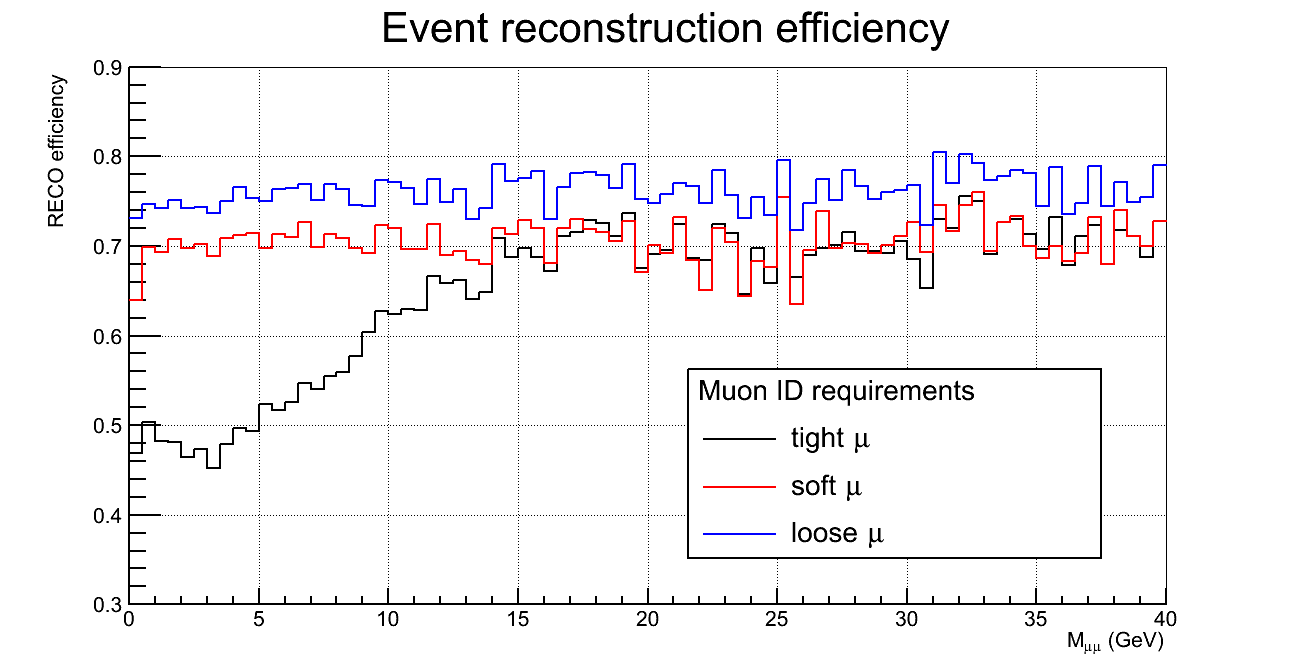}
  \caption[Event reconstruction efficiency as a function of dimuon invariant mass.]
  {Comparison of three common muon IDs:
    event reconstruction efficiency as a function of dimuon invariant mass.
    This efficiency is calculated with the full events selection, described in Section~\ref{sec:sel}.}
  \label{fig:muID_eff}
\end{figure}

For this analysis, the muon candidates must be selected by the PF algorithm and satisfy
the following requirements. The fit of the resulting track in the pixel detector must pass
$\chi^2/n_{DoF} < 3$ criterion.  In CMS, this ID criteria is considered \textit{loose} and
the motivation for this choice is driven by the dedicated studies of the event
reconstruction efficiency.  Due to the properties of the signal, described in
Section~\ref{sec:features}, it is important to maintain the reconstruction efficiency for
events with small $\Delta R(\mu\mu)$ separation, which corresponds to low $m_{\mu\mu}$.
Figure~\ref{fig:muID_eff} shows the event reconstruction efficiency vs. dimuon invariant
mass for different muon IDs.  From this figure one can see that a more commonly used
\textit{tight} ID would be inefficient for the region with $m_{\mu\mu}<10\,\GeV$. However
with the selected \textit{loose} ID we maintain a high efficiency, which is independent of
the dimuon invariant mass. Figure~\ref{fig:muID_eff} includes additional kinematic
requirements: muon with the highest-\pt (called \textit{leading} lepton) has to have
$\pt>23\GeV$, and the next to highest-\pt muon (called \textit{subleading}) has to have
$\pt>4\GeV$; photon with $\pt>25\,\GeV$ is also selected.  Full event selection of the
analysis is described in Section~\ref{sec:sel}.

Additionally, the PF isolation in the cone $R_0=0.4$ is then calculated for the leading
muon:
\begin{equation}
  \label{eq:pfiso}
  I_{PF} = I_{ch} + max(0, I_{nue} + I_{pho} - 0.5\cdot I_{PU}).
\end{equation}
The isolation is required to be less than 0.4 for the leading muon. No isolation
requirement is applied for the subleading muon since those muons are already within the
isolation cone of the leading muon in most events.  The isolation requirement rejects
misidentified leptons and background arising from hadronic jets.  Full dimuon
identification and isolation efficiency of about 80\% is obtained.

The energy scale (in data events) and resolution (in MC) of the muons are corrected using
$\Z\to\mu\mu$ events.

\subsection{Electrons}
\label{sec:reco-el}

Similar to the photon reconstruction, electrons in CMS are built from the superclusters in
ECAL. The shape of the supercluster is different from the photons, because electrons bend
in the magnetic field along $\phi$ direction.  The superclusters are then matched to
tracks in the silicon tracker~\cite{ele-CMS,ele-7TeV}.

In the electron channel of the $\PH\to\gamma^*\gamma\to \ell\ell\gamma$ decay, the two
electrons produced by $\gamma^*$ are rather close to each other.  Even more so than in the
muon channel, since the $m_{\ell\ell}$ is smaller in $\gamma^*\to\Pe\Pe$ process, see
Fig.~\ref{fig:mll-sig}.  Therefore, their energy deposits in the electromagnetic
calorimeter are merged into one supercluster by the CMS reconstruction algorithms, giving
rise to a very special signature.  In order to identify these merged electrons, at least
two tracks reconstructed with Gaussian Sum Filter (GSF) algorithm~\cite{ele-GSF}
associated to the supercluster are required. Also at least two basic ECAL clusters within
a supercluster are required.  The supercluster of the reconstructed \textit{merged}
electrons must have $\pt>30\GeV$, $\abs{\eta}<1.44$, and $\PT^{\Pe_1} + \PT^{\Pe_2} >
44\,\GeV$ for the corresponding two GSF tracks.  Both GSF tracks must have no more than
one missing hit in the pixel detector in order to reduce the background from photon
converting into $\Pe^+\Pe^-$ induced by interactions with the detector material.  These
criteria remove 92\% of the QCD $\gamma\gamma$ events, 80\% of the QCD dijet events and
36\% of the $\Z\to\Pe\Pe$ background events, while losing 19\% of signal events.

Furthermore, a multivariate discriminator is trained to separate the $\gamma^*\to\Pe\Pe$
objects from jets and single electrons.  The MVA used for the ID is Boosted Decision Tree
(BDT), implemented in TMVA~\cite{TMVA}.  The variables used as inputs to the BDT training
include lateral shower shape variables, the energy median density per unit area in the
event ($\rho$), and the kinematic information of supercluster energy and GSF tracks.

There are three kinds of major backgrounds for the \textit{merged} electrons signature:
\begin{itemize}
\item prompt photon conversion, which is suppressed by requiring missing hits and conversion veto, mentioned above;
\item fake photons from jets, that fragment to neutral mesons and then decay to the photon pair;
\item prompt electron with a second (fake) GSF track close to the real GSF track, or
  a photon from bremsstrahlung which converts into GSF tracks.
\end{itemize}

\begin{figure}[t]
  \centering
  \includegraphics[width=0.46\textwidth]{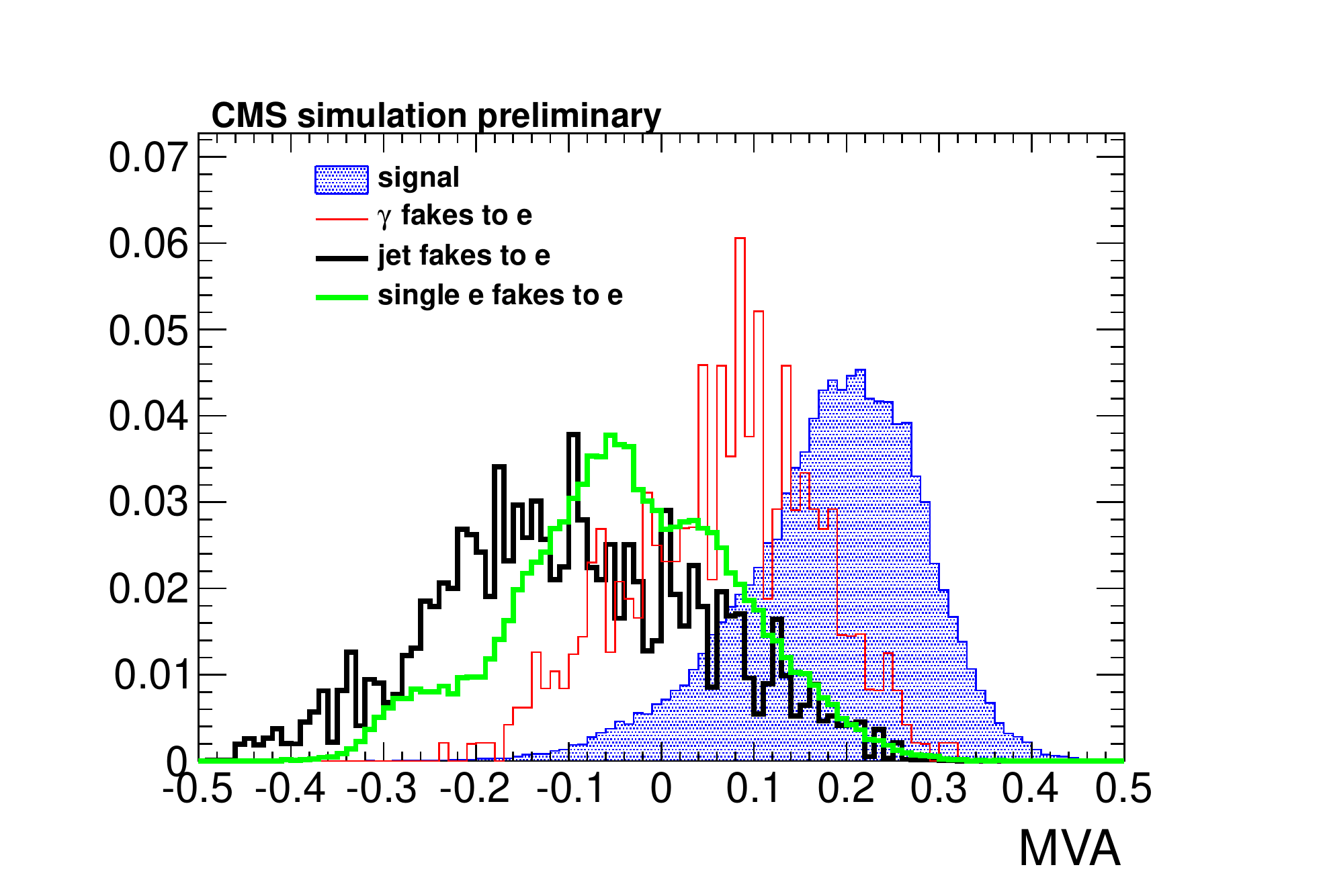}~
  \includegraphics[width=0.46\textwidth]{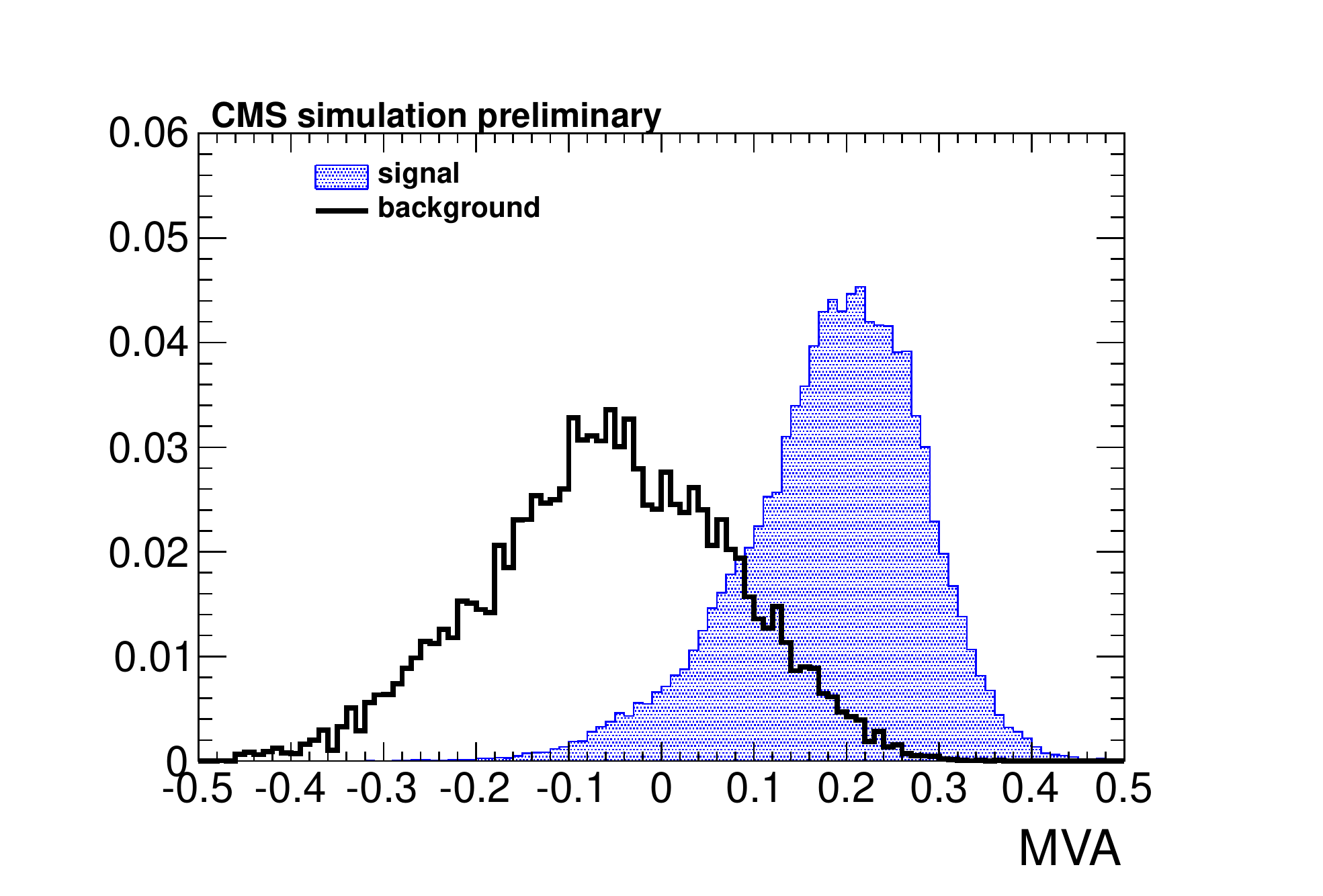}
  \caption{MVA response for signal and individual backgrounds (left), and  combined background objects (right).}
  \label{fig:eID_MVA_output}
\end{figure}

The BDT output discriminator is trained with simulated samples, where the signal objects
of $\gamma^*\to\Pe\Pe$ are taken from the Higgs signal samples and background objects are
taken from various MC background samples: $\gamma+jet$, QCD processes, and DY+jet.  As an
input to the BDT training, both signal and background electrons must pass the selection
described in the beginning of this section. In order to validate the results of the MVA
training the events are divided evenly between the \textit{training} and \textit{testing}
sub-samples. Comparing the BDT output of the two samples we've concluded that there is no
overtraining and the quality of the BDT discriminant is good.
Figure~\ref{fig:eID_MVA_output} shows the MVA response for the signal and backgrounds.
The final cut on the BDT output of 0.12 is used in the analysis.

\subsection{Data/MC scale factors}
Efficiency of certain object ID selection or trigger may not be properly simulated in the
MC samples.  Hence, event scale factors are applied to the MC samples in order to
compensate for those differences.  Specifically, the scale factors are applied due to the
photon ID.  Those corrections are \textit{standard} in CMS and derived using a
tag-and-probe method from $\Z\to \Pe\Pe$ events on both data and MC, see
Table~\ref{tab:phoSF} for a summary.  On the other hand, the muon scale factors needed to
be estimated for our event topology.  This was done using MC signal samples and
$\JPsi\to\mu\mu$ events in data.  It was found that the uncertainty on the scale factors
is larger than the correction itself.  Hence, no correction is applied, and the systematic
uncertainty is assigned instead.  Those uncertainties are summarized in
Section~\ref{sec:syst}.  Similarly, efficiency in the electron channel is not measures in
data, instead the uncertainty is assigned based on the studies of the simulated signal
samples.

\begin{table}[ht]
  \caption{Photon ID scale factors applied per photon.}
  \label{tab:phoSF}
  \centering
  \begin{tabular}{c|cc}
    \multicolumn{1}{r}{$\eta/\PT$ range } & 40--50 \GeV &$>50 \GeV$\\
    \hline
    \multicolumn{3}{c}{ID}\\
    \hline
    $0  <|\eta|<0.8$  & 0.9804 $\pm$ 0.0005 & 0.9787 $\pm$ 0.0009\\
    $0.8<|\eta|<1.5$  & 0.9840 $\pm$ 0.0006 & 0.9822 $\pm$ 0.0011\\
    \hline
    \multicolumn{3}{c}{Conv. electron veto}\\
    \hline
    $0 < |\eta|<1.44$ & 0.993 $\pm$ 0.029 & 1.0 $\pm$ 0.0\\
  \end{tabular}
\end{table}

\clearpage

\section{Event selection}
\label{sec:sel}

Initial event selection is performed during the data-taking by the HLT.  In the muon
channel, the trigger requires a muon and a photon, both with $\pt>22\GeV$.  In the
electron channel, the $\gamma^* \to \Pe\Pe$ process at low dielectron invariant mass
mimics a photon at the trigger level.  For this reason, a diphoton trigger is used in the
electron channel, to select the $\gamma^*\gamma$ final state events. The trigger requires
a leading (subleading) photon with \pt greater than 26\,(18)\GeV.  The diphoton trigger is
inefficient for events with high dielectron invariant mass ($m_{\Pe\Pe} > 2\GeV$) due to
the isolation and shower shape requirements. The available dielectron triggers cannot be
used to select events with $2 < m_{\Pe\Pe} < 20\,\GeV$ either, because they require
isolation, and their \pt threshold on the subleading lepton is too stringent.

The efficiency of the triggers for the signal events after the selection requirements
described bellow is 85\%\,(90\%) in the muon (electron) channel, as obtained from the
simulated samples.

In the offline selection, the events are required to have at least one primary vertex, as
described in Section~\ref{sec:PV}.  The lepton tracks from $\gamma^*\to\mu\mu\,(\Pe\Pe)$
are required to originate from the primary vertex, and to have transverse and longitudinal
impact parameters with respect to that vertex smaller than $2.0\,(0.2)\mm$ and $5\,(1)\mm$,
respectively.

The muons (electrons) are required to be within $\abs{\eta} < 2.4\,(1.44)$, while the
photon must have $\abs{\eta} < 1.44$. The three-body invariant mass is required to satisfy
$110 < m_{\ell\ell\gamma}< 170\,\GeV$.  The photon and dilepton momenta must satisfy
$\pt^{\gamma}> 0.3\cdot m_{\ell\ell\gamma}$ and $\pt^{\ell\ell}>0.3\cdot
m_{\ell\ell\gamma}$ requirements, which are optimized for high signal efficiency and
background rejection.  The muons must be oppositely charged, and have \PT greater than
$23\,(4)\GeV$ for the leading (subleading) muon.  The \PT requirement on the leading muon
is driven by the trigger threshold, and on the subleading muon by the minimum energy
needed for a particle to reach the muon system, while maintaining high reconstruction
efficiency.  In the electron channel, no additional selection on \PT of the GSF tracks is
necessary, beyond those described in Section~\ref{sec:reco-el}.  Finally, in both muon and
electron channels, the separation between each lepton and the photon is required to
satisfy $\DR>1$ in order to suppress Drell--Yan background events with FSR.

\begin{figure}[t]
  \begin{center}
    \includegraphics[width=0.6\textwidth]{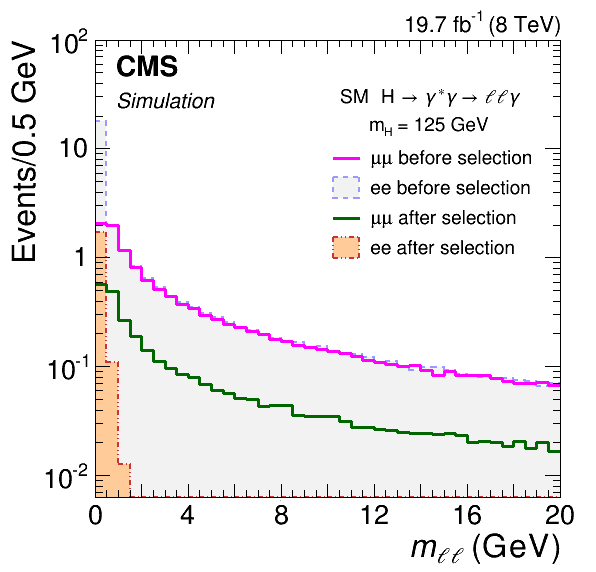}~
    \caption[The invariant mass of the dilepton system in signal simulation for
    $m_\PH=125\GeV$.]
    {The invariant mass of the dilepton system in signal simulation for $m_\PH=125\GeV$.
      Distributions are shown for muon and electron channels, before and after selection.
      The invariant mass before selection is obtained from the leptons at the generator
      level, while after selection the reconstructed invariant mass is used.}
    \label{fig:mll-sig}
  \end{center}
\end{figure}

\begin{figure}[t]
  \begin{center}
    \includegraphics[width=0.6\textwidth]{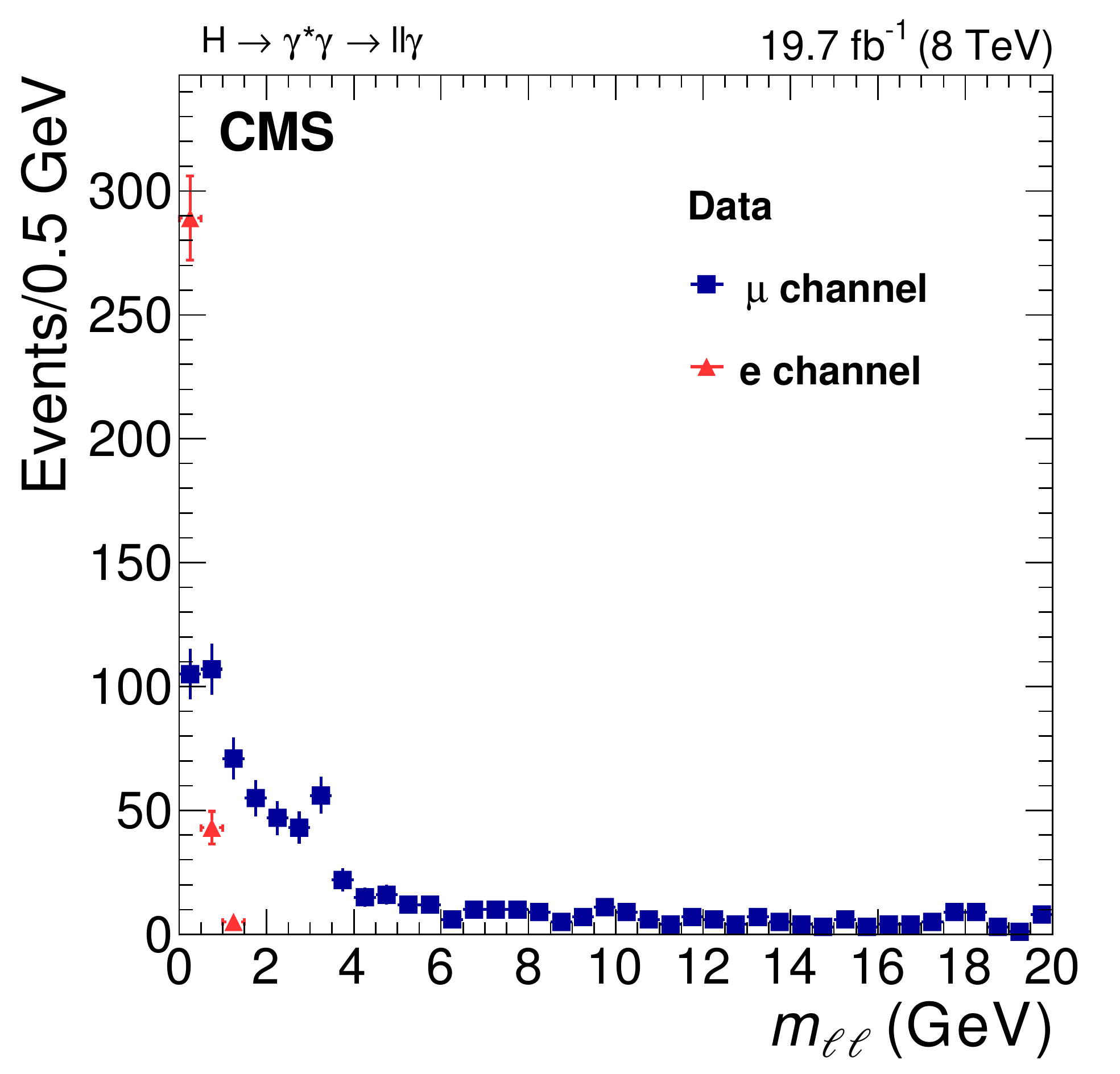}~
    \caption[The invariant mass of the dilepton system in data for muon and electron
    channels.]
    {The invariant mass of the dilepton system in data for muon and electron channels.
      The distributions produced after the selection described in the text, but without
      rejecting the $V\to\mu\mu$ contributions in data.}
    \label{fig:mll-data}
  \end{center}
\end{figure}

The dilepton invariant mass in the muon channel is required to be less than 20\GeV to
reject contributions from $\Pp\Pp\to\Z\gamma$ and to suppress interference effects from
$\PH\to\Z\gamma$ process and box diagrams shown in Fig.~\ref{fig:dia-dalitz}.  Events with
a dimuon mass $2.9< m_{\mu\mu}< 3.3\,\GeV$ and $9.3 < m_{\mu\mu} < 9.7\,\GeV$ are rejected
to avoid the $\JPsi\to\mu\mu$ and $\Upsilon\to\mu\mu$ contamination.  In the electron
channel the invariant mass, constructed from the two GSF tracks, is required to satisfy
$m_{\Pe\Pe} < 1.5\GeV$.  There are almost no events in the electron channel for
$m_{\Pe\Pe}>1.5\GeV$ due to the trigger requirement mentioned before.  The $m_{\ell\ell}$
distributions for the simulated signal events are shown in Fig.~\ref{fig:mll-sig} in the
muon and electron channels.  The $m_{\ell\ell}$ distributions in data are shown in
Fig.~\ref{fig:mll-data}. The distributions of the data events should be thought of as the
background, thus the shapes of the background and signal events in Fig.~\ref{fig:mll-sig}
are to be compared.

In the special case of the search for $\PH\to(\JPsi)\gamma\to\mu\mu\gamma$, both
$\pt^{\gamma}>40\GeV$ and $\pt^{\mu\mu}>40\GeV$ are required, and the events are selected
within $2.9 < m_{\mu\mu}<3.3\,\GeV$.

The expected signal yields for $m_\PH=125\GeV$ and the observed yield in the 10\GeV mass
bins after the full event selection are listed in Table~\ref{tab:yield}. Additionally,
Table~\ref{tab:yield-data} shows the event yield in data, broken down into four
data-taking periods of CMS. The beam conditions (\eg pileup) and the integrated luminosity
were different in those periods. The data show the yields statistically consistent with
the integrated luminosity per period, mostly independent of the different conditions.

\begin{table}[t]
  \caption[The expected signal yields and number of events in data.]{The expected signal yield and the number of events in data,
    for an integrated luminosity of $19.7\fbinv$.
    Signal events are presented before and after applying the full selection criteria described in the text.
    In the $(\JPsi)\gamma$ sub-category only $\JPsi\to\mu\mu$ decay is considered,
    and the signal yield is a sum of two contributions:
    $\PH\to(\JPsi)\gamma\to\mu\mu\gamma$ and $\PH\to\gamma^*\gamma\to\mu\mu\gamma$, where dimuon mass distribution is non-resonant.}
  \label{tab:yield}
  \begin{center}
    \begin{tabular}{l c c|c}
      & \multicolumn{2}{c}{Signal events $m_\PH=125\GeV$}              & Number of events in data  \\
      Sample& before selection  & after selection       & $120 < m_{\ell\ell\gamma} <130\,\GeV$  \\
      \hline
      $\mu\mu\gamma$           & 13.9                       & 3.3                                & 151 \\
      $\Pe\Pe\gamma$           & 25.8                       & 1.9                                & 65 \\
      \hline
      $(\JPsi\to\mu\mu)\gamma$ & 0.065 (\JPsi) + & 0.014 (\JPsi) +  & 12  \\
      &   0.32 \text{(non-res.)} & 0.078 \text{(non-res.)} &  \\
    \end{tabular}
  \end{center}
\end{table}

\begin{table}[t]
  \begin{center}
    \begin{tabular}{ cr|cccc|c}
      \multicolumn{2}{r|}{$m_{\ell\ell\gamma}$ range,\GeV} & A & B & C & D & Total \\
      \hline
      $\mu$ channel      &110--120& 14&  40&  81&  69& 204  \\
      &120--130&  6&  33&  49&  63& 151  \\
      &130--140& 13&  29&  39&  36& 117  \\
      &140--150&  3&  18&  41&  24&  86  \\
      &150--160&  5&  15&  19&  28&  67  \\
      &160--170&  1&  12&  17&  13&  43  \\
      \hline
      \multicolumn{2}{l|}{Total in 110--170}& 42& 147& 246& 233& 668  \\
      \hline
      \hline
      $\Pe$ channel&110--120& 7 & 23 & 37 & 34 & 101  \\
      &120--130& 2 & 13 & 21 & 29 &  65  \\
      &130--140& 3 & 16 & 24 & 17 &  60 \\
      &140--150& 2 & 15 & 14 &  9 &  40 \\
      &150--160& 2 &  4 & 12 & 14 &  32 \\
      &160--170& 2 &  2 & 10 &  8 &  22 \\
      \hline
      \multicolumn{2}{l|}{Total in 110--170}&18 & 73 &118 & 111 & 320  \\
      \hline
      \multicolumn{2}{c|}{Integrated luminosity, \fbinv}& 0.88 & 4.41 & 7.06 & 7.36 & 19.7 \\
      &&\multicolumn{4}{c|}{}&  \\
    \end{tabular}
  \end{center}
  \caption{Events break down by the  data-taking periods and bins of $m_{\ell\ell\gamma}$.}
  \label{tab:yield-data}
\end{table}

After the full selection one can look at the distributions of interest: $m_{\ell\ell}$,
$\PT^{\ell_1}$, $\PT^{\ell_2}$, $\PT^{\gamma}$, $\Delta R(\ell,\gamma)$, etc.  These
figures are presented in Appendix~\ref{sec:app-extra}.  The shapes of all these
distributions in data end up looking very similar to those in simulated signal sample,
which suggests no further kinematic separation can be achieved using these variables.

In the muon channel, about 3.4 signal events are expected for $m_\PH =125\,\GeV$ Higgs
boson, while the background is about $92\pm9$ events within $122 <m_{\ell\ell\gamma}<
128\GeV$ (estimated from the fit to the data, as discussed in Section~\ref{sec:stat}).  In
the electron channel the signal-to-background ratio is much smaller, which results in a
weaker limit on the SM signal strength. It should be also mentioned that in the electron
channel there is a contribution from the $\PH\to\gamma\gamma$ process due to unidentified
conversions, which is about 15\% of the $\PH\to\gamma^*\gamma$ signal (0.2 events at
$m_\PH = 125\GeV$).  This contribution is considered as a background to
$\PH\to\gamma^*\gamma$, and it is negligible compared to the continuum background
estimated from the fit to data described in the next section.

From Table~\ref{tab:yield}, the total signal efficiency is $\varepsilon_{tot} =
3.3/13.9\approx0.24$, \ie 24\%, in the muon channel for $m_\PH = 125\GeV$. It rises to
${\sim}25\%$ for $m_\PH=150\GeV$. Correspondingly, in the electron channel,
$\varepsilon\approx7\%$, which rises to ${\sim}9\%$ for $m_\PH=150\GeV$. For the
$\PH\to(\JPsi)\gamma$ signal, $\varepsilon\approx22\%$.

\begin{figure}[ht]
  \begin{center}
    \includegraphics[width=0.48\textwidth]{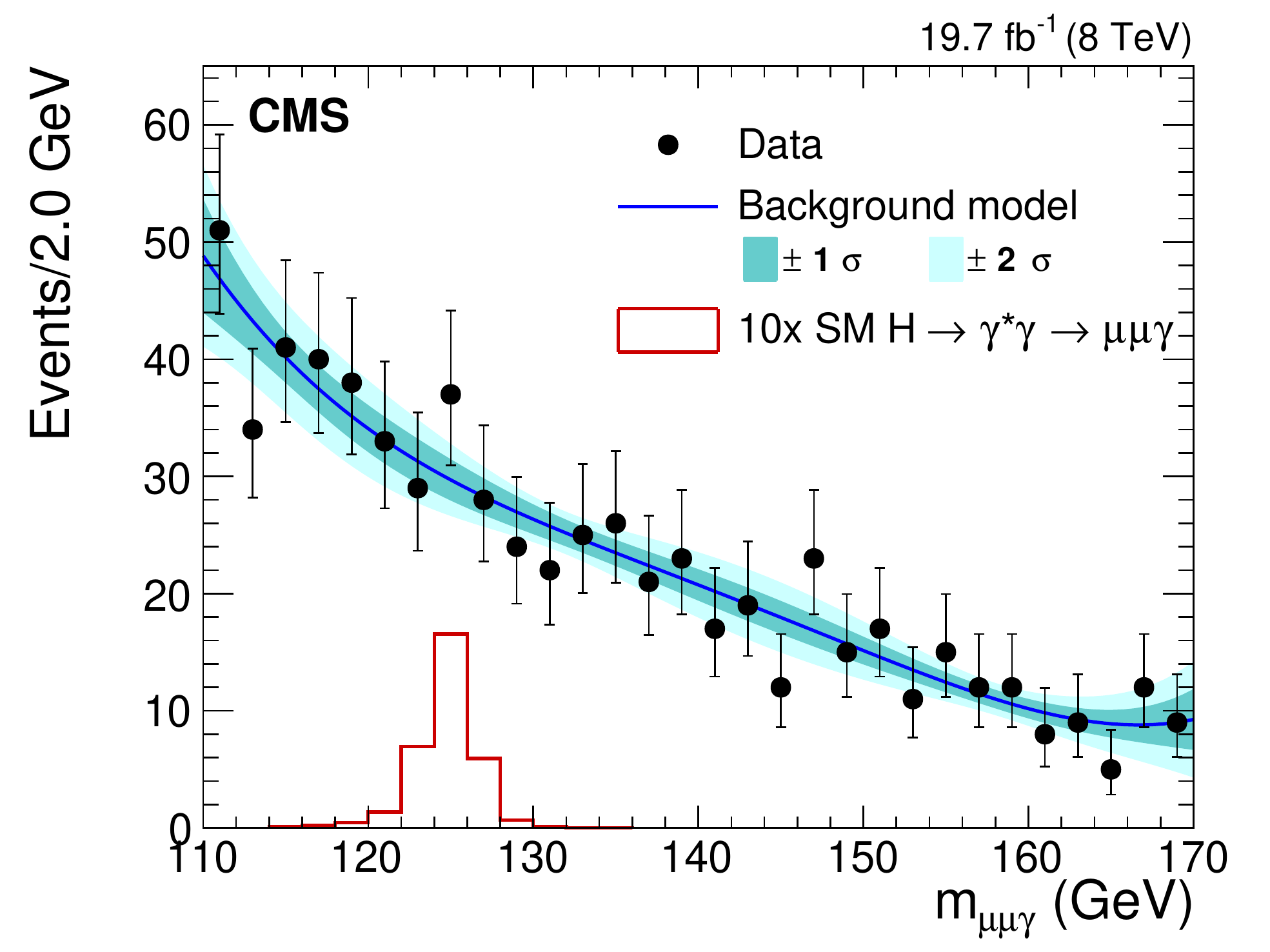}
    \includegraphics[width=0.48\textwidth]{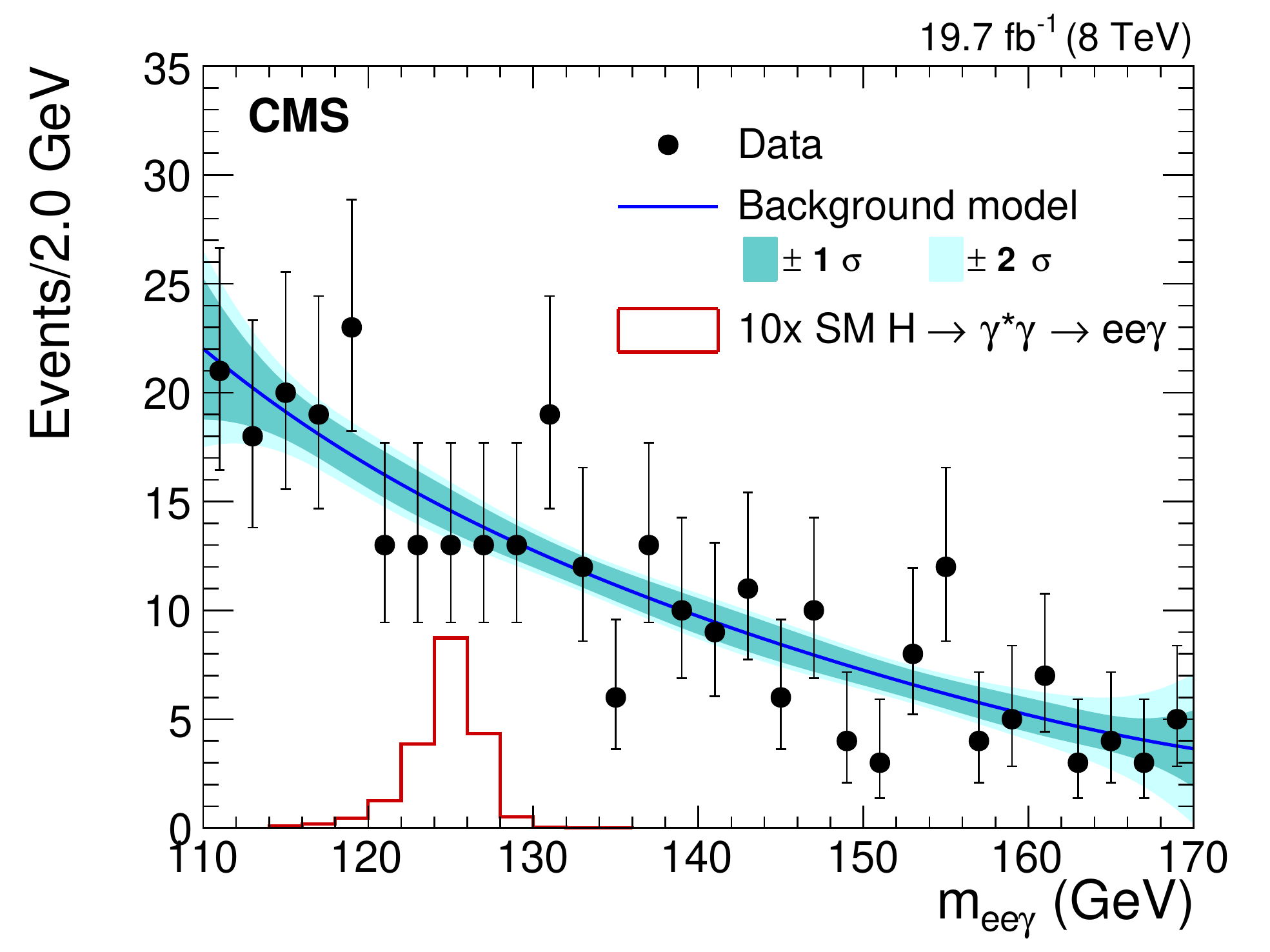}
    \caption[The $m_{\mu\mu\gamma}$ and $m_{\Pe\Pe\gamma}$ spectra in Dalitz analysis.]  {The
      $m_{\mu\mu\gamma}$ (left) and $m_{\Pe\Pe\gamma}$ (right) spectra for 8\TeV data
      (points with error bars), together with the result of a background-only fit to the
      data. The 1$\sigma$ and 2$\sigma$ uncertainty bands represent the uncertainty in the
      parameters of the fitted function.  The expected contribution from the SM Higgs
      boson signal with $m_\PH=125\GeV$, scaled up by a factor of~10, is shown as a
      histogram.}
    \label{fig:fit}
  \end{center}
\end{figure}

\begin{figure}[ht]
  \begin{center}
    \includegraphics[width=0.48\textwidth]{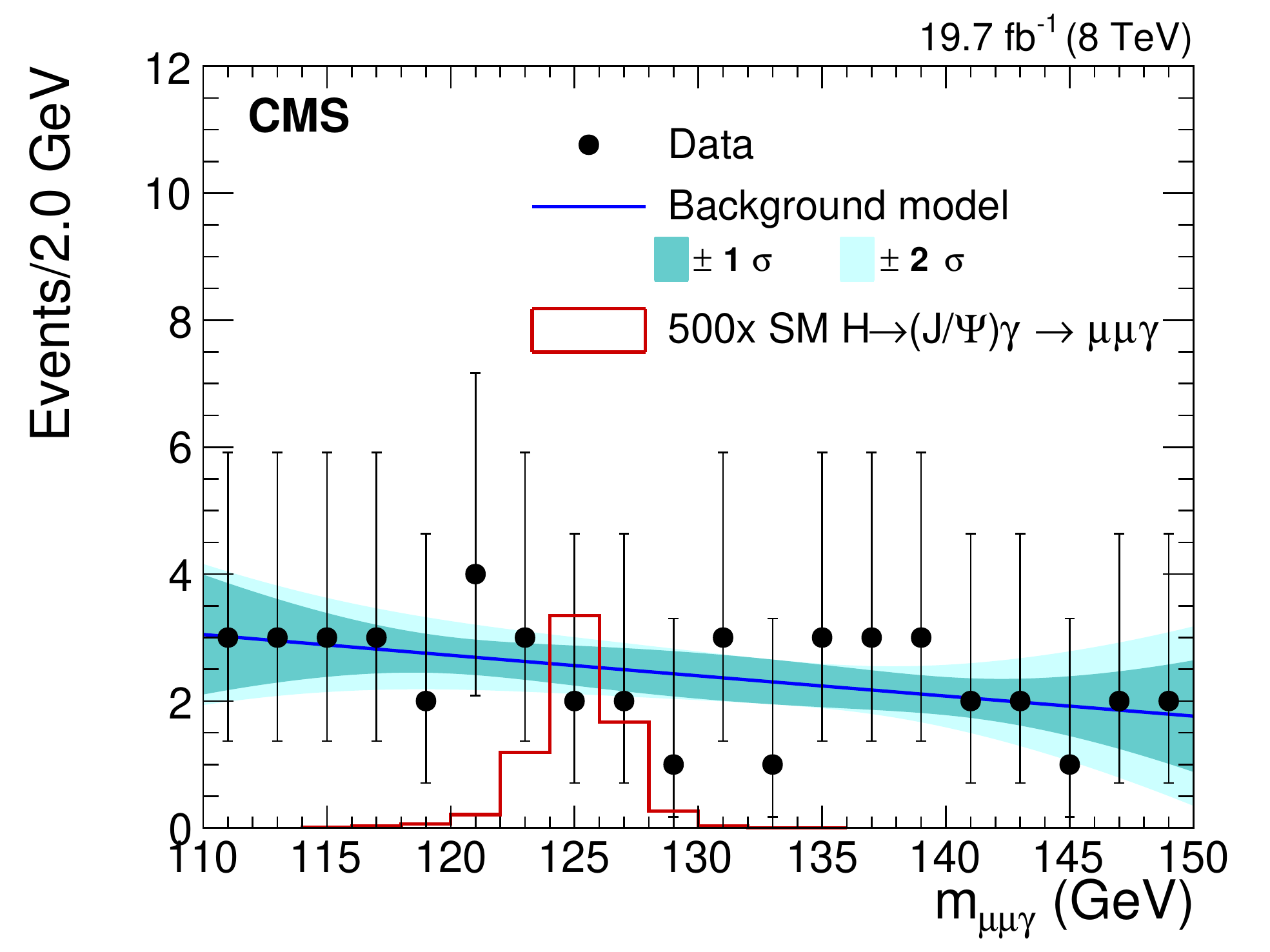}
    \caption[The $m_{\mu\mu\gamma}$ distribution for events within $2.9 < m_{\mu\mu} <
      3.3\GeV$.]  {The $m_{\mu\mu\gamma}$ distribution for events within $2.9 < m_{\mu\mu}
        < 3.3\GeV$ for 8\TeV data (points with error bars), together with the result of a
        background-only fit to the data. The 1$\sigma$ and 2$\sigma$ uncertainty bands
        represent the uncertainty in the parameters of the fitted function.  The expected
        contribution from the $\PH\to(\JPsi)\gamma\to\mu\mu\gamma$ process of the SM Higgs
        boson, $m_\PH=125\GeV$, scaled up by a factor of~500, is shown as a histogram.
      \label{fig:fit-jp}
    }
  \end{center}
\end{figure}

Finally, the $m_{\ell\ell\gamma}$ distributions are shown in Fig.~\ref{fig:fit} for the
Dalitz search and in Fig.~\ref{fig:fit-jp} for the $\PH\to(\JPsi)\gamma$ search.  A smooth
polynomial fit to these spectra in data is used as a background prediction.  An excess of
events in data above the background curve, at any particular mass point, could indicate a
presence of a signal peak. The strength of the signal is then determined using the
statistical methods described in Section~\ref{sec:stat}.

\begin{figure}[ht]
  \begin{center}
    \includegraphics[width=0.48\textwidth]{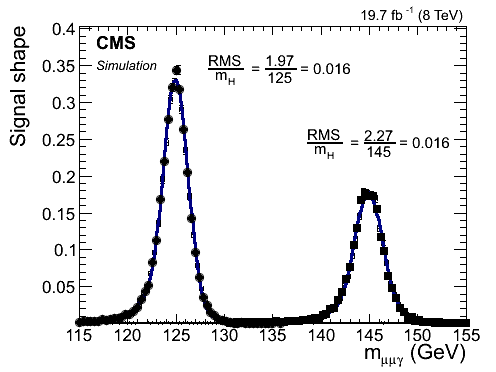}~
    \includegraphics[width=0.48\textwidth]{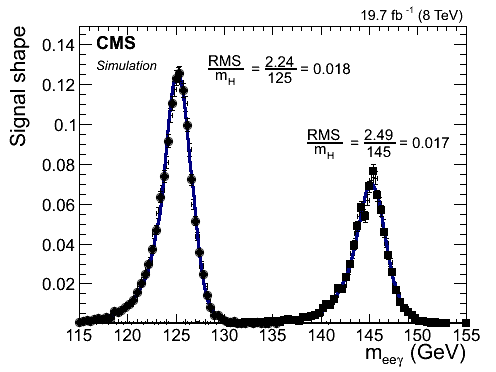}
    \caption[Reconstructed three-body mass distributions from the signal MC samples for
    $m_\PH=125$ and 145\GeV in muon and electron channels.]
    {Reconstructed three-body mass distributions from the signal MC samples for
      $m_\PH=125$ and 145\GeV in muon (left) and electron (right) channels. This
      distributions are clearly non-Gaussian. The blue curves represent the fits to a
      Crystal Ball plus a Gaussian function, discussed in the next section.}
      \label{fig:sig-mass-res}
  \end{center}
\end{figure}

Resolution of the $m_{\ell\ell\gamma}$ variable plays a crucial role in the analysis
sensitivity. Table~\ref{tab:sig-mass-res} shows the effective width of the
$m_{\ell\ell\gamma}$ distributions obtained from the MC signal samples.  It is calculated
as an RMS of the $m_{\ell\ell\gamma}$ dataset obtained from the MC signal sample,
considering only the points within $0.9\times m_\PH < m_{\ell\ell\gamma} < 1.1\times
m_\PH$.  Examples of the mass distributions for $m_\PH = 125$ and 145\GeV are shown in
Fig.~\ref{fig:sig-mass-res}. The resolution of ${\sim}1.6\%$ is achieved in the muon
channel and ${\sim}1.8\%$ in the electron channel.  For comparison, in the
$\PH\to\gamma\gamma$ search the resolution of the $m_{\gamma\gamma}$ varies from 0.9\% to
2\% for $m_\PH=125\GeV$, depending on the event category~\cite{cms-Hgg-Legacy}.

\begin{table}[ht]
  \begin{center}
    \begin{tabular}{c|cc}
      & \multicolumn{2}{c}{$\sigma_{\ell\ell\gamma}^{eff}$, \GeV; ($\sigma_{\ell\ell\gamma}^{eff}/m_{\PH}$)} \\
      $m_\PH$& $\mu$ channel & $\Pe$ channel \\
      \hline
      120& 1.79 (1.5\%)& 2.13 (1.8\%)\\
      125& 1.97 (1.6\%)& 2.24 (1.8\%)\\
      130& 2.09 (1.6\%)& 2.28 (1.8\%)\\
      135& 2.12 (1.6\%)& 2.40 (1.8\%)\\
      140& 2.22 (1.6\%)& 2.43 (1.7\%)\\
      145& 2.27 (1.6\%)& 2.50 (1.7\%)\\
      150& 2.35 (1.6\%)& 2.54 (1.7\%)\\
\end{tabular}
  \end{center}
  \caption{Effective width of the Higgs boson candidate mass distribution obtained from a gluon fusion MC signal sample.}
  \label{tab:sig-mass-res}
\end{table}

\clearpage
\subsection{VBF tag in the muon channel}
An importance of the VBF/VH event tagging for the purpose of increasing the sensitivity of
an analysis was mentioned in Sec.~\ref{sec:prod}.  At the current stage, however, such
tagging does not bring much to the sensitivity.  For example, after a basic selection for
the VBF tag, very little signal is expected.  In addition, there are too few events left
in the data to perform the fit for the background estimation.  I describe this result in
the Appendix~\ref{sec:app-vbf}.  Due to these two reasons, no VBF/VH tagging is used
throughout the analysis at present, although it will be useful in the future data-taking
at 13\TeV, with larger data sample.

\subsection{Additional categories in the muon channel}
The selection described in the previous section can be extended in various ways, in order
to increase the sensitivity of the search.  Let me remind that the main analysis only
includes the photons in the barrel, $|\eta^{\gamma}_{SC}| < 1.4$, and low dilepton
invariant mass, $m_{\mu\mu} < 20\,\GeV$ (labeled \textit{EB} further in the text). Hence,
the two obvious choices to extend the analysis selection, are:
\begin{itemize}
\item To extend the pseudorapidity of the photons to the endcap: $1.6 <
  |\eta^{\gamma}_{SC}| < 2.5$ (labeled \textit{EE})
\item To extend the $m_{\mu\mu}$ range to $20 < m_{\mu\mu} < 50\,\GeV$, while
  $|\eta^{\gamma}_{SC}| < 1.4$ (labeled \textit{mll50})
\end{itemize}

\begin{figure}[ht]
  \begin{center}
    \includegraphics[width=0.48\textwidth]{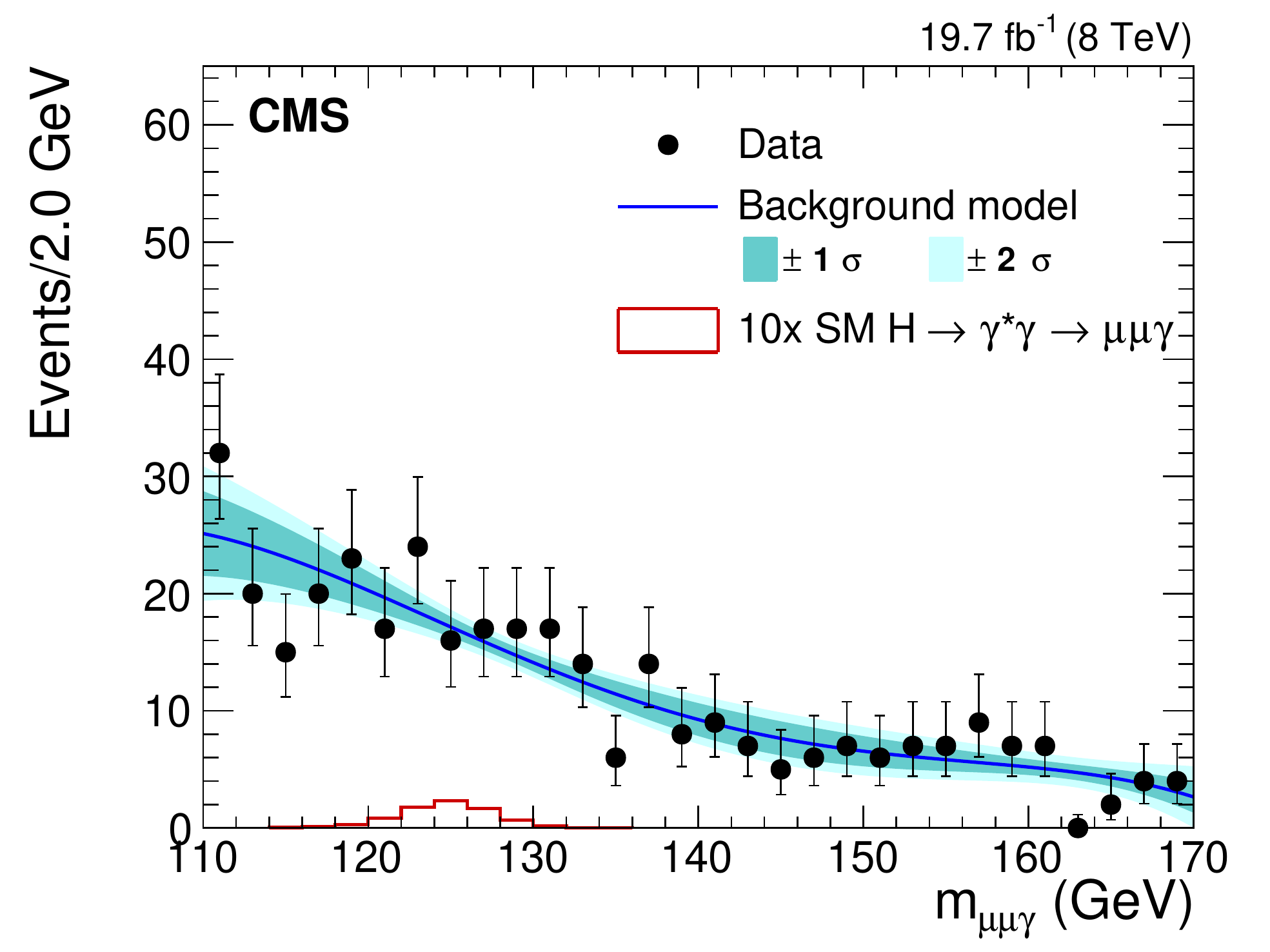}
    \includegraphics[width=0.48\textwidth]{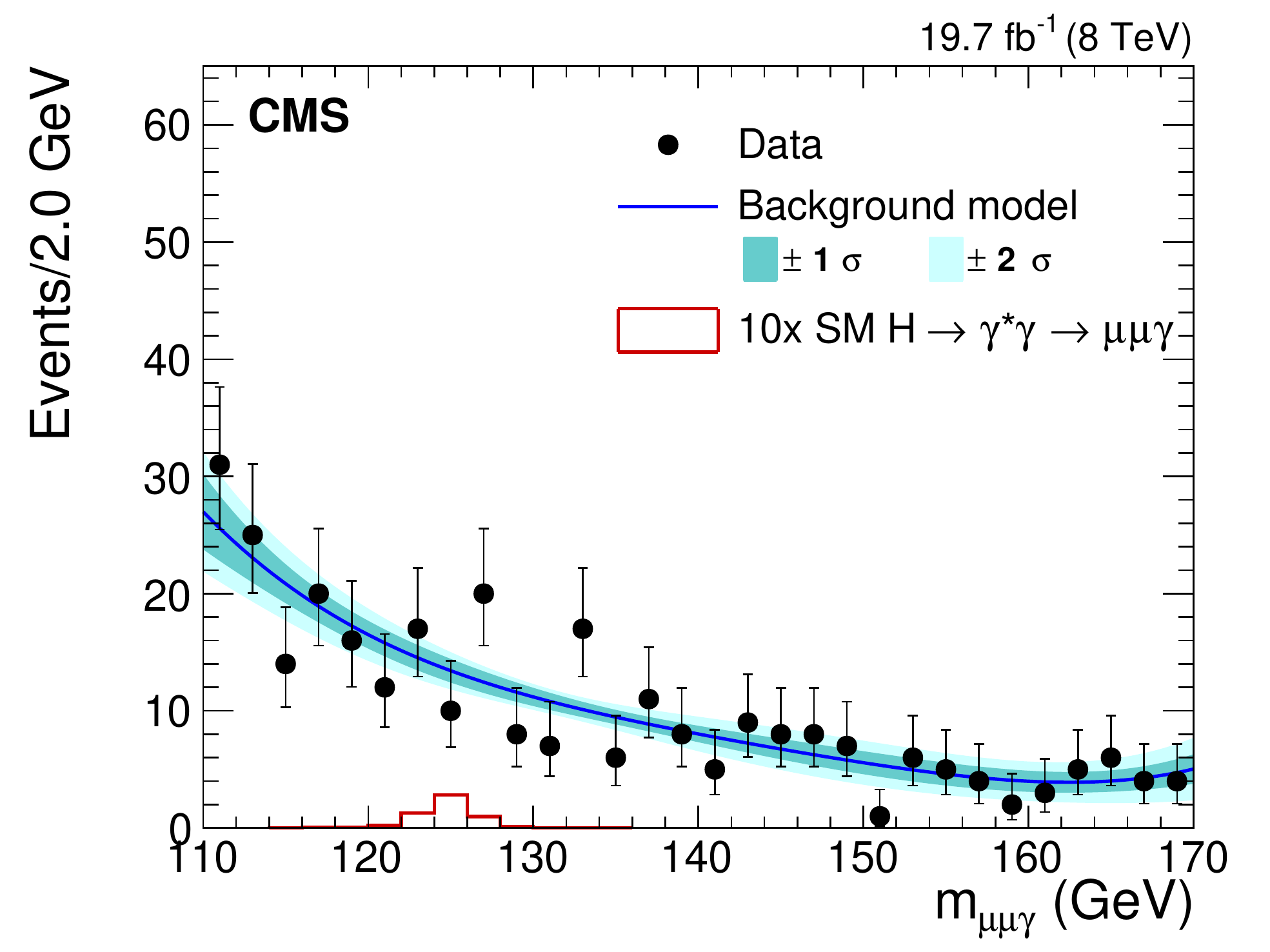}
    \caption[The $m_{\ell\ell\gamma}$ distributions in the muon channel for additional
    event categories.]
    {The $m_{\ell\ell\gamma}$ distributions in muon channel for two additional event
      categories, \textit{EE} (left) and \textit{mll50} (right), described in the text.}
      \label{fig:best-fits-mu-extra}
  \end{center}
\end{figure}

These two categories can be included in the analysis and the number of events in each
category are shown in Table~\ref{tab:yield-mu-cat}.  Already from this table once can see
that the signal-to-background ratios in the two extra categories is much lower than in the
main category. This will manifest itself in a lower sensitivity.
Figure~\ref{fig:best-fits-mu-extra} shows the final distributions of the three-body
invariant mass in the two additional categories.

\begin{table}[t]
\caption[Events per category in the muon channel after the selection.]
{Events per category in the muon channel after the selection described in the text.}
  \label{tab:yield-mu-cat}
  \begin{center}

    \begin{tabular}{l||c|c}
      Category           & Total signal   & Data events in \\
      &  $m_\PH=125\GeV$           & $120<m_{\mu\mu\gamma}<130\GeV$\\
      \hline
      (1) \textit{EB}    & 3.25 & 151\\
      (2) \textit{EE}    & 0.80 & 91\\
      (3) \textit{mll50} & 0.56 & 67\\
    \end{tabular}

  \end{center}
\end{table}

\clearpage

\section{Statistical Methods}
\label{sec:stat}

The $m_{\ell\ell\gamma}$ distributions are used to obtain the background prediction with a
fit to the data events.  An un-binned fitting to a polynomial is performed, over the range
of $110 < m_{\mu\mu\gamma} < 170\,\GeV$ for the Dalitz search (Fig.~\ref{fig:fit}) and the
range of $110 < m_{\mu\mu\gamma} < 150\,\GeV$ for the $\PH\to(\JPsi)\gamma$ search
(Fig.~\ref{fig:fit-jp}).  The fit chosen for the background model is a \textit{Bernstein}
polynomial of degree 4, and its probability density function (pdf) is:
\begin{equation}
  b = Bern(p_1, p_2, p_3),
\end{equation}
with 3 free parameters, $p_1, p_2, p_3$.  The degree of the polynomial is the lowest that
gives an unbiased fit in the full mass range and for the most of the MC toy models (see
Section~\ref{sec:bias} below regarding the bias studies).

\begin{figure}[t]
  \centering
  \includegraphics[width=0.4\textwidth]{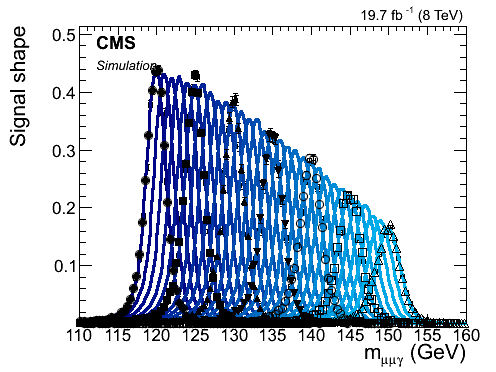}~
  \includegraphics[width=0.4\textwidth]{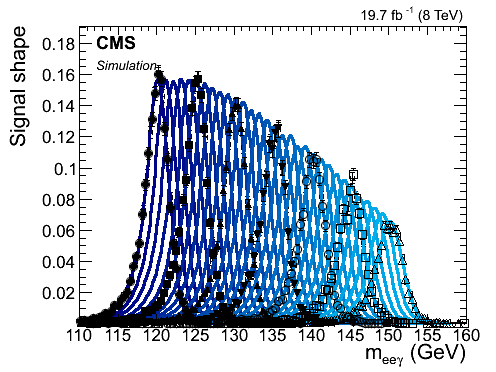}~
  \caption[Signal model fits.]
    {Signal model fits are shown for ggH production mode in muon channel (left) and
      electron channel (right).  The simulated samples are produced for masses at every
      $5\GeV$ and indicated by marker points.  The functions in between are obtained by an
      interpolation of the two near-by mass points at 1\GeV intervals.}
  \label{fig:fit-sig}
\end{figure}

Similarly, the pdf of the signal model is obtained from the un-binned fit of the
three-body mass in the signal MC sample. The fit function is \textit{Crystal
  Ball~\cite{CB-Oreglia} plus a Gaussian}:
\begin{equation}
  s = CB(m, \sigma_1, n, \alpha) + f_G\cdot\mathcal{G}(m, \sigma_2),
\end{equation}
with the same mean, $m$.  The same pdf form is used for the ggF, VBF and VH samples, as
well as the $\PH\to(\JPsi)\gamma$ sample. Figure~\ref{fig:fit-sig} shows the distributions
of the signal pdfs obtained from the ggF MC sample. Once the fit for a particular signal
mass is obtained, the parameters of the fit are frozen, and two nuisance parameters,
$\kappa_m$ and $\kappa_\sigma$, are introduced, as multiplicative factors to the mean and
width of the signal peaks:
\begin{equation}
  m' = \kappa_m \cdot m, \qquad \sigma' = \kappa_\sigma \cdot \sigma.
\end{equation}
These nuisances are used to incorporate the systematic uncertainties, described in the
Section~\ref{sec:syst}. That is, $\kappa_m = 1\pm\delta\kappa_m$ $\kappa_\sigma =
1\pm\delta\kappa_\sigma$, where $\delta\kappa_m$ and $\delta\kappa_\sigma$ are the
one-sigma uncertainties on the scale and resolution of the signal peak.

Then, following the limit setting procedure described in Ref.~\cite{CMS-AN-2011-298}, the
Likelihood function can be written as:
\begin{equation}
  \label{eq:likeli}
  \mathcal{L}(\mu, \theta) = Poisson( \mu \cdot s(\kappa_m,\kappa_\sigma) + b(p_1,p_2,p_3)) \cdot p(\tilde{\theta}|\theta),
\end{equation}
where $\mu$ is a signal strength, and $\theta$ represents all other nuisance parameters in
the model, $\theta = \{\kappa_m, \kappa_\sigma, p_1, p_2, p_3\}$.  In this expression
$p(\tilde{\theta}|\theta)$ is the pdf for the nuisance parameters $\theta$, from a
fictional auxiliary measurement $\tilde{\theta}$. Including the nuisance pdf in this way
allows us to constrain the likelihood using a pure frequentist
calculation~\cite{CMS-AN-2011-298}.  The function in eq.~(\ref{eq:likeli}) can be used to
represent the \textit{background only} hypothesis, $\mathcal{L}_b = \mathcal{L}(\mu=0)$,
and \textit{backgrounds plus signal} hypothesis, $\mathcal{L}_{s+b} = \mathcal{L}(\mu=1)$,
where $\mu$ is equal to 1 for the nominal SM Higgs boson hypothesis.

For the purpose of hypothesis testing, the Likelihood ratio
$\mathcal{L}_{s+b}/\mathcal{L}_b$ provides the most powerful test, according to
Neyman\&Pearson~\cite{NP}.  This Likelihood ratio is called \textit{test~statistic} and
can be written as:
\begin{equation}
  t(\mu) = \frac{\mathcal{L}_{s+b}}{\mathcal{L}_{b}}
  = \frac{ \mathcal{L}(\mu, \theta)}{\mathcal{L}(0, \theta)}.
\end{equation}

However, given that the expected signal from the SM Higgs boson is quite small, we are not
sensitive to determine the presence of the signal (that is $\mathcal{L}_{s+b}$ and
$\mathcal{L}_b$ hypotheses can not be strictly separated).  Hence, the statistical
analysis is instead set up to place an upper limit on the signal strength parameter,
$\mu$. For this purpose, according to Ref.~\cite{Cowan}, a different test statistic is
used:
\large
\begin{equation}
\label{eq:lambda}
\tilde{\lambda}(\mu) =
\begin{cases}
  \frac{ \mathcal{L}(\mu, \hat{\theta}_\mu)}{\mathcal{L}(\hat{\mu}, \hat{\theta})}  & \text{if~~} \hat{\mu} \geq 0\\
  \frac{ \mathcal{L}(\mu, \hat{\theta}_\mu)}{\mathcal{L}(0, \hat{\theta}_{\mu=0})}  & \text{if~~} \hat{\mu} < 0\\
\end{cases},
\end{equation}
\normalsize
where $\hat{\theta}_\mu$ denotes the value of $\theta$ that maximizes $\mathcal{L}$ for
the specified $\mu$ (thus it is a function of $\mu$), and the denominator is an
unconditional maximum likelihood function, i.e.  $\hat{\mu}$ and $\hat{\theta}$ are their
maximum likelihood estimators for $\mathcal{L}$. The second part of this definition
restricts negative signals, which is the case in our situation.

Furthermore, for the purpose of setting an upper limit on $\mu$, one should not regard the
data with $\hat{\mu} > \mu$ as representing less compatibility with $\mu$ than the data
obtained. Thus, we define:
\begin{equation}
  \tilde{q}_\mu =
  \begin{cases}
    -2\ln\tilde{\lambda}(\mu) & \text{if~~} \hat{\mu} \leq \mu\\
    0                 & \text{if~~} \hat{\mu}  >  \mu
    \end{cases}.
\end{equation}
Using this test statistic and setting the confidence level at 95\% with CL$_s$
criterion~\cite{CMS-AN-2011-298,Read,Junk}, the upper limit on $\mu$ is obtained given the
observed data.  Moreover, in order to evaluate the expected limits based on the pdfs of
the signal and background (not looking at the data) the \textit{asymptotic methods} are
used as described in Ref~\cite{Cowan}.

The results of this statistical approach are presented in Section~\ref{sec:results}, after
describing the treatment of systematic uncertainties, $\theta$, in Section~\ref{sec:bias}
and \ref{sec:syst}.

%
%
%


\subsection{Background Fit Bias Study}
\label{sec:bias}

The true form of the background $m_{\ell\ell\gamma}$ distribution is unknown.  Hence the
analysis described in the previous section can suffer from a mis-modeling of this
distribution obtained from the fit in data.  The effect of this mis-modeling can lead to
biases in the analysis sensitivity.  These biases can be quantified with a Monte Carlo toy
study, which I describe in this section.

\begin{figure}[t]\begin{center}
    \includegraphics[width=0.45\textwidth]{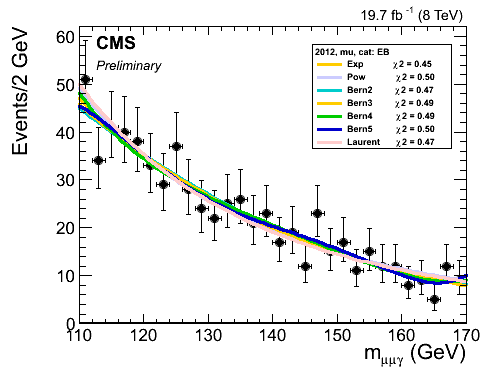}~
    \includegraphics[width=0.45\textwidth]{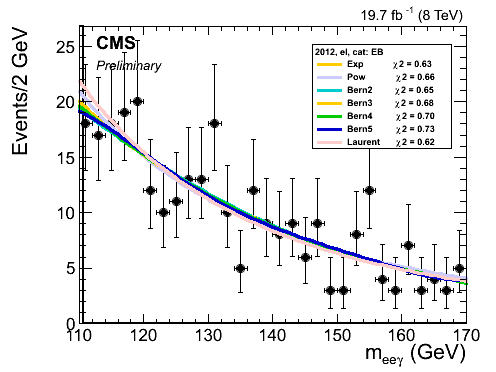}~
    \caption{Data events after the final selection and the fits of $m_{\mu\mu\gamma}$
      distribution, in the range $110 < m_{\mu\mu\gamma} < 170\,\GeV$ for three
      categories: \textit{EB, EE, mll50}.}
    \label{fig:fits-data}\end{center}
\end{figure}

As can be seen from Fig.~\ref{fig:fits-data}, many different functions would result in a
good fit to the data. Some of these functions are chosen for this test and are used as a
truth model when generating the toy events. The following list of functions is used:
\begin{itemize}
\item Exponential,  $e^{-ax}$;
\item Power law,  $ax^{-b}$;
\item Laurent polynomials of the form $ax^{-4}+bx^{-6}$;
\item Bernstein polynomials of degrees 2 to 5.
\end{itemize}

\begin{figure}[ht]
  \begin{center}
    \includegraphics[width=0.45\textwidth]{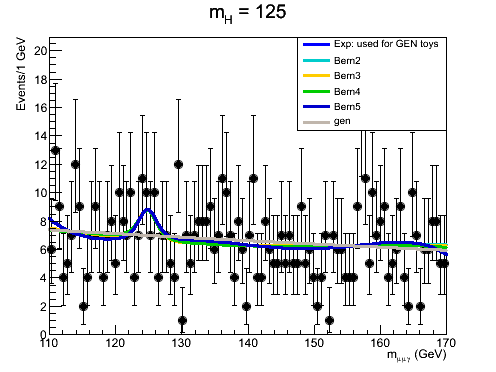}~
    \includegraphics[width=0.45\textwidth]{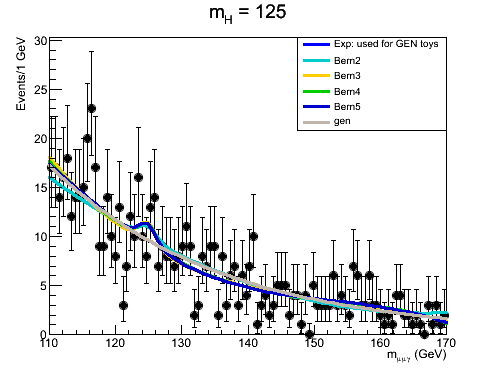}\\
    \includegraphics[width=0.45\textwidth]{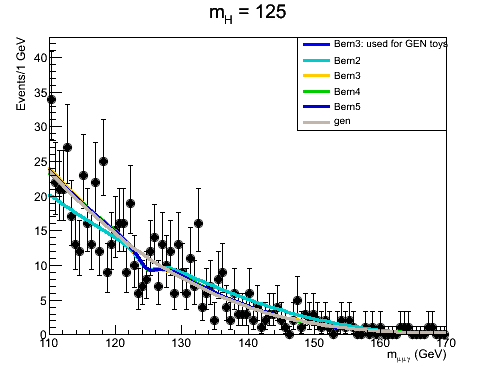}~
    \includegraphics[width=0.45\textwidth]{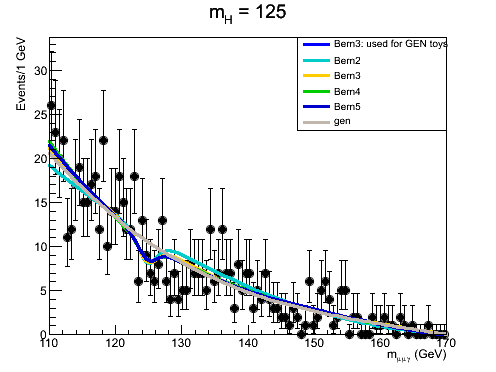}\\
    \includegraphics[width=0.45\textwidth]{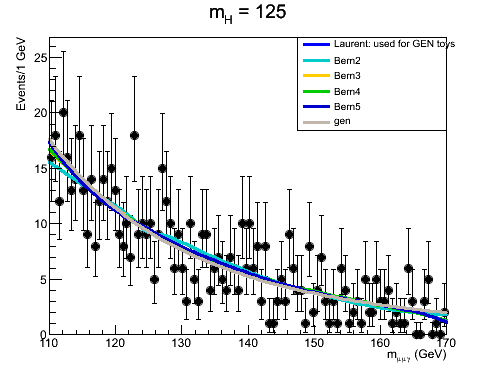}~
    \includegraphics[width=0.45\textwidth]{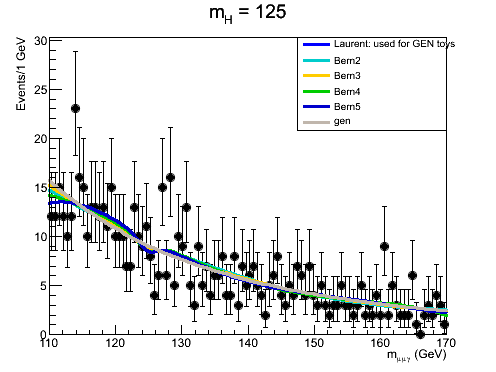}\\
    \caption[Examples of the toy data generated from Exp, Bernstein $3^d$ order and
    Laurent functions.]
    {Examples of the toy data generated from Exp (top), Bernstein $3^d$ order (middle) and
      Laurent (bottom) functions. Generator function is indicated by gray colored line
      (background only) and light-blue (for background plus signal).  The fits of
      Bernstein functions of degrees 2 to 5 with a signal at $m_\PH=125\,\GeV$ are shown
      in other colors.  Negative fluctuations of the signal are allowed.  On average we
      expect zero signal within 1 sigma of the statistical fluctuation of the background.}
    \label{fig:toys-125}
  \end{center}
\end{figure}

First, the chosen function is fit to the data events, then a toy dataset is generated from
the fit. No signal is introduced.  Then, the resulting toy dataset is fit to the Bernstein
polynomials \textit{plus} the signal pdf (which can be negative).  Fig.~\ref{fig:toys-125}
shows a few examples of the toy data and the fits to it for $m_\PH = 125\GeV$ signal.
Repeating this toy experiment many times, we expect on average zero signal events
predicted by the fit.  To quantify if that is the case, two pull distributions are
obtained, $N_{Sig}^{FIT}/\sigma_{Sig}^{FIT}$ and $N_{Sig}^{FIT}/\sigma_{Bkg}^{FIT}$, and
the following criteria are used to identify an unbiased fit:
\begin{itemize}
\item The pull distribution of $N_{Sig}^{FIT}/\sigma_{Sig}^{FIT}$ have to be Gaussian with
  mean zero and width~one.  Here, $N_{Sig}^{FIT}$ is the number of signal events predicted
  by the fit and $\sigma_{Sig}^{FIT}$ is the error on that number.  This distribution is
  constructed from 50\,000 toys. If its mean is less than~0.2, for a particular background
  function, then that fit function is considered unbiased.  This criterion ensures that a
  possible bias is at least five times smaller than the statistical fluctuation.
\item A modified pull distribution, $N_{Sig}^{FIT}/\sigma_{Bkg}^{FIT}$, should also have
  the mean less than~0.2 for an unbiased fit. Here $\sigma_{Bkg}^{FIT}$ is the error on
  the number of background events from the fit.
\end{itemize}

\begin{figure}[ht]
  \begin{center}
    \includegraphics[width=0.45\textwidth]{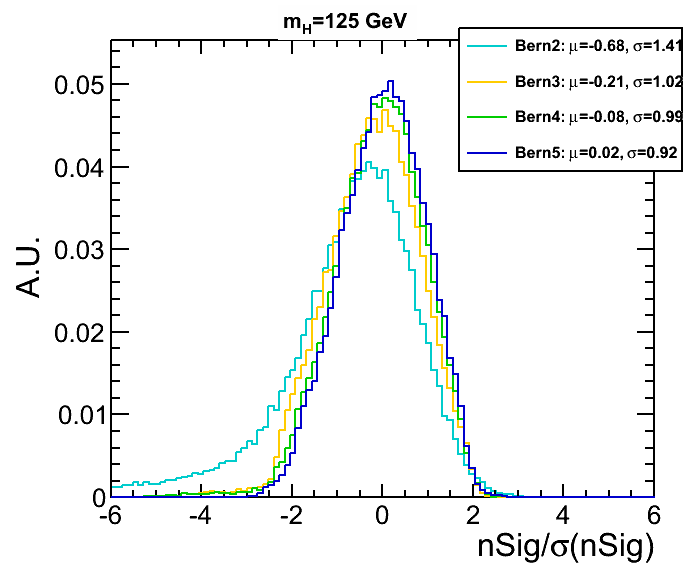}
    \includegraphics[width=0.45\textwidth]{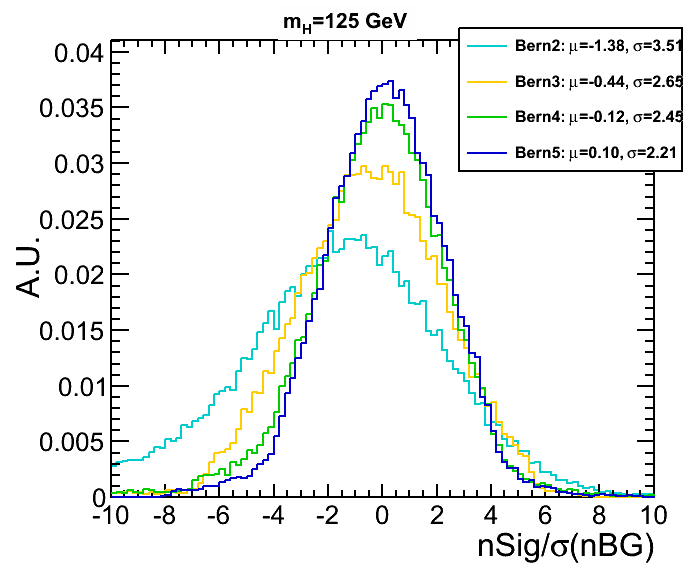}
    \caption{Examples of pull distribution obtained from the toy data with $m_\PH=125\GeV$ signal
      and the Exponential function as the true background model. Category 1 in muon channel.}
    \label{fig:pulls-125}\end{center}
\end{figure}

As an example, the pull distributions for $m_\PH=125\GeV$, obtained with the
\textit{Exponential} function as true model are shown on Figure~\ref{fig:pulls-125}.  A
complete set of the means of the distributions are presented in
Tables~\ref{tab:pulls-1-mu}, \ref{tab:pulls-2-mu} for the muon channel and
\ref{tab:pulls-1-el},~\ref{tab:pulls-2-el} for the electron channel.  One can see that the
Bernstein polynomials of degree 3 do not satisfy the criteria of the mean being less
than~0.2.  On the other hand, the degree 4 polynomial does pass this condition (except for
a few cases), thus it is chosen as the background model, both in the muon and electron
channels.

\clearpage
\begin{table}[t]
  \caption[Mean values of the $N_{Sig}^{FIT}/\sigma_{Sig}^{FIT}$ pull distributions in the muon channel.]
  {Mean values of the $N_{Sig}^{FIT}/\sigma_{Sig}^{FIT}$ pull distributions in the muon channel.
    Three true models (gen. functions) are used, each is fit to a polynomial of degrees 3, 4 and 5.}
  \label{tab:pulls-1-mu}
  \centering
  \begin{tabular}{|c|c|c|c||c|c|c||c|c|c|}
    \hline
    Gen func: & \multicolumn{3}{c|}{Exp}& \multicolumn{3}{c|}{Pow}& \multicolumn{3}{c|}{Laurent}\\
    \hline
    & \multicolumn{9}{c|}{Mean of the pull, using Bernstein polynomial of degree:}\\
    $m_\PH$   & 3&4&5   &3&4&5   &3&4&5\\
    \hline
    120 & -0.19 & -0.13 & -0.08 & -0.29 & -0.18 & -0.10 & -0.06 & 0.01 & 0.04 \\
    125 & -0.21 & -0.08 &  0.02 & -0.38 & -0.15 & 0.03 & -0.10 & -0.01 & 0.01 \\
    130 & -0.13 &  0.01 &  0.04 & -0.26 & 0.01 & 0.08 & -0.02 & 0.03 & 0.03 \\
    135 & -0.11 &  0.03 &  0.01 & -0.12 & 0.05 & 0.03 & 0.02 & -0.00 & -0.01 \\
    140 & -0.04 &  0.03 & -0.01 & 0.06 & 0.09 & 0.01 & 0.06 & -0.01 & -0.02 \\
    145 &  0.05 & -0.01 & -0.06 & 0.15 & 0.05 & -0.05 & 0.05 & -0.04 & -0.05 \\
    150 &  0.08 & -0.01 & -0.05 & 0.19 & 0.00 & -0.04 & -0.01 & -0.07 & -0.05 \\
    \hline
  \end{tabular}
\end{table}

\begin{table}[ht]
  \caption{Mean values of the $N_{Sig}^{FIT}/\sigma_{Bkg}^{FIT}$ pulls in the muon channel.}
  \label{tab:pulls-2-mu}
  \centering
  \begin{tabular}{|c|c|c|c||c|c|c||c|c|c|}
    \hline
    Gen func: & \multicolumn{3}{c|}{Exp}& \multicolumn{3}{c|}{Pow}& \multicolumn{3}{c|}{Laurent}\\
    \hline
    & \multicolumn{9}{c|}{Mean of the pull, using Bernstein polynomial of degree:}\\
    $m_\PH$   & 3&4&5   &3&4&5   &3&4&5\\
    \hline
    120 & -0.41 & -0.20 & -0.07 & -0.66 & -0.30 & -0.08 & -0.08 & 0.08 & 0.12 \\
    125 & -0.44 & -0.12 & 0.10 & -0.87 & -0.22 & 0.14 & -0.17 & 0.07 & 0.06 \\
    130 & -0.31 & 0.12 & 0.21 & -0.67 & 0.13 & 0.28 & 0.09 & 0.17 & 0.13 \\
    135 & -0.04 & 0.19 & 0.12 & -0.25 & 0.26 & 0.18 & 0.19 & 0.10 & 0.08 \\
    140 & 0.11 & 0.18 & 0.07 & 0.31 & 0.33 & 0.11 & 0.26 & 0.10 & 0.04 \\
    145 & 0.31 & 0.12 & -0.04 & 0.61 & 0.26 & -0.01 & 0.24 & 0.05 & -0.00 \\
    150 & 0.41 & 0.12 & 0.00 & 0.65 & 0.17 & 0.04 & 0.11 & -0.03 & 0.01 \\
    \hline
  \end{tabular}
\end{table}

\clearpage
\begin{table}[t]
  \caption{Mean values of the $N_{Sig}^{FIT}/\sigma_{Sig}^{FIT}$ pulls in the electron channel.}
  \label{tab:pulls-1-el}
  \centering
  \begin{tabular}{|c|c|c|c||c|c|c||c|c|c|}
    \hline
    Gen func: & \multicolumn{3}{c|}{Exp}& \multicolumn{3}{c|}{Pow}& \multicolumn{3}{c|}{Laurent}\\
    \hline
    & \multicolumn{9}{c|}{Mean of the pull, using Bernstein polynomial of degree:}\\
    $m_H$   & 3&4&5   &3&4&5   &3&4&5\\
    \hline
    120 & -0.29 & -0.20 & -0.14 & -0.39 & -0.21 & -0.09 & -0.09 & -0.02 & 0.03 \\
    125 & -0.40 & -0.25 & -0.16 & -0.35 & -0.08 & 0.02 & -0.08 & 0.01 & 0.04 \\
    130 & -0.47 & -0.28 & 0.02 & -0.27 & -0.07 & 0.05 & -0.06 & -0.01 & -0.01 \\
    135 & -0.09 & 0.02 & 0.00 & -0.15 & 0.03 & 0.03 & -0.00 & 0.00 & -0.01 \\
    140 & -0.07 & -0.02 & -0.05 & -0.07 & 0.03 & -0.03 & -0.01 & -0.04 & -0.06 \\
    145 & 0.02 & -0.02 & -0.07 & 0.09 & 0.04 & -0.03 & 0.01 & -0.05 & -0.05 \\
    150 & 0.09 & -0.04 & -0.09 & 0.16 & -0.01 & -0.06 & -0.06 & -0.12 & -0.09 \\
    \hline
  \end{tabular}
\end{table}

\begin{table}[ht]
  \caption{Mean values of the $N_{Sig}^{FIT}/\sigma_{Bkg}^{FIT}$ pulls in the electron channel.}
  \label{tab:pulls-2-el}
  \centering
  \begin{tabular}{|c|c|c|c||c|c|c||c|c|c|}
    \hline
    Gen func: & \multicolumn{3}{c|}{Exp}& \multicolumn{3}{c|}{Pow}& \multicolumn{3}{c|}{Laurent}\\
    \hline
    & \multicolumn{9}{c|}{Mean of the pull, using Bernstein polynomial of degree:}\\
    $m_H$   & 3&4&5   &3&4&5   &3&4&5\\
    \hline
    120 & -0.62 & -0.41 & -0.26 & -0.85 & -0.40 & -0.10 & -0.09 & 0.08 & 0.21 \\
    125 & -1.00 & -0.51 & -0.19 & -0.85 & -0.05 & 0.21 & -0.05 & 0.20 & 0.21 \\
    130 & -0.68 & -0.25 & 0.19 & -0.65 & 0.00 & 0.28 & 0.04 & 0.15 & 0.10 \\
    135 & -0.02 & 0.23 & 0.16 & -0.22 & 0.27 & 0.25 & 0.23 & 0.16 & 0.12 \\
    140 & 0.10 & 0.14 & 0.04 & 0.14 & 0.29 & 0.10 & 0.21 & 0.10 & 0.00 \\
    145 & 0.32 & 0.18 & -0.01 & 0.55 & 0.37 & 0.11 & 0.22 & 0.09 & 0.05 \\
    150 & 0.54 & 0.17 & -0.01 & 0.71 & 0.25 & 0.07 & 0.11 & -0.07 & 0.01 \\
    \hline
  \end{tabular}
\end{table}

\section{Systematic uncertainties}
\label{sec:syst}

Systematic uncertainties are propagated to the final results through the nuisance
parameters, $\theta$, in the likelihood function and test statistic of
eq.~(\ref{eq:lambda}).  These uncertainties are caused by the incomplete understanding of
the detector, and theoretical uncertainties on the signal production and decay mechanisms.
The background prediction is taken from a fit to the data with no systematic uncertainties
assigned to its estimation. Only the uncertainty on the fit itself, provided by
RooFit~\cite{RooFit} for each parameter of the fit function, and the statistical
uncertainty of the background prediction are considered. The procedure to ensure that the
fits are unbiased is followed as described in Section~\ref{sec:bias}.

In this section I discuss the uncertainties on the simulated signal.  We account for them
by propagating every uncertainty to the estimation of the signal yield and/or its shape,
using the MC samples.  There are three distinct classes of uncertainties that are assigned
to the signal modeling:
\begin{enumerate}
\item \textbf{Uncertainty on the predicted yield.} The main source for it is the
  theoretical uncertainty on the cross section due to PDFs and scale (up to 8\%)
  ~\cite{pdf4lhc0,pdf4lhc1,pdf4lhc2,pdf-heavy-Q}, and the branching fraction of the Higgs
  decay, 10\%~\cite{YR3,Passarino}.

  Second source of this uncertainty is due to the detector simulation of the reconstructed
  objects, which leads to a different reconstruction and ID efficiencies in the simulated
  events and data.  The uncertainty due to the dimuon reconstruction efficiency, 11\%, is
  obtained from data using $\JPsi\to\mu\mu$ events. It is dominated by the statistical
  uncertainty of the data sample (see Sec.~\ref{sec:mu-unc} for more details).  In the
  electron channel, the corresponding uncertainty 3.5\%, is obtained from simulation,
  because no data-driven methods are available for the unique object of merged electrons
  (see Sec.~\ref{sec:ele-unc}).  The 11\% uncertainty estimated for the muons is
  sufficiently small that it has no impact on the result, thus although it is probable
  that a simulation study could greatly reduce the uncertainty, no such study was
  attempted.

  The uncertainty due to the photon ID is quite small and comes from the errors on the
  scale factors presented in Table~\ref{tab:phoSF}.

\item \textbf{Uncertainty on the energy scale.}  It arises from the uncertainties on the energy
  scale of the muons, electrons and photons, and propagated to the uncertainty on the mean
  of the Higgs boson mass peak in the simulated samples.  Technically, it is implemented as
  a multiplicative nuisance parameter, $\kappa_{\mu}$ on the mean of the signal fit
  function (see Section~\ref{sec:stat}). The uncertainty of $\delta\kappa_{\mu} =
  0.1\,(0.5)\%$ is assigned in the muon (electron) channel.

\item \textbf{Uncertainty on the energy resolution.}  It comes from the same sources as the scale
  uncertainty and implemented as a multiplicative nuisance parameter $\kappa_{\sigma}$ on
  the width of the signal model function. Conservatively, a $\delta\kappa_{\sigma} = 10\%$
  is assigned in both muon and electron channels.

  Details on the photon, muon and electron scale and resolution can be found in
  Sections~\ref{sec:pho-unc}, \ref{sec:jpsi-unc} and \ref{sec:ele-unc} respectively.
\end{enumerate}

The full list of uncertainties are listed in Table~\ref{tab:syst}, while more details on
some of them are presented in the next subsections.

\begin{table}[ht]
  \begin{center}
    \caption{Sources of the systematic uncertainties.}
    \label{tab:syst}
    \begin{tabular}{l|c}
      ~~~~~ Source & Uncertainty \\
      \hline 
      Integrated luminosity (ref.~\cite{CMS-PAS-LUM-13-001}) & 2.6\%\\
      \hline  
      Theoretical uncertainties: &\\
      ~~~ PDF                                       & 2.6--7.5\%\\
      ~~~ Scale                                     & 0.2--7.9\%\\
      ~~~ $\PH\to\gamma^*\gamma\to\ell\ell\gamma$ branching fraction  &  10\% \\
      \hline 
      Signal modeling: &\\
      ~~~~~~~ Pilup reweighting         & 0.8\%\\
      ~~~~~~~ Trigger efficiency, muon (electron) channel  & 4\,(2)\%\\
      ~~~~~~~ Muon reconstruction efficiency       & 11\%\\
      ~~~~~~~ Electron reconstruction efficiency  & 3.5\%\\
      ~~~~~~~ Photon reconstruction efficiency    & 0.6\%\\
      ~~~~~~~ $m_{\ell\ell\gamma}$ scale, muon (electron) channel & 0.1 (0.5)\%\\
      ~~~~~~~ $m_{\ell\ell\gamma}$ resolution, muon (electron) channel & 10 (10)\%\\
    \end{tabular}
  \end{center}
\end{table}

\subsection{Photons}
\label{sec:pho-unc}
Photons in this analysis are well identified, isolated and have large transverse energy.
Calibration for such photons was well understood in CMS for the purpose of
$\PH\to\gamma\gamma$ search.  The uncertainty due to energy scale and resolution of the
photons is discussed in detail in $\PH\to\gamma\gamma$ legacy paper~\cite{cms-Hgg-Legacy},
and here we use those results.  The uncertainties taken from Ref.~\cite{cms-Hgg-Legacy}
and propagated to the signal MC sample result in ${<}\,1\%$ uncertainty on the width of
the Higgs peak and ${<}\,0.06\%$ on its mean (scale), which are quite small compared to
the other uncertainties of the analysis.

\clearpage
\subsection{Dimuon ID efficiency from \texorpdfstring{$\JPsi\to\mu\mu$}{JPsi to mumu} events}
\label{sec:mu-unc}
In order to derive the uncertainty due to muon reconstruction efficiency we need a way to
determine this efficiency in data. Usually, $\Z\to\mu\mu$ events are used to do this job,
but in our case there is no $\Z$-peak, since $m_{\mu\mu} < 20\,\GeV$.  Hence, we make use
of $\JPsi\to\mu\mu$ peak it data events in order to extract these efficiencies.  Moreover,
the two muons are close to each other and anti-correlated in \PT, due to the selection
requirements applied. Therefore, we do not attempt to derive the efficiency per-muon,
instead we get them per-event.

\begin{figure}[b]
  \begin{center}
    \includegraphics[width=0.45\textwidth]{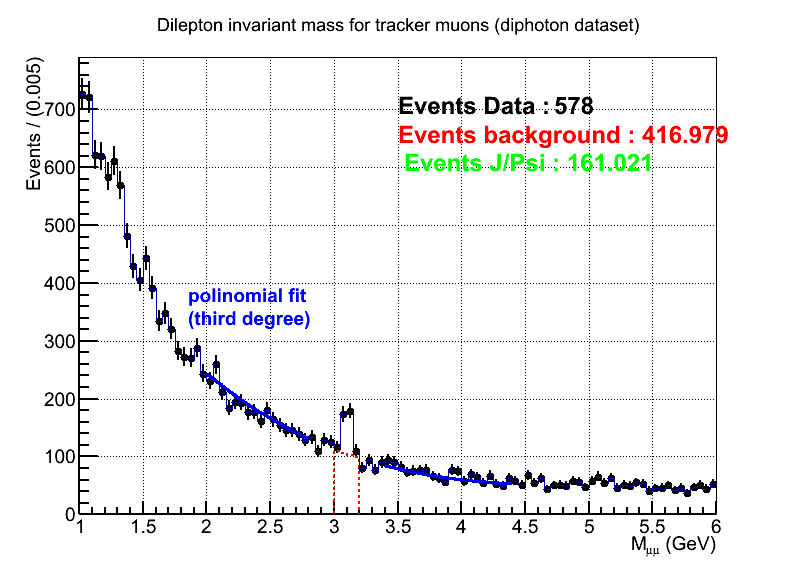}~
    \includegraphics[width=0.45\textwidth]{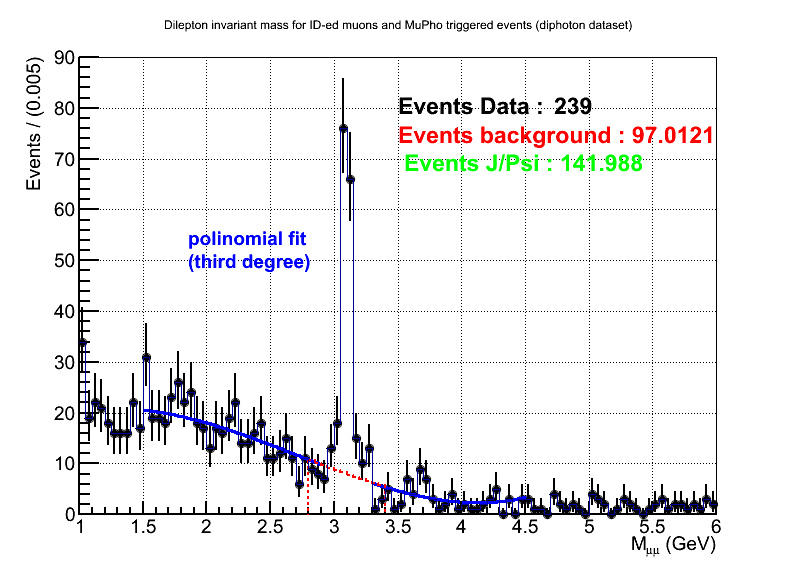}\\
    \includegraphics[width=0.45\textwidth]{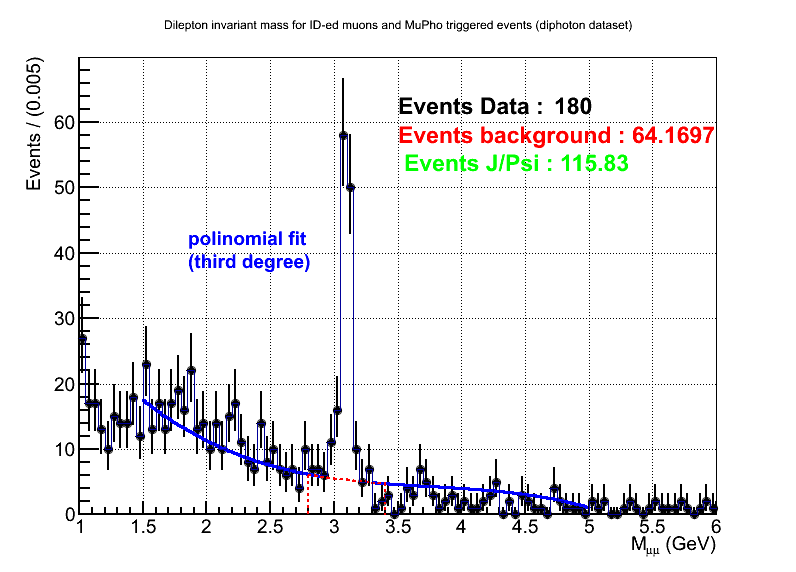}~
    \caption[Dimuon invariant mass distributions with different muon IDs.]
    {Dimuon invariant mass distributions with different muon IDs from the Double
      Photon dataset; Top-left: tracker ID muons; top-right: loose ID (\ie analysis muon);
      bottom: ID plus MuEG trigger.  A polynomial fit is applied to the side-bands in
      order to estimate the background in the region of the \JPsi resonance, thus
      extracting the \JPsi yield.}
    \label{fig:JPsi_diMuEff_data}
  \end{center}
\end{figure}

In order to obtain unbiased results we use a statistically independent dataset triggered
by the double-photon trigger for this study.  From this dataset we select events that have
a photon with $\PT > 40\,\GeV$ and two muons with $\PT^1 > 23\,\GeV$ and $\PT^2 >
4\,\GeV$, using the most trivial muon ID, the \textbf{tracker} (TR) ID.  \textit{No ID
  criteria is applied to the photon in order to factorize its efficiency.}  We also
require $\DR(\mu,\gamma)>0.4$ for each selected muon.  We then plot the dimuons mass
distributions, as shown in Fig.~\ref{fig:JPsi_diMuEff_data} (left).  After that, we apply
the muon ID criteria used in the analysis, \ie \textbf{loose} ID, described in
Section~\ref{sec:reco}; and finally we select events, which pass the \verb|Mu22_Pho22|
trigger (TRIG).  After both those selections we also plot the $m_{\mu\mu}$ distribution,
shown in Fig.~\ref{fig:JPsi_diMuEff_data}.

The $\JPsi\to\mu\mu$ peak is clearly seen in all three distributions and we extract the
number of $\JPsi$ events from them.  This is done by subtracting the background within the
\JPsi region, which in turn is estimated by a fit to a third order polynomial in the
side-bands.  The result is the three numbers, $N_{TR}$, $N_{ID}$ and $N_{TRIG}$, presented
in Table~\ref{tab:JPsi_diMuEff}.  The dimuon reconstruction efficiency is now determined
as the ratios, $\varepsilon_{ID} = \frac{N_{ID}}{N_{TR}}$ and $\varepsilon_{TRIG} =
\frac{N_{TRIG}}{N_{ID}}$, and it is also shown in Table~\ref{tab:JPsi_diMuEff}.  Notice
that this is a per-event efficiency, not a per-muon efficiency.  The uncertainty on these
efficiencies are statistical and come from the uncertainty on the number of \JPsi events
extracted from the fits.

\begin{figure}[b]
  \begin{center}
    \includegraphics[width=0.45\textwidth]{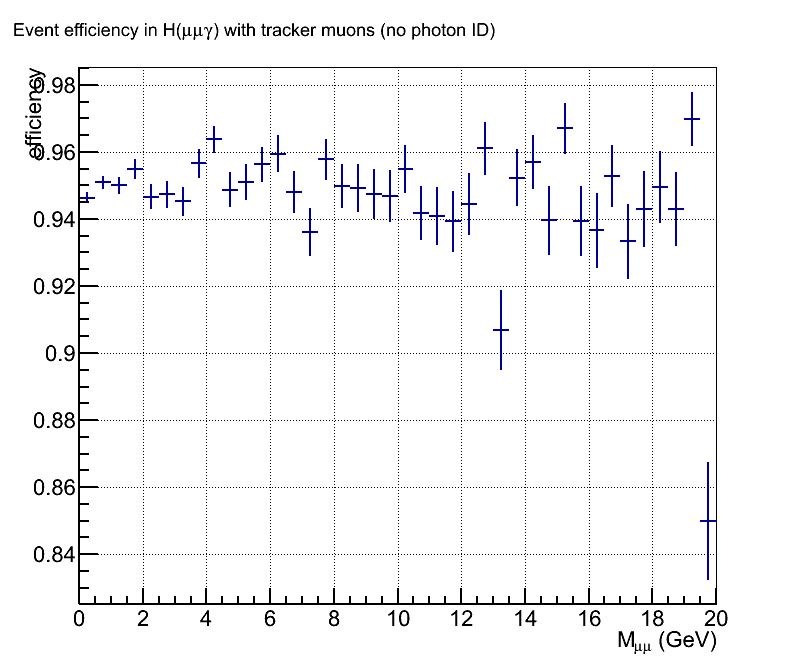}~
    \includegraphics[width=0.45\textwidth]{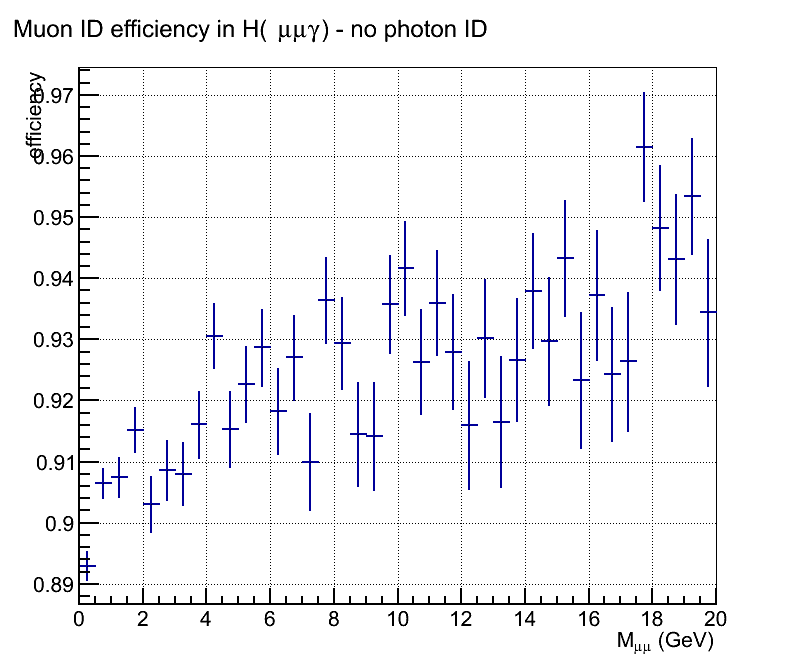}\\
    \includegraphics[width=0.45\textwidth]{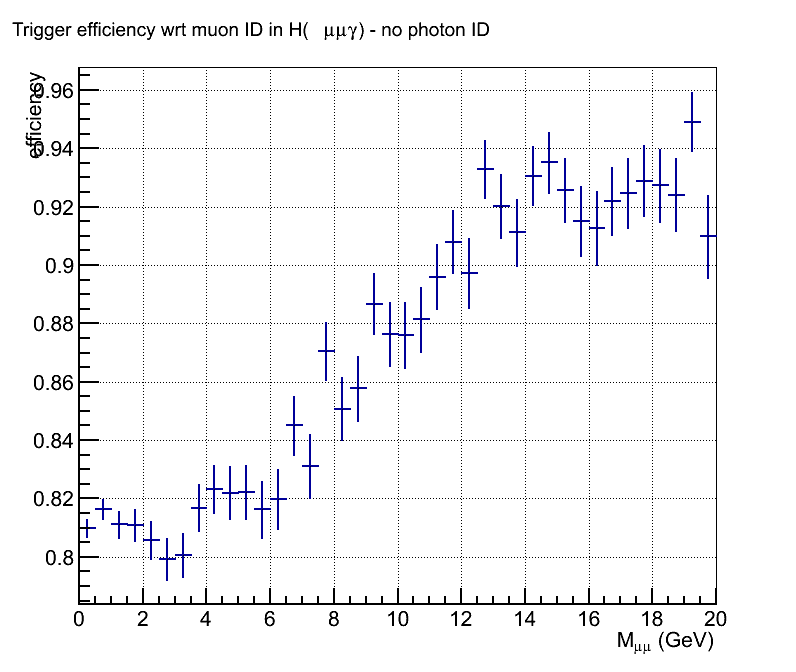}
    \caption[Event efficiency associated with a specific muon ID selection vs dimuon
    invariant mass in MC signal sample.]  {Event efficiency associated with a specific
      muon ID selection (other sources are factored out) vs dimuon invariant mass in MC
      signal sample.  Top-left: tracker muon ID; top-right: loose muon ID, as adopted in
      the analysis (the tracker muon is implicitly a part of the full ID).  Bottom: event
      trigger efficiency associated with the muon part of the trigger used in the
      analysis.}
    \label{fig:diMu_eff_MC}
  \end{center}
\end{figure}

In the MC signal sample the determination of efficiencies is straightforward, since we
have true information at generator level.  Similar to what was described above, we use two
muon IDs: \textbf{tracker} and \textbf{loose}, and the trigger selection at the end.  For
each selection we plot directly the efficiencies vs $m_{\mu\mu}$ using the MC sample of
$m_\PH=125\GeV$, as shown in Fig.~\ref{fig:diMu_eff_MC}.  Here efficiencies are defined
as: $\varepsilon_{TR} = \frac{N_{TR}}{N_{Acc}}$, $\varepsilon_{ID} =
\frac{N_{ID}}{N_{TR}}$, $\varepsilon_{TRIG} = \frac{N_{TRIG}}{N_{ID}}$, where ${N_{Acc}}$
is the number of events with generator level particles in kinematic acceptance (this can
only be obtained in MC).  We can see a dependence of those efficiencies on the dimuon
invariant mass.  In order to be consistent with the results obtained in data we should
take the numbers at around $m_{\mu\mu} = 3.1\,\GeV$, which are presented in
Table~\ref{tab:JPsi_diMuEff}.

We should note that the \textbf{loose ID} efficiency is obtained with respect to the
\textit{tracker} ID.  We don't have a way to determine the tracker ID efficiency in
data. However, we do know it in the MC sample, it is $\sim95\%$ and mostly independent of
$m_{\mu\mu}$, as shown in Fig.~\ref{fig:diMu_eff_MC} (left). Thus, being conservative, we
assign another 5\% uncertainty to the dimuon ID efficiency.

In conclusion, for the dimuon ID from data we obtain $\varepsilon_{ID}=0.88\pm0.10$, which
we declare consistent with the efficiency in MC, $\varepsilon_{ID}=0.96$, and do not apply
any MC/data scale factors. Instead we assign an 11\% systematic uncertainty on the MC
signal yield.

\begin{table}[t]
  \caption[Per-event efficiency due to Muon ID.]{Per-event efficiency due to Muon ID.
    \JPsi yield in data is extracted from the Double Photon dataset, as described in text.}
    \label{tab:JPsi_diMuEff}
    \centering
    \begin{tabular}{ l | l | c || c }
      & & \multicolumn{2}{c}{Per-event efficiency, \%}\\
      \hline
      Muon ID & \JPsi yield & Data & MC at $m_{\mu\mu} = 3.1\,\GeV$ \\
      \hline
      Tracker muon       & 161 $\pm$ 16 & - & 95\\
      \hline
      + \textit{Loose}  & 142 $\pm$ 10 & 88 $\pm$ 9 & 96\\
      \hline
      + \verb|Mu22_Pho22| trigger      &116 $\pm$ 5 & 82 $\pm$ 4 & 80\\
      \hline
    \end{tabular}
\end{table}

There is a caveat with the trigger efficiency obtained with this method: in the data it
only accounts for the \verb|Mu22| leg of the \verb|Mu22_Pho22| trigger, used in the main
analysis, because the double-photon trigger has tighter photon ID.  In order to estimate
the efficiency of the Pho22 leg we make use of yet another dataset, triggered by the
single muon trigger \verb|Single_MuIso|, and select events that pass the full analysis
selection (without the trigger requirement).  Then, the ratio of the number of events
triggered with (\verb|Single_MuIso| + \verb|Mu22_Pho22|) to the number of events triggered
only by \verb|Single_MuIso| gives the trigger efficiency for the \verb|Pho22| part of the
\verb|Mu22_Pho22| trigger (that is what we want). We can do the same in the signal MC.
Finally we get: $\varepsilon_{Data}^{Pho22} = 0.975 \pm 0.007$ and
$\varepsilon_{MC}^{Pho22} = 0.9992 \pm 0.0002$.  The 2\% difference is assigned as another
systematic uncertainty due to the \verb|Mu22_Pho22| trigger efficiency.

\clearpage
\subsection{Muon momentum scale and resolution from \texorpdfstring{$\JPsi\to\mu\mu$}{JPsi to mumu}}
\label{sec:jpsi-unc}
\begin{figure}[t]
  \centering
  \includegraphics[width=0.45\textwidth]{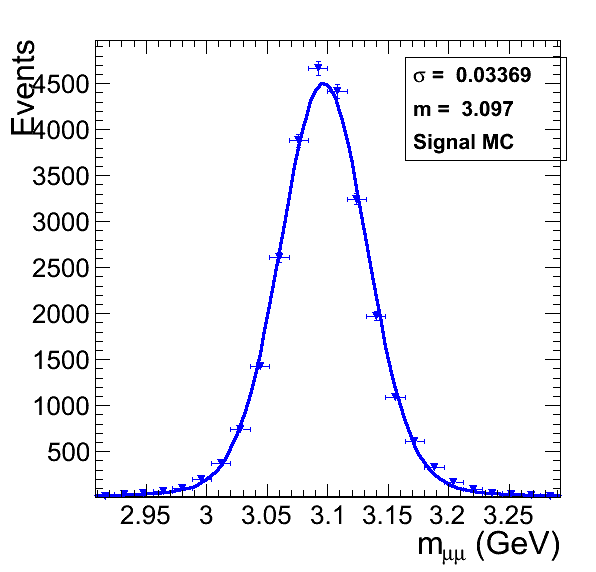}~
  \includegraphics[width=0.45\textwidth]{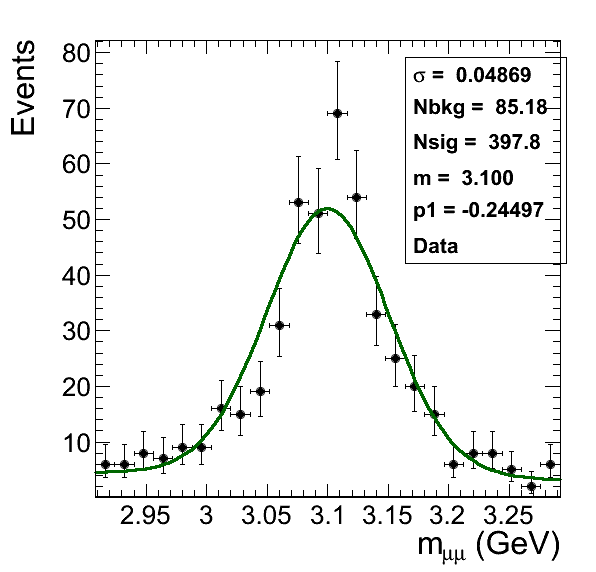}
  \caption{$\JPsi\to\mu\mu$ peak and the fits in the $\PH\to\JPsi\gamma\to\mu\mu\gamma$ MC
    sample (left) and data events (right).}
  \label{fig:mu-jpsi}
\end{figure}

It is common for CMS simulation that the resolution of Monte Carlo samples is better than
it is in real data. Hence, a smearing correction is applied to muons and photons.  In
addition to this correction, a corresponding uncertainty is assigned.  The uncertainty
provided with the standard muon momentum corrections are small -- when propagated to Higgs
mass in MC sample it yields to less than 1\% difference in the width and less than 0.05\%
in the mean.  However, those corrections were derived from $\Z\to\mu\mu$ events, in mass
window, $60 < m_{\mu\mu} < 120\,\GeV$, and for the muons with $p_T^{\mu} > 20\,\GeV$,
hence they do not fully cover our kinematics (low dimuon invariant mass and low trailing
muon $\PT$.)  Therefore, in order to make sure that the scale and resolution of the muons
are good, we once again make use of the $\JPsi\to\mu\mu$ events in data.  (And we also use
the $\PH\to(\JPsi)\gamma\to\mu\mu\gamma$ MC signal sample for this study).  Indeed,
looking at Fig.~\ref{fig:mu-jpsi}, one can see that $\JPsi\to\mu\mu$ peak is narrower in
the MC sample than in data.  Events in those plots have a photon with $\PT>40 \GeV$, two
muons with $\PT^1>23\,\GeV, \PT^2>4\,\GeV$, and $\PT^{\mu\mu}>40\,\GeV$, which is close to
the selection in the main analysis.  The muon momentum corrections have been applied.

To quantify the scale and resolution, we fit the \JPsi peak in MC with
a Breit-Wigner function convoluted with the Guassian:
\begin{equation*}
  f_{\JPsi}(m,\sigma) = BW(m, \Delta) \otimes \mathcal{G}(0,\sigma),
\end{equation*}
where parameter $\Delta$ is fixed to 0.01, while $m$ and $\sigma$ are subject to the fit.
For the data events, the fit contains a linear background contribution:
\begin{equation*}
  f_{data} =  N_{sig}\times f_{\JPsi}(m,\sigma) + N_{bkg}\times (1+p_1m).
\end{equation*}
The following resolution from the fit are obtained: $\sigma_{MC} = 34\,\MeV$,
$\sigma_{data} = 49\,\MeV$.  This difference suggests that there is indeed a residual
resolution difference not accounted for in the MC sample.  We do not attempt to derive a
new corrections for this. Instead we assign an uncertainty on the resolution of the Higgs
boson candidate mass, of 10\%, which covers any possible differences in resolution between
data and simulation (including the uncertainties due to photon energy resolution).

\subsection{Electron ID, Energy Scale and Resolution}
\label{sec:ele-unc}
Unfortunately, there is no $\JPsi\to \Pe\Pe$ peak available in the electron channel.  This
is due to the trigger and electron ID implemented in the analysis, which effectively lead
to $m_{ee}<1.5\GeV$ selection, see Fig.~\ref{fig:mll-sig}.  Hence, we don't have a method
of estimating electron ID efficiency in data.  Instead, we make use of various simulation
samples to evaluate the relevant systematic uncertainties.  We follow the approach
developed in \cite{cms-Hgg-Legacy} and produce the MC signal samples, varying parameters
of the simulation:
\begin{itemize}
 \item the tracker material budget;
 \item the underlying event modeling;
 \item the pile-up simulation.
\end{itemize}
In each of those samples we can measure the electron ID efficiency and assign the largest
difference as systematic.  The total uncertainty on the dielectron ID obtained from these
studies is 3.5\%.

As for the energy scale and resolution, we use a 10\% uncertainty on the width and 0.5\%
on the scale, same as in muon channel analysis.

\clearpage
\chapter{Results}
\label{sec:results}

Due to the absence of a signal, the data are used to derive upper limits on the Higgs
boson production cross section times the branching fraction,
$\sigma(\Pp\Pp\to\PH)\times\mathcal{B}(\PH\to\gamma^*\gamma\to\ell\ell\gamma)$, divided by
that expected for a SM Higgs boson, for $m_{\ell\ell} < 20\GeV$.  No significant excess
above background is observed in the full mass range, $120<m_\PH<150\,\GeV$, with a maximum
excess of less than two standard deviations.  In the electron channel a correction is made
to account for the events that are removed by the requirement of $m_{\Pe\Pe} < 1.5\GeV$
due to the trigger and reconstruction inefficiencies described above.  The exclusion
limits are calculated using the modified frequentist CL$_s$ method, as described in
Sec.~\ref{sec:stat}. An un-binned evaluation over the full mass range of data is used, as
shown in Figs.~\ref{fig:fit} and \ref{fig:best-fits-mu-extra}.  The uncertainty in the
limit is dominated by the size of the data sample. The systematic uncertainties have a
small impact, which effect is further quantified in Appendix~\ref{sec:lim-syst}.

\section{Muon channel}

\begin{figure}[ht]
  \begin{center}
    \includegraphics[width=0.32\textwidth]{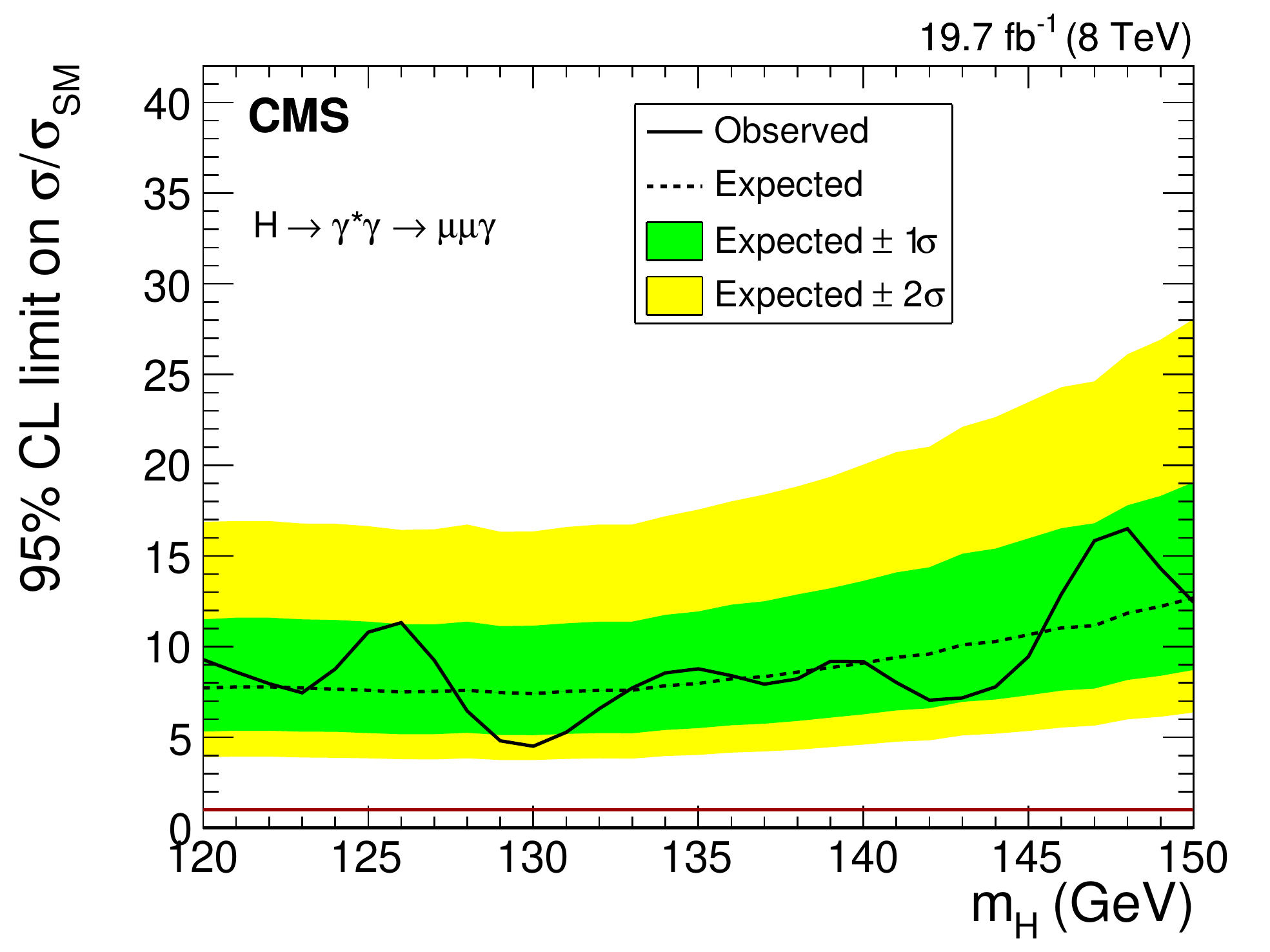}
    \includegraphics[width=0.32\textwidth]{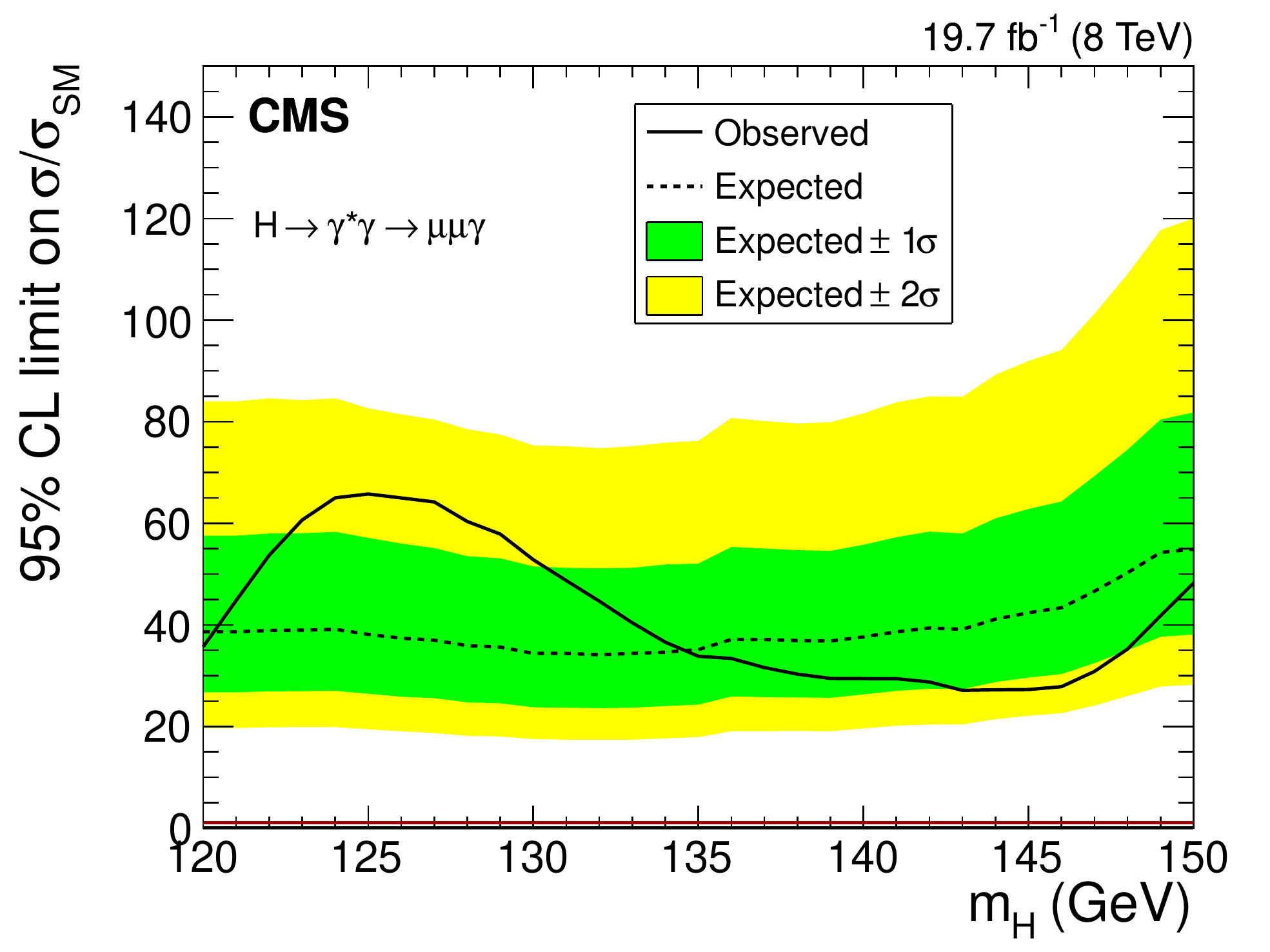}
    \includegraphics[width=0.32\textwidth]{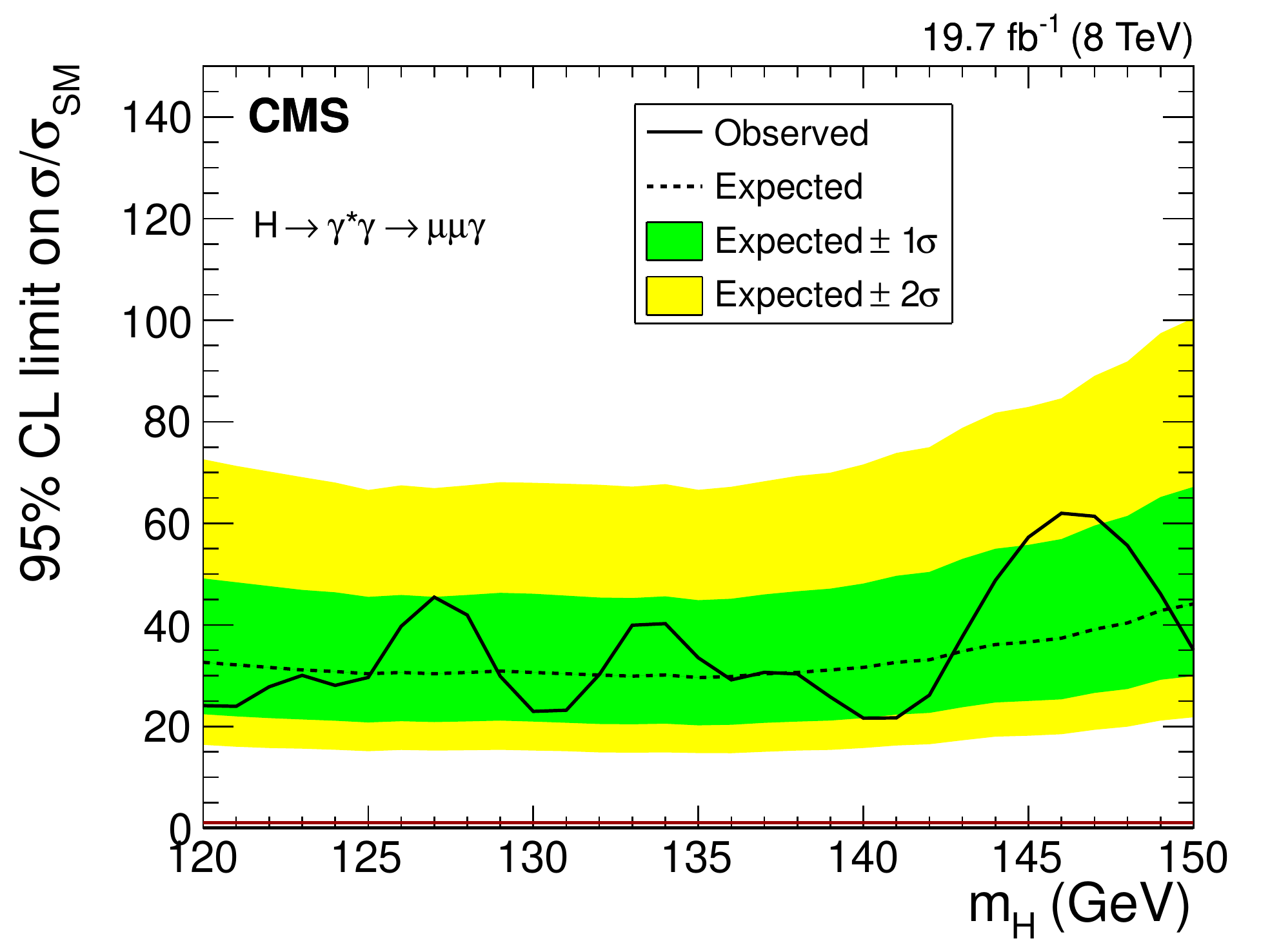}\\
    \includegraphics[width=0.65\textwidth]{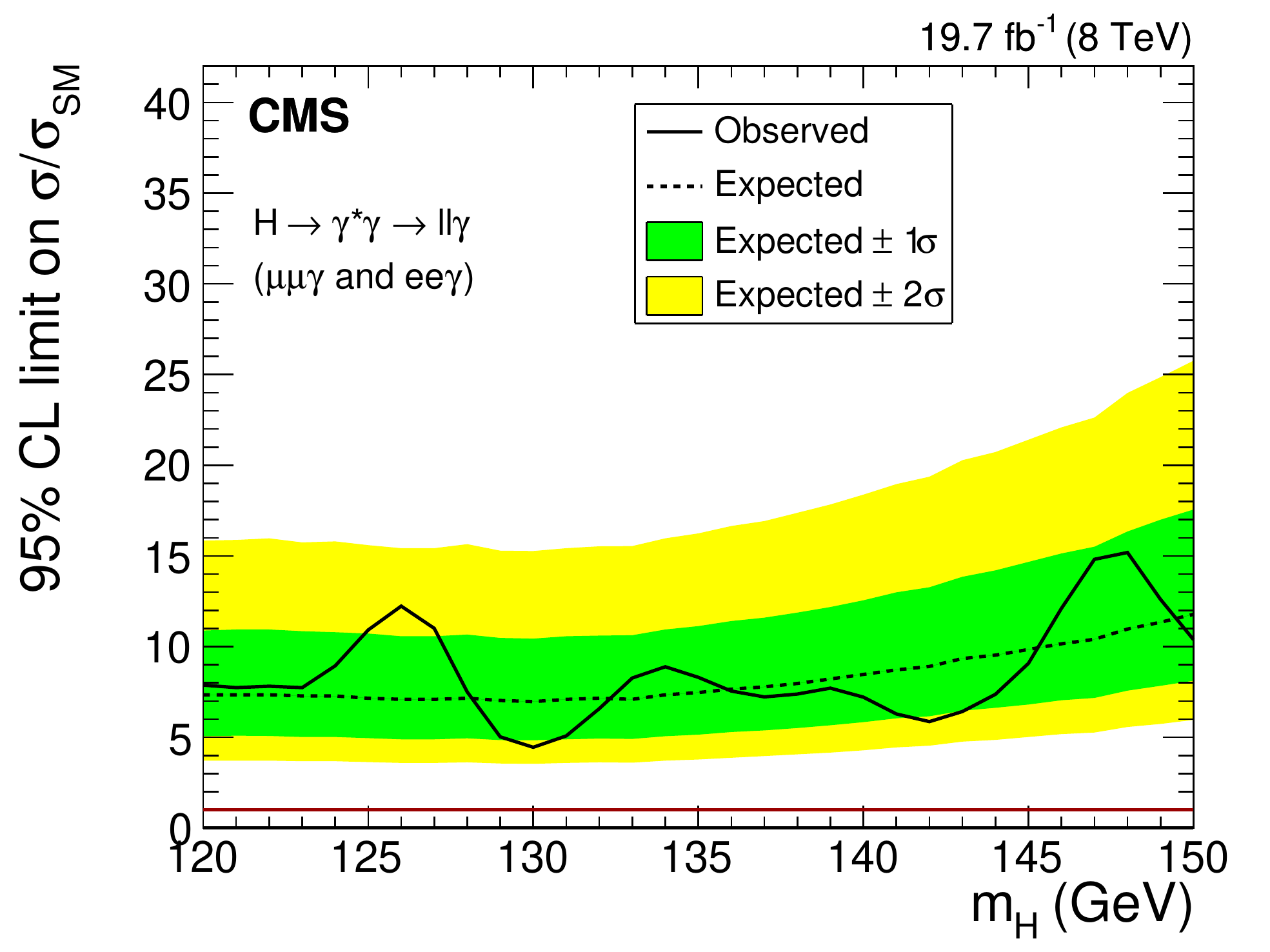}
    \caption[Exclusion upper limit on the $\mu$-value for $H\to\mu\mu\gamma$ decay of a
    Higgs boson, in the muon channel.]  {Exclusion upper limit at 95\% CL on the cross
      section times the branching fraction for $H\to\mu\mu\gamma$ decay of a Higgs boson
      divided by the SM prediction ($\mu$-value). Top plots show three categories:
      \textit{EB, EE, mll50}.  Bottom plot shows the combination of them. The result is
      dominated by the most sensitive, \textit{EB} category.}
    \label{fig:limit-mu}
  \end{center}
\end{figure}

In the muon channel, the 95\% CL exclusion upper limits are shown in
Fig.~\ref{fig:limit-mu}, separately for three event categories, and their combination.
The limits are calculated for $m_\PH$ hypotheses in the 120--150\GeV range with 1\GeV
intervals. In the main \textit{EB} category the expected exclusion limits are between 7
and 13 times the SM prediction depending on $m_\PH$. Combination with two other categories
improves the limit by about 6\%.  For instance, at $m_\PH=125\GeV$ the median expected
upper limit of the \textit{EB} category alone is ${\sim}7.6$ times the SM prediction and
it improves to ${\sim}7.2\times$SM when the three categories are combined.  The observed
limit for $m_\PH=125\GeV$ is ${\sim}10.8\times$SM prediction for \textit{EB} category
alone and it degrades a little, to ${\sim}10.9\times$SM for the combination.

\section{Combination with the Electron Channel}
\begin{figure}[ht]
  \begin{center}
    \includegraphics[width=0.45\textwidth]{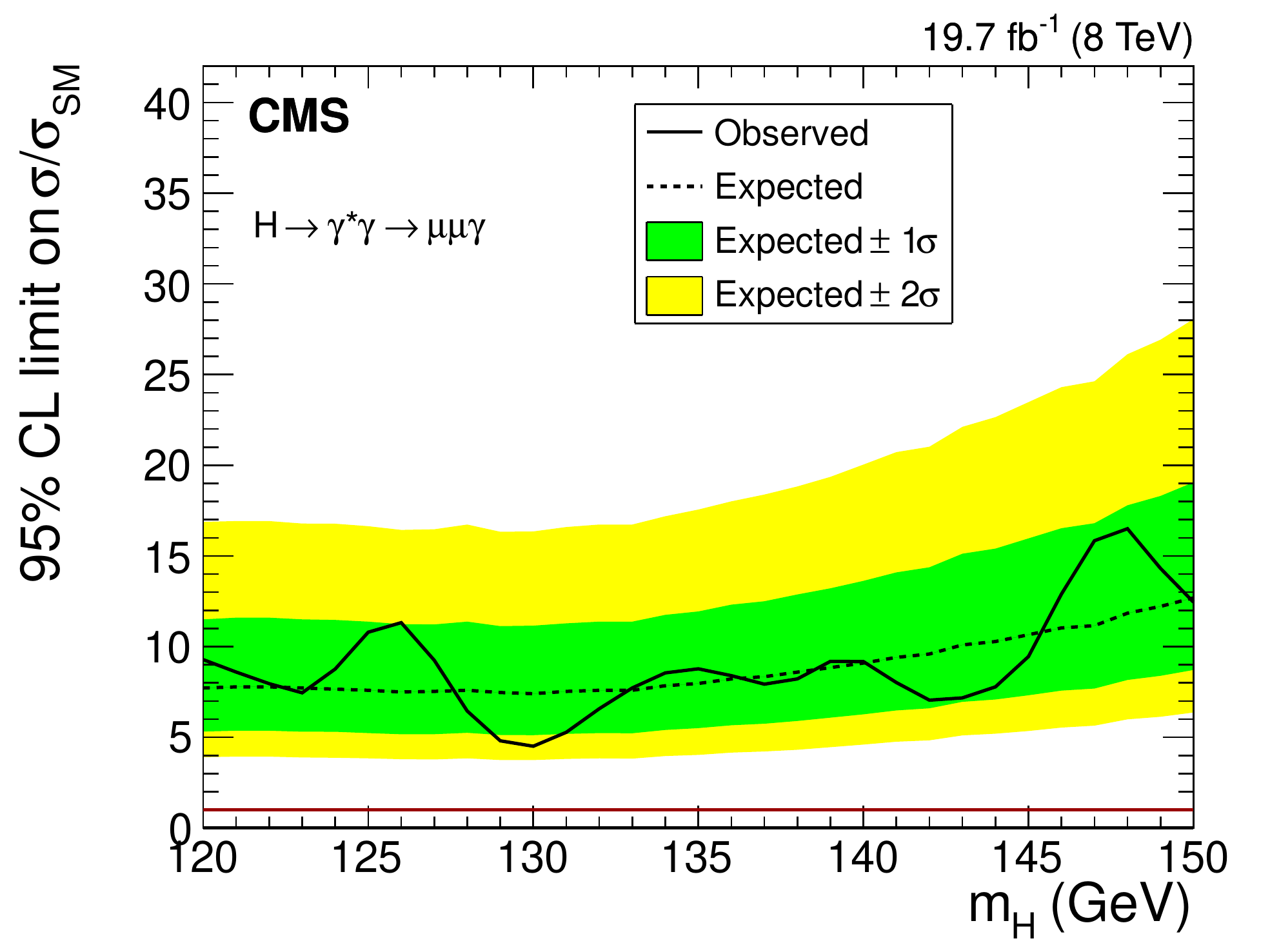}~
    \includegraphics[width=0.45\textwidth]{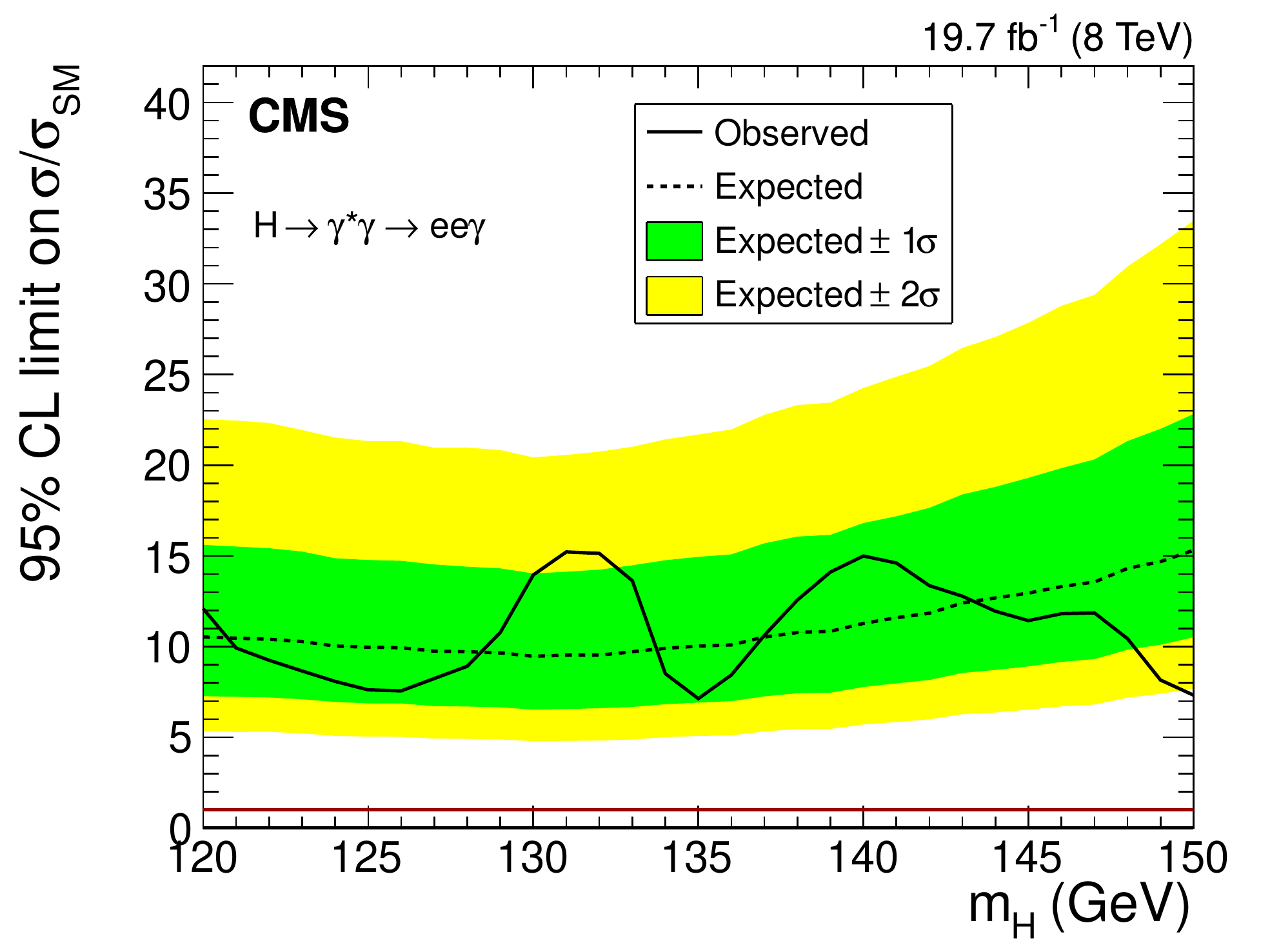}\\
    \includegraphics[width=0.85\textwidth]{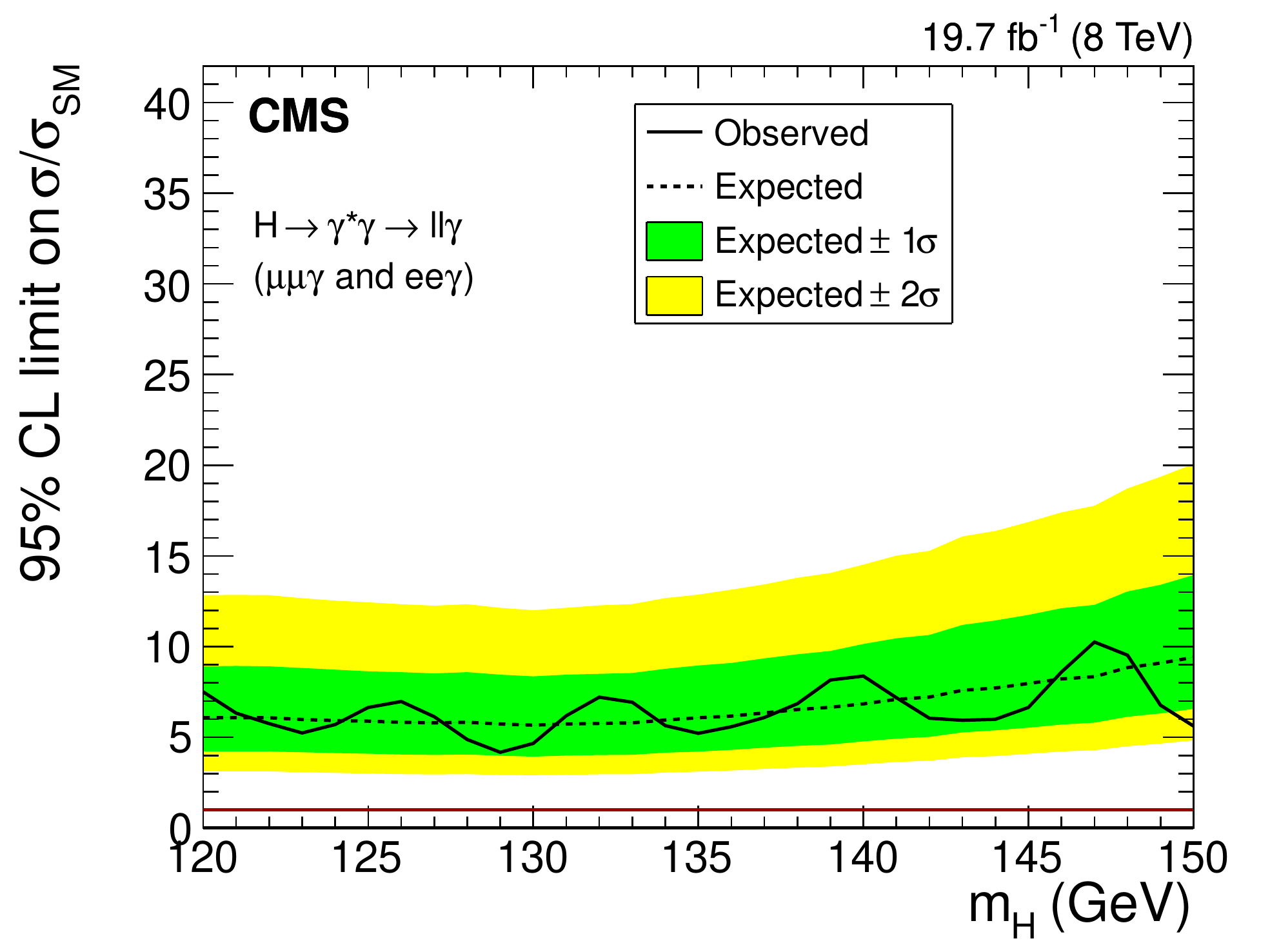}~
    \caption[Exclusion upper limit on the $\mu$-value for $H\to\ell\ell\gamma$ decay of a
    Higgs boson, in the muon and electron channels.]
    {The 95\% CL exclusion upper limit, as a function of the mass hypothesis, $m_\PH$, on
      the $\sigma/\sigma_{SM}$, the cross section times the branching fraction of a
      Higgs-like particle decaying into a photon and a lepton pair with $m_{\ell\ell} <
      20\GeV$, divided by the SM value.  (Upper) left: muon, right: electron channels;
      (lower) a statistical combination of the results in the two channels
      \label{fig:limit-comb}}
  \end{center}
\end{figure}

As expected, the sensitivity of the electron channel is weaker than of the muon channel.
On the top-right of the Fig.~\ref{fig:limit-comb} the 95\% CL exclusion limit is shown for
the electron channel alone. For comparison, the limits in the muon channel of the
\textit{EB} category with the same $y$-axis scale is shown on the left of the same figure.
For the combination of the muon and electron channels only the \textit{EB} category of the
muon channel is used.  The resulting upper limit plot is presented on the bottom of the
Fig.~\ref{fig:limit-comb}.  The observed (expected) 95\% CL upper limit for $m_\PH =
125\GeV$ is 6.7 ($5.9^{+.2.8}_{-1.8}$) times the SM prediction.

\clearpage

\section{Model independent limits}
\label{sec:res-model-ind}
\begin{figure}[b]
  \begin{center}
    \includegraphics[width=0.45\textwidth]{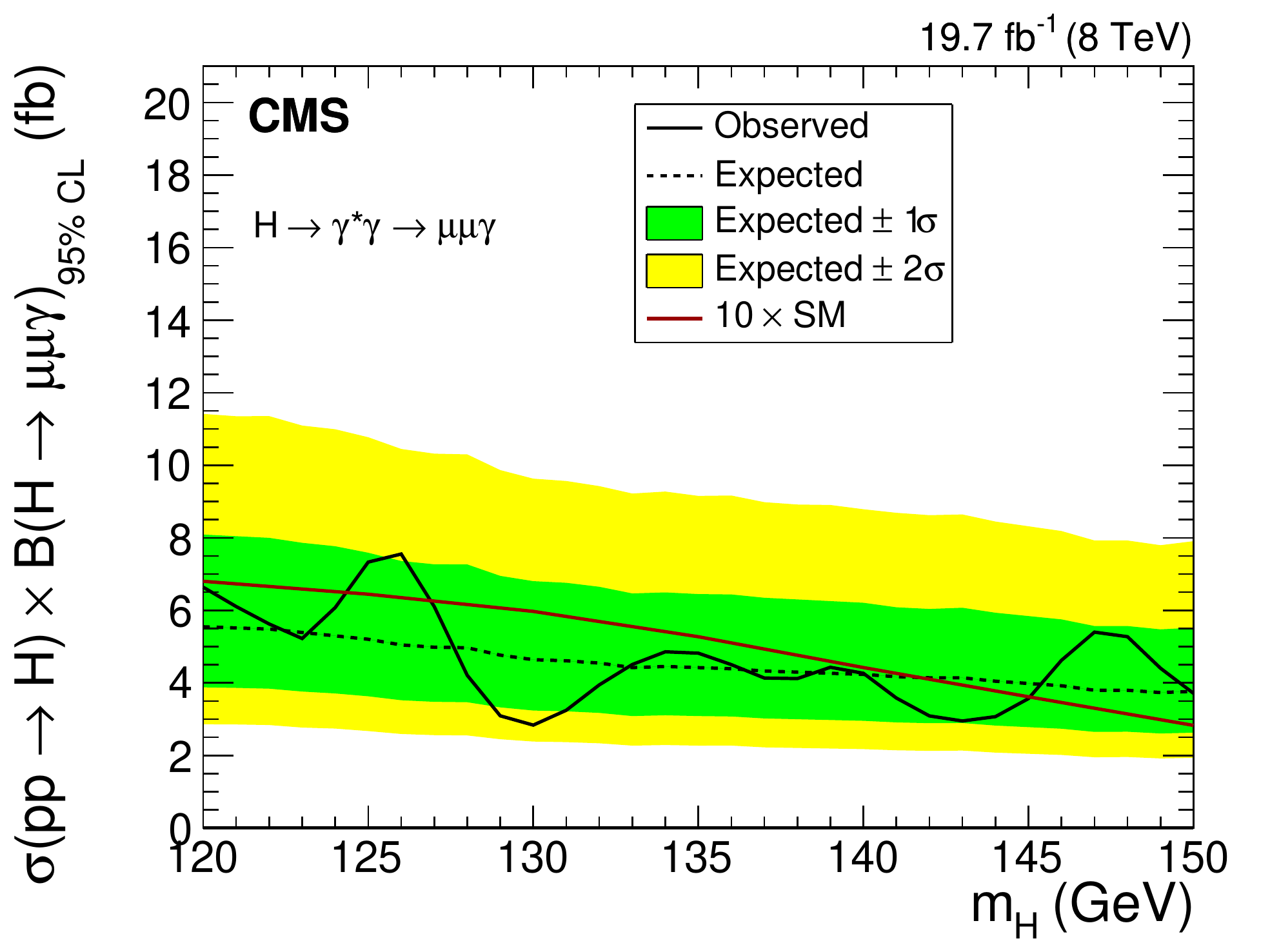}
    \includegraphics[width=0.45\textwidth]{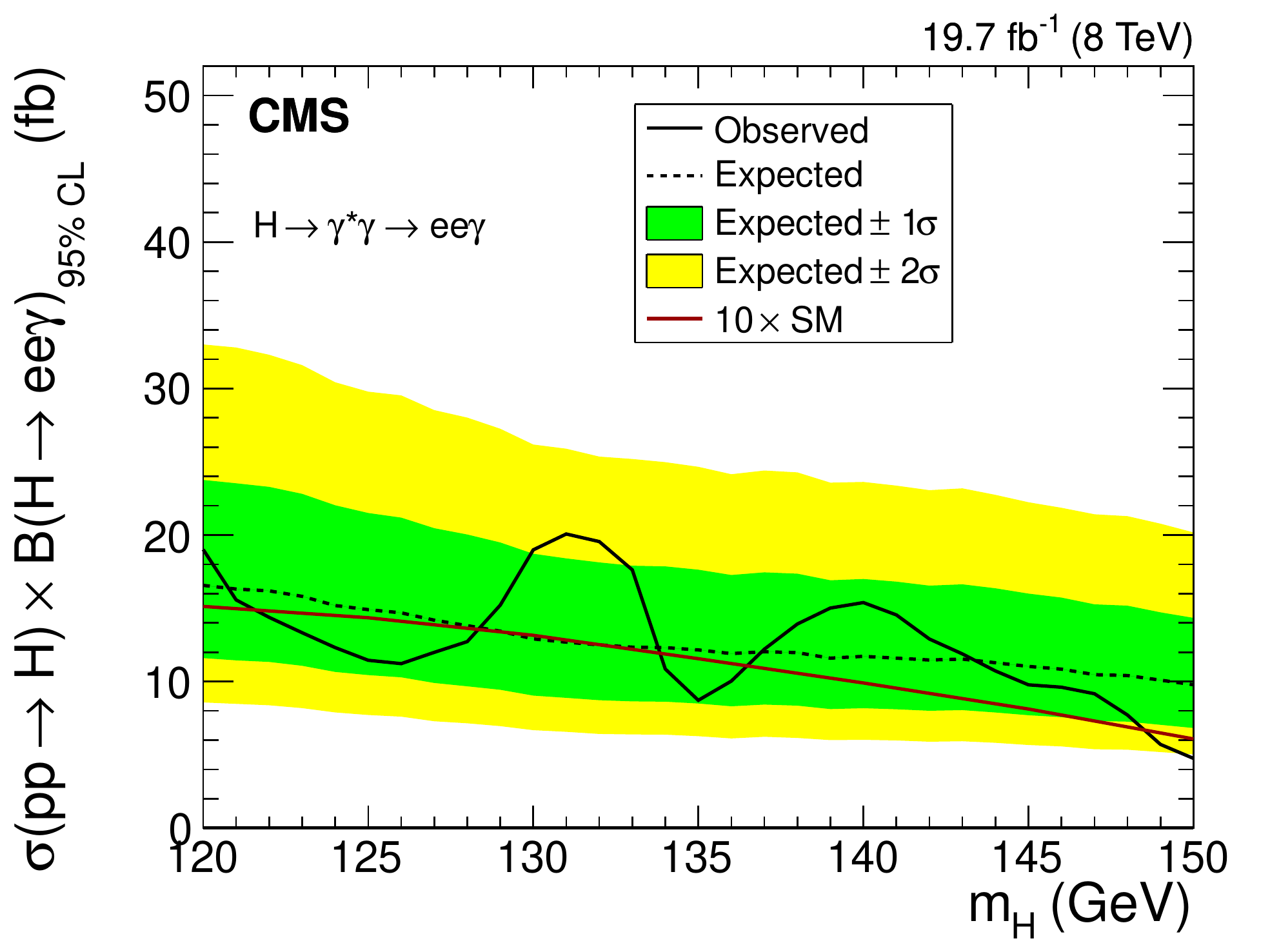}
    \caption{The 95\% CL exclusion limit on $\sigma(\Pp\Pp\to \PH)\times
      \mathcal{B}(\PH\to\ell\ell\gamma)$, with $m_{\ell\ell} < 20\GeV$, for a Higgs-like
      particle, as a function of the mass hypothesis, $m_\PH$.
      \label{fig:limit-xsBR}}
  \end{center}
\end{figure}

In addition to the limits on the SM process one can re-interpret the results in a more
general way to obtain a limit on the inclusive cross section times the branching fraction
of the $\PH\to\ell\ell\gamma$ decay, where $\PH$ now denotes a Higgs-like scalar particle
of any BSM theory.  No theoretical uncertainties of the Higgs boson production cross
sections are needed for this limit.  The result of the 95\% CL upper limits is now
expressed in femtobarns, and shown in Fig.~\ref{fig:limit-xsBR}.  One should use these
results with care though. In the muon channel, the total signal efficiency is about 24\%
and almost independent of the dimuon invariant mass. In the electron channel, efficiency
depends on the dielectron mass, since it is strongly shaped by the selection. For this
reason, the result in the electron channel is really \textit{not} model independent.  In
the muon channel, however, it can be interpreted as such. The observed (expected) 95\% CL
upper limit of 7.3 ($5.2^{+2.4}_{-1.6}$)\unit{fb} is obtained at $m_\PH = 125\GeV$ for
$\PH\to\mu\mu\gamma$ decay.

\section{Differential limits in bins of  \texorpdfstring{$m_{\mu\mu}$}{m(mumu)}}
\begin{figure}[ht]
  \begin{center}
    \includegraphics[width=0.32\textwidth]{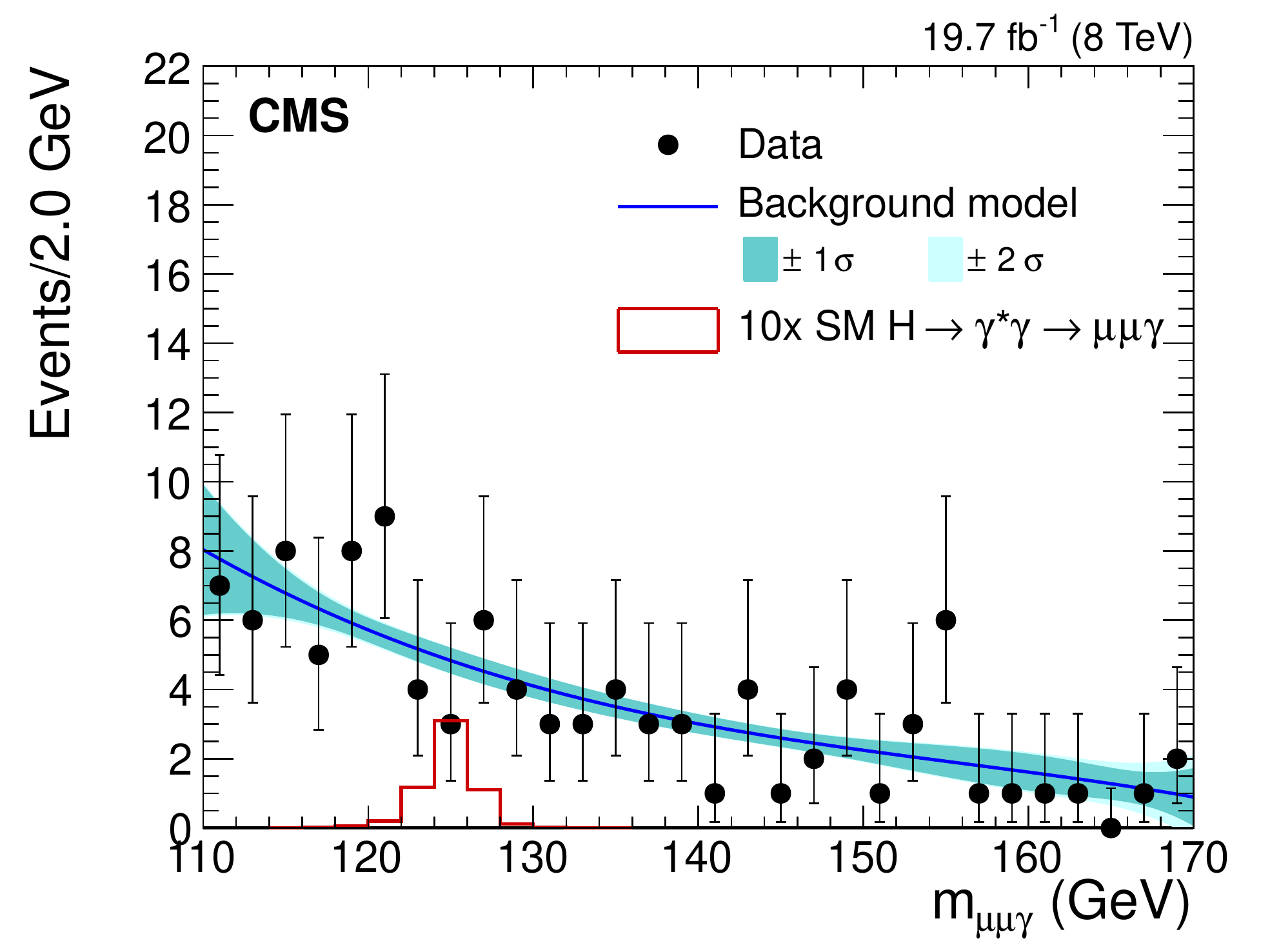}~
    \includegraphics[width=0.32\textwidth]{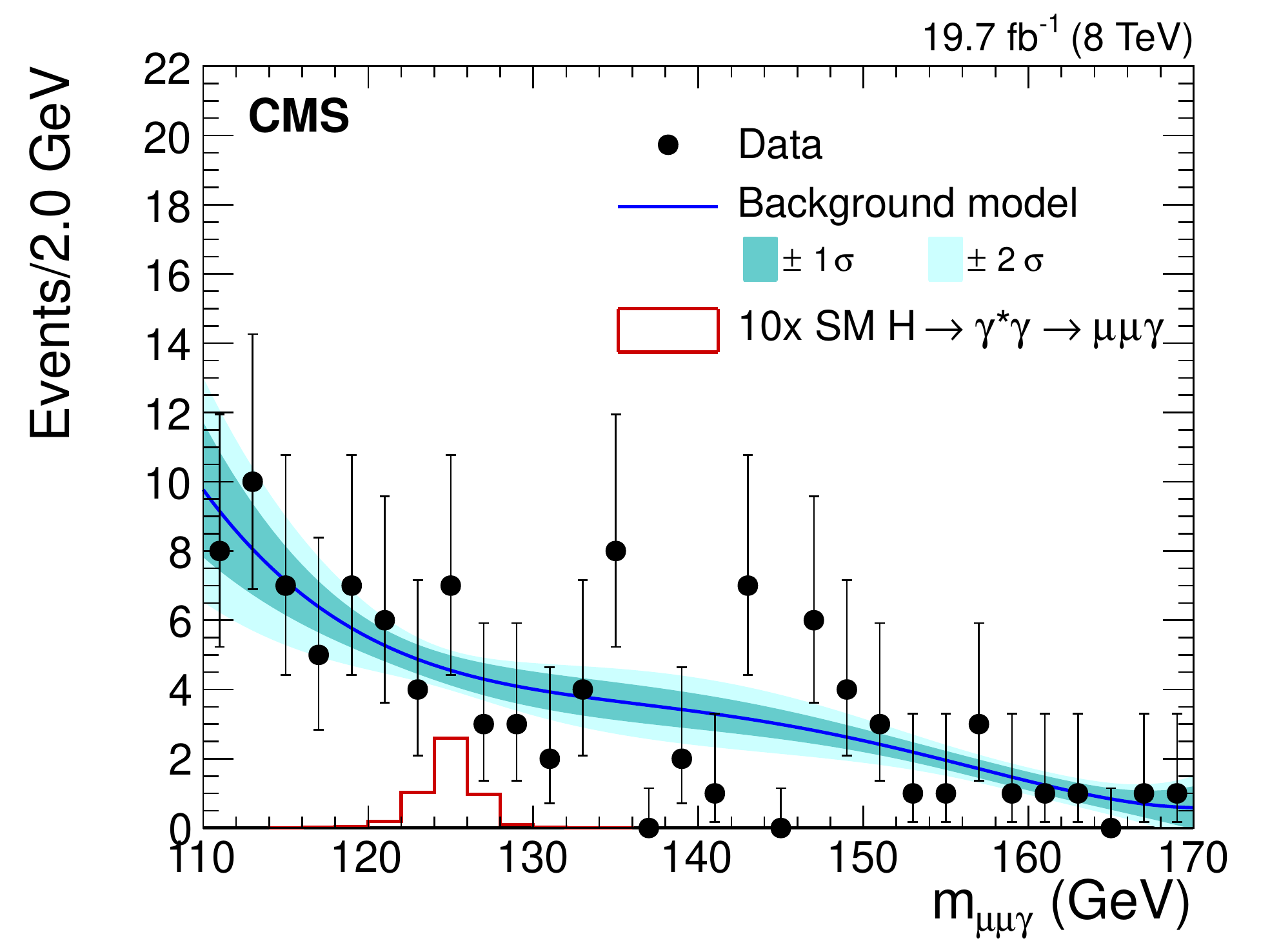}~
    \includegraphics[width=0.32\textwidth]{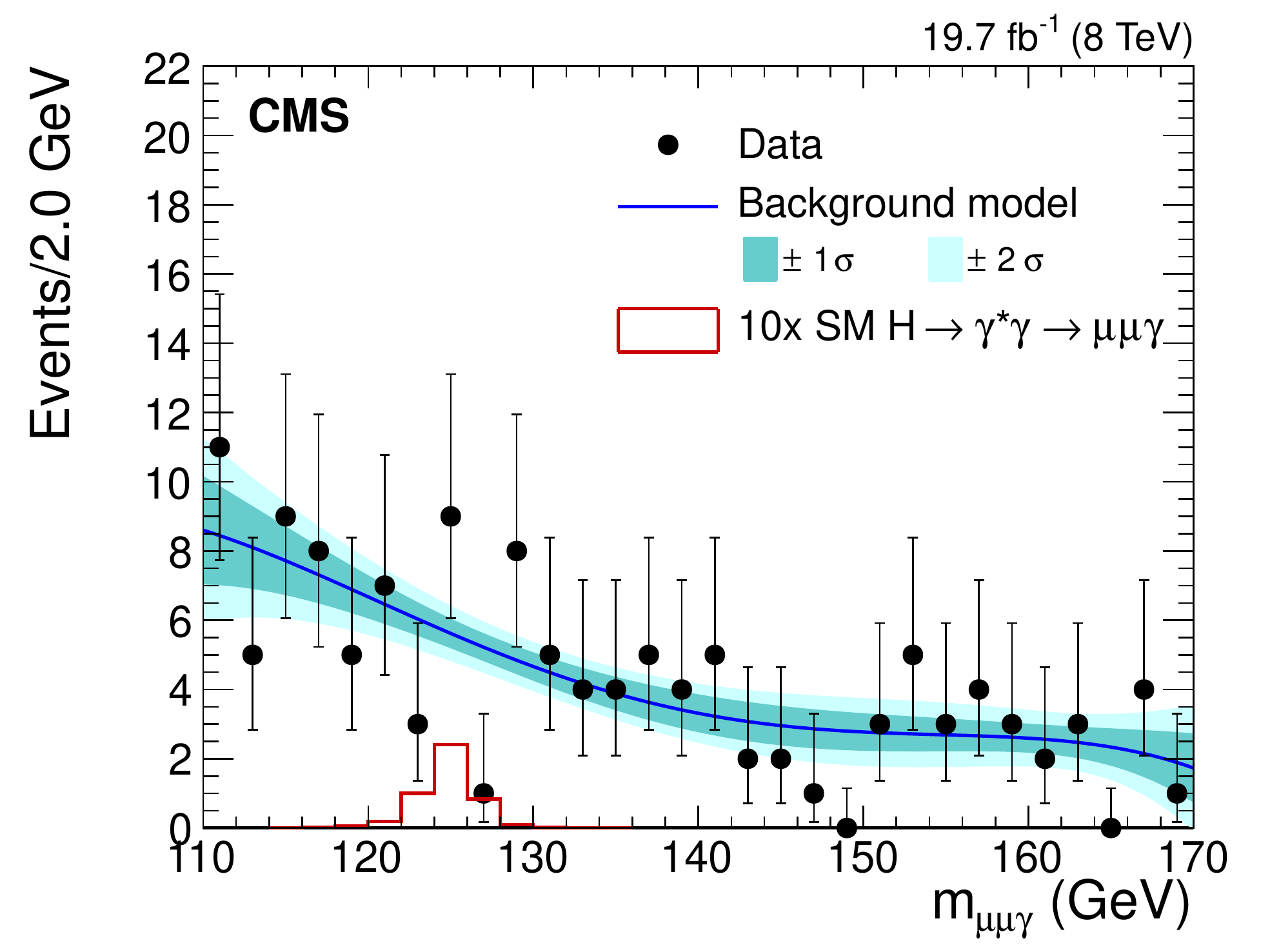}\\
    \includegraphics[width=0.32\textwidth]{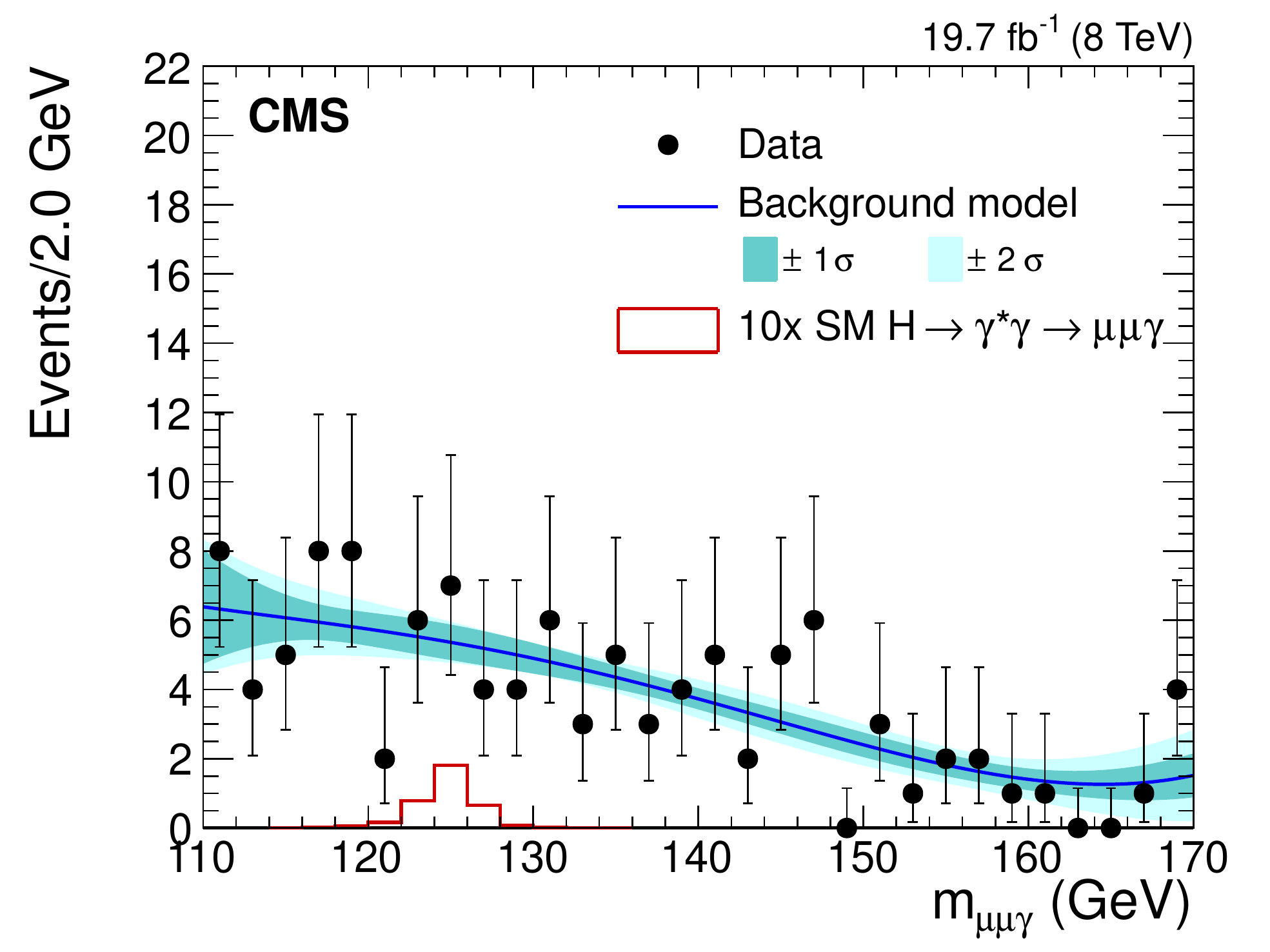}~
    \includegraphics[width=0.32\textwidth]{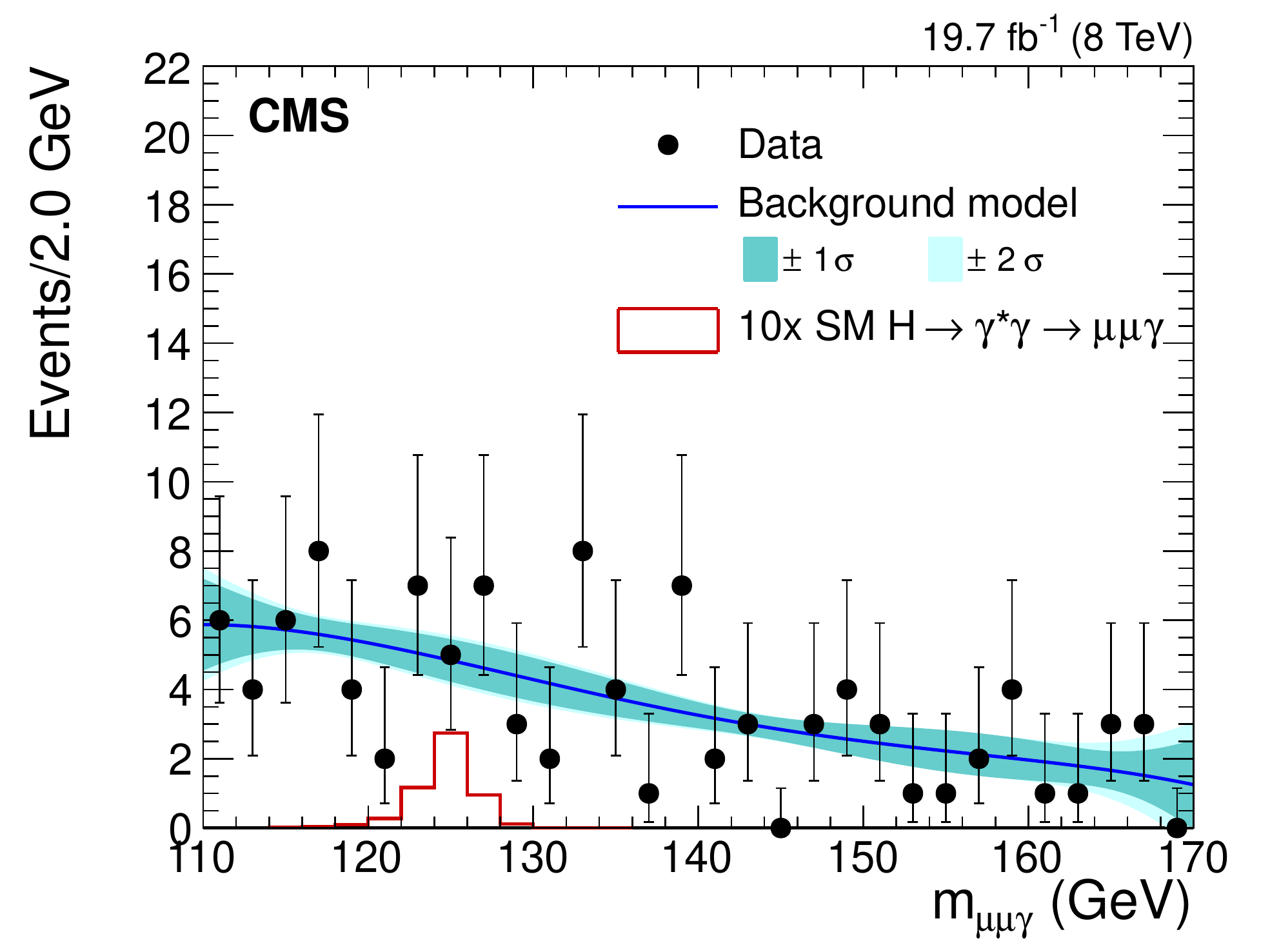}~
    \includegraphics[width=0.32\textwidth]{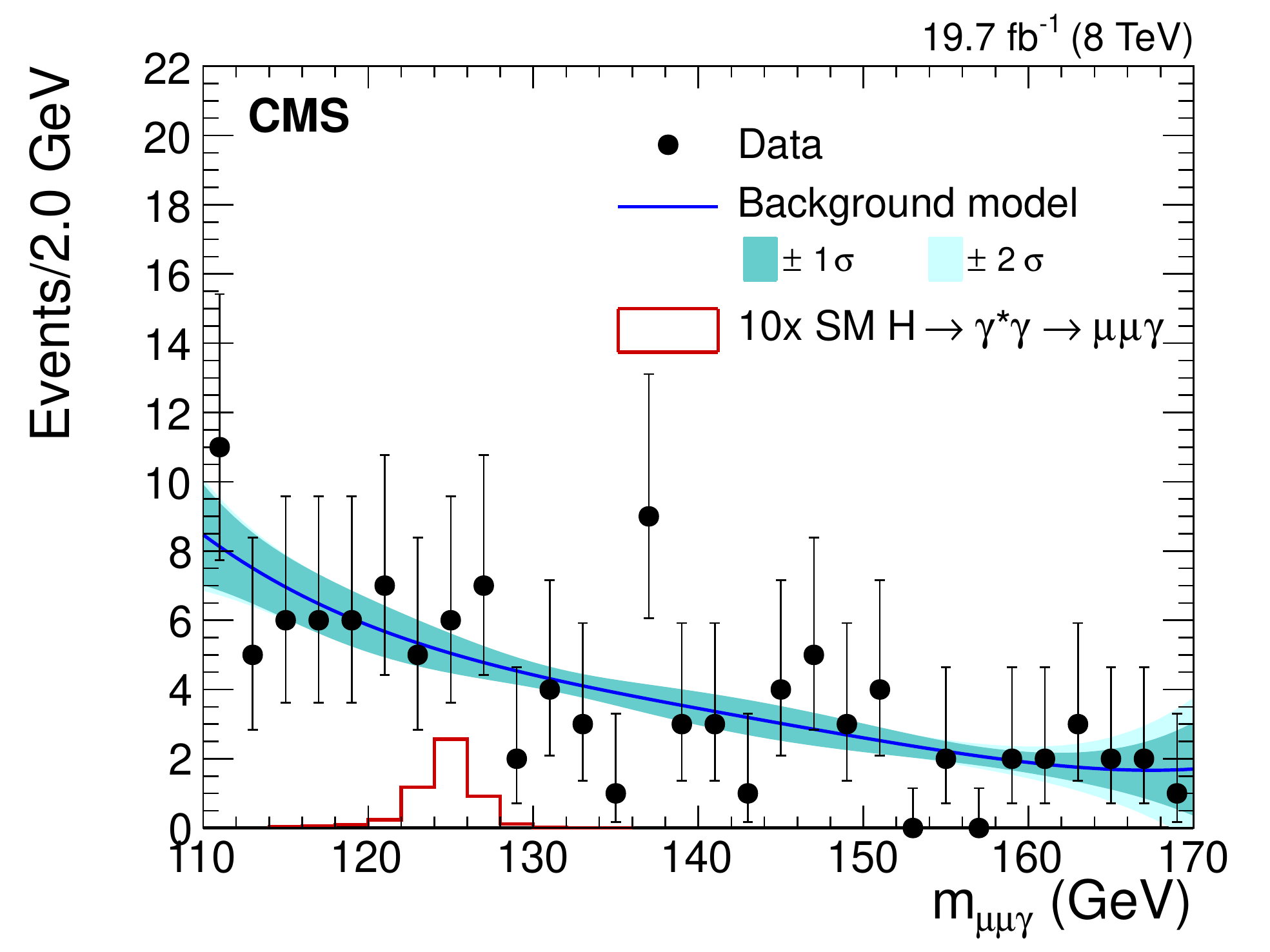}\\
    \includegraphics[width=0.32\textwidth]{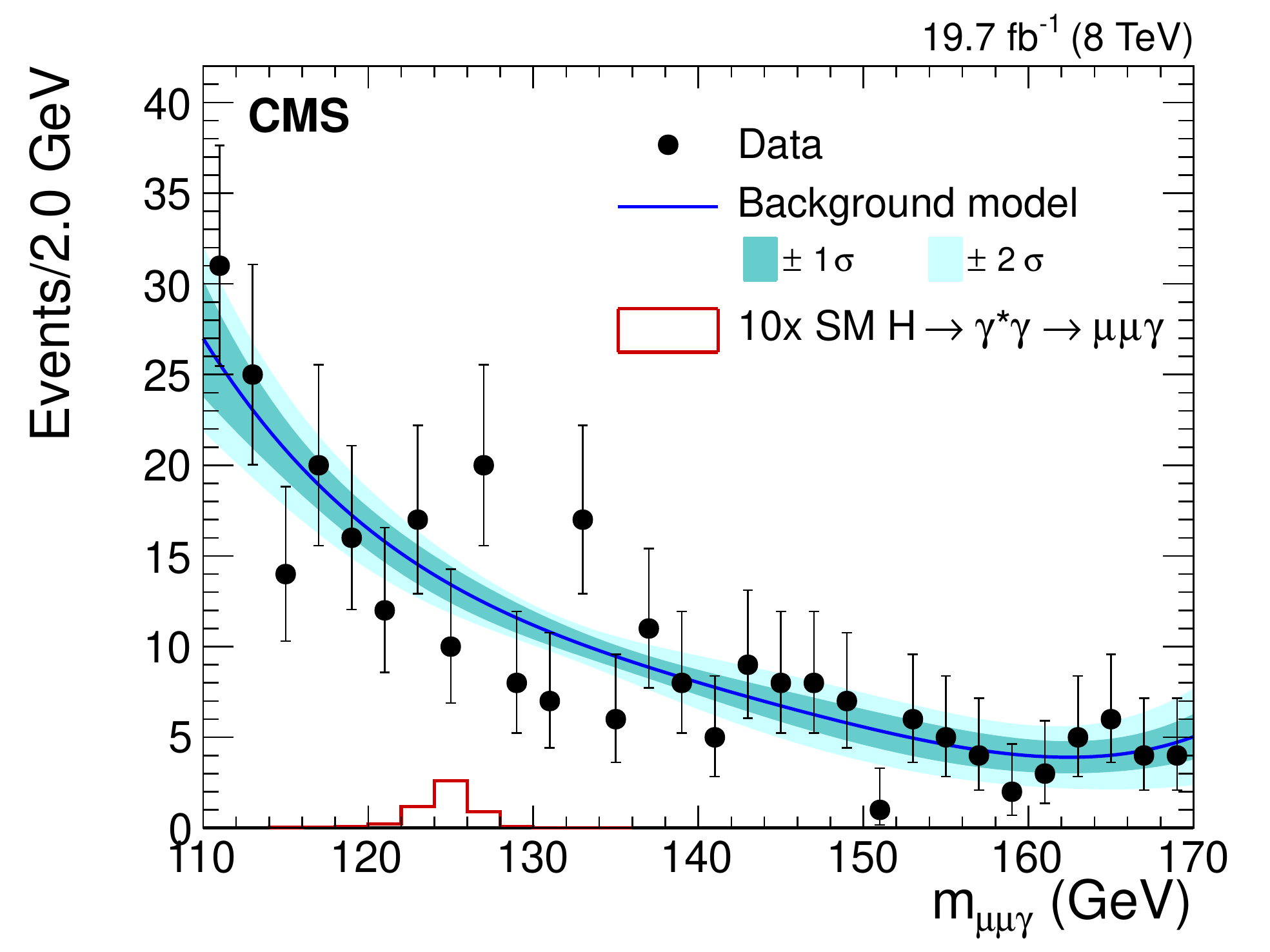}
    \caption[Fit to the data events of $m_{\mu\mu\gamma}$, where the plots correspond to
    one of the 7 bins in $m_{\mu\mu}$.]
    {Fit to the data events of $m_{\mu\mu\gamma}$, where the plots correspond to one of
      the 7 bins in $m_{\mu\mu}$ (described in the text and ordered from left to right).}
    \label{fig:best-fits-mll}
  \end{center}
\end{figure}

\begin{figure}[ht]
  \begin{center}
    \includegraphics[width=0.65\textwidth]{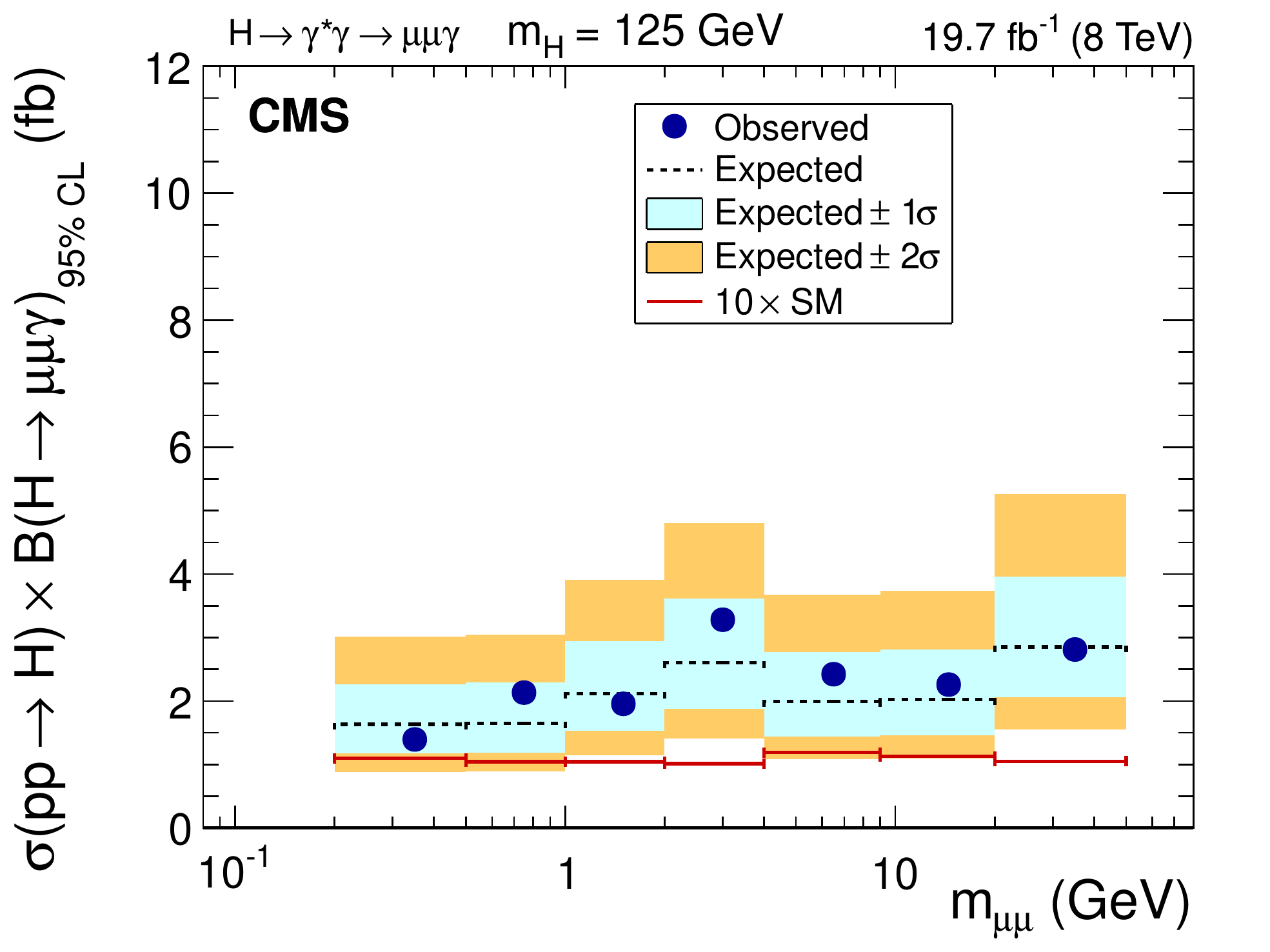}\\
    \includegraphics[width=0.65\textwidth]{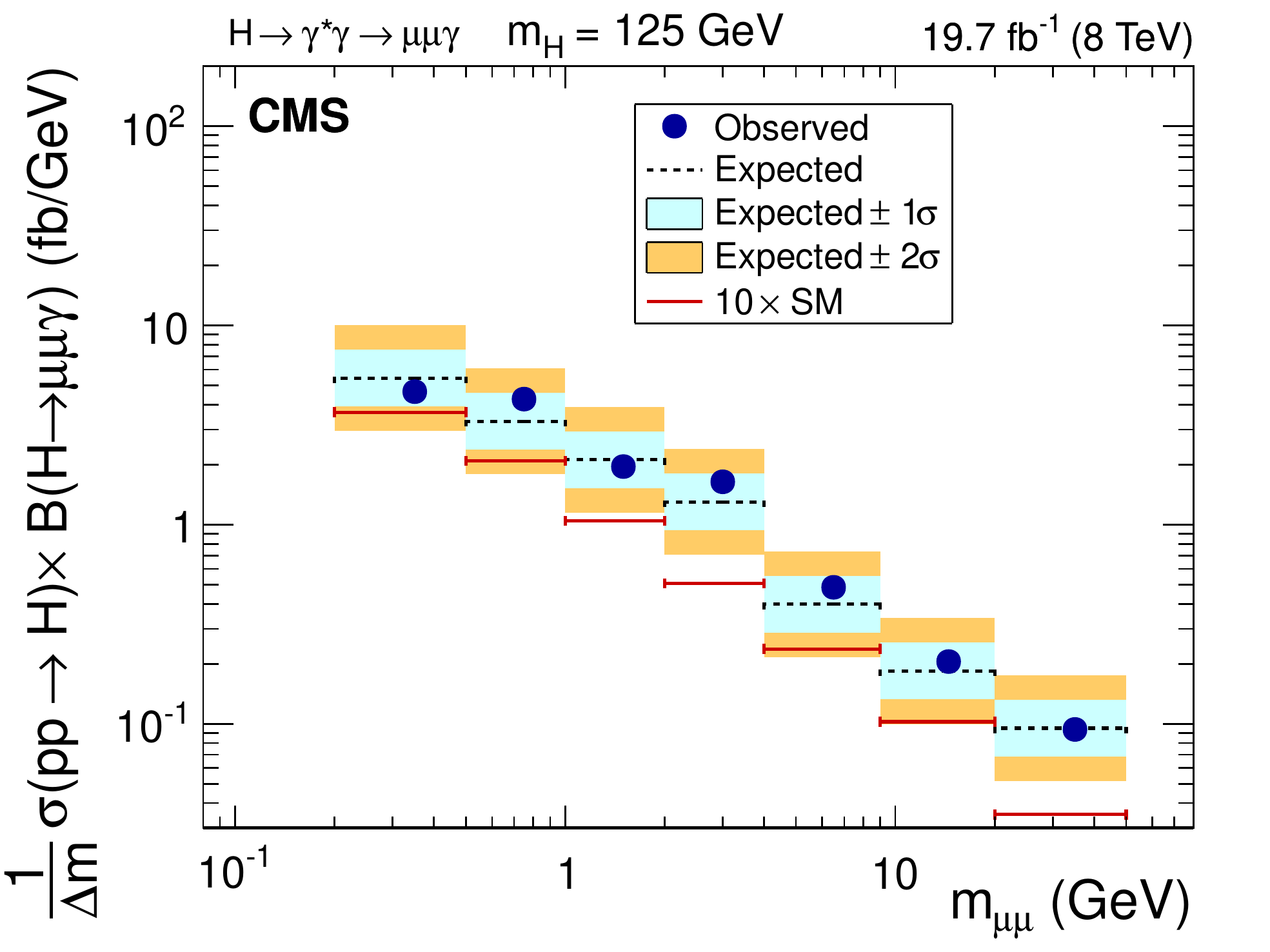}
    \caption[Differential limit on the cross section times the branching fraction of a
    Higgs-like particle H, with $m_\PH = 125\GeV$,] {Differential limit on the cross
      section times the branching fraction of a Higgs-like particle H, with $m_\PH =
      125\GeV$, decaying into a photon and a muon pair within bins of $m_{\mu\mu}$.  Two
      plots show the same result but the second one is scaled by the bin size, $\Delta m$,
      and shown in the logarithmic scale.}
    \label{fig:limit-mu-mll}
  \end{center}
\end{figure}

Since we know of the existence of the Higgs-like particle with 125\GeV mass, it is
interesting to look at $m_\PH = 125\GeV$ specifically and ask for the differential cross
section measurements.  For example, in $\PH\to\gamma\gamma$ decay the differential cross
sections are measured for many kinematic variables, including ${\PT}^{\gamma\gamma}$,
$\abs{y_{\gamma\gamma}}$~\cite{atl-Hgg-diff}.  In our channel, among other variables, it
is interesting to perform the differential measurement of the dilepton invariant
mass. Currently we are not yet sensitive to the signal, thus instead of measuring the
cross section I present a differential limit for it in bins of $m_{\ell\ell}$.  This is
done only in the muon channel because of the shaped $m_{\Pe\Pe}$ distribution due to
selection in the electron channel. In order to produce this result I introduce 7 bins in
$m_{\mu\mu}$ variable, with the edges: 0.2 -- 0.5 -- 1 -- 2 -- 4 -- 9 -- 20 --
50\,\GeV. These bins are chosen so that they approximately contain the same number of
signal events (corresponding to ${\sim}\,0.1\unit{fb}$ of $\sigma\cdot\mathcal{B}$).  In
each of those bins, the fit of the data events to a Bernstein polynomial of degree 4 is
performed (see Fig.~\ref{fig:best-fits-mll}).  And the upper limits in each bin are
determined in the same manner as it is done for the limits presented in the previous
section. The result is shown in Figure~\ref{fig:limit-mu-mll}.

It is important to note that there is no migration of events between $m_{\mu\mu}$ bins,
therefore no unfolding is necessary.  In Appendix~\ref{sec:app-mu-res} I show the
resolution of the dimuon mass for every mass bin.  The resolution is good, varying from
1.1\% in high $m_{\mu\mu}$ bins to 2.4\% in the lowest bin.

\clearpage
\section{Result for \texorpdfstring{$\PH\to(\JPsi)\gamma$}{Higgs to JPsi gamma}}
\label{sec:res-Hjp}

As it is described in Section~\ref{sec:ana}, a region with $2.9 < m_{\ell\ell} <
3.3\,\GeV$ in the muon channel is selected for a special case -- a search for the
$\PH\to(\JPsi)\gamma\to\mu\mu\gamma$ process.  After the complete event selection, just
like in the muon Dalitz search analysis, the $m_{\mu\mu\gamma}$ distribution in data is
fitted to a Bernstein polynomial of degree 2, in the range
$110<m_{\mu\mu\gamma}<150\,\GeV$, see Figure~\ref{fig:fit-jp}. The signal model function
is Crystal-Ball plus a Gaussian, and only $m_\PH=125\GeV$ is considered.

The 95\% CL upper limit is placed on the cross section times the branching fraction,
\begin{equation}
  \sigma(\Pp\Pp\to \PH)\times\mathcal{B}(\PH\to\mu\mu\gamma) < 1.80\unit{fb},
\end{equation}
while the expected limit is $1.90~\pm~0.97$\unit{fb}.  One can interpret this result as an
upper limit on $\sigma(\Pp\Pp\to \PH)\times \mathcal{B}(\PH\to
(\JPsi)\gamma\to\mu\mu\gamma)$ and obtain for the branching fraction,
\begin{equation}
  \mathcal{B}(\PH\to(\JPsi)\gamma) < 1.5\times10^{-3}
\end{equation}
at 95\% CL, which is about 540 times the prediction in Ref.~\cite{hToJPsiGamma-2014}. At
90\% CL, $\mathcal{B}(\PH\to(\JPsi)\gamma) < 1.2\times10^{-3}$. The number of events
present in the 2.9--3.3\GeV mass window coming from the
$\PH~\to~\gamma^*\gamma~\to~\mu\mu\gamma$ process is large compared to the $\PH\to
(\JPsi)\gamma\to\mu\mu\gamma$ (see Table~\ref{tab:yield}). On the other hand it is small
compared to the total background, hence it is considered as a part of the background when
extracting the limit on $\mathcal{B}(\PH\to(\JPsi)\gamma)$.  The interference between the
$\PH\to\gamma^*\gamma\to\mu\mu\gamma$ and $\PH\to(\JPsi)\gamma~\to~\mu\mu\gamma$ signal
processes is negligible due to small $\Gamma_{\JPsi}/m_{\JPsi}$ ratio.


\section{Conclusion and outlook}
In this dissertation I presented the search analyses of the rare decays of the Higgs boson
at CMS: $\PH\to\gamma^*\gamma\to\ell\ell\gamma$, where $\ell=\mu,\Pe$, and
$\PH\to(\JPsi)\gamma\to\mu\mu\gamma$.  No signal is observed due to insufficient
statistical power of the recorded data.  The upper limits are set on the decay rate of
these processes.  For the $\PH\to\gamma^*\gamma$ process the observed limit is 6.7 times
the SM prediction, which comes from the combination of the muon and electron channels.
This result is dominated by the sensitivity in the muon channel. The sensitivity in the
electron channel is suppressed due to the difficulty of reconstructing two close-by
electrons that are merged into a single shower in the electromagnetic
calorimeter. Nevertheless, it seems important to present the result of the electron
channel in the hopes that in the future analysis the techniques and the reconstruction
algorithms can be improved and better sensitivity will be achieved.  Furthermore, this
challenge in the electrons reconstruction can lead to a solution in the design of the
future particle detectors.

In the next data taking of the LHC operation, in addition to a higher collision energy,
the increase of the integrated luminosity is anticipated. In the Run-2, the LHC plans to
deliver $300\fbinv$ of data at $\sqrt{s}=13\TeV$, while in the high luminosity (HL) run,
$\mathcal{L}_{int}=3000\fbinv$ is expected.  With such luminosity the sensitivity to the
SM Higgs boson decay, $\PH\to\gamma^*\gamma$, will be greatly increased. At $300\fbinv$
one expects to achieve the signal significance greater than $2\sigma$, while at
$3000\fbinv$ the Higgs boson signal with ${>}5\sigma$ significance can be observed.  This
would allow us to determine the rate of this decay and its compatibility with the SM
predictions.

For the $\PH\to(\JPsi)\gamma\to\mu\mu\gamma$ decay the obtained limit on the branching
ratio is $1.5\times10^{-3}$, which is 540 times higher than the SM prediction.  This means
that even after the HL run one would not be sensitive to this decay at the SM rates. It is
possible, however, that in some BSM models the Hcc coupling is larger than it is in the
SM.  In that case the $\PH\to(\JPsi)\gamma$ could be interesting to look at.

I think further improvements to this analysis can be done. First of all, a better
simulation of the background processes is needed.  This will allow us to better understand
the background composition and could lead to an optimization of the event selection,
including the MVA techniques. Secondly, the developments in the electron channel for a
better identification of the merged electrons topology could boost the sensitivity. The
improvements here could come from further exploitation of the results obtained by CMS for a
photon conversion process $\gamma\to\Pe\Pe$. Also, the development of a dedicated trigger
for this channel is anticipated, and it is expected to improve the total signal selection
efficiency.

I hope you enjoyed the reading of this dissertation.  If you have any questions or
comments, please send them to me at
\href{mailto:Andrey.Pozdnyakov@cern.ch}{Andrey.Pozdnyakov@cern.ch}.

%

\renewcommand\refname{\begin{centering}References\end{centering}}
\bibliography{thesis}

\providecommand{\href}[2]{#2}\begingroup\raggedright\begin{thebibliography}{10}%
\makeatletter
\providecommand{\hrefCMSnoop }[0]{\@secondoftwo}%
\makeatother
\providecommand{\doi}{\texttt{doi:}\begingroup \urlstyle{tt}\Url}

\bibitem{Peskin}
M.~E. Peskin and D.~V. Schroeder, ``An introduction to quantum field theory''.
\newblock Addison-Wesley,
1995.
\newblock

\bibitem{PDG}
\hrefCMSnoop {}{{Particle Data Group} Collaboration, ``Review of Particle
  Physics'',} \textit{ Chin. Phys. C} \textbf{ 38} (2014) 090001,
\href{http://dx.doi.org/10.1088/1674-1137/38/9/090001}{\doi{10.1088/1674-1137/38/9/090001}}.

\bibitem{Rubakov}
V.~A. Rubakov, ``Classical theory of gauge fields''.
\newblock Princeton University Press, Princeton, N.J,
2002.
\newblock

\bibitem{atlas_h}
\hrefCMSnoop {}{{ATLAS} Collaboration, ``Observation of a new particle in the
  search for the {S}tandard {M}odel {H}iggs boson with the {ATLAS} detector at
  the {LHC}'',} \textit{ Phys. Lett. B} \textbf{ 716} (2012) 1,
  \href{http://dx.doi.org/10.1016/j.physletb.2012.08.020}{\doi{10.1016/j.physletb.2012.08.020}},
\href{http://www.arXiv.org/abs/1207.7214}{\texttt{arXiv:1207.7214}}.

\bibitem{cms_h}
\hrefCMSnoop {}{{CMS} Collaboration, ``Observation of a new boson at a mass of
  125\GeV with the {CMS} experiment at the {LHC}'',} \textit{ Phys. Lett. B}
  \textbf{ 716} (2012) 30,
  \href{http://dx.doi.org/10.1016/j.physletb.2012.08.021}{\doi{10.1016/j.physletb.2012.08.021}},
\href{http://www.arXiv.org/abs/1207.7235}{\texttt{arXiv:1207.7235}}.

\bibitem{LHC}
\hrefCMSnoop {}{O.~Br{\"u}ning, H.~Burkhardt, and S.~Myers, ``The {L}arge
  {H}adron {C}ollider'',} \textit{ Prog. Part. Nucl. Phys.} \textbf{ 67}
  (2012), no.~CERN-ATS-2012-064, 705.

\bibitem{pdf4lhc0}
\hrefCMSnoop {}{A.~D. Martin, W.~J. Stirling, R.~S. Thorne, and G.~Watt,
  ``Parton distributions for the {LHC}'',} \textit{ Eur. Phys. J. C} \textbf{
  63} (2009) 189,
  \href{http://dx.doi.org/10.1140/epjc/s10052-009-1072-5}{\doi{10.1140/epjc/s10052-009-1072-5}},
\href{http://www.arXiv.org/abs/0901.0002}{\texttt{arXiv:0901.0002}}.

\bibitem{higgs-theory}
\hrefCMSnoop {}{A.~Djouadi, ``Higgs Physics: Theory'',} \textit{ Pramana}
  \textbf{ 79} (2012) 513--539,
  \href{http://dx.doi.org/10.1007/s12043-012-0361-y}{\doi{10.1007/s12043-012-0361-y}},
\href{http://www.arXiv.org/abs/1203.4199}{\texttt{arXiv:1203.4199}}.

\bibitem{cs-ATL}
\hrefCMSnoop {}{{ATLAS} Collaboration, ``{Measurement of the Inelastic
  Proton-Proton Cross-Section at $\sqrt{s}=7$ TeV with the ATLAS Detector}'',}
  \textit{ Nature Commun.} \textbf{ 2} (2011) 463,
  \href{http://dx.doi.org/10.1038/ncomms1472}{\doi{10.1038/ncomms1472}},
\href{http://www.arXiv.org/abs/1104.0326}{\texttt{arXiv:1104.0326}}.

\bibitem{cs-CMS}
\hrefCMSnoop {}{{CMS} Collaboration, ``{Measurement of the inelastic
  proton-proton cross section at $\sqrt{s}=7$ TeV}'',} \textit{ Phys. Lett.}
  \textbf{ B722} (2013) 5--27,
  \href{http://dx.doi.org/10.1016/j.physletb.2013.03.024}{\doi{10.1016/j.physletb.2013.03.024}},
\href{http://www.arXiv.org/abs/1210.6718}{\texttt{arXiv:1210.6718}}.

\bibitem{CMS-cs-summary}
\href
  {https://twiki.cern.ch/twiki/bin/view/CMSPublic/PhysicsResultsCombined}{{CMS
  Collaboration}, ``{{Summaries of {CMS} cross section measurements}}''.}
  \url{https://twiki.cern.ch/twiki/bin/view/CMSPublic/PhysicsResultsCombined}.

\bibitem{pz-fermions}
\hrefCMSnoop {}{A.~Pozdnyakov, ``Fermionic decays of {SM} {H}iggs'',} in
  \textit{ XXXIV Physics in Collision}.
\newblock SLAC eCONF, 2014.
\newblock
\href{http://www.arXiv.org/abs/1411.1981}{\texttt{arXiv:1411.1981}}.
\newblock

\bibitem{Abba96}
\hrefCMSnoop {}{A.~Abbasabadi, D.~Bowser-Chao, D.~A. Dicus, and W.~W. Repko,
  ``Radiative {H}iggs boson decays $H \to ff \gamma$'',} \textit{ Phys. Rev. D}
  \textbf{ 55} (1997) 5647,
  \href{http://dx.doi.org/10.1103/PhysRevD.55.5647}{\doi{10.1103/PhysRevD.55.5647}},
\href{http://www.arXiv.org/abs/hep-ph/9611209}{\texttt{arXiv:hep-ph/9611209}}.

\bibitem{Dicus13}
\hrefCMSnoop {}{D.~A. Dicus and W.~W. Repko, ``Calculation of the decay $H \to
  e\bar{e}\gamma$'',} \textit{ Phys. Rev. D} \textbf{ 87} (2013) 077301,
  \href{http://dx.doi.org/10.1103/PhysRevD.87.077301}{\doi{10.1103/PhysRevD.87.077301}},
\href{http://www.arXiv.org/abs/1302.2159}{\texttt{arXiv:1302.2159}}.

\bibitem{Chen12}
\hrefCMSnoop {}{L.~B. Chen, C.~F. Qiao, and R.~L. Zhu, ``Reconstructing the
  125\GeV {SM} {H}iggs boson through $\ell\bar{\ell}\gamma$'',} \textit{ Phys.
  Lett. B} \textbf{ 726} (2013) 306,
  \href{http://dx.doi.org/10.1016/j.physletb.2013.08.050}{\doi{10.1016/j.physletb.2013.08.050}},
\href{http://www.arXiv.org/abs/1211.6058}{\texttt{arXiv:1211.6058}}.

\bibitem{Passarino}
\hrefCMSnoop {}{G.~Passarino, ``Higgs Boson Production and Decay: {Dalitz}
  Sector'',} \textit{ Phys. Lett. B} \textbf{ 727} (2013) 424--431,
  \href{http://dx.doi.org/10.1016/j.physletb.2013.10.052}{\doi{10.1016/j.physletb.2013.10.052}},
\href{http://www.arXiv.org/abs/1308.0422}{\texttt{arXiv:1308.0422}}.

\bibitem{Htollg-FB-Sun}
\hrefCMSnoop {}{Y.~Sun, H.~Chang, and D.~Gao, ``Higgs decays to $\gamma
  \ell^+\ell^-$ in the standard model'',} \textit{ JHEP} \textbf{ 05} (2013)
  061,
  \href{http://dx.doi.org/10.1007/JHEP05(2013)061}{\doi{10.1007/JHEP05(2013)061}},
\href{http://www.arXiv.org/abs/1303.2230}{\texttt{arXiv:1303.2230}}.

\bibitem{atl-HZG}
\hrefCMSnoop {}{{ATLAS} Collaboration, ``{Search for Higgs boson decays to a
  photon and a Z boson in pp collisions at $\sqrt{s}$=7 and 8 {TeV} with the
  {ATLAS} detector}'',} \textit{ Phys. Lett. B} \textbf{ 732} (2014) 8,
  \href{http://dx.doi.org/10.1016/j.physletb.2014.03.015}{\doi{10.1016/j.physletb.2014.03.015}},
\href{http://www.arXiv.org/abs/1402.3051}{\texttt{arXiv:1402.3051}}.

\bibitem{cms-HZG}
\hrefCMSnoop {}{{CMS} Collaboration, ``Search for a {Higgs} boson decaying into
  a {Z} and a photon in pp collisions at $\sqrt{s}$ = 7 and 8 {TeV}'',}
  \textit{ Phys. Lett. B} \textbf{ 726} (2013) 587,
  \href{http://dx.doi.org/10.1016/j.physletb.2013.09.057}{\doi{10.1016/j.physletb.2013.09.057}},
\href{http://www.arXiv.org/abs/1307.5515}{\texttt{arXiv:1307.5515}}.

\bibitem{atl-Hmm}
\hrefCMSnoop {}{{ATLAS} Collaboration, ``Search for the {S}tandard {M}odel
  {H}iggs boson decay to $\mu^{+}\mu^{-}$ with the {ATLAS} detector'',}
  \textit{ Phys. Lett. B} \textbf{ 738} (2014) 68,
  \href{http://dx.doi.org/10.1016/j.physletb.2014.09.008}{\doi{10.1016/j.physletb.2014.09.008}},
\href{http://www.arXiv.org/abs/1406.7663}{\texttt{arXiv:1406.7663}}.

\bibitem{cms-Hmm}
\hrefCMSnoop {}{{CMS} Collaboration, ``Search for a standard model-like {H}iggs
  boson in the $\mu^+\mu^-$ and $\mathrm{e^+e^-}$ decay channels at the
  {LHC}'',}
\href{http://www.arXiv.org/abs/1410.6679}{\texttt{arXiv:1410.6679}}.

\bibitem{Firan07}
\hrefCMSnoop {}{A.~Firan and R.~Stroynowski, ``Internal conversions in {H}iggs
  decays to two photons'',} \textit{ Phys. Rev. D} \textbf{ 76} (2007) 057301,
  \href{http://dx.doi.org/10.1103/PhysRevD.76.057301}{\doi{10.1103/PhysRevD.76.057301}},
\href{http://www.arXiv.org/abs/0704.3987}{\texttt{arXiv:0704.3987}}.

\bibitem{Dicus14}
\hrefCMSnoop {}{D.~A. Dicus and W.~W. Repko, ``Dalitz decay $H\to
  f\bar{f}\gamma$ as a background for $H\to\gamma\gamma$'',} \textit{ Phys.
  Rev. D} \textbf{ 89} (2014) 093013,
  \href{http://dx.doi.org/10.1103/PhysRevD.89.093013}{\doi{10.1103/PhysRevD.89.093013}},
\href{http://www.arXiv.org/abs/1402.5317}{\texttt{arXiv:1402.5317}}.

\bibitem{YR3}
\hrefCMSnoop {}{{LHC Higgs Cross Section Working Group} Collaboration,
  ``{Handbook of LHC Higgs Cross Sections: 3. Higgs Properties}'',}
  \href{http://dx.doi.org/10.5170/CERN-2013-004}{\doi{10.5170/CERN-2013-004}},
\href{http://www.arXiv.org/abs/1307.1347}{\texttt{arXiv:1307.1347}}.

\bibitem{MCFM}
\hrefCMSnoop {}{J.~Campbell and R.~Ellis, ``MCFM for the Tevatron and the
  LHC'',} \textit{ Nuclear Physics B - Proceedings Supplements} (2010)
  \href{http://dx.doi.org/10.1016/j.nuclphysbps.2010.08.011}{\doi{10.1016/j.nuclphysbps.2010.08.011}}.

\bibitem{Htollg-FB-Kor}
\hrefCMSnoop {}{A.~Korchin and V.~Kovalchuk, ``Angular distribution and
  forward-backward asymmetry of the Higgs-boson decay to photon and lepton
  pair'',} \textit{ Eur. Phys. J. C} \textbf{ 74} (2014)
  \href{http://dx.doi.org/10.1140/epjc/s10052-014-3141-7}{\doi{10.1140/epjc/s10052-014-3141-7}}.

\bibitem{eeHg}
\hrefCMSnoop {}{A.~Abbasabadi, D.~Bowser-Chao, D.~A. Dicus, and W.~W. Repko,
  ``Higgs-boson\char21{}photon associated production at \textit{e\ifmmode
  \bar{e}\else \={e}\fi{}} colliders'',} \textit{ Phys. Rev. D} \textbf{ 52}
  (Oct, 1995) 3919--3928,
  \href{http://dx.doi.org/10.1103/PhysRevD.52.3919}{\doi{10.1103/PhysRevD.52.3919}}.

\bibitem{hToJPsiGamma-2013}
\hrefCMSnoop {}{G.~T. Bodwin, F.~Petriello, S.~Stoynev, and M.~Velasco, ``Higgs
  boson decays to quarkonia and the $H\bar{c}c$ coupling'',} \textit{ Phys.
  Rev. D} \textbf{ 88} (2013) 053003,
  \href{http://dx.doi.org/10.1103/PhysRevD.88.053003}{\doi{10.1103/PhysRevD.88.053003}},
\href{http://www.arXiv.org/abs/1306.5770}{\texttt{arXiv:1306.5770}}.

\bibitem{hToJPsiGamma-2014}
G.~T. Bodwin\hrefCMSnoop {}{ {et~al.}, ``Relativistic corrections to {Higgs}
  boson decays to quarkonia'',} \textit{ Phys. Rev. D} \textbf{ 90} (2014)
  1130,
  \href{http://dx.doi.org/10.1103/PhysRevD.90.113010}{\doi{10.1103/PhysRevD.90.113010}},
\href{http://www.arXiv.org/abs/1407.6695}{\texttt{arXiv:1407.6695}}.

\bibitem{atl-hjp}
\hrefCMSnoop {}{{ATLAS} Collaboration, ``Search for {Higgs} and {Z} boson
  decays to {$J/\Psi\gamma$} and {$\Upsilon(nS)\gamma$} with the {ATLAS}
  detector'',} \textit{ Phys. Rev. Lett.} \textbf{ 114} (2015) 121801,
  \href{http://dx.doi.org/10.1103/PhysRevLett.114.121801}{\doi{10.1103/PhysRevLett.114.121801}},
\href{http://www.arXiv.org/abs/1501.03276}{\texttt{arXiv:1501.03276}}.

\bibitem{Perez15}
\hrefCMSnoop {}{G.~Perez, Y.~Soreq, E.~Stamou, and K.~Tobioka, ``Constraining
  the Charm {Y}ukawa and {H}iggs-quark Universality'',}
\href{http://www.arXiv.org/abs/1503.00290}{\texttt{arXiv:1503.00290}}.

\bibitem{cms-jinst}
\hrefCMSnoop {}{{CMS} Collaboration, ``{The CMS experiment at the CERN LHC}'',}
  \textit{ JINST} \textbf{ 3} (2008) S08004,
\href{http://dx.doi.org/10.1088/1748-0221/3/08/S08004}{\doi{10.1088/1748-0221/3/08/S08004}}.

\bibitem{CMS-PAS-TRK-10-003}
\href {https://cds.cern.ch/record/1279138}{{CMS} Collaboration, ``{Studies of
  Tracker Material}'',} Technical Report CMS-PAS-TRK-10-003, 2010.

\bibitem{TRK-11-001}
\hrefCMSnoop {}{{CMS} Collaboration, ``Description and performance of track and
  primary-vertex reconstruction with the {CMS} tracker'',} \textit{ JINST}
  \textbf{ 9} (2014) P10009,
  \href{http://dx.doi.org/10.1088/1748-0221/9/10/P10009}{\doi{10.1088/1748-0221/9/10/P10009}},
\href{http://www.arXiv.org/abs/1405.6569}{\texttt{arXiv:1405.6569}}.

\bibitem{HCAL-calib}
\hrefCMSnoop {}{P.~Goldenzweig, ``Operational Experience with the CMS Hadronic
  Calorimeter in the 2011 LHC run'',} \textit{ Journal of Physics: Conference
  Series} \textbf{ 404} (2012), no.~1, 012005,
  \href{http://dx.doi.org/10.1088/1742-6596/404/1/012005}{\doi{10.1088/1742-6596/404/1/012005}}.

\bibitem{LHC-BEAM}
\href {http://accelconf.web.cern.ch/accelconf/p07/PAPERS/THXC01.PDF}{O.~Jones,
  ``{LHC} beam instrumentation'',} in \textit{ Particle Accelerator Conference
  (PAC07)}.
\newblock IEEE, 2007.

\bibitem{BPM}
\href {http://www-bd.gsi.de/uploads/paper/cas\_bpm\_main.pdf}{P.~Forck,
  P.~Kowina, and D.~Liakin, ``Beam Position Monitors'',} in \textit{ CERN
  Accelerator School on Beam Diagnostics}.
\newblock CERN, 2008.

\bibitem{LHC-design}
O.~Br{\"u}ning\href {http://cds.cern.ch/record/782076?ln=en}{ {et~al.}, ``{LHC}
  Design Report Vol.1: The {LHC} Main Ring'',} technical report, 2004.

\bibitem{CFD935}
``EG\&G ORTEC Model 935 Quad Constant-Fraction 200 MHz Discriminator'', \url
  {http://www.ortec-online.com/download/935.pdf}.

\bibitem{WaveRunner}
\href
  {http://teledynelecroy.com/support/techlib/registerpdf.aspx?documentid=6594}{{Lecroy},
  ``{{WaveRunner Xi-A/MXi-A oscilloscopes}}''.}
  \url{http://teledynelecroy.com}.

\bibitem{dip}
\href {http://j2eeps.cern.ch/wikis/display/EN/DIP+and+DIM}{``{{DIP and
  DIM}}''.} http://j2eeps.cern.ch/wikis/display/EN/DIP+and+DIM.

\bibitem{vistar}
\href
  {http://op-webtools.web.cern.ch/op-webtools/vistar/vistars.php?usr=LHCCONFIG}{{LHC},
  ``{{LHC Configuration Vistar}}''.} \url{http://op-webtools.web.cern.ch}.

\bibitem{MAD5}
J.~Alwall\hrefCMSnoop {}{ {et~al.}, ``{MadGraph 5 : Going Beyond}'',} \textit{
  JHEP} \textbf{ 06} (2011) 128,
  \href{http://dx.doi.org/10.1007/JHEP06(2011)128}{\doi{10.1007/JHEP06(2011)128}},
\href{http://www.arXiv.org/abs/1106.0522}{\texttt{arXiv:1106.0522}}.

\bibitem{ANO-HEFT}
\hrefCMSnoop {}{T.~Corbett, O.~J.~P. \'Eboli, J.~Gonzalez-Fraile, and M.~C.
  Gonzalez-Garcia, ``Constraining anomalous Higgs boson interactions'',}
  \textit{ Phys. Rev. D} \textbf{ 86} (Oct, 2012) 075013,
  \href{http://dx.doi.org/10.1103/PhysRevD.86.075013}{\doi{10.1103/PhysRevD.86.075013}}.

\bibitem{pythia6}
\hrefCMSnoop {}{T.~Sj{\"o}strand, S.~Mrenna, and P.~Z. Skands, ``{PYTHIA} 6.4
  physics and manual'',} \textit{ JHEP} \textbf{ 05} (2006) 026,
  \href{http://dx.doi.org/10.1088/1126-6708/2006/05/026}{\doi{10.1088/1126-6708/2006/05/026}},
\href{http://www.arXiv.org/abs/hep-ph/0603175}{\texttt{arXiv:hep-ph/0603175}}.

\bibitem{CTEQ6L}
J.~Pumplin\hrefCMSnoop {}{ {et~al.}, ``{New generation of parton distributions
  with uncertainties from global QCD analysis}'',} \textit{ JHEP} \textbf{
  0207} (2002) 012,
  \href{http://dx.doi.org/10.1088/1126-6708/2002/07/012}{\doi{10.1088/1126-6708/2002/07/012}},
\href{http://www.arXiv.org/abs/hep-ph/0201195}{\texttt{arXiv:hep-ph/0201195}}.

\bibitem{pythia8}
\hrefCMSnoop {}{T.~Sj{\"o}strand, S.~Mrenna, and P.~Z. Skands, ``A brief
  introduction to {PYTHIA} 8.1'',} \textit{ Comput. Phys. Commun.} \textbf{
  178} (2008) 852,
  \href{http://dx.doi.org/10.1016/j.cpc.2008.01.036}{\doi{10.1016/j.cpc.2008.01.036}},
\href{http://www.arXiv.org/abs/0710.3820}{\texttt{arXiv:0710.3820}}.

\bibitem{MG-jets}
\href
  {https://cp3.irmp.ucl.ac.be/projects/madgraph/wiki/IntroMatching}{``Introduction
  to jet-parton matching in {MG/ME}''.}
  \url{https://cp3.irmp.ucl.ac.be/projects/madgraph/wiki/IntroMatching}.

\bibitem{PF09}
\href {http://cdsweb.cern.ch/record/1194487}{{CMS} Collaboration, ``Particle
  flow Event Reconstruction in {CMS} and Performance for Jets, Taus, and
  {\MET}'',} CMS Physics Analysis Summary CMS-PAS-PFT-09-001, 2009.

\bibitem{PF10}
\href {http://cdsweb.cern.ch/record/1247373}{{CMS} Collaboration,
  ``Commissioning of the Particle-flow Event Reconstruction with the first
  {LHC} collisions recorded in the {CMS} detector'',} CMS Physics Analysis
  Summary CMS-PAS-PFT-10-001, 2010.

\bibitem{EGM-14-001}
\hrefCMSnoop {}{{CMS} Collaboration, ``Performance of photon reconstruction and
  identification with the {CMS} detector in proton-proton collisions at
  $\sqrt{s}$ = 8 {TeV}'',} (2015).
\href{http://www.arXiv.org/abs/1502.02702}{\texttt{arXiv:1502.02702}}.

\bibitem{cms-Hgg-Legacy}
\hrefCMSnoop {}{{CMS} Collaboration, ``Observation of the diphoton decay of the
  {Higgs} boson and measurement of its properties'',} \textit{ Eur. Phys. J.}
  \textbf{ C74} (2014), no.~10, 3076,
  \href{http://dx.doi.org/10.1140/epjc/s10052-014-3076-z}{\doi{10.1140/epjc/s10052-014-3076-z}},
\href{http://www.arXiv.org/abs/1407.0558}{\texttt{arXiv:1407.0558}}.

\bibitem{Marinelli}
\href {https://cds.cern.ch/record/927374}{N.~Marinelli, ``Track finding and
  identification of converted photons'',} Technical Report CMS-NOTE-2006-005,
  CERN, Geneva, 2006.

\bibitem{cms-mu-7TeV}
\hrefCMSnoop {}{{CMS} Collaboration, ``Performance of {CMS} muon reconstruction
  in $pp$ collision events at $\sqrt{s}=7$ {\TeV}'',} \textit{ JINST} \textbf{
  7} (2012) P10002,
  \href{http://dx.doi.org/10.1088/1748-0221/7/10/P10002}{\doi{10.1088/1748-0221/7/10/P10002}},
\href{http://www.arXiv.org/abs/1206.4071}{\texttt{arXiv:1206.4071}}.

\bibitem{ele-CMS}
S.~Baffioni\hrefCMSnoop {}{ {et~al.}, ``Electron reconstruction in {CMS}'',}
  \textit{ Eur. Phys. J. C} \textbf{ 49} (2007) 1099,
\href{http://dx.doi.org/10.1140/epjc/s10052-006-0175-5}{\doi{10.1140/epjc/s10052-006-0175-5}}.

\bibitem{ele-7TeV}
\href {http://cdsweb.cern.ch/record/1299116}{{CMS} Collaboration, ``Electron
  reconstruction and identification at $\sqrt{s}$ = 7 {TeV}'',} CMS Physics
  Analysis Summary CMS-PAS-EGM-10-004, CERN, 2010.

\bibitem{ele-GSF}
\hrefCMSnoop {}{W.~Adam, R.~Fruhwirth, A.~Strandlie, and T.~Todor,
``{Reconstruction of Electrons with the Gaussian-Sum Filter in the CMS Tracker
  at the LHC}'',}.

\bibitem{TMVA}
A.~Hoecker\hrefCMSnoop {}{ {et~al.}, ``{TMVA: Toolkit for Multivariate Data
  Analysis}'',} \textit{ PoS} \textbf{ ACAT} (2007) 040,
\href{http://www.arXiv.org/abs/physics/0703039}{\texttt{arXiv:physics/0703039}}.

\bibitem{CB-Oreglia}
\hrefCMSnoop {}{M.~Oreglia, ``A study of the reactions $\psi^\prime \to \gamma
  \gamma \psi$''}.
\newblock PhD thesis, Stanford University,
1980.
\newblock

\bibitem{CMS-AN-2011-298}
\hrefCMSnoop {}{{CMS} Collaboration, ``Procedure for the {LHC} {H}iggs boson
  search combination in summer 2011'',} {Analysis Note}, CERN, 2011.

\bibitem{NP}
\hrefCMSnoop {}{J.~{Neyman} and E.~S. {Pearson}, ``On the problem of the most
  efficient tests of statistical hypotheses'',} \textit{ Royal Society of
  London Philosophical Transactions Series A} \textbf{ 231} (1933) 289,
  \href{http://dx.doi.org/10.1098/rsta.1933.0009}{\doi{10.1098/rsta.1933.0009}}.

\bibitem{Cowan}
\hrefCMSnoop {}{G.~Cowan, K.~Cranmer, E.~Gross, and O.~Vitells, ``Asymptotic
  formulae for likelihood-based tests of new physics'',} \textit{ Eur. Phys. J.
  C} \textbf{ 71} (2011) 1554,
  \href{http://dx.doi.org/10.1140/epjc/s10052-011-1554-0}{\doi{10.1140/epjc/s10052-011-1554-0}},
\href{http://www.arXiv.org/abs/1007.1727}{\texttt{arXiv:1007.1727}}.

\bibitem{Read}
\hrefCMSnoop {}{A.~L. Read, ``Presentation of search results: the $CL_s$
  technique'',} \textit{ J. Phys. G} \textbf{ 28} (2002) 2693,
\href{http://dx.doi.org/10.1088/0954-3899/28/10/313}{\doi{10.1088/0954-3899/28/10/313}}.

\bibitem{Junk}
\hrefCMSnoop {}{T.~Junk, ``Confidence level computation for combining searches
  with small statistics'',} \textit{ Nucl. Instrum. Meth. A} \textbf{ 434}
  (1999) 435,
  \href{http://dx.doi.org/10.1016/S0168-9002(99)00498-2}{\doi{10.1016/S0168-9002(99)00498-2}},
\href{http://www.arXiv.org/abs/hep-ex/9902006}{\texttt{arXiv:hep-ex/9902006}}.

\bibitem{RooFit}
\hrefCMSnoop {}{W.~Verkerke and D.~P. Kirkby, ``The {RooFit} toolkit for data
  modeling'',} \textit{ eConf} \textbf{ C0303241} (2003) MOLT007,
\href{http://www.arXiv.org/abs/physics/0306116}{\texttt{arXiv:physics/0306116}}.

\bibitem{pdf4lhc1}
\hrefCMSnoop {}{S.~Alekhin {et~al.}, ``{The PDF4LHC Working Group Interim
  Report}'',} (2011).
\href{http://www.arXiv.org/abs/1101.0536}{\texttt{arXiv:1101.0536}}.

\bibitem{pdf4lhc2}
M.~Botje\hrefCMSnoop {}{ {et~al.}, ``{The PDF4LHC Working Group Interim
  Recommendations}'',} (2011).
\href{http://www.arXiv.org/abs/1101.0538}{\texttt{arXiv:1101.0538}}.

\bibitem{pdf-heavy-Q}
R.~D. Ball\hrefCMSnoop {}{ {et~al.}, ``Impact of Heavy Quark Masses on Parton
  Distributions and {LHC} Phenomenology'',} \textit{ Nucl. Phys. B} \textbf{
  849} (2011) 296,
  \href{http://dx.doi.org/10.1016/j.nuclphysb.2011.03.021}{\doi{10.1016/j.nuclphysb.2011.03.021}},
\href{http://www.arXiv.org/abs/1101.1300}{\texttt{arXiv:1101.1300}}.

\bibitem{CMS-PAS-LUM-13-001}
\hrefCMSnoop {}{{CMS} Collaboration, ``{CMS} luminosity based on pixel cluster
  counting - Summer 2013 update'',} CMS Physics Analysis Summary
  CMS-PAS-LUM-13-001, CERN, 2013.

\bibitem{atl-Hgg-diff}
\hrefCMSnoop {}{{ATLAS} Collaboration, ``{Measurements of fiducial and
  differential cross sections for Higgs boson production in the diphoton decay
  channel at $\sqrt{s}=8$ $\TeV$ with ATLAS}'',} \textit{ JHEP} \textbf{ 1409}
  (2014) 112,
  \href{http://dx.doi.org/10.1007/JHEP09(2014)112}{\doi{10.1007/JHEP09(2014)112}},
\href{http://www.arXiv.org/abs/1407.4222}{\texttt{arXiv:1407.4222}}.

\end{thebibliography}\endgroup
\bibliographystyle{lucas_unsrt}

\appendix		

\clearpage
\chapter{\texorpdfstring{$\JPsi$}{JPsi} polarization}
\label{sec:app-jpsi}

\begin{figure}[hbtp]
  \centering
  \includegraphics[width=0.4\textwidth]{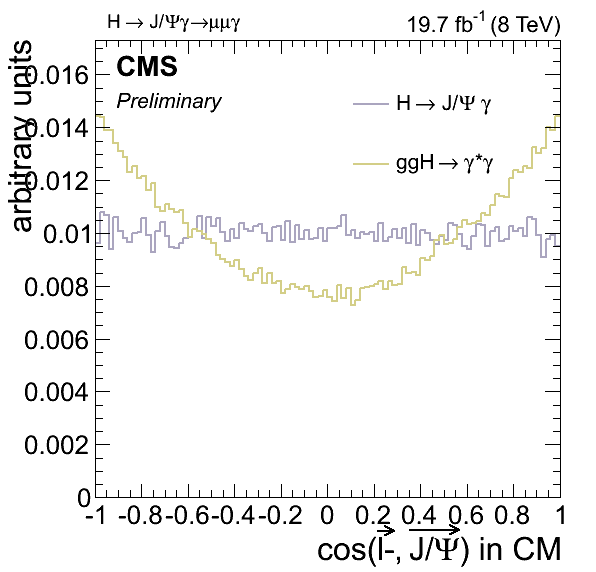}~
  \includegraphics[width=0.4\textwidth]{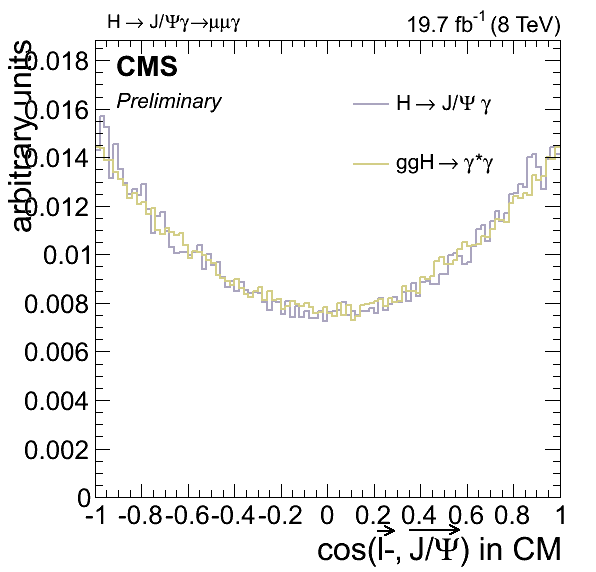}~
  \caption[Distribution of polarization angle from $\JPsi\to\mu\mu$ and
  $\gamma^*\to\mu\mu$.]  {Distribution of polarization angle from $\JPsi\to\mu\mu$ and
    $\gamma^*\to\mu\mu$.  Left: before reweighting of the $\PH\to(\JPsi)\gamma$ sample;
    right: after reweighting.}
  \label{fig:jpsi-pol}
\end{figure}

In the MC sample of the $\PH\to(\JPsi)\gamma$ process, the \JPsi is expected to be 100\%
polarized, since both $\gamma$ and $\JPsi$ are spin-one particles and \PH has spin-zero.
This polarization of \JPsi is not taken into account in the MC sample (\PYTHIA8), which is
produced for the analysis.  It can be checked by looking at the distribution of
$\cos{\theta}$, where $\theta$ is the angle between the positive (negative) lepton and the
direction of the \JPsi. The angle has to be obtained at the generator level, before
selection and calculated in the rest frame of the \JPsi, while the direction of \JPsi is
taken from the centre-of-mass frame of the Higgs boson (\ie $\JPsi+\gamma$
system). Figure~\ref{fig:jpsi-pol} (left) shows this distribution, compared with
$\PH\to\gamma^*\gamma\to\mu\mu\gamma$ (Dalitz) sample. The $\gamma^*$ in the Dalitz sample
is polarized while the \JPsi is not.  This issue could result in a difference in the event
acceptance. In order to estimate this effect, I reweight the MC sample based on that
$cos{\theta}$ distribution with per-event weight
$w=(3/4)\times(1+\cos{\theta}^2)$. Figure~\ref{fig:jpsi-pol} (right) shows the distributions
after reweighting of the $\PH\to(\JPsi)\gamma$ sample.  It confirms the correct
implementation of the re-weighting.  The signal acceptance deceases by 5.5\% once the
reweighting is performed, which leads to ${\sim}\,6.5\%$ decrease in the sensitivity (\ie
an upper limit).

\clearpage
\chapter{Background Simulation study in Muon Channel}
\label{sec:app-mu-bkg}

In order to better understand the background composition in the signal region, I have used
Monte Carlo simulation and generated two main background processes: $\gamma^*\gamma$ and
$\gamma^* + jet$.  The first consists of the ISR ($\Pp\Pp \to \gamma^* + \gamma \to
\mu\mu\gamma$) and FSR ($\Pp\Pp \to \gamma^*/\Z \to \mu\mu\gamma$) of the Drell--Yan
process, with low di-muon invariant mass, $m_{\mu\mu} < 20\,\GeV$.  The second is an ISR
process, $\Pp\Pp \to \gamma^* + jet \to \mu\mu\ + jet$, where a jet in the final state is
mis-identified as a photon. Further in the text I will refer to those processes as
\textit{DYG}amma and \textit{DYJ}et, where DY means $\gamma^*/\Z^*\to\mu\mu$ conversion.
Only $\mu\mu\gamma$ final state is considered for this study.

Both of the above samples are produced starting with \MADGRAPH for the tree diagrams
generation and then hadronized with \PYTHIA6.  In order to avoid double-counting of FSR
photons from \PYTHIA, the FSR process was disabled during hadronization.  for the
production of the \textit{DYG} sample. For the \textit{DYJ} sample, for the same reason of
avoiding double-counting of the jets, I applied jet-matching settings prescribed in
Ref.~\cite{MG-jets}.  This however may not have worked properly (see further discussion).
Event pre-selection is applied at the generator level for both samples, which is close to
the selection used in the analysis.  This is done in order to reduce the number of events
produced. Particularly, for the Jet in \textit{DYJ} sample, only $p_T^j > 35\,\GeV$ and
$|\eta^j| < 1.5$ jets are generated.

\begin{figure}[b]
  \centering
  \includegraphics[width=0.4\textwidth]{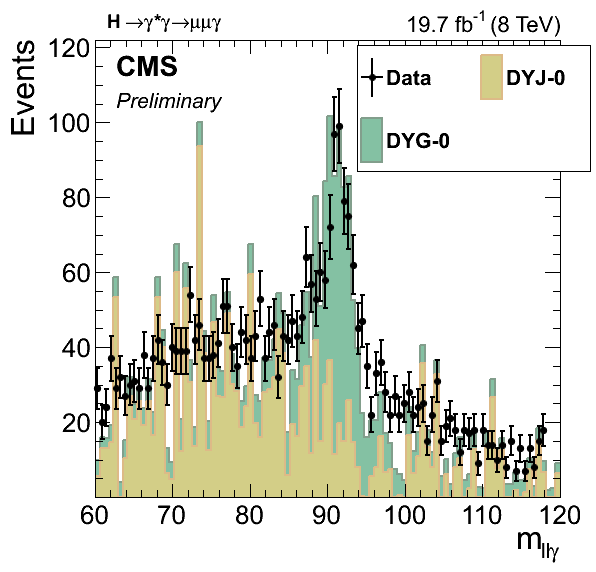}~
  \includegraphics[width=0.4\textwidth]{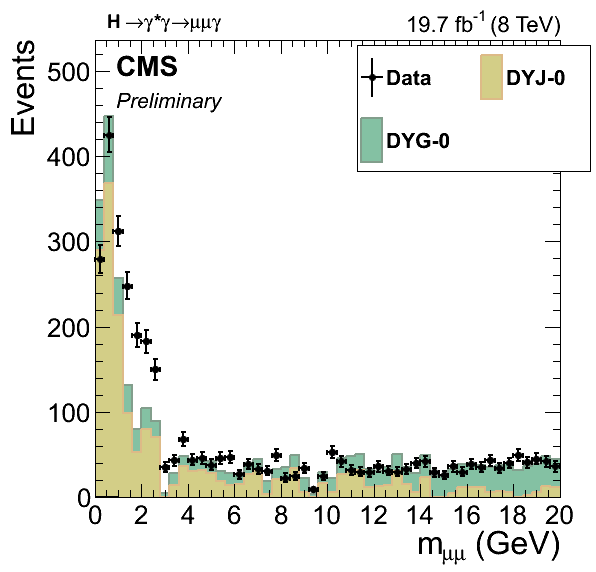}~
  \caption[$m_{\mu\mu\gamma}$ and $m_{\mu\mu}$ distributions in the $\Z$ peak control
  region.]
  {$m_{\mu\mu\gamma}$ and $m_{\mu\mu}$ distributions in the \Z peak control region.  Background MC
    samples are normalized in order to simultaneously fit these two distributions.}
  \label{fig:bkgCR-m}
\end{figure}

The normalization of the MC samples are determined from the fit to the data in a control
region (CR), defined as $60 < m_{\mu\mu\gamma} < 120\GeV$ (while all the other cuts are
the same as described in Section~\ref{sec:sel}).  For this fitting I make use of the
Z-peak in $m_{\mu\mu\gamma}$ distribution from the FSR events, and normalize the MC
samples to match the data.  At the same time we want a good description of the
$m_{\mu\mu}$ distribution, see Fig.~\ref{fig:bkgCR-m}. Simultaneous fit to these two
distributions leads to an effective cross sections of the samples reported in
Table~\ref{tab:datasets-bkg}.

\begin{table}[t]
  \caption[Simulated samples of the background processes for $\mu\mu\gamma$ final state.]
  {Simulated samples of the background processes for $\mu\mu\gamma$ final state.
    Effective cross sections of the samples are determined from the fit to the data in the control region (see text).}
  \label{tab:datasets-bkg}
\begin{center}
\begin{tabular}{c|c| r}
  Process & tag & $\sigma_{eff}$, pb \\
  \hline
  $\Pp\Pp\to\gamma^*\gamma$ & DYG & 1.1\\
  $\Pp\Pp\to\gamma^*+jet$ & DYJ & 180\\
\end{tabular}
\end{center}
\end{table}

\begin{figure}[t]
  \centering
  \includegraphics[width=0.4\textwidth]{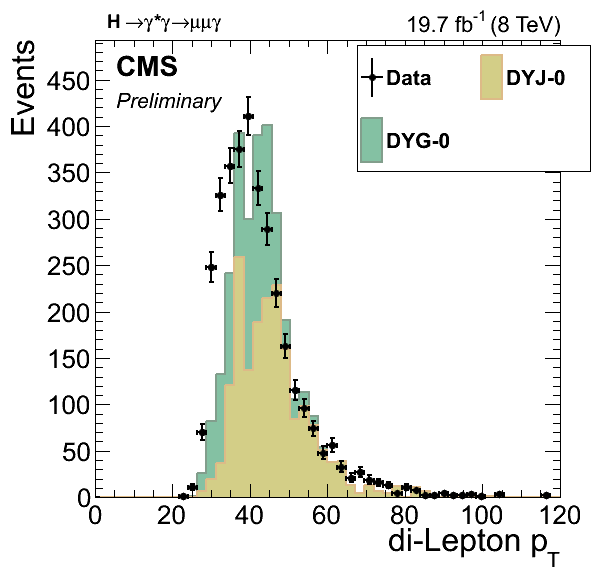}~
  \includegraphics[width=0.4\textwidth]{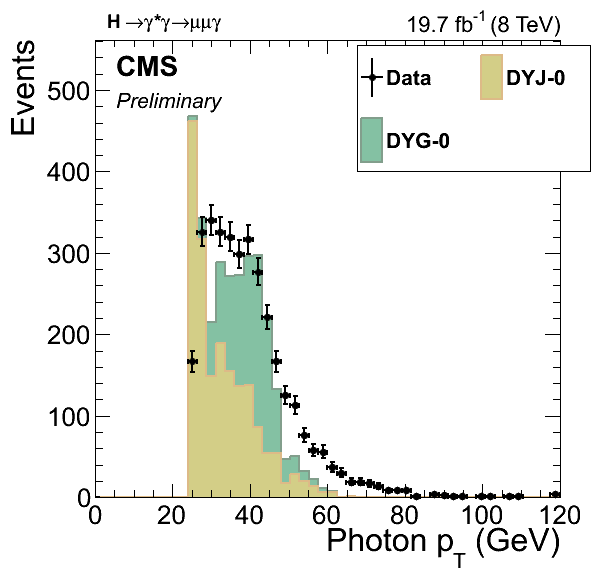}\\
  \includegraphics[width=0.4\textwidth]{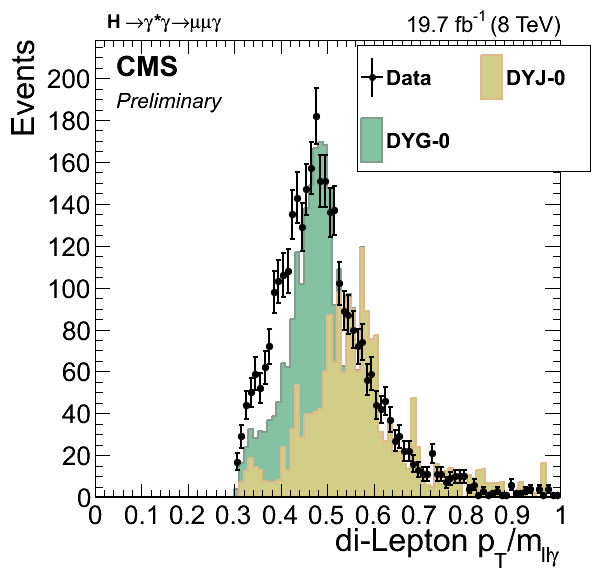}~
  \includegraphics[width=0.4\textwidth]{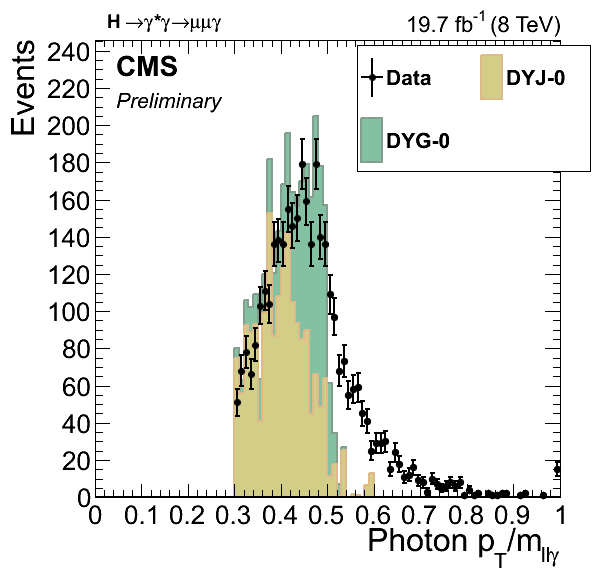}
  \caption{Dimuon and photon \pt distributions of events in the \Z peak control region.}
  \label{fig:bkgCR-pt}
\end{figure}

\begin{figure}[b]
  \centering
  \includegraphics[width=0.32\textwidth]{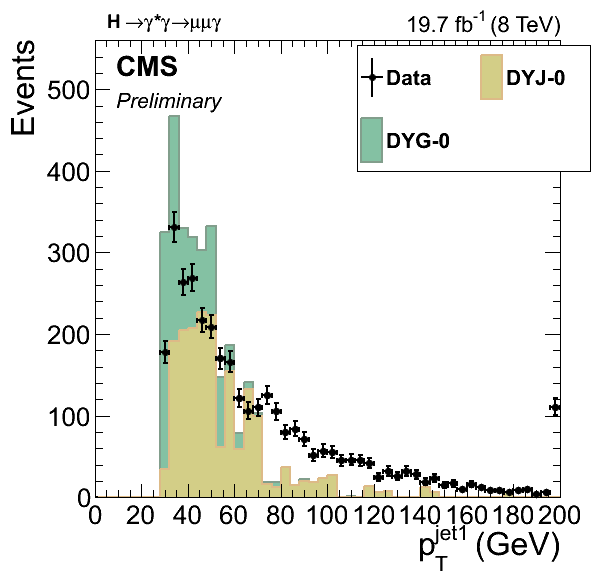}~
  \includegraphics[width=0.32\textwidth]{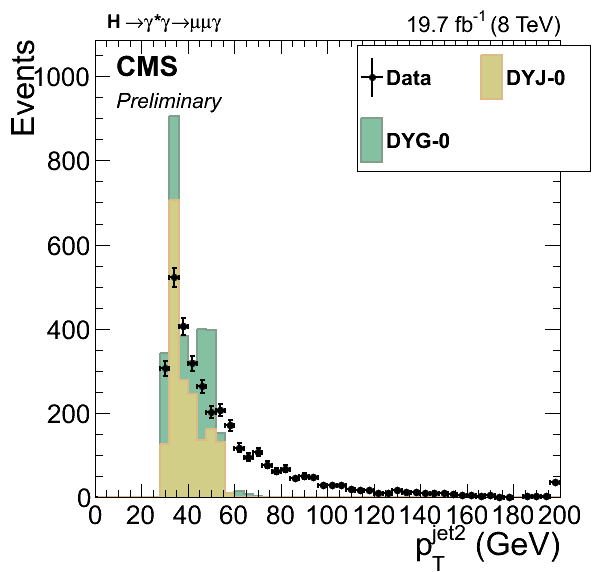}~
  \includegraphics[width=0.32\textwidth]{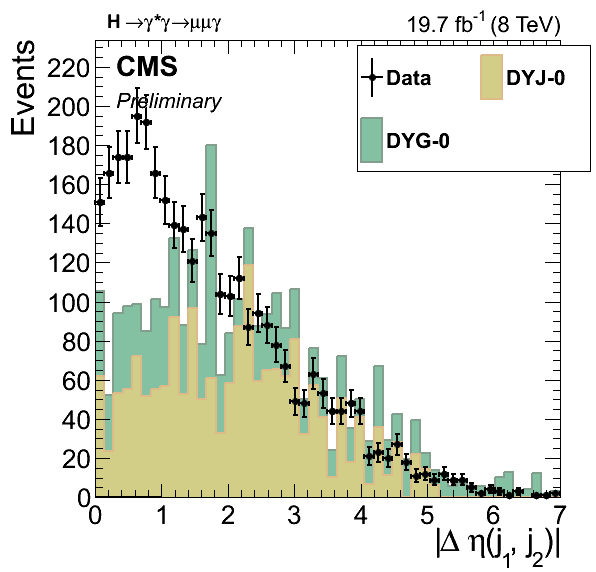}~
  \caption{Distributions of jets' \pt and $\Delta\eta$ in the \Z peak control region.}
  \label{fig:bkgCR-jets}
\end{figure}

Figures~\ref{fig:bkgCR-pt} and \ref{fig:bkgCR-jets} show other relevant distributions,
from which one can see that the two MC samples do not account for all of the events in
data.  Overall, about 10\% of events are missing and this problem probably comes from the
mis-modeling of jets in the \textit{DYJ} sample, see jet distributions on
Fig.~\ref{fig:bkgCR-jets}.

\begin{figure}[b]
  \centering
  \includegraphics[width=0.4\textwidth]{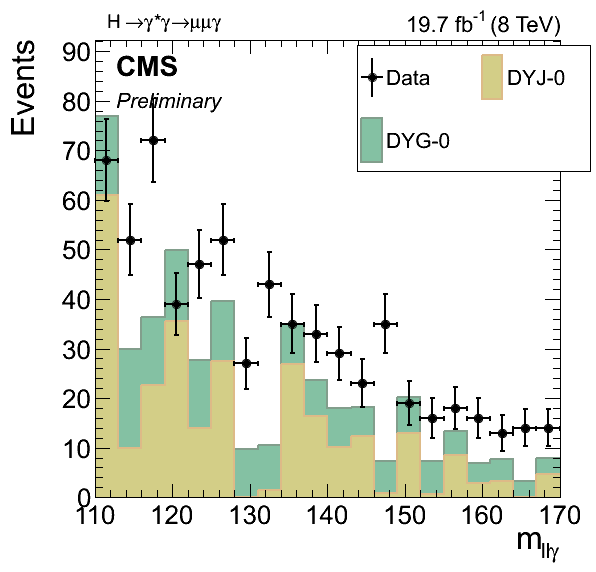}~
  \includegraphics[width=0.4\textwidth]{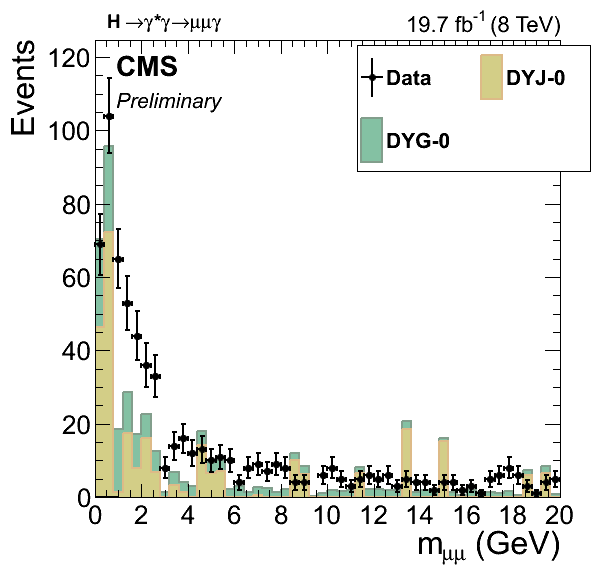}~
  \caption{$m_{\mu\mu\gamma}$ and $m_{\mu\mu}$ distributions in the SR, where the
    background normalization is taken from the CR.}

  \label{fig:bkgSR-m}
\end{figure}

\begin{figure}[hbtp]
  \centering
  \includegraphics[width=0.4\textwidth]{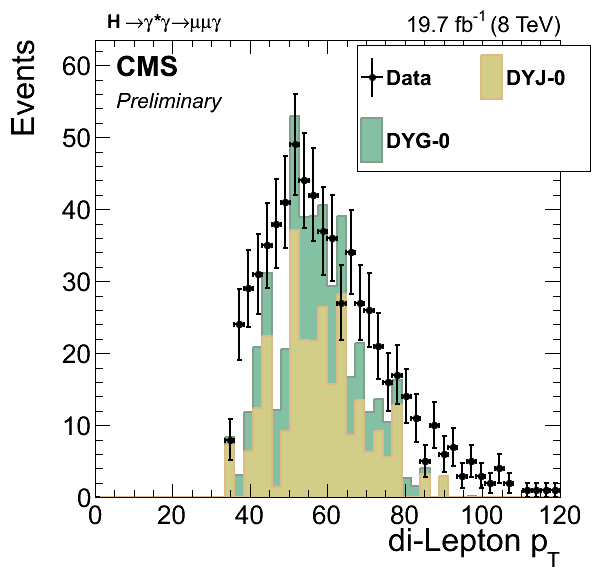}~
  \includegraphics[width=0.4\textwidth]{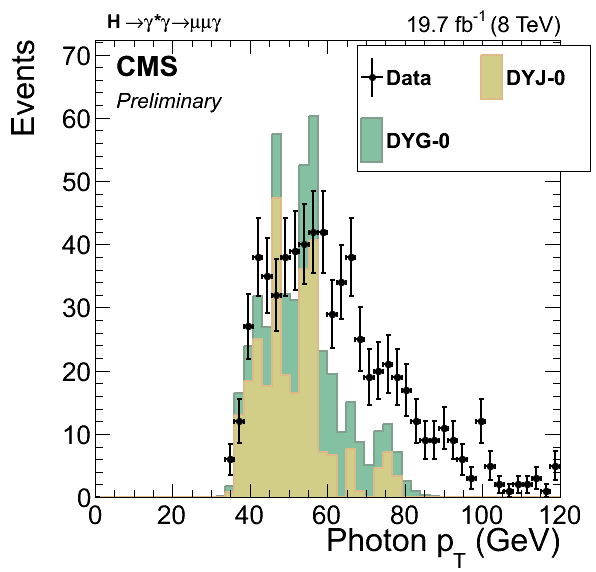}\\
  \includegraphics[width=0.4\textwidth]{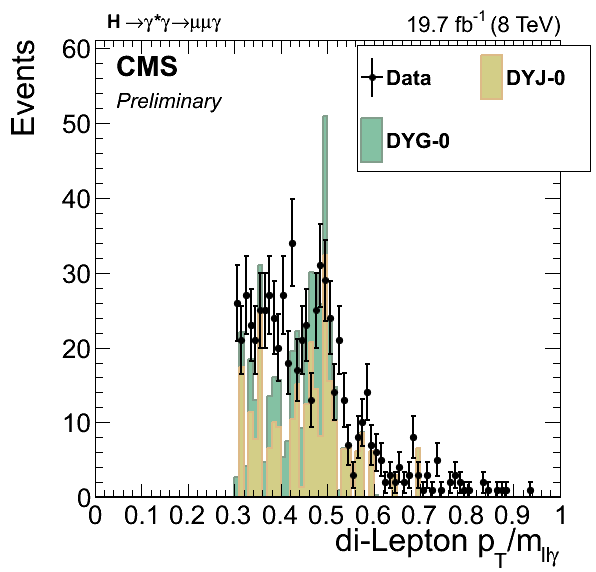}~
  \includegraphics[width=0.4\textwidth]{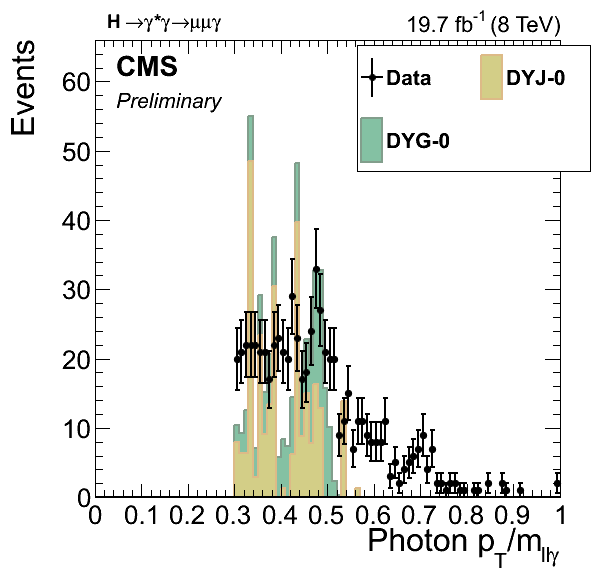}
  \caption{Dimuon and photon \pt and $p_T/m_{\mu\mu\gamma}$ distributions of the events in
    the signal region.}
  \label{fig:bkgSR-pt}
\end{figure}

Nevertheless, we can use this normalization from the CR and see the predictions of the
background in the signal region (SR, $110 < m_{\mu\mu\gamma} < 170\GeV$). The discrepancy
between the data and MC becomes larger in SR: about 35\% of events are not described by
the MC. I think this discrepancy comes from the \textit{DYJ} sample, while \textit{DYG}
gives a reliable prediction of the background. Hence, I assume that about 40\% of the
total background is from \textit{DYG} process. See Figs.~\ref{fig:bkgSR-m} and
\ref{fig:bkgSR-pt} for the relevant distributions in the SR and Table~\ref{tab:bkg-events}
for the event yields in the CR and SR.

\begin{table}[ht]
  \caption{Number of events from data and MC backgrounds in the control and signal regions.}
  \label{tab:bkg-events}
  \begin{center}
    \begin{tabular}{r|cccc}
      &  & \multicolumn{3}{c}{Events, (\% of total)}\\
      & data & DYG& DYJ & Other\\
      \hline
      CR, $60 < m_{\mu\mu\gamma} < 120\GeV$       & 3372 & 1833 (54) & 1146 (34) & 393 (12) \\
      \hline
      SR, $110< m_{\mu\mu\gamma} < 170\GeV$       & 665  &  272 (41) & 177  (27) & 216 (32) \\
    \end{tabular}
  \end{center}
\end{table}

\clearpage
\chapter{Vector Bosons Fusion selection in muon channel}
\label{sec:app-vbf}
In the muon channel of the analysis, a separate category for vector bosons fusion (VBF)
event topology was considered.  The jets used in the VBF tag are reconstructed with PF
algorithm, described in Sec.~\ref{sec:pf}, and their energy is corrected using the
techniques described in Ref.~\cite{PF09}.  These jets are required to be within $|\eta| <
4.7$, and have $p_T^j > 30\,\GeV$.  Two such jets have to be present in an event.

\begin{figure}[hbtp]
  \centering
  \includegraphics[width=0.32\textwidth]{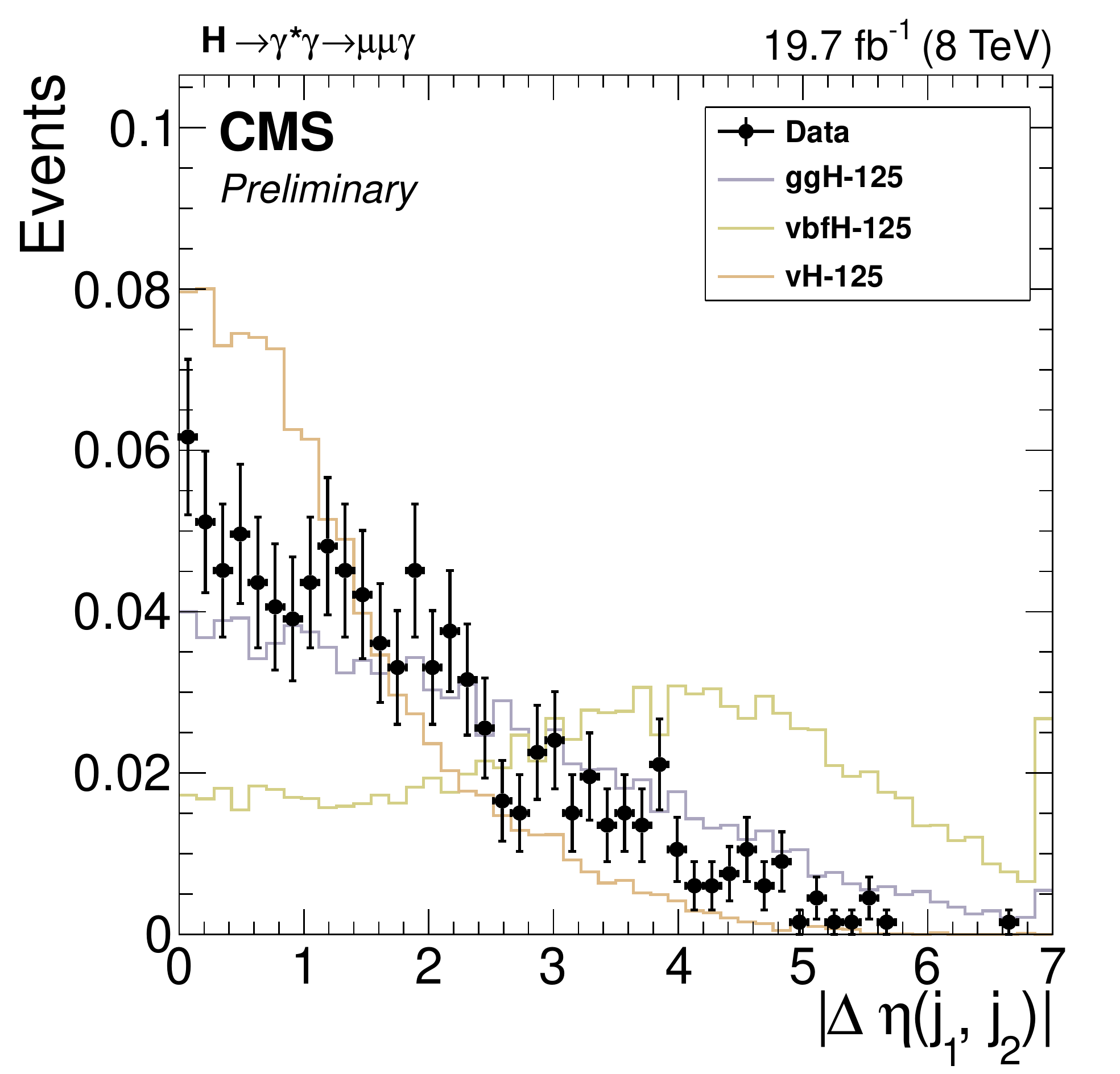}~
  \includegraphics[width=0.32\textwidth]{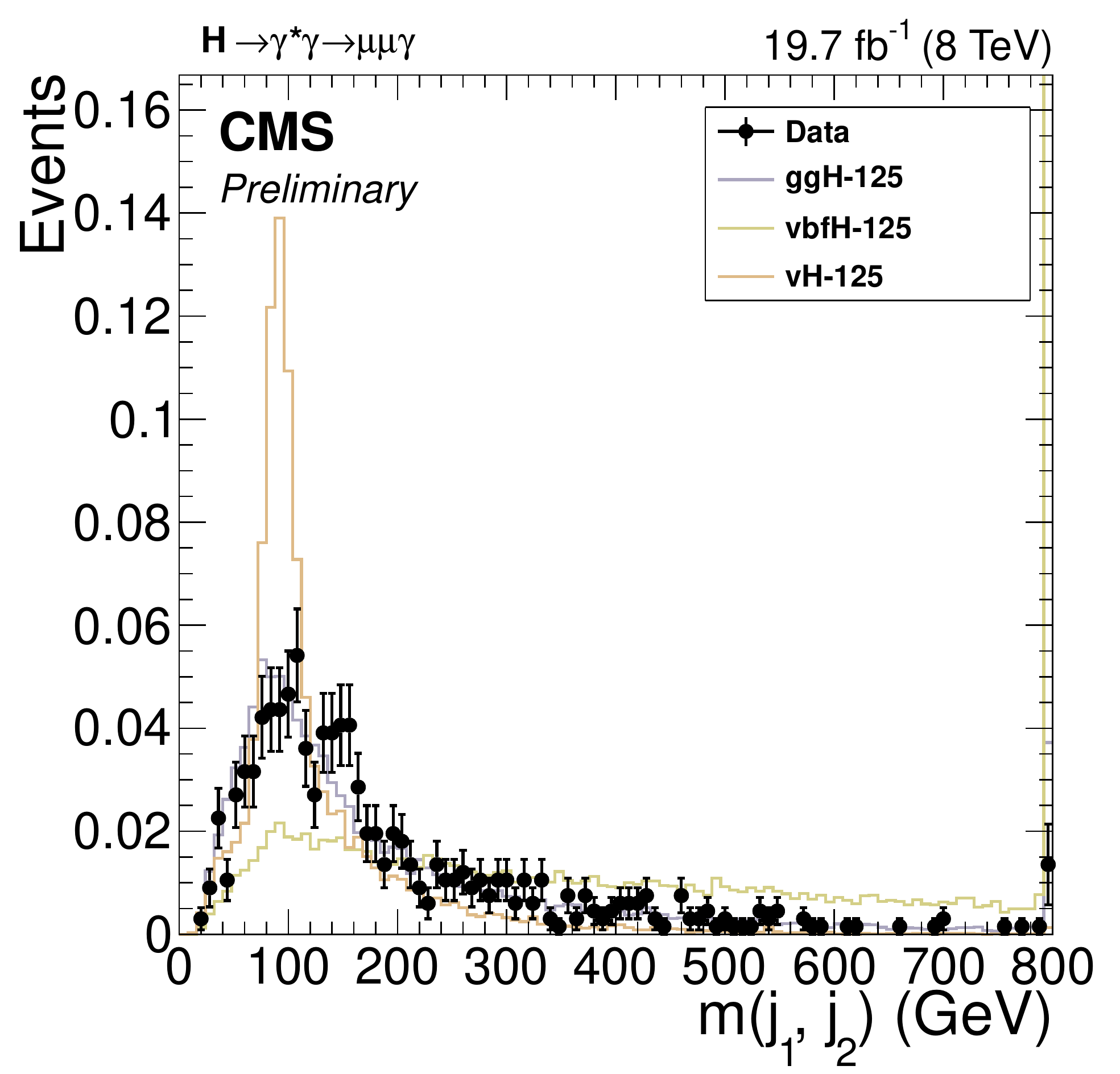}~
  \includegraphics[width=0.32\textwidth]{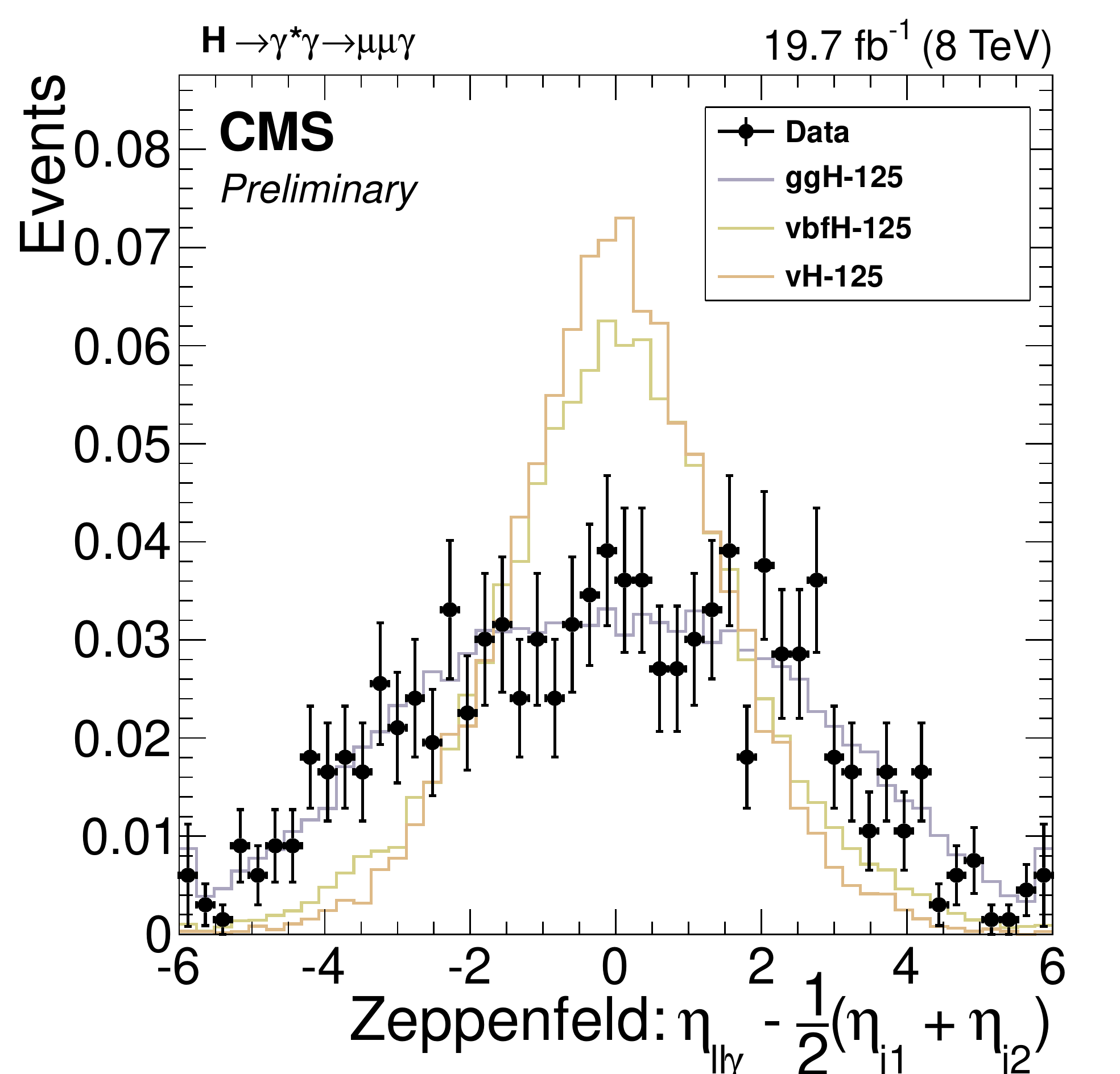}~
  \caption[Input variables for the VBF selection in the muon channel.]  {Input variables
    for the VBF selection. Signals of three production channels are shown.  Signal
    distributions are independently normalized to the total number of events in data.}
  \label{fig:vbf}
\end{figure}

Three variables are considered to identify the VBF events: difference
in $\eta$, di-jet invariant mass and a Zeppenfeld variable.  The
distributions of those variables for the events in data and simulated
signal samples are shown in Fig.~\ref{fig:vbf}.

The following selection for VBF-like events was chosen:
\begin{itemize}
\item $|\eta^{j_1} - \eta^{j_2}| > 3$;
\item $m_{jj} > 450\,\GeV$;
\item $|\eta^{\mu\mu\gamma} - \frac12 (\eta^{j_1} + \eta^{j_2})| < 4$.
\end{itemize}

\begin{figure}[ht]
  \centering
  \includegraphics[width=0.45\textwidth]{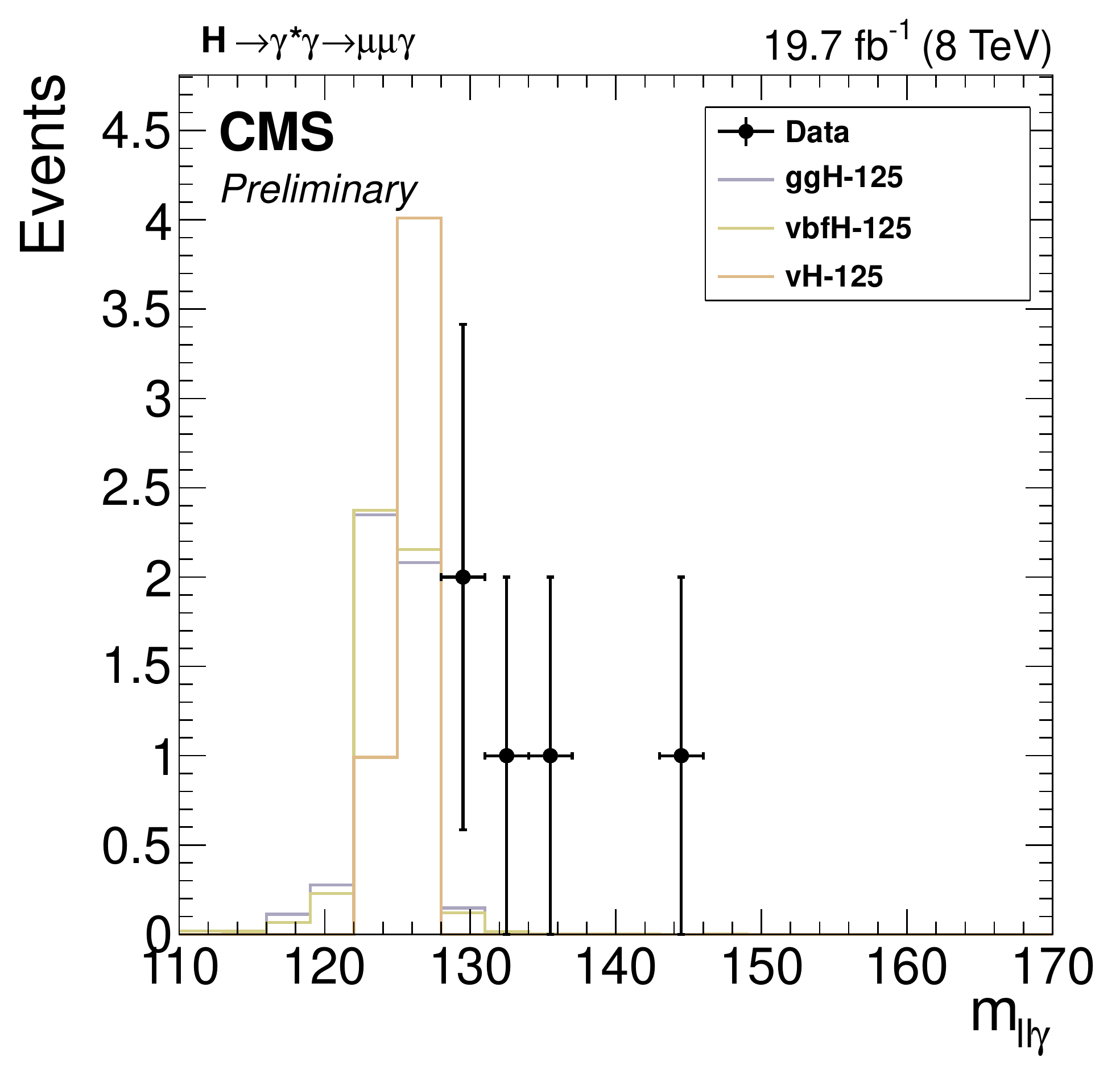}~
  \caption[Three body mass distribution in the VBF category.]  {Three body mass
    distribution in the VBF category.  Signal histograms are independently normalized to
    the number of events in data.}
  \label{fig:vbf-mass}
\end{figure}

After this selection we get 0 events in data within a $122 < m_{\mu\mu\gamma} < 128\,\GeV$
mass window, while expecting 0.08 signal events, see Table~\ref{tab:yield-vbf} for the
yields and Fig.~\ref{fig:vbf-mass} for the three-body mass distributions.  Due to the lack
of statistics we are unable to estimate the background by fitting the data, and we don't
have a MC background sample either.  Because of that and the fact that very little signal
is expected, we don't use the VBF category throughout the analysis.  Even if this category
is used, it would give an insignificant improvement in the limit.  One can estimate this
with a simple counting experiment as follows.  Assuming that the background prediction is
$0.1\pm 1$ events in the signal region and observed number of events is zero, we can place
an upper limit on $\mu$ at $\sim30\times$ SM prediction for the signal.

Although it is not used at present analysis, this category will be
useful in the future data-taking at 13\TeV, with larger statistics.

\begin{table}[ht]
  \caption{Event yield after full selection in VBF category.}
  \label{tab:yield-vbf}
  \begin{center}
    \begin{tabular}{rc|c|c|ccc}
       & $m_{\mu\mu\gamma}$ selection, GeV  & Data & Total signal & ggH & vbfH & VH \\ \hline
       & [110, 170] & 5 & 0.09 & 0.02 & 0.07 & $<$0.005 \\
       & [122, 128] & 0 & 0.08 & 0.02 & 0.06 & $<$0.005 \\
    \end{tabular}
  \end{center}
\end{table}

\chapter{Auxiliary information and plots}
\label{sec:app-extra}
In this Section I include additional information, which is not necessary for the results
presented in the main part of this dissertation, but relevant for a better understanding
of the analysis.

\section{Muon channel}
\label{sec:app-mu}

The rejection and efficiency of the selection presented in the analysis can be looked at
through Table~\ref{tab:yield-mu}, where the event yields in data and signal are shown
after each selection.

\begin{figure}[t]
  \centering
  \includegraphics[width=0.40\textwidth]{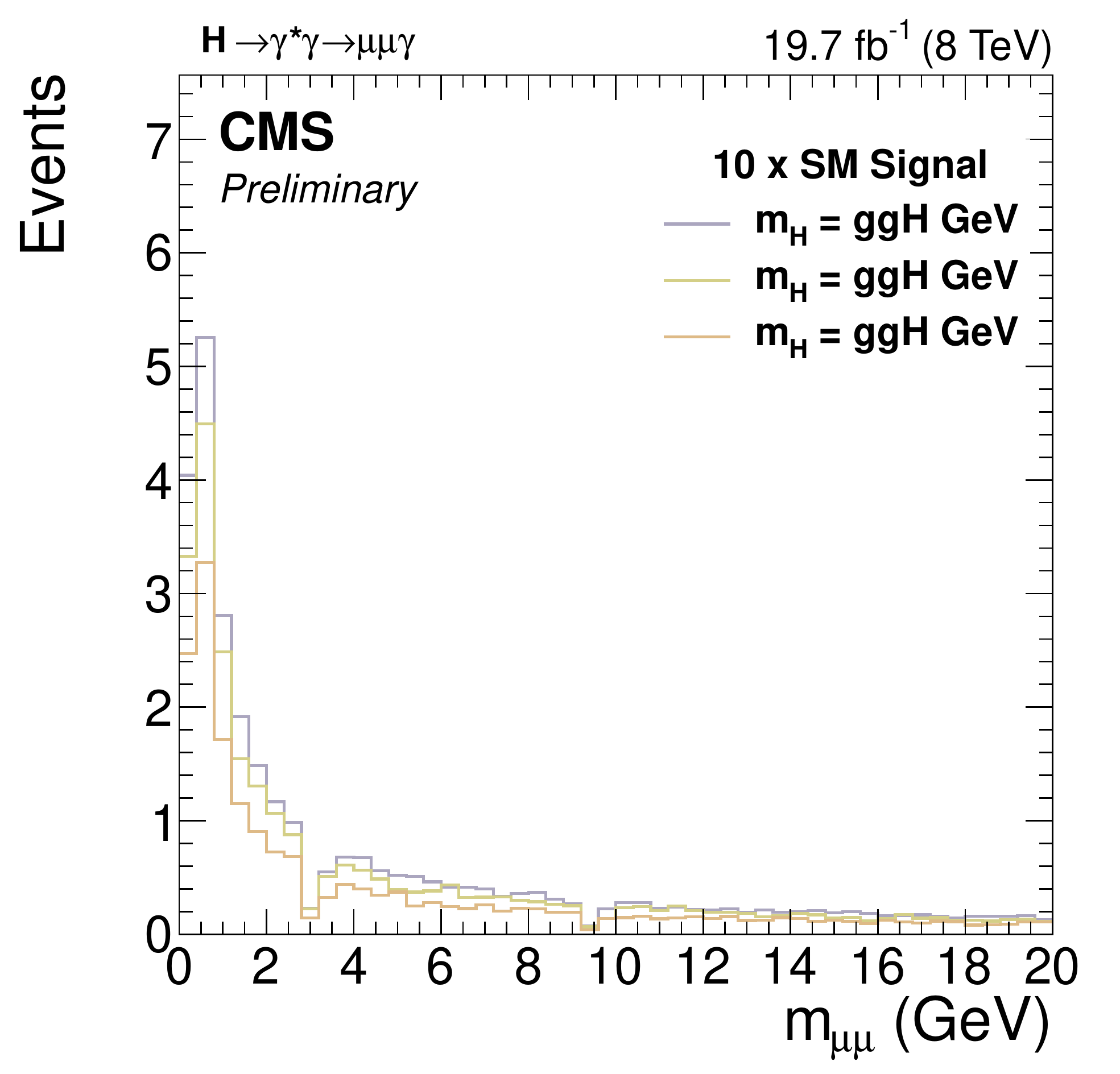}~
  \includegraphics[width=0.40\textwidth]{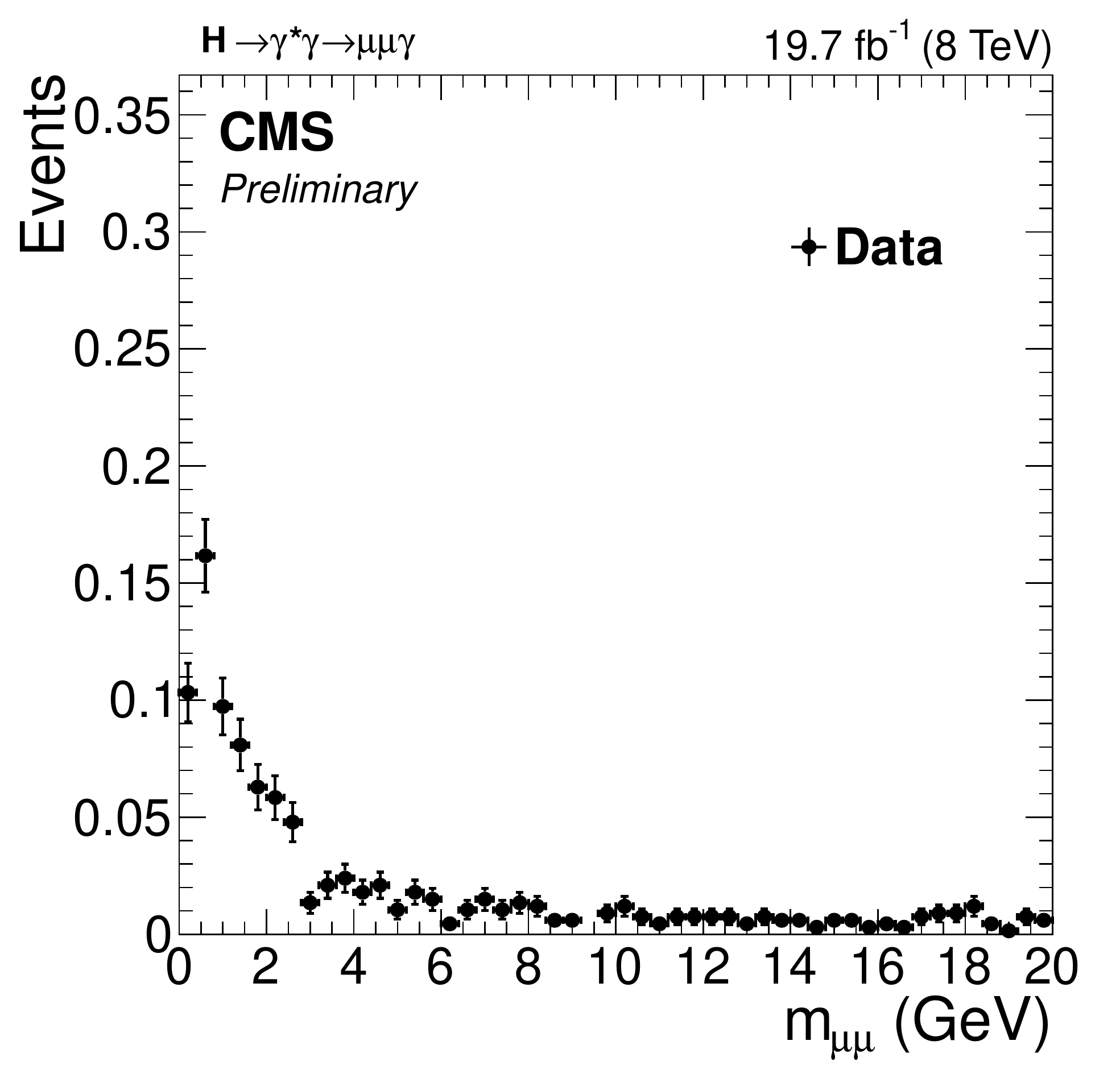}\\
  \includegraphics[width=0.40\textwidth]{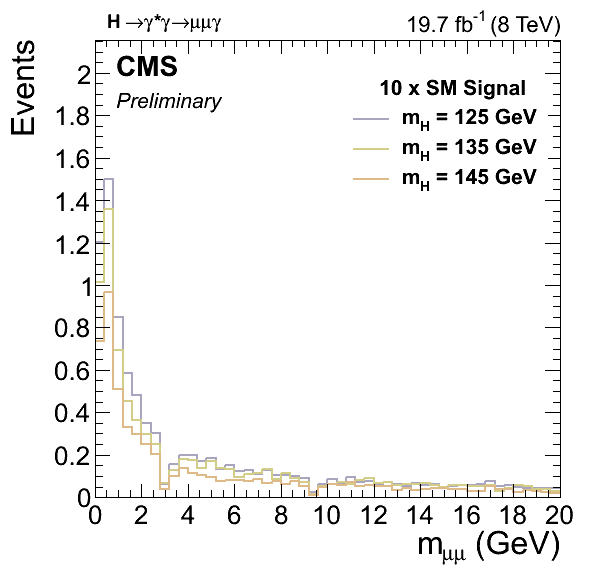}~
  \includegraphics[width=0.40\textwidth]{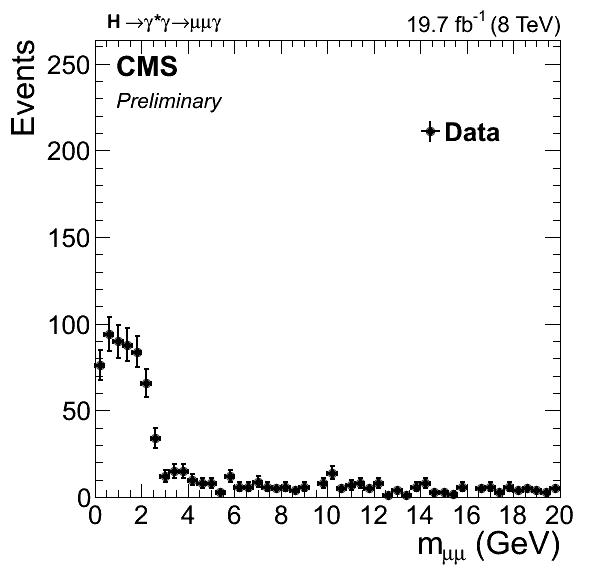}\\
  \includegraphics[width=0.40\textwidth]{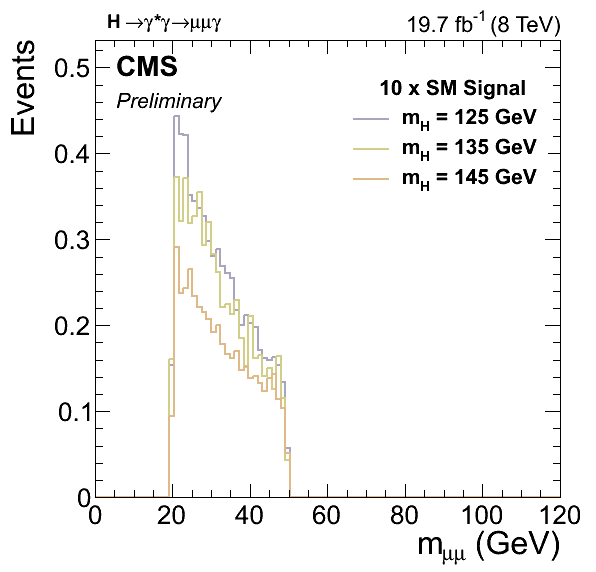}~
  \includegraphics[width=0.40\textwidth]{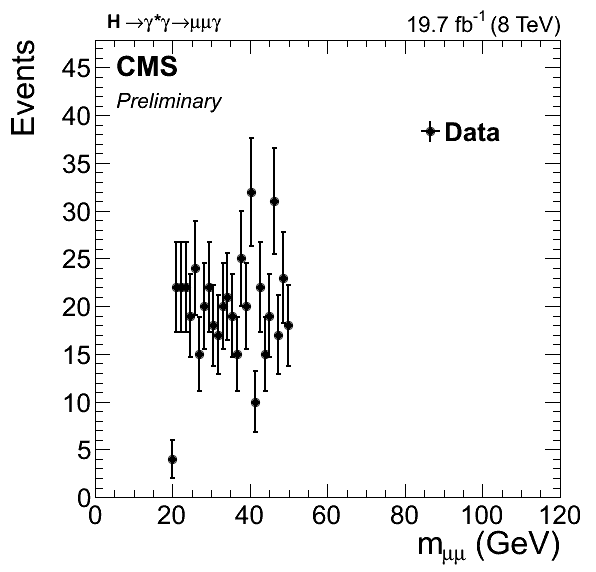}
  \caption[Distributions of the $m_{\mu\mu}$, after full selection in $110 <
  m_{\mu\mu\gamma} < 170\,\GeV$ window.]  {Distributions of $m_{\mu\mu}$, after full
    selection in $110 < m_{\mu\mu\gamma} < 170\,\GeV$ window.  Rows from top to bottom
    correspond to categories 1 (top), 2 (middle), 3 (bottom) as described in the text.
    The ggF signal distributions are shown on the left and scaled by 10. Data is on the
    right.}
  \label{fig:mu-mll}
\end{figure}

\begin{figure}[t]
  \centering
  \includegraphics[width=0.42\textwidth]{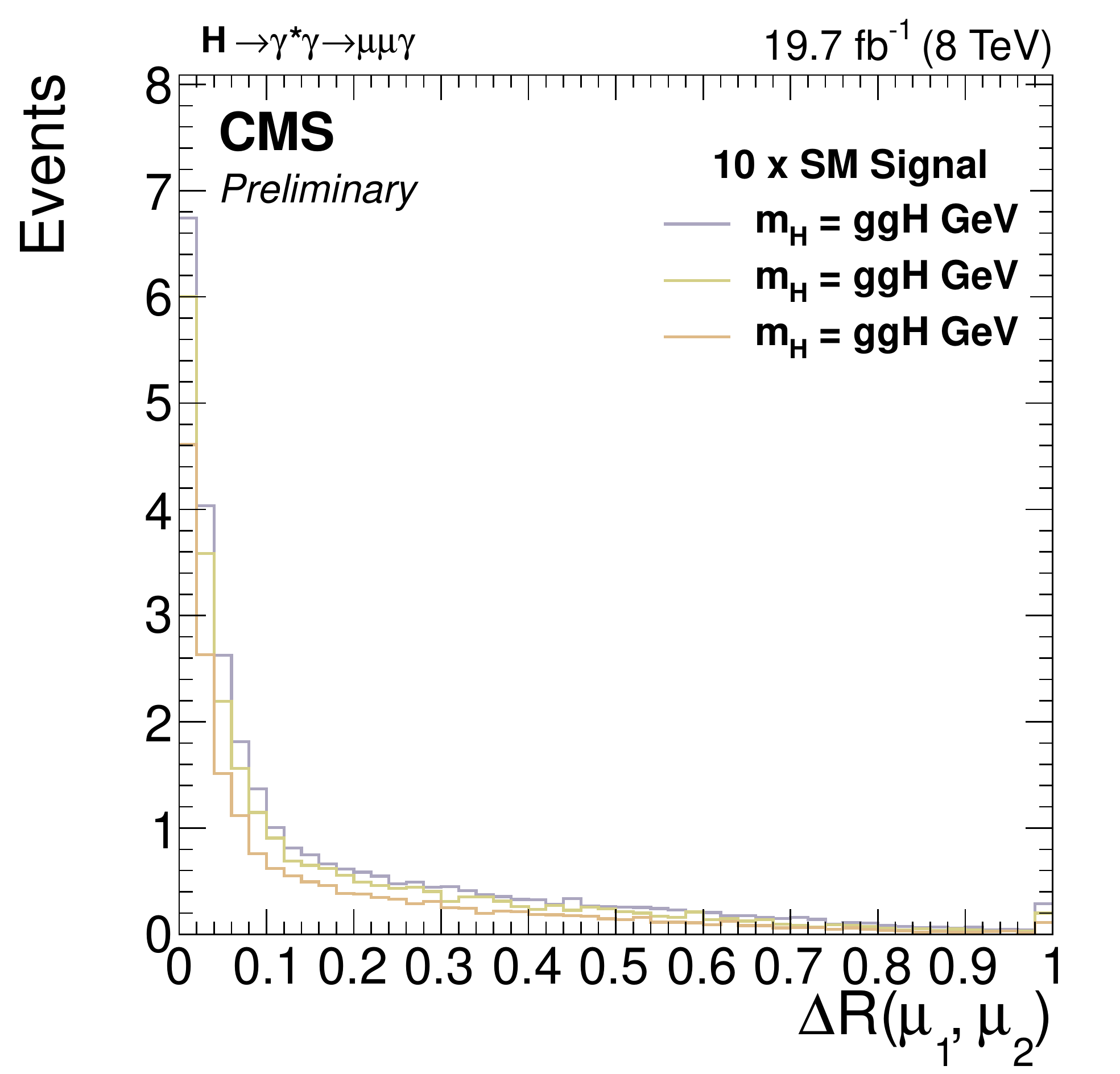}~
  \includegraphics[width=0.42\textwidth]{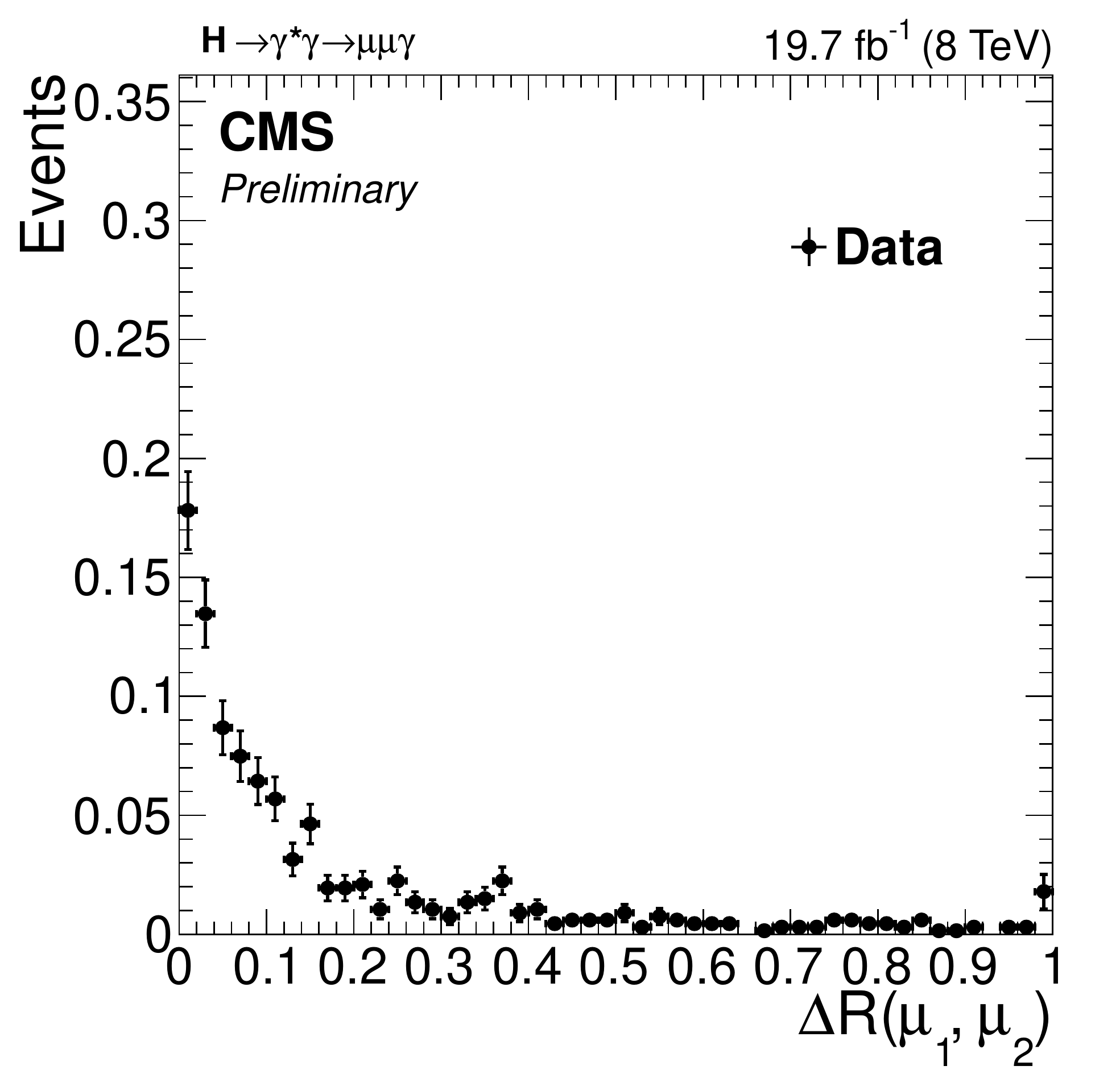}\\
  \includegraphics[width=0.42\textwidth]{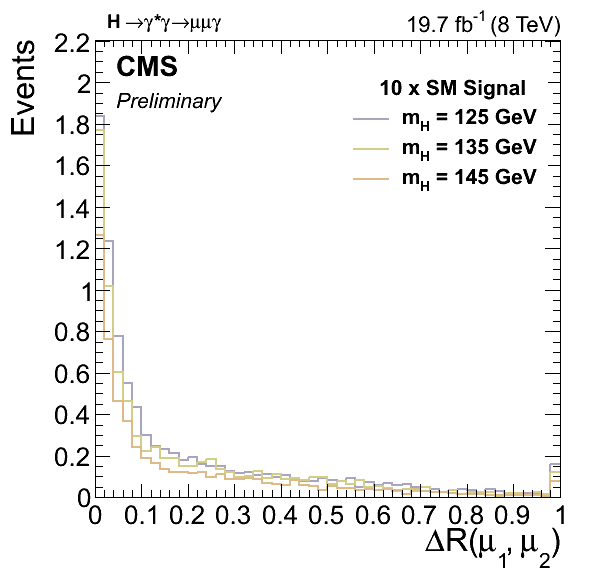}~
  \includegraphics[width=0.42\textwidth]{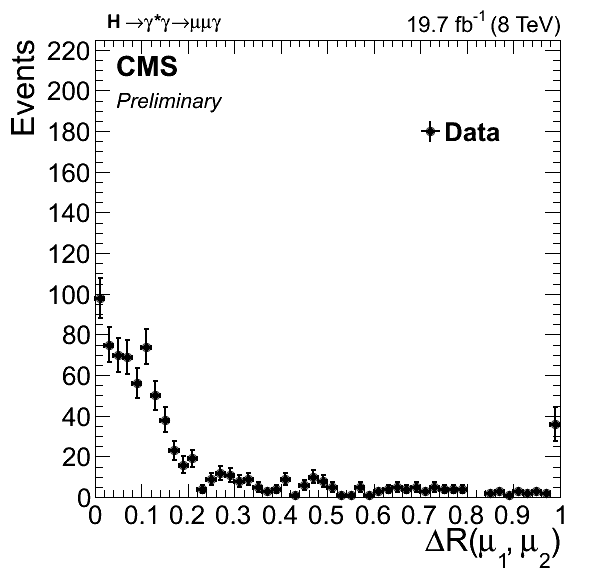}\\
  \includegraphics[width=0.42\textwidth]{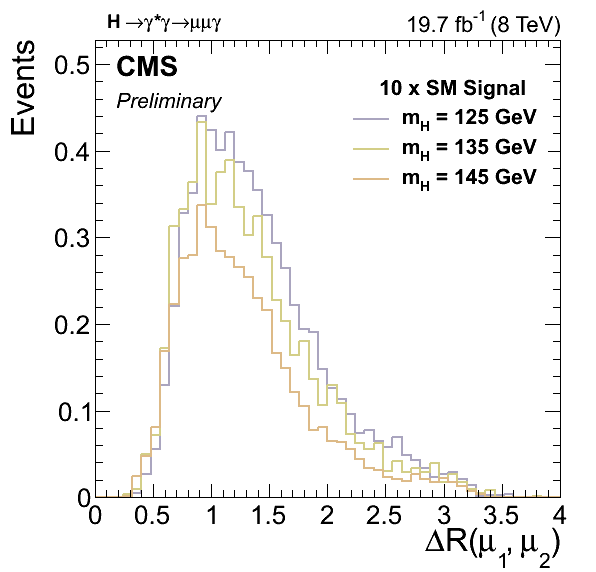}~
  \includegraphics[width=0.42\textwidth]{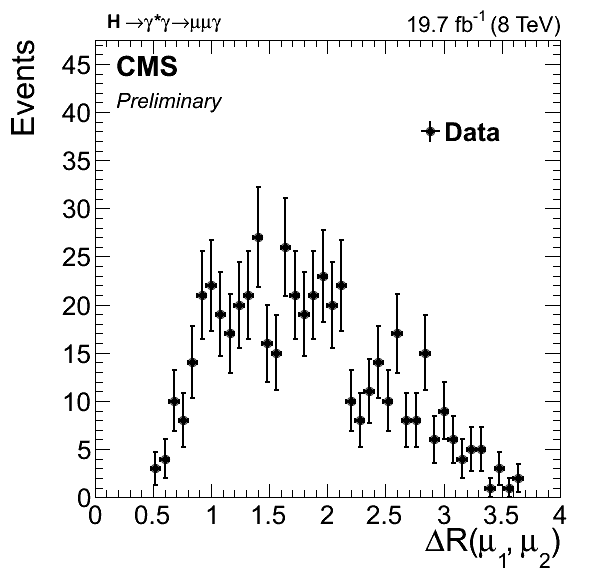}
  \caption[Distributions of $\Delta R(\mu_1,\mu_2)$ for signal and data.]{Distributions of
    $\Delta R(\mu_1,\mu_2)$ for ggF signal (left) and data (right) in 3 categories (top to
    bottom).}
  \label{fig:mu-dR}
\end{figure}

\begin{figure}[t]
  \centering
  \includegraphics[width=0.42\textwidth]{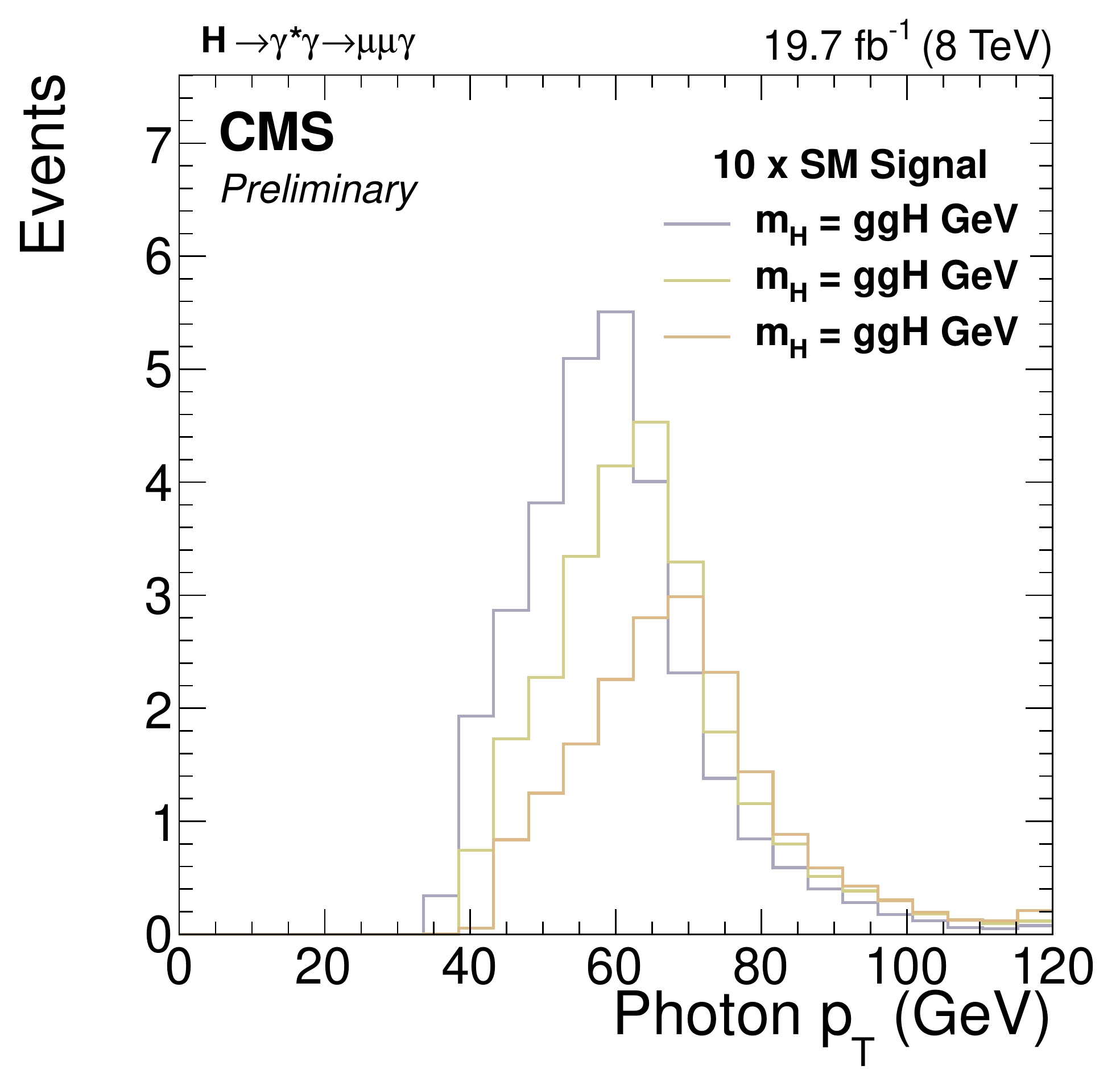}~
  \includegraphics[width=0.42\textwidth]{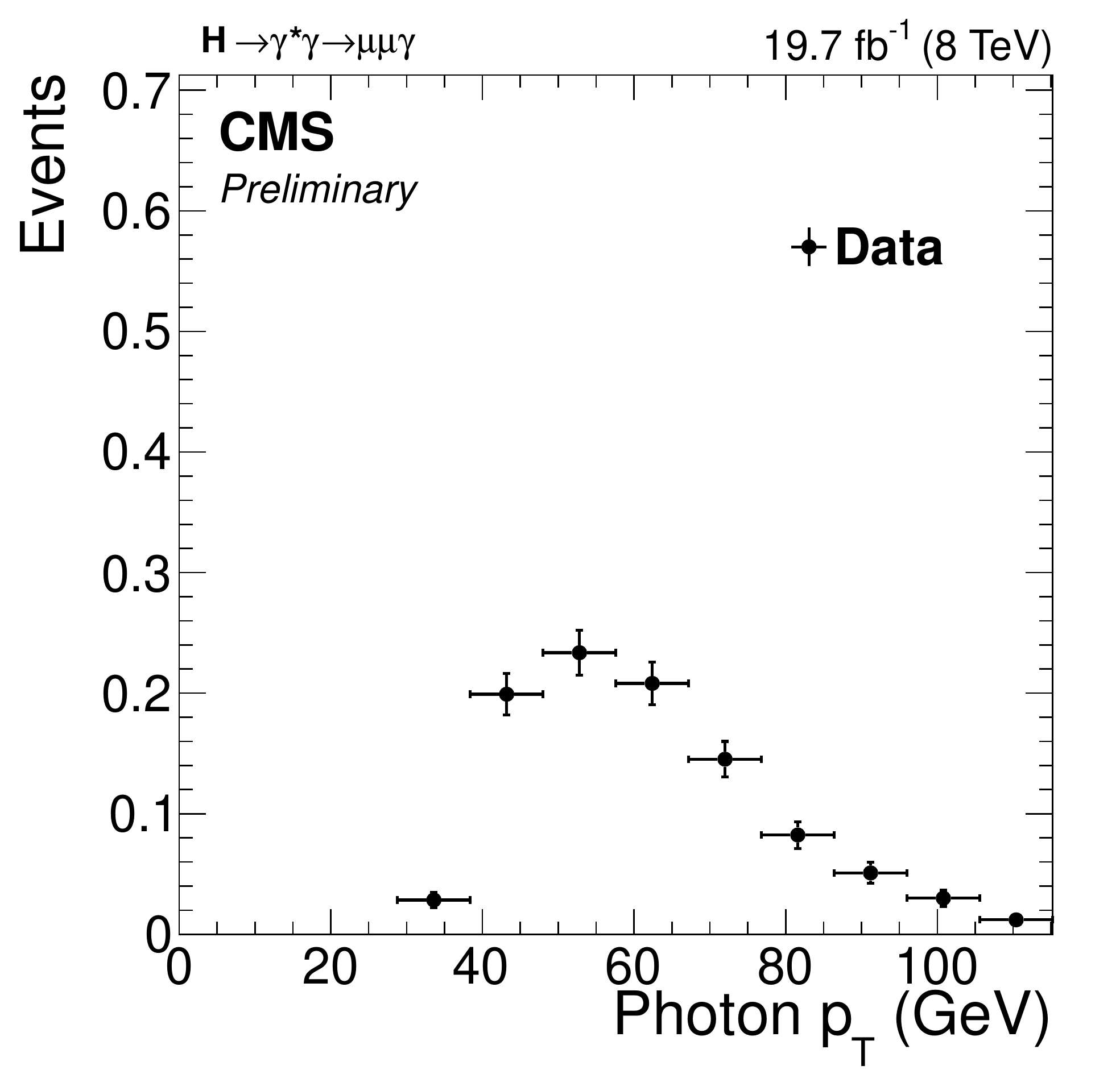}\\
  \includegraphics[width=0.42\textwidth]{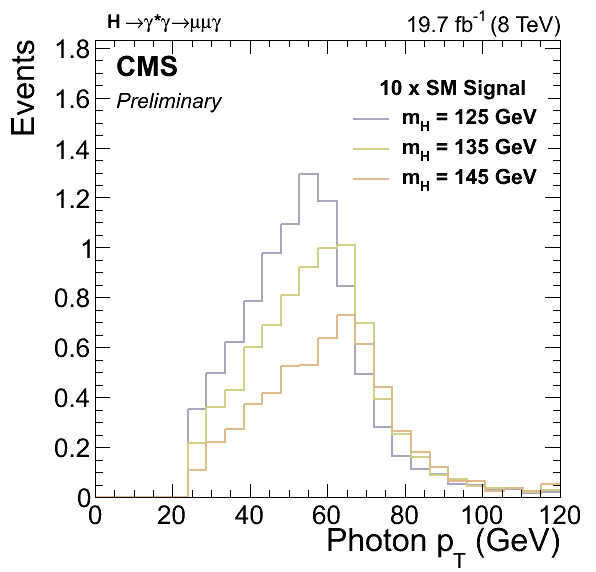}~
  \includegraphics[width=0.42\textwidth]{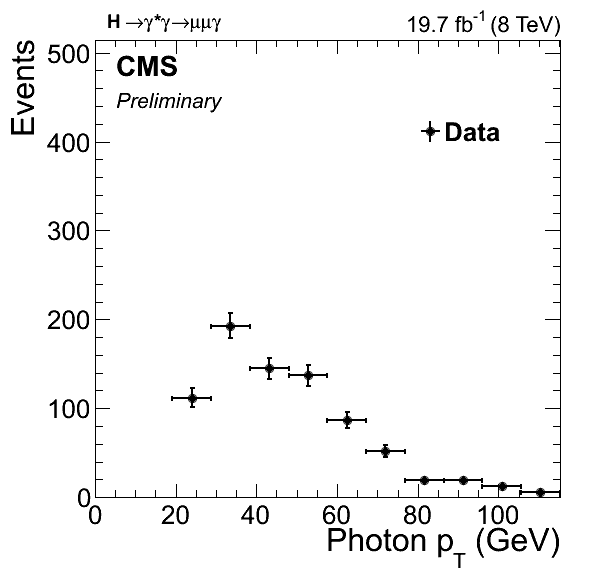}\\
  \includegraphics[width=0.42\textwidth]{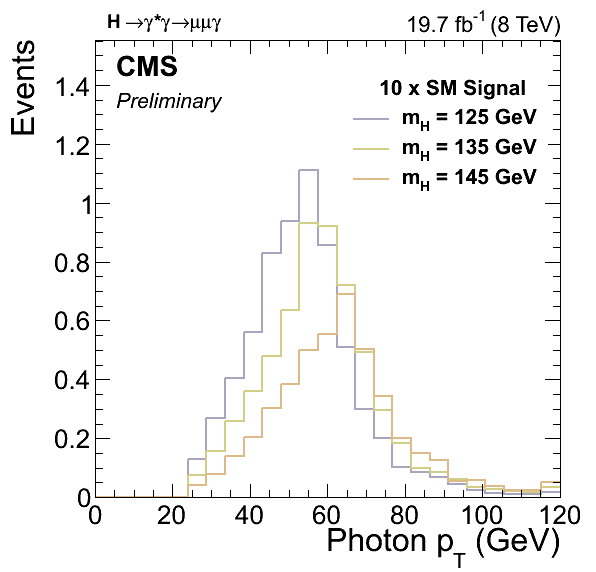}~
  \includegraphics[width=0.42\textwidth]{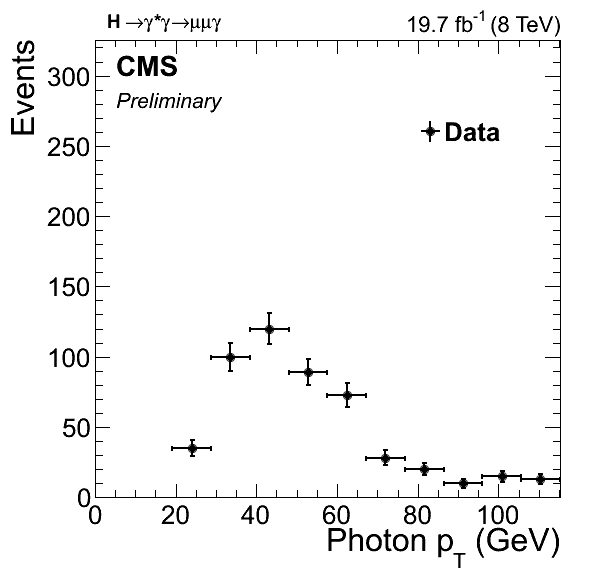}
  \caption[Distributions of $p_T^{\gamma}$ for signal and data.] {Distributions of
    $p_T^{\gamma}$ for ggF signal (left) and data (right) in 3 categories (top to
    bottom).}
  \label{fig:mu-gamma-pt}
\end{figure}

\begin{figure}[hbtp]
  \centering
  \includegraphics[width=0.42\textwidth]{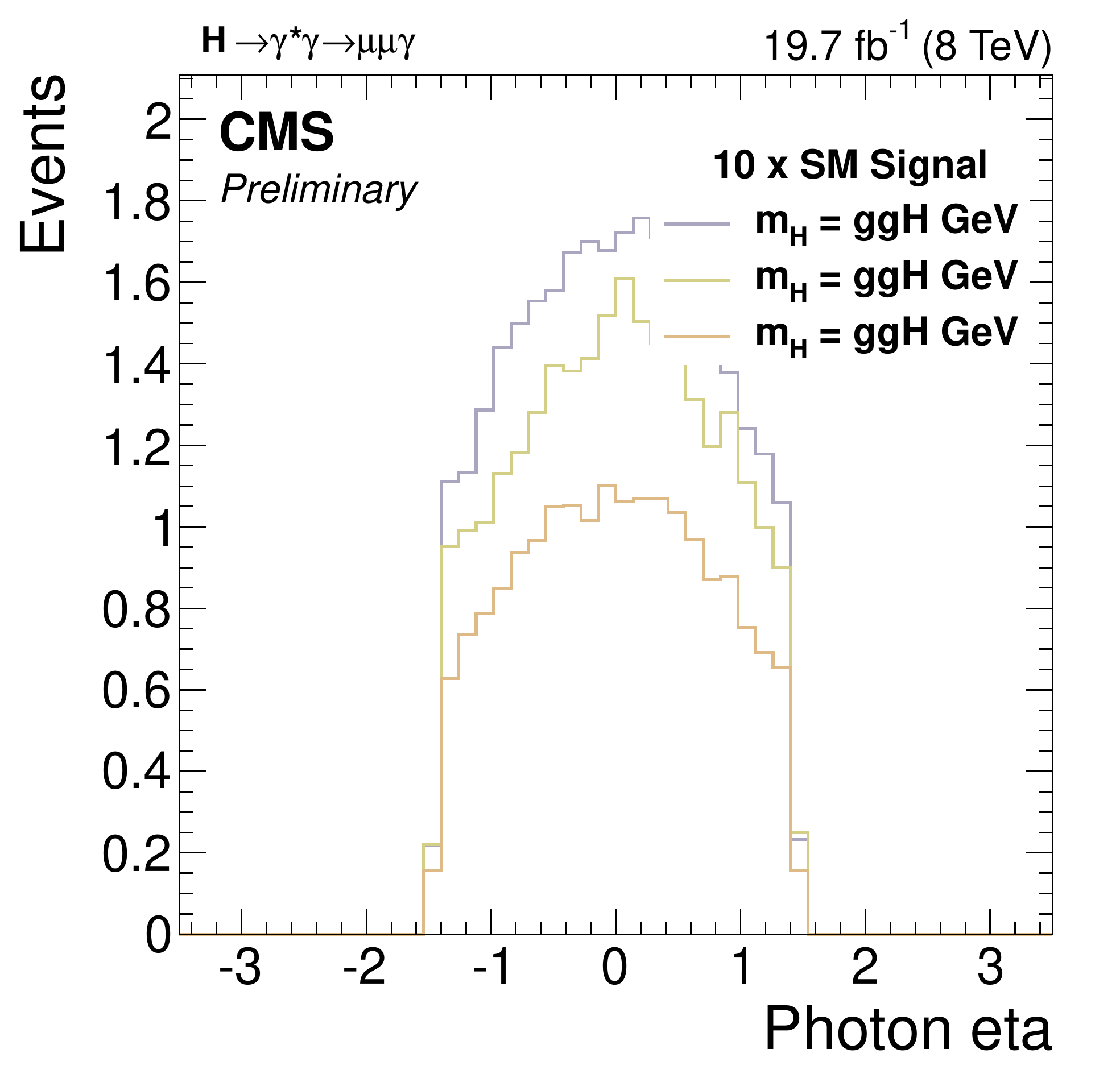}~
  \includegraphics[width=0.42\textwidth]{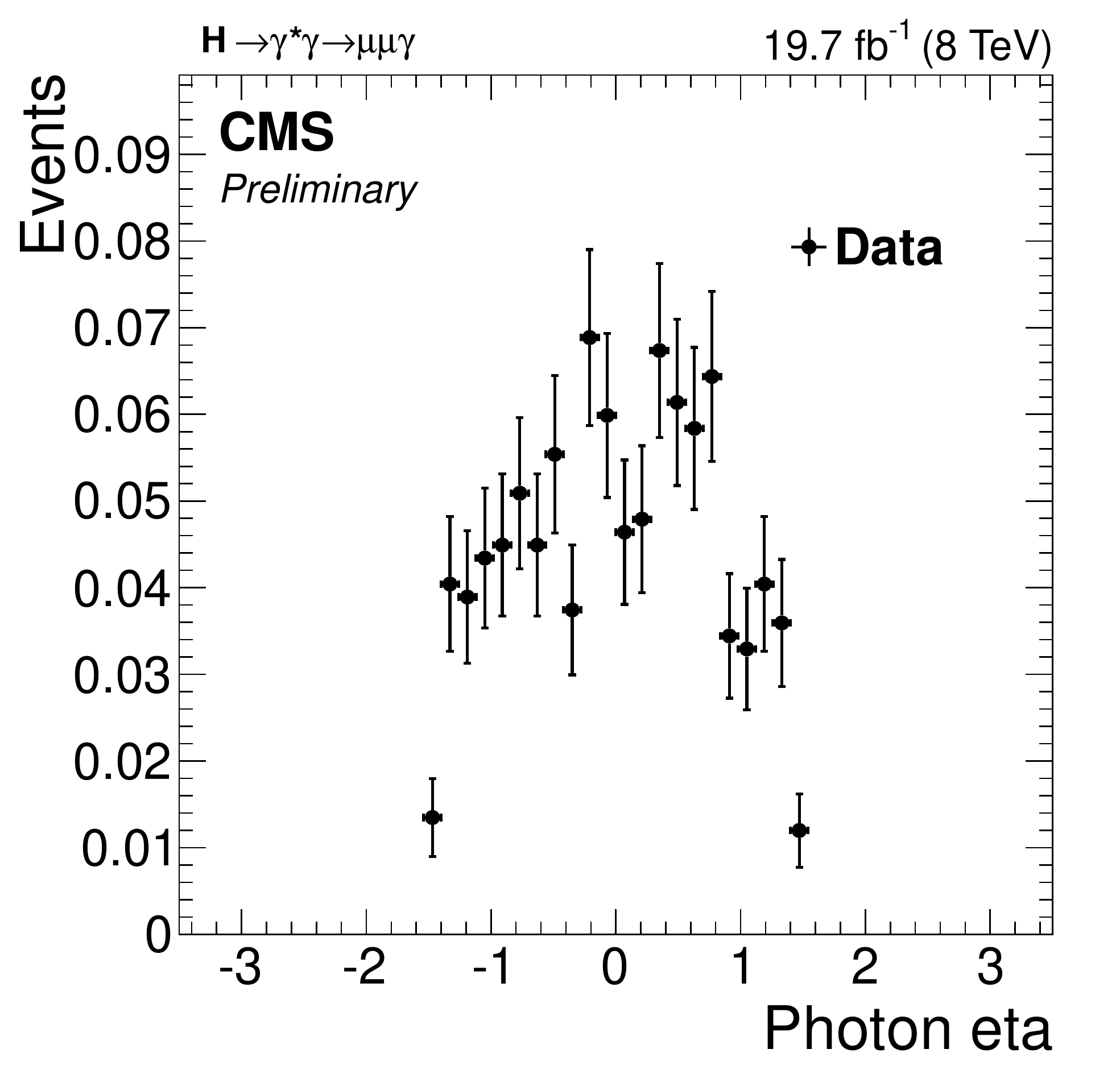}\\
  \includegraphics[width=0.42\textwidth]{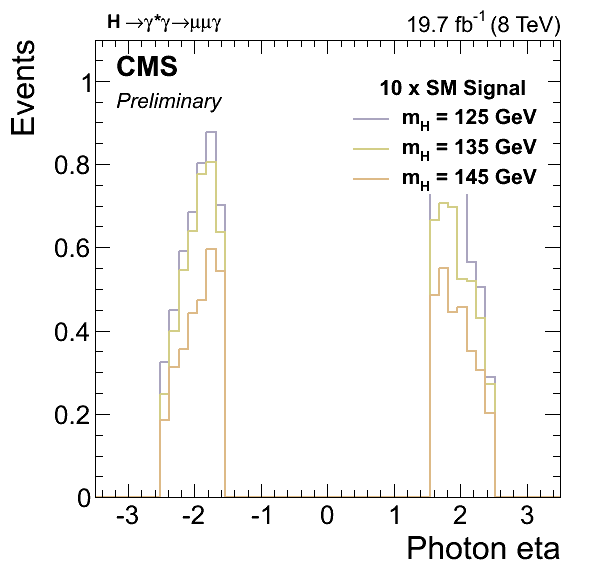}~
  \includegraphics[width=0.42\textwidth]{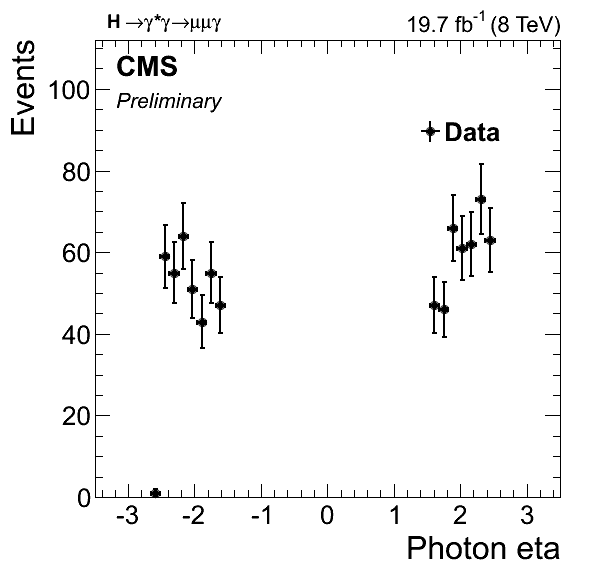}\\
  \includegraphics[width=0.42\textwidth]{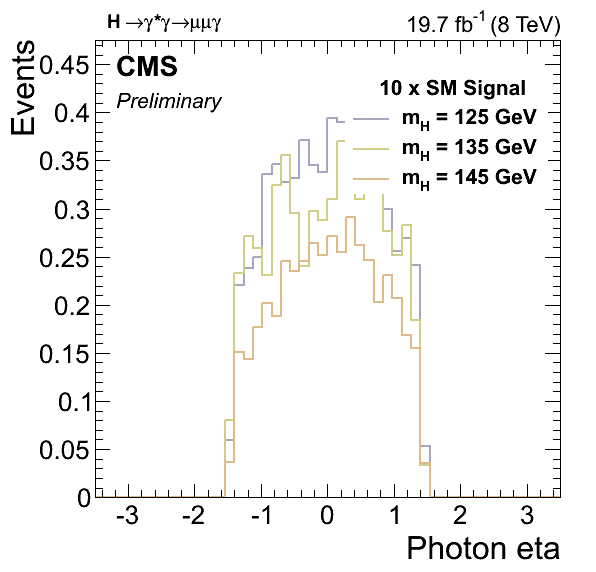}~
  \includegraphics[width=0.42\textwidth]{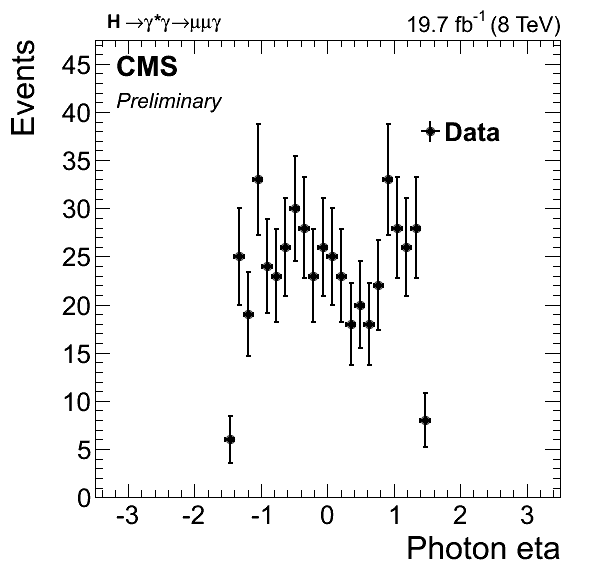}
  \caption[Distributions of $\eta^\gamma$ for signal and data.]{Distributions of
    $\eta^\gamma$ for ggF signal (left) and data (right) in 3 categories (top to bottom).}
  \label{fig:mu-gamma-eta}
\end{figure}

\begin{figure}[t]
  \centering
  \includegraphics[width=0.42\textwidth]{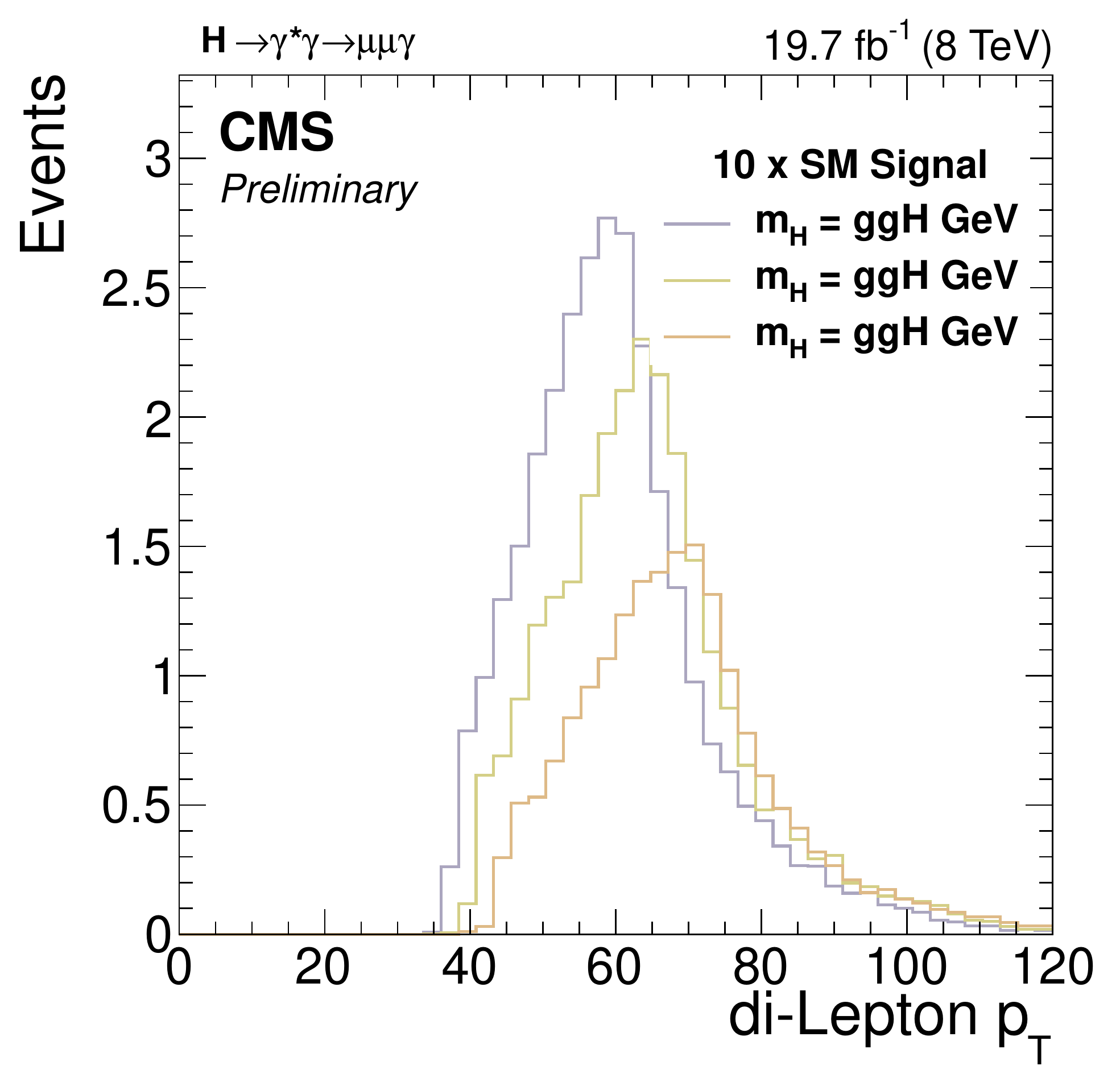}~
  \includegraphics[width=0.42\textwidth]{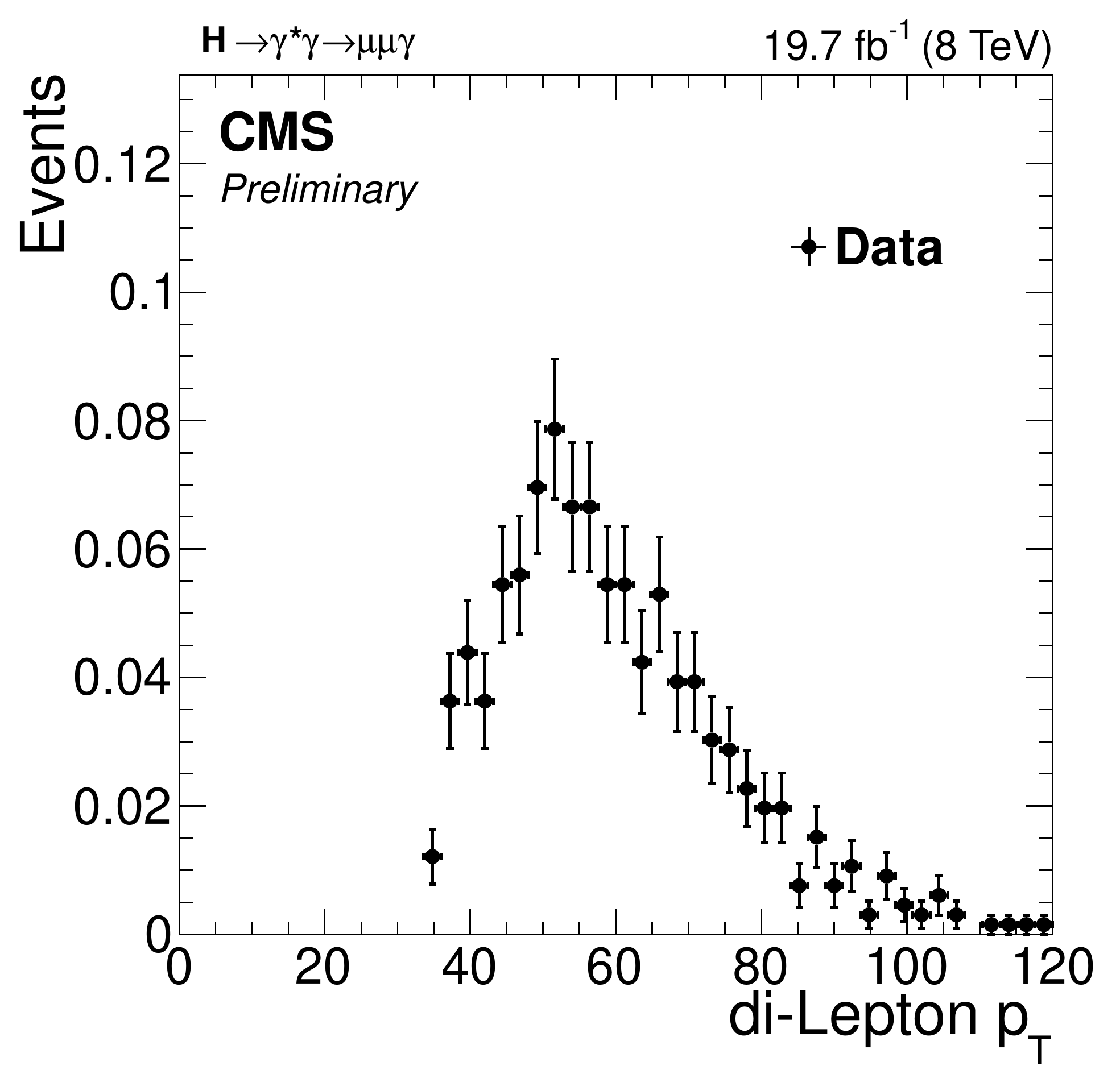}\\
  \includegraphics[width=0.42\textwidth]{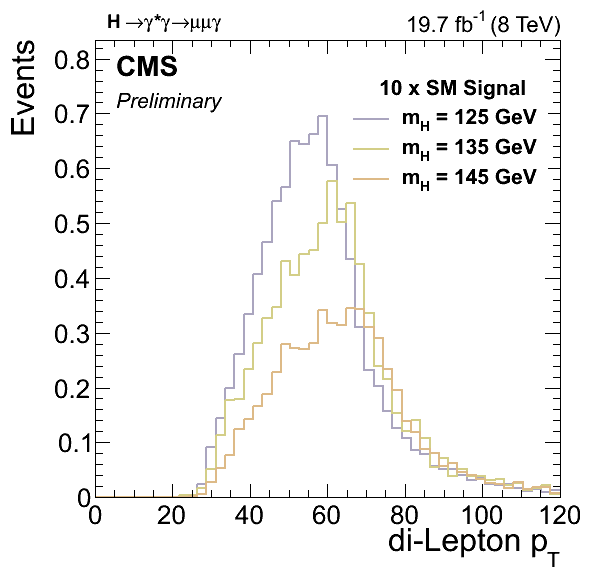}~
  \includegraphics[width=0.42\textwidth]{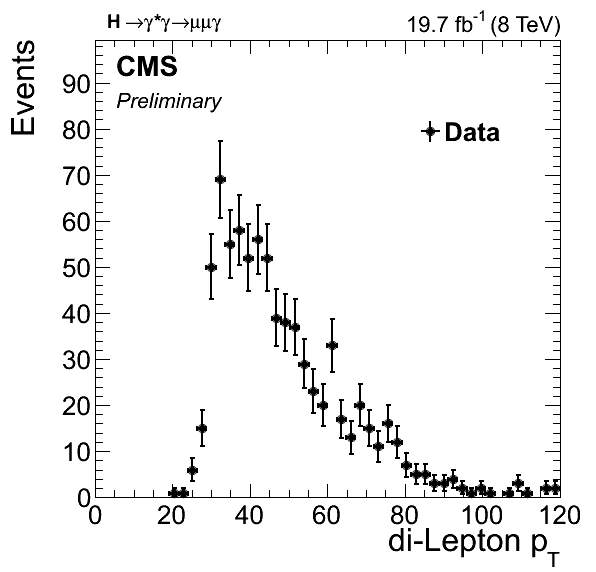}\\
  \includegraphics[width=0.42\textwidth]{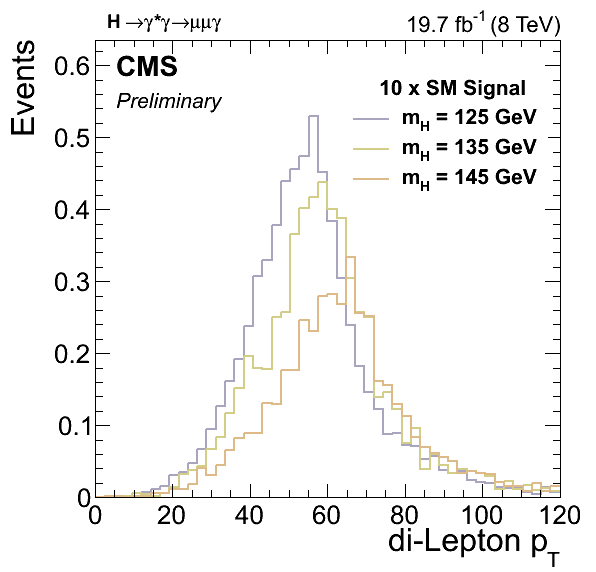}~
  \includegraphics[width=0.42\textwidth]{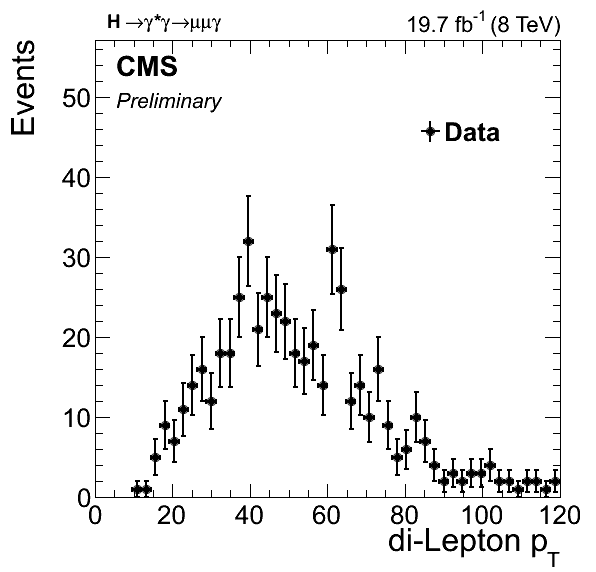}
  \caption[Distributions of the $p_T^{\mu\mu}$.]{Distributions of the dimuon transverse
    momentum, $p_T^{\mu\mu}$, in 3 categories (top to bottom).}
  \label{fig:mu-di}
\end{figure}

\begin{figure}[t]
  \centering
  \includegraphics[width=0.42\textwidth]{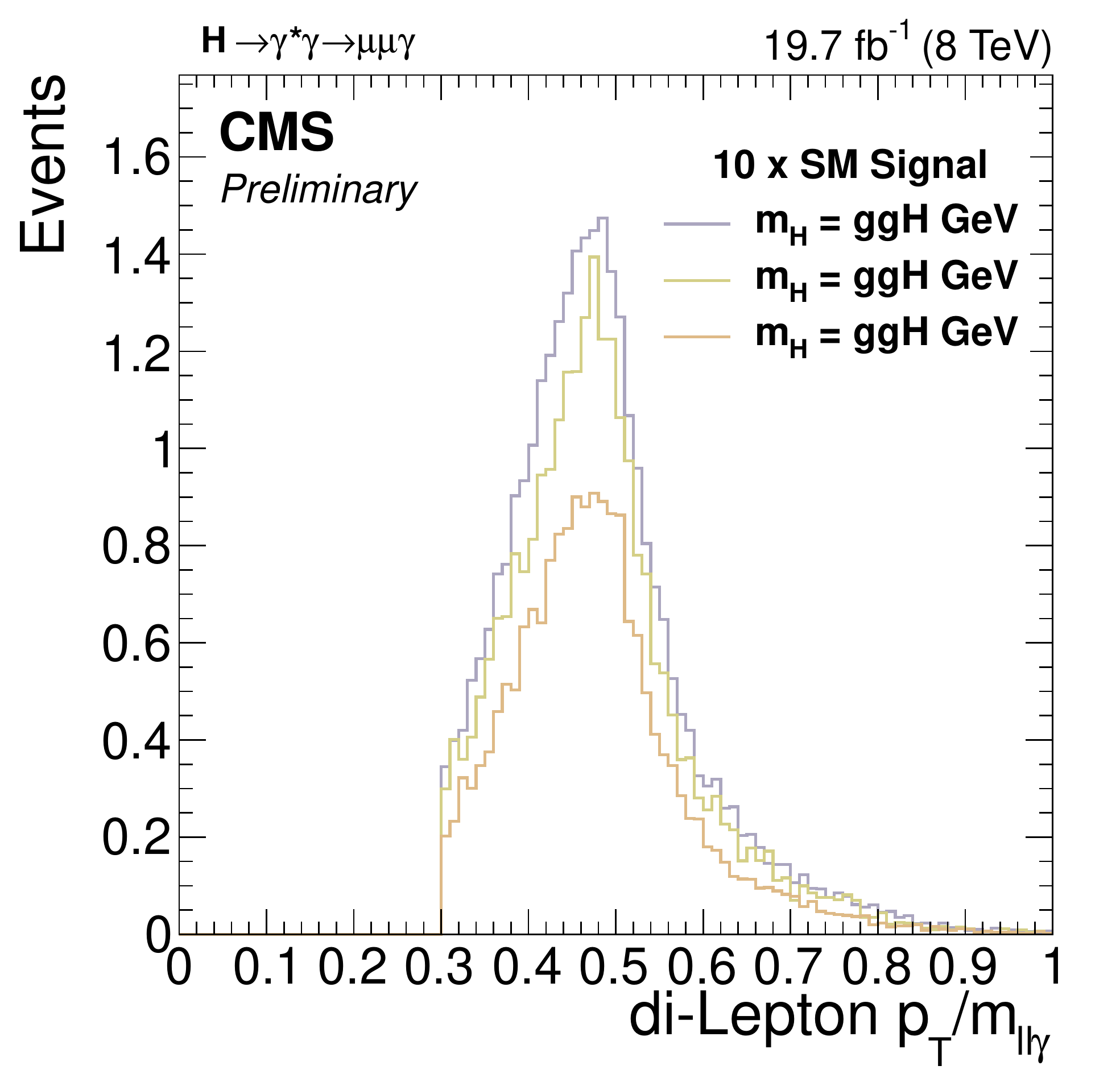}~
  \includegraphics[width=0.42\textwidth]{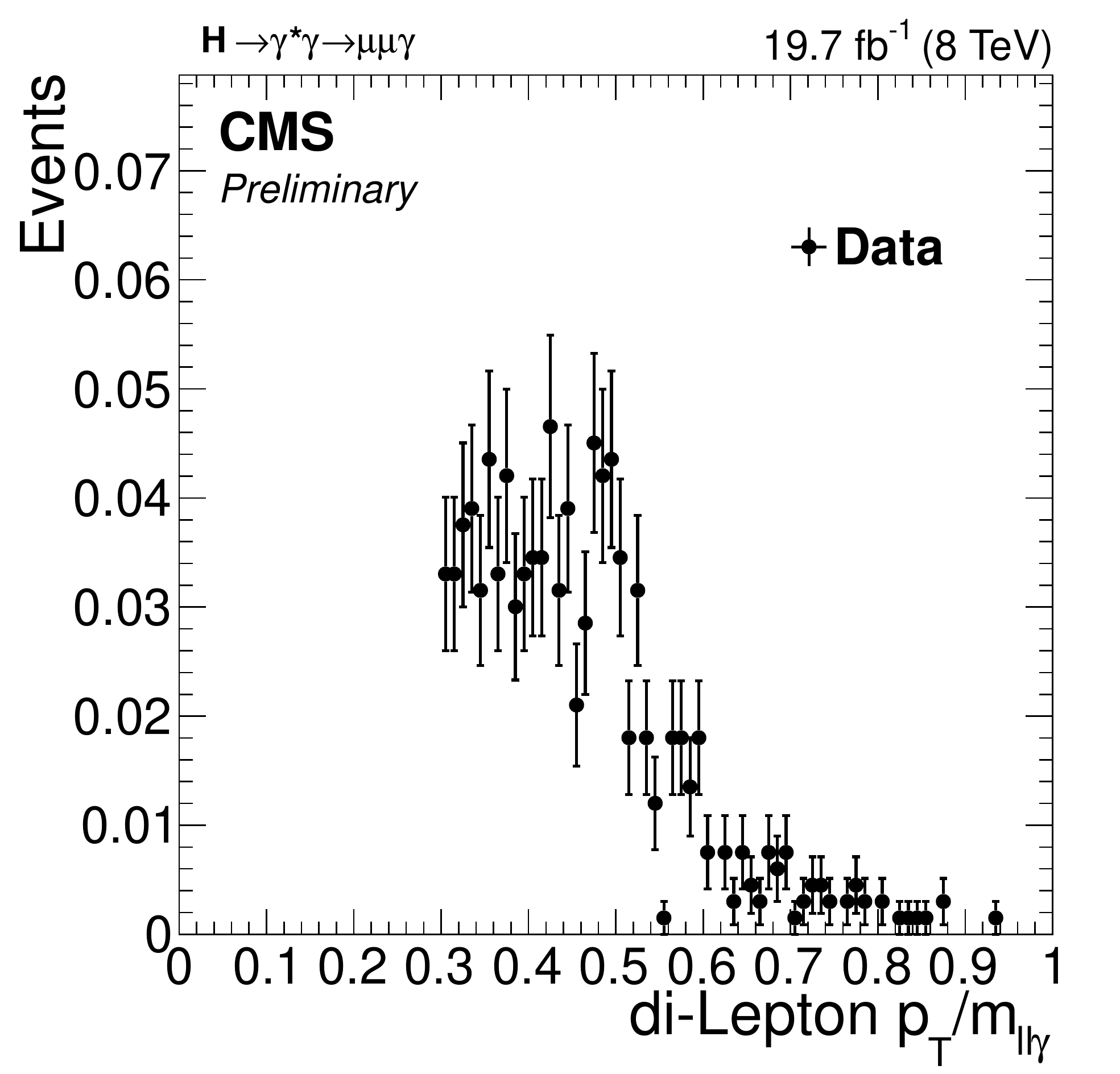}\\
  \includegraphics[width=0.42\textwidth]{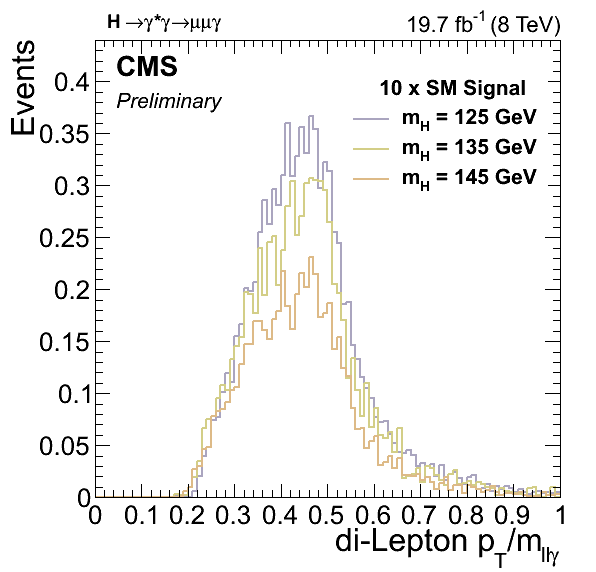}~
  \includegraphics[width=0.42\textwidth]{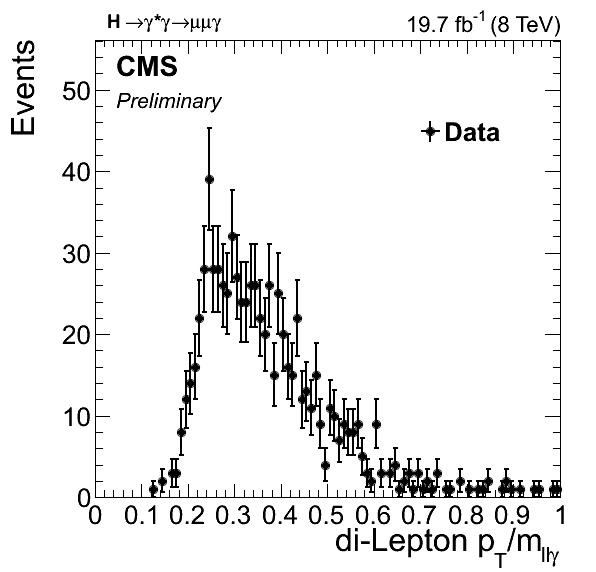}\\
  \includegraphics[width=0.42\textwidth]{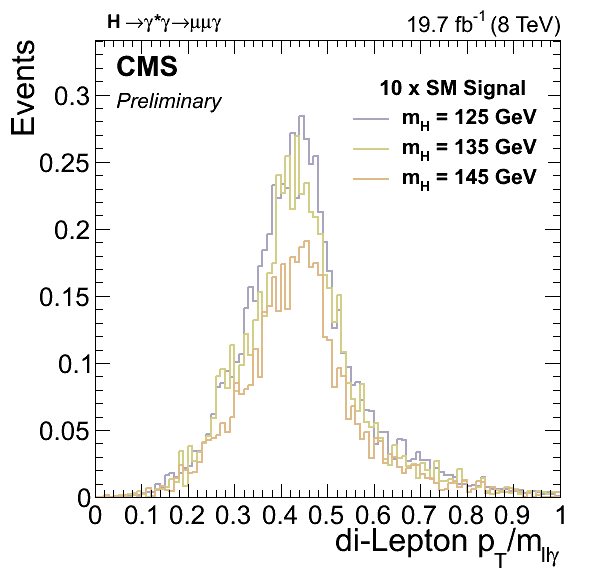}~
  \includegraphics[width=0.42\textwidth]{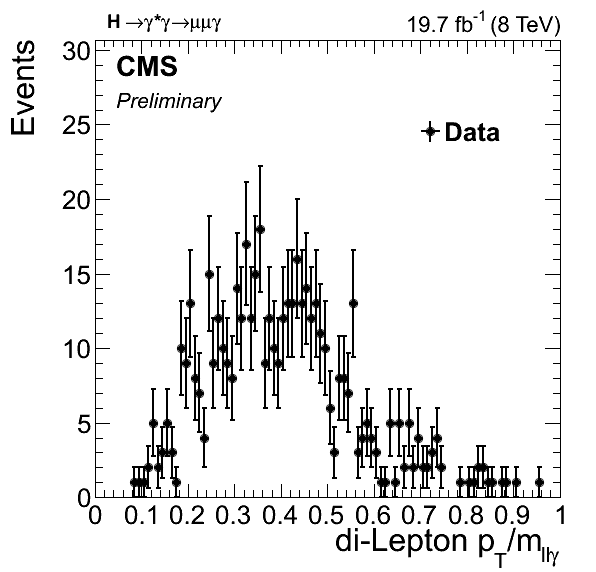}
  \caption[Distributions of the $p_T^{\mu\mu}/m_{\mu\mu\gamma}$.]{Distributions of the
    $p_T^{\mu\mu}/m_{\mu\mu\gamma}$, in 3 categories (top to bottom).}
  \label{fig:mu-di-scaled}
\end{figure}

\begin{figure}[t]
  \centering
  \includegraphics[width=0.42\textwidth]{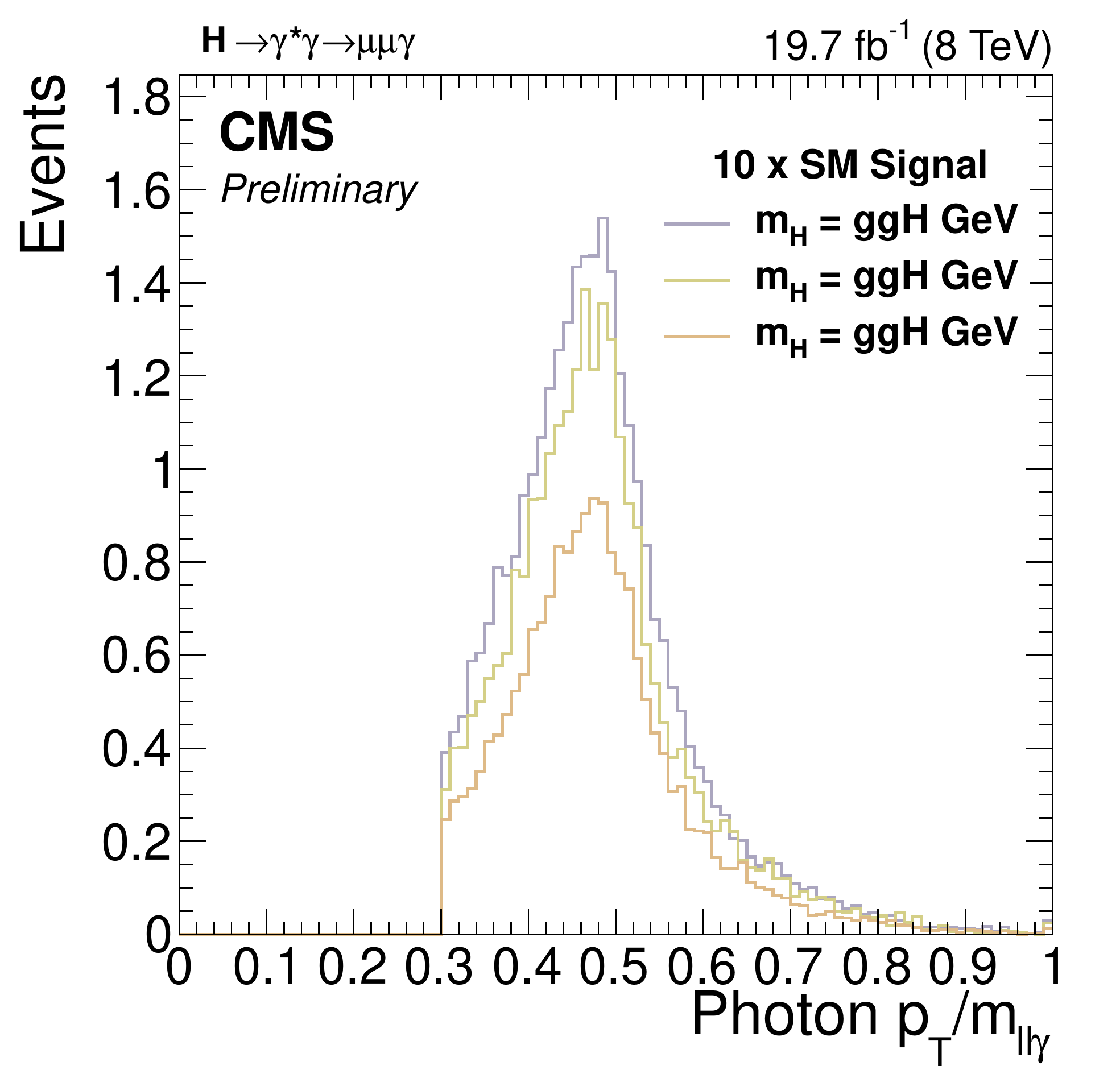}~
  \includegraphics[width=0.42\textwidth]{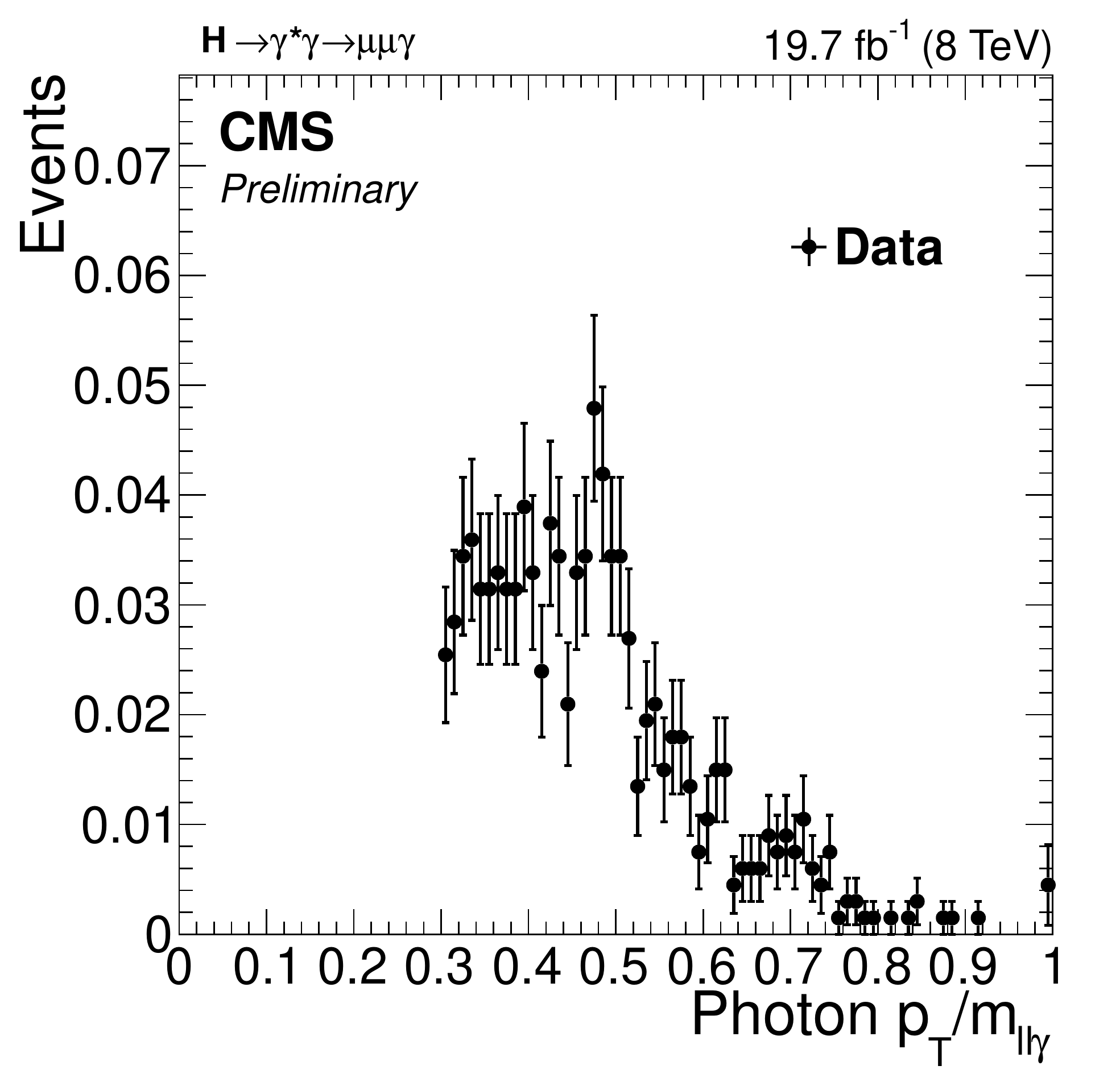}\\
  \includegraphics[width=0.42\textwidth]{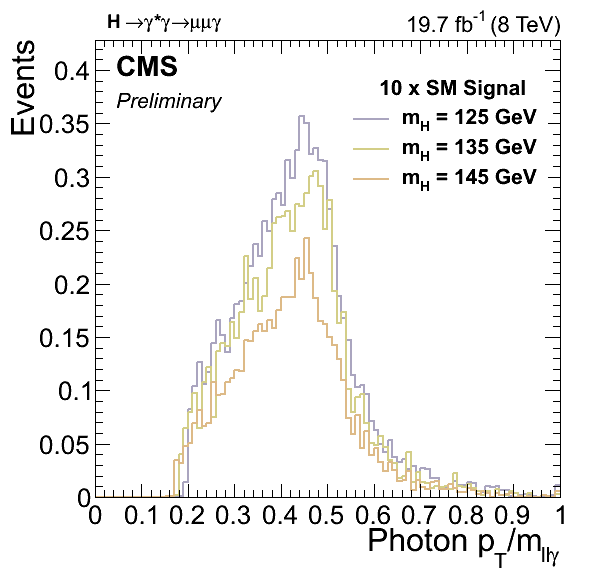}~
  \includegraphics[width=0.42\textwidth]{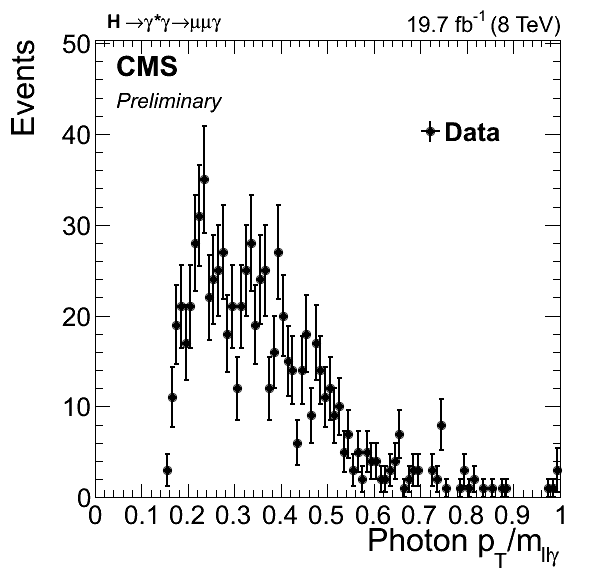}\\
  \includegraphics[width=0.42\textwidth]{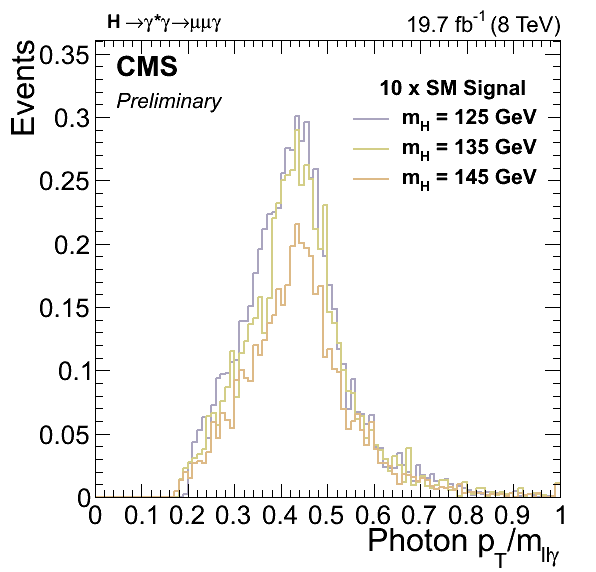}~
  \includegraphics[width=0.42\textwidth]{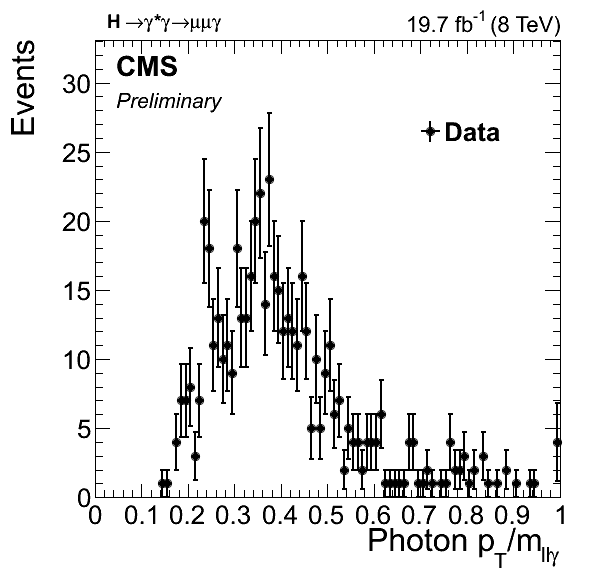}
  \caption[Distributions of the $p_T^{\gamma}/m_{\mu\mu\gamma}$.]{Distributions of the
    $p_T^{\gamma}/m_{\mu\mu\gamma}$, in 3 categories (top to bottom).}
  \label{fig:mu-gamma-scaled}
\end{figure}

\begin{figure}[t]
  \centering
  \includegraphics[width=0.42\textwidth]{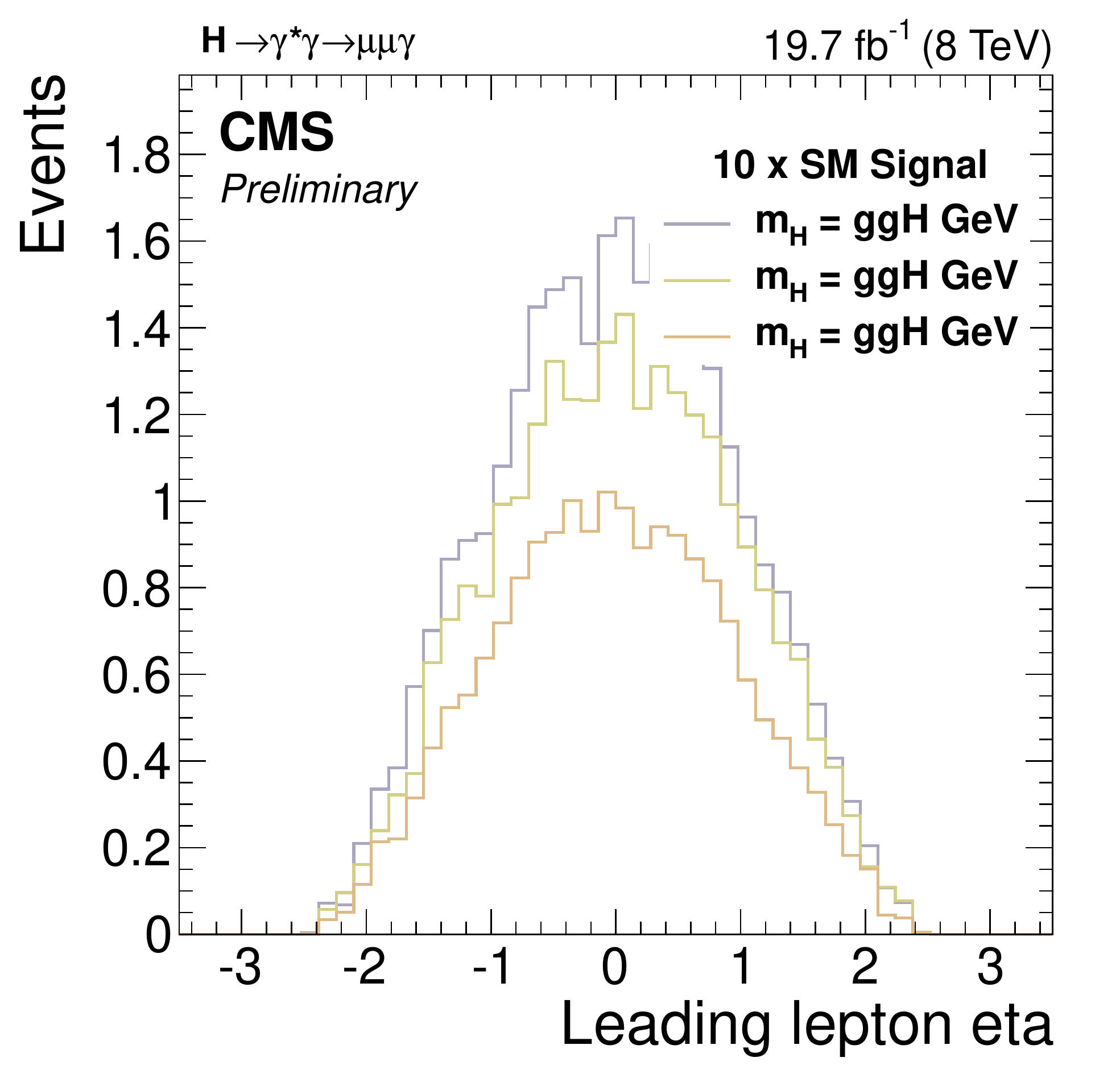}~
  \includegraphics[width=0.42\textwidth]{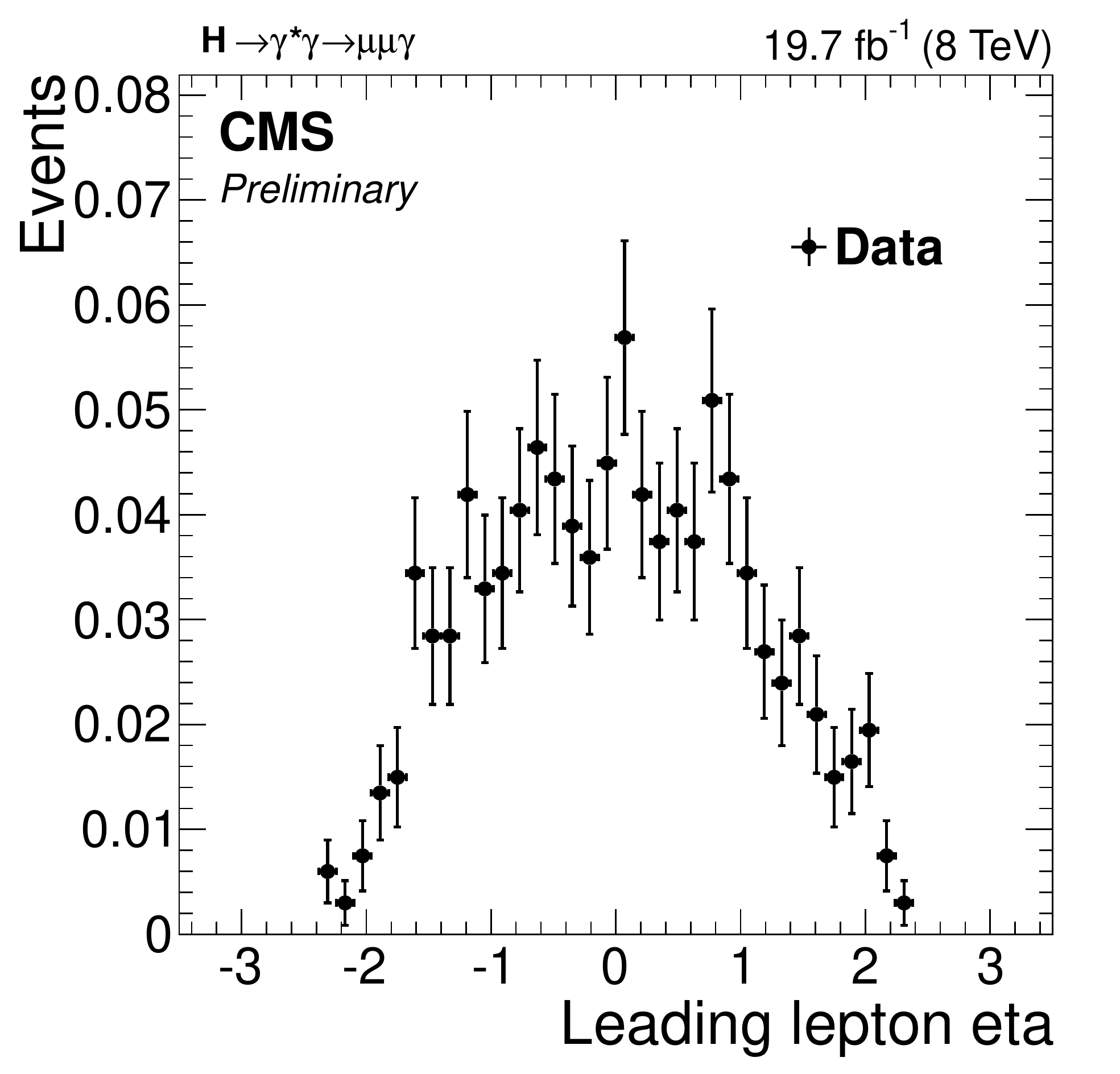}\\
  \includegraphics[width=0.42\textwidth]{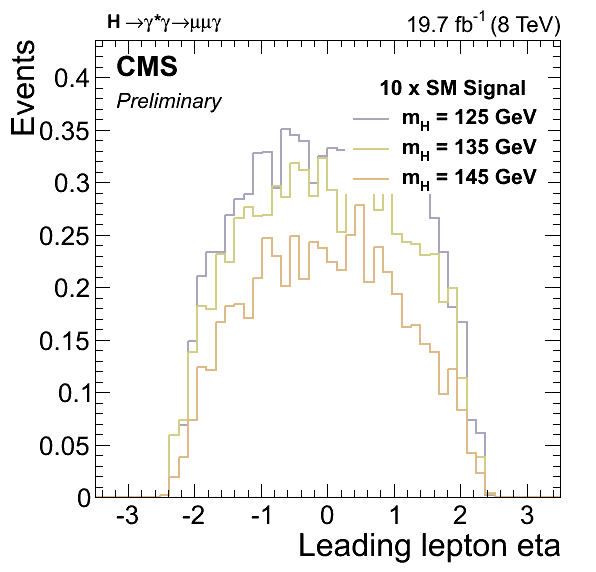}~
  \includegraphics[width=0.42\textwidth]{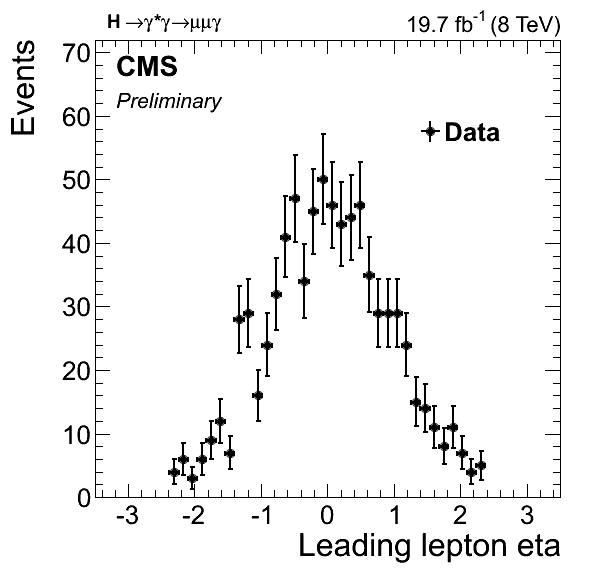}\\
  \includegraphics[width=0.42\textwidth]{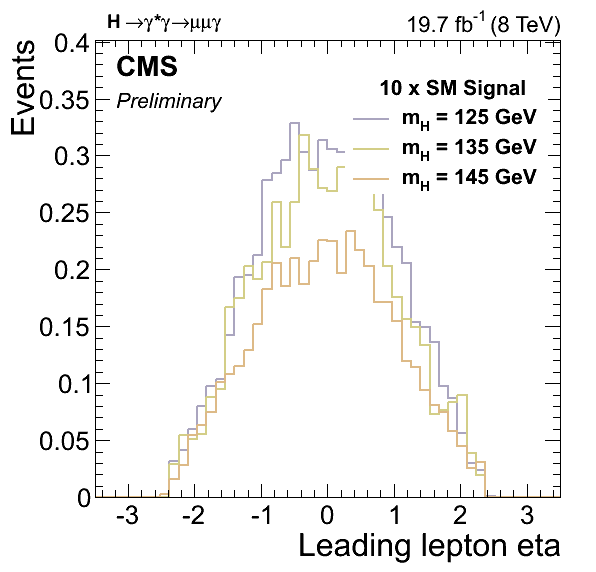}~
  \includegraphics[width=0.42\textwidth]{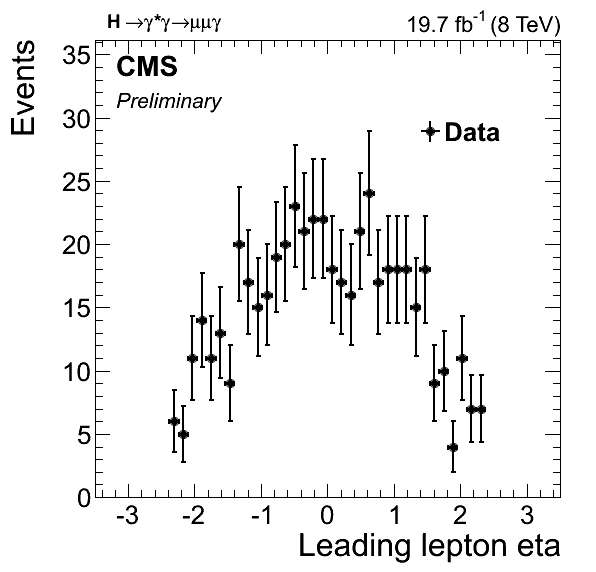}
  \caption[Distributions of the leading muon $\eta$.]{Distributions of the leading muon
    $\eta$, in 3 categories (top to bottom).}
  \label{fig:mu-leta1}
\end{figure}

\begin{figure}[t]
  \centering
  \includegraphics[width=0.42\textwidth]{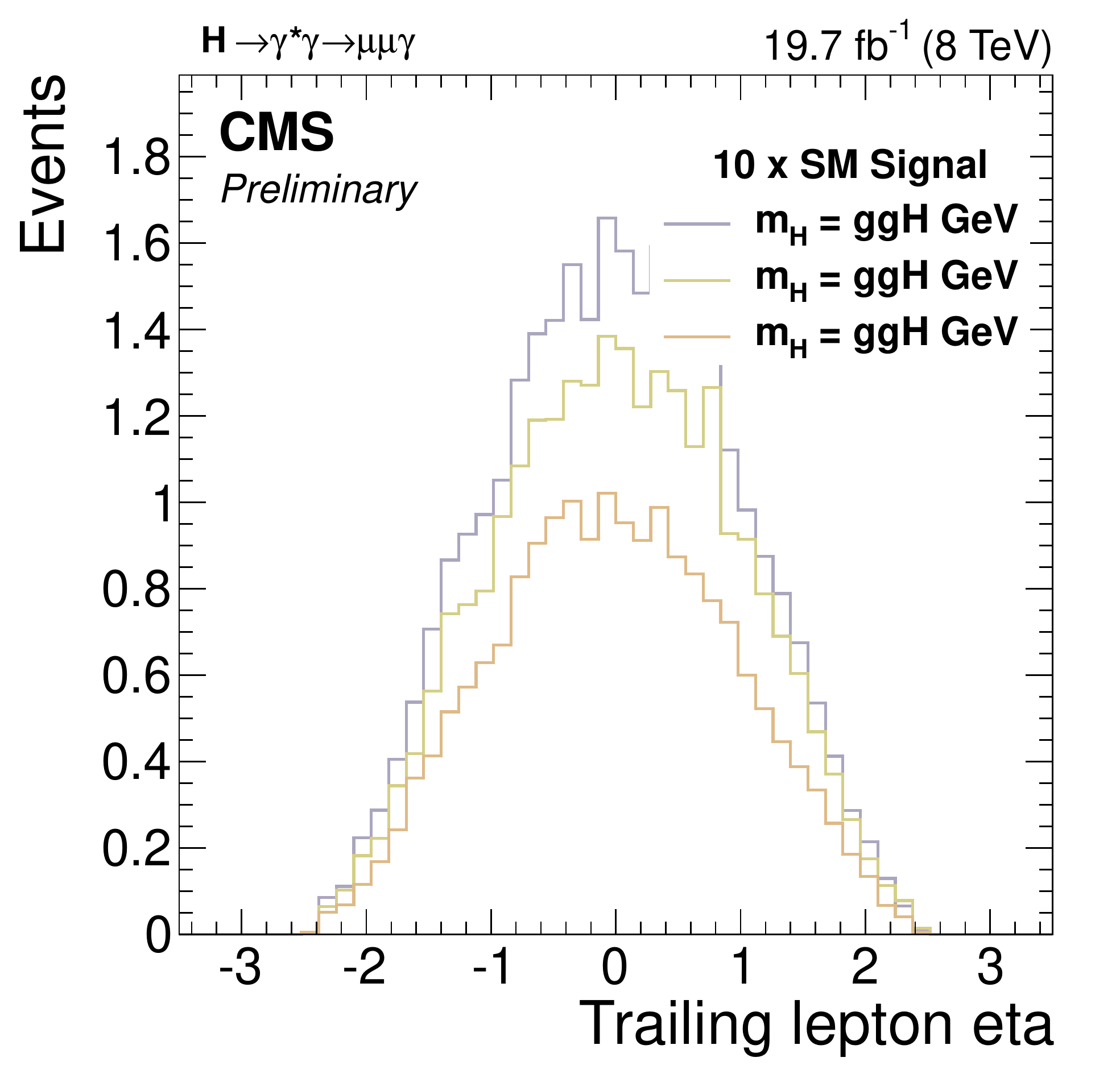}~
  \includegraphics[width=0.42\textwidth]{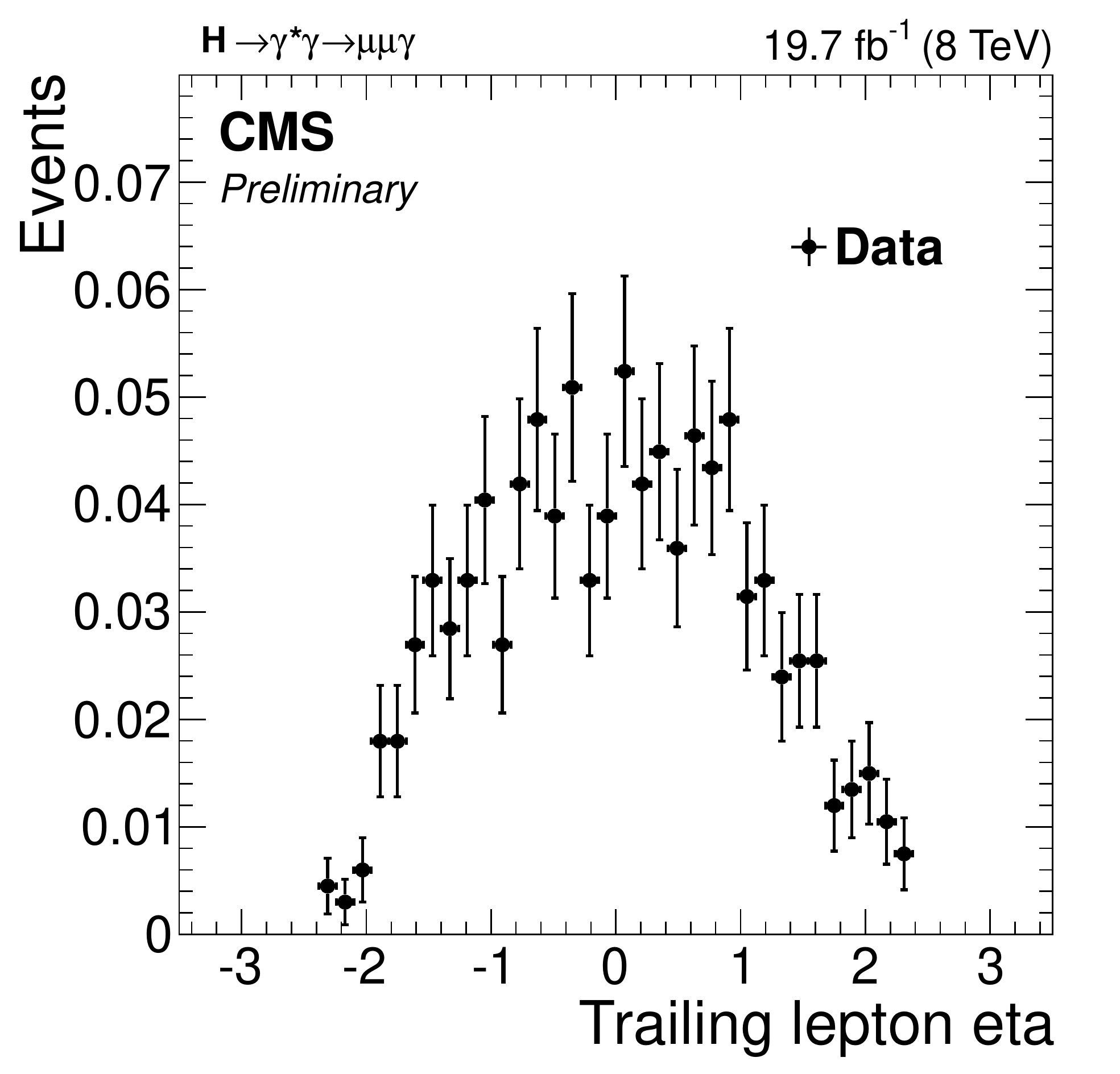}\\
  \includegraphics[width=0.42\textwidth]{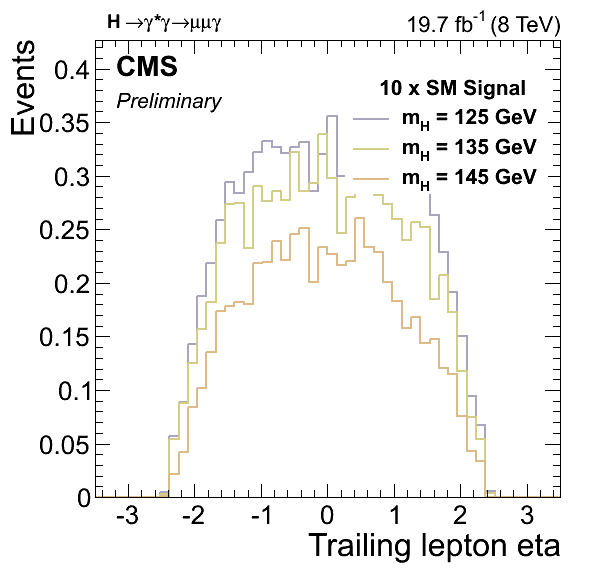}~
  \includegraphics[width=0.42\textwidth]{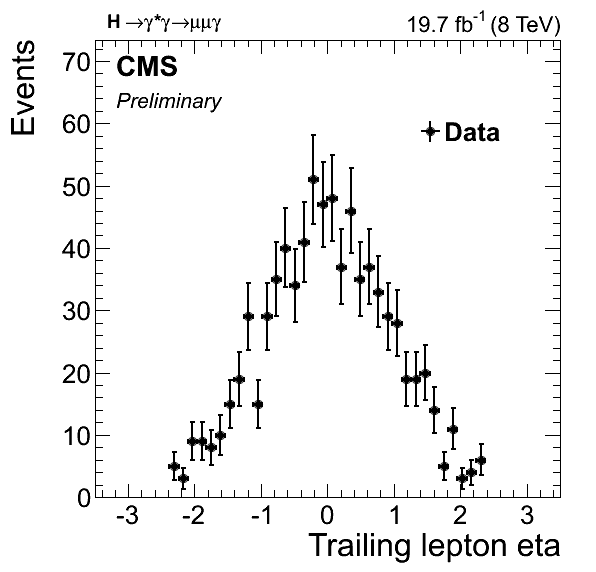}\\
  \includegraphics[width=0.42\textwidth]{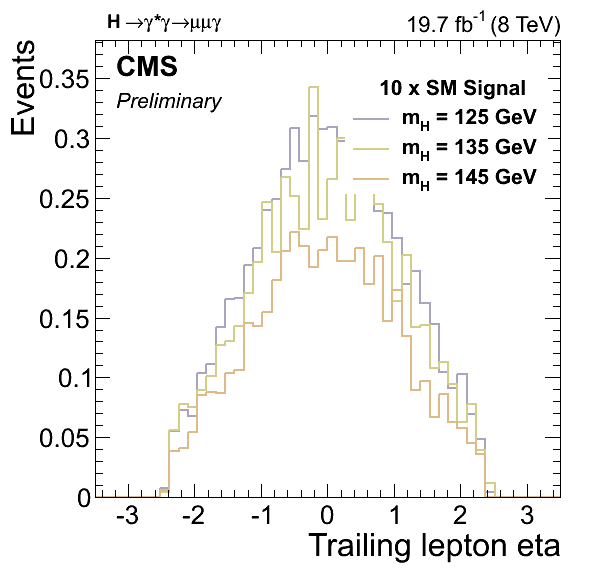}~
  \includegraphics[width=0.42\textwidth]{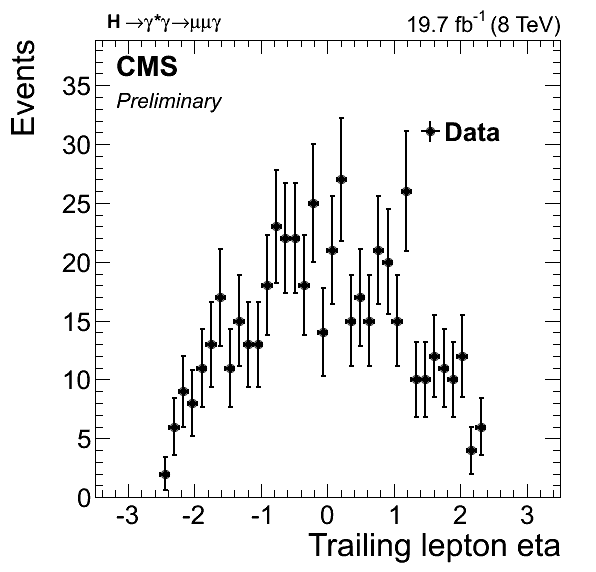}
  \caption[Distributions of the sub-leading muon $\eta$.]{Distributions of the sub-leading
    muon $\eta$, in 3 categories (top to bottom).}
  \label{fig:mu-leta2}
\end{figure}

\begin{figure}[t]
  \centering
  \includegraphics[width=0.42\textwidth]{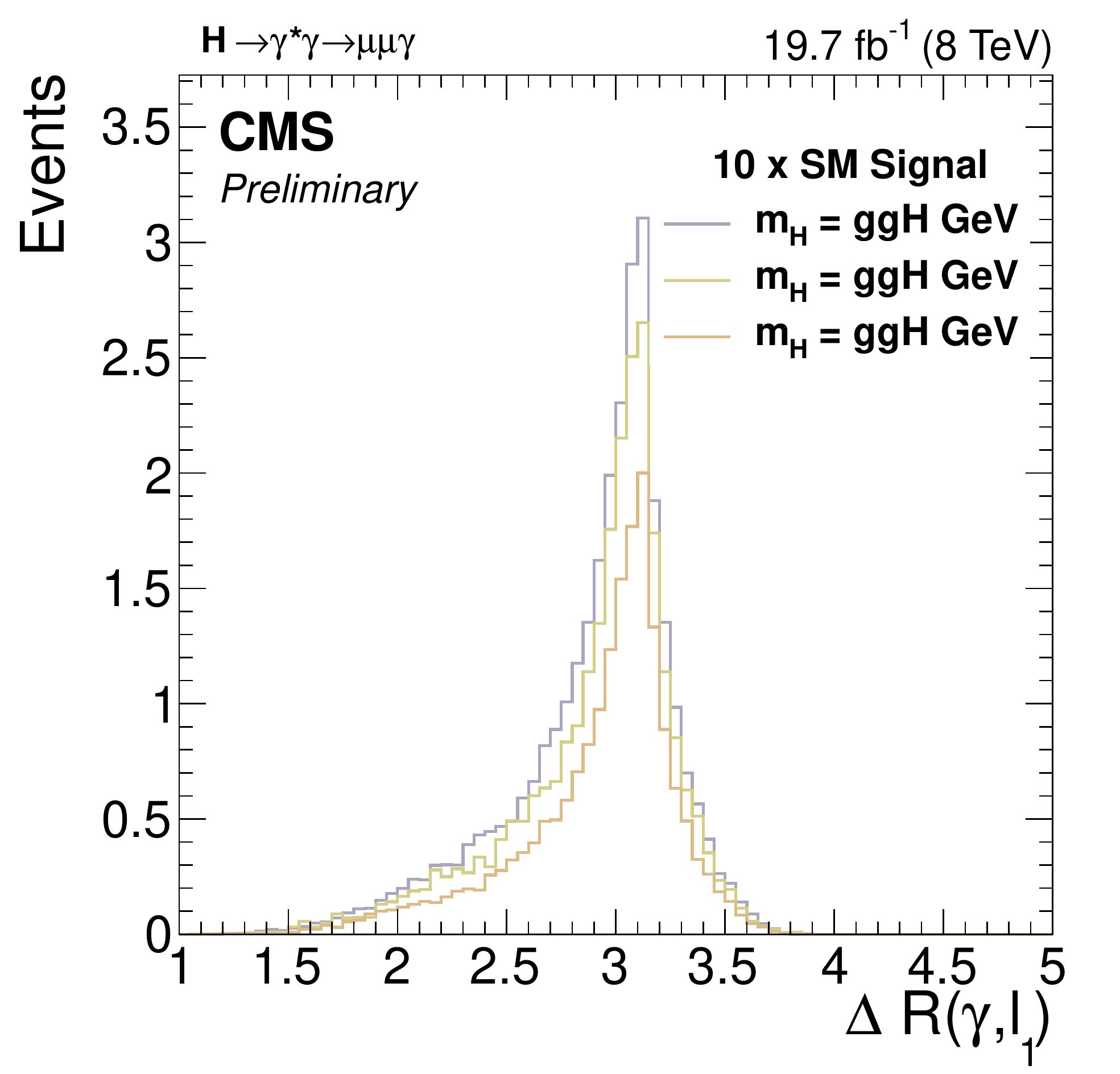}~
  \includegraphics[width=0.42\textwidth]{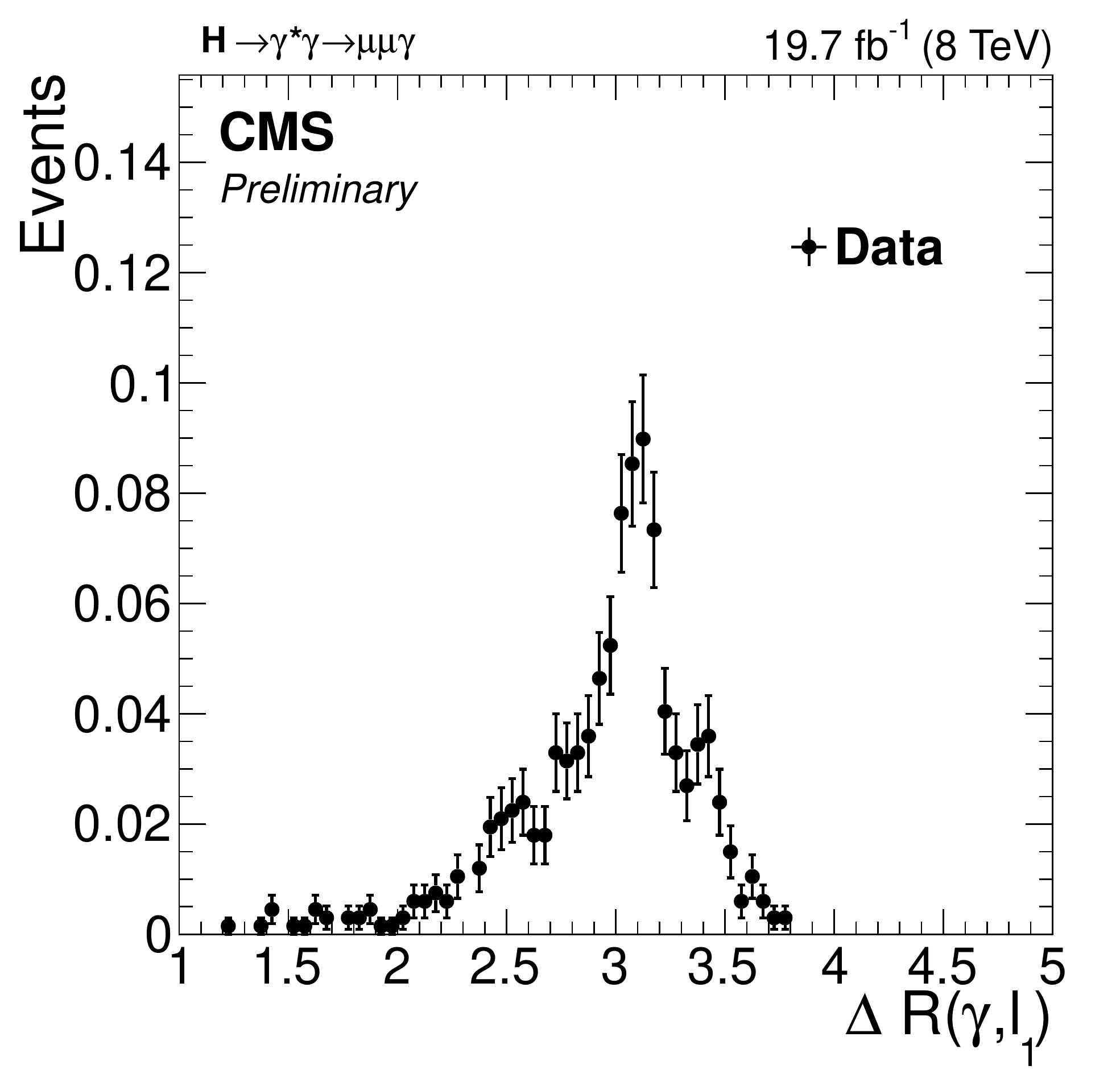}\\
  \includegraphics[width=0.42\textwidth]{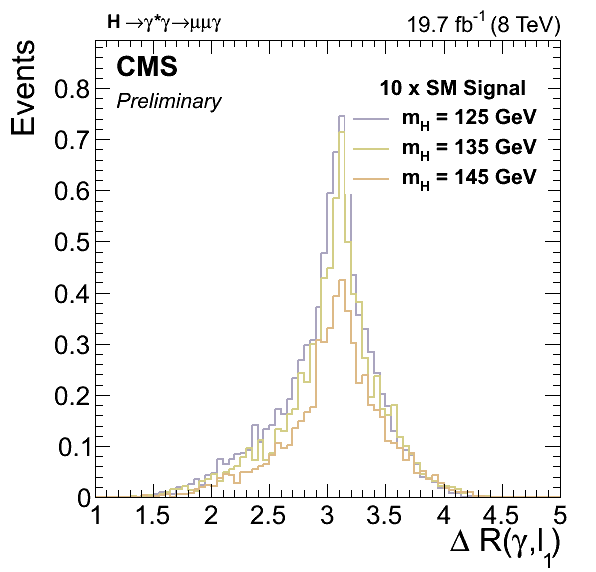}~
  \includegraphics[width=0.42\textwidth]{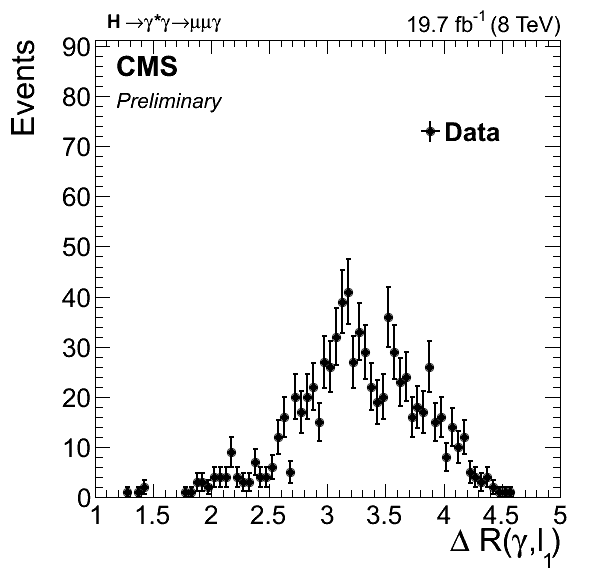}\\
  \includegraphics[width=0.42\textwidth]{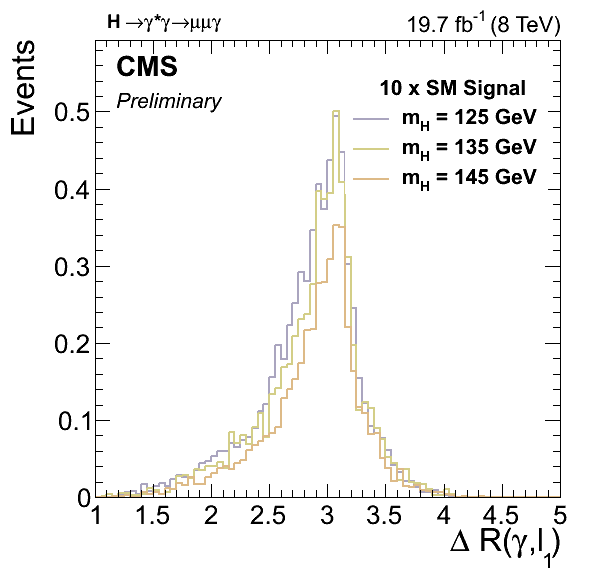}~
  \includegraphics[width=0.42\textwidth]{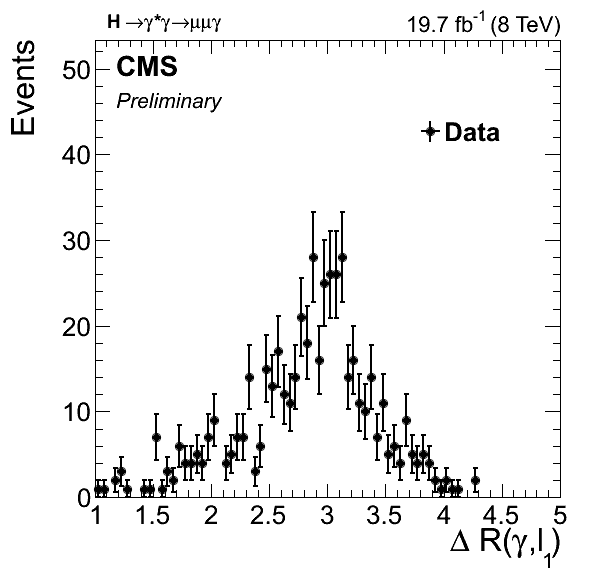}
  \caption{Distributions of the $\Delta R_{\eta\phi}$ between the leading muon and the
    photon.}
  \label{fig:mu-dR-lg}
\end{figure}

Figures~\ref{fig:mu-mll} show the key $m_{\mu\mu}$ distributions.  They are plotted after
the full selection in three categories denoted by 1 -- \textit{EB}, 2 -- \textit{EE}, 3 --
\textit{mll50} in the Table~\ref{tab:yield-mu}.  Figures
\ref{fig:mu-dR}--\ref{fig:mu-dR-lg} show additional kinematic distributions relevant in
the analysis also split in those three categories.  Signal MC distributions (left) from
gluon fusion sample are to be compared with data (right), which should be thought of as
background.

\begin{table}[ht]
  \caption[Event yield after each selection criteria for data and signal with $m_\PH = 125\GeV$ for $L = 19.7\fbinv$ in the muon channel.]
  {Event yield after each selection criteria for data and signal with $m_\PH = 125\,\GeV$
    for $L = 19.7\,\fbinv$ in the muon channel. Three independent categories of events used in the analysis are also marked.}
  \label{tab:yield-mu}
\begin{center}
\newcommand{\cutOne}{}
\newcommand{\cutTwo}{cut2}
\newcommand{\cutThree}{Pass Trigger and Photon selection}
\newcommand{\cutFour}{Two muons selected}
\newcommand{\cutFive}{$110 < m_{\mu\mu\gamma} < 170$; $m_{\mu\mu} < 50~\GeV$}
\newcommand{\cutSix}{$\Delta R(\mu, \gamma) > 1$; removed \JPsi, $\Upsilon$}
\newcommand{\cutSeven}{\multicolumn{2}{c|}{$m_{\mu\mu} < 20~\GeV$}}
\newcommand{\cutEight}{$|\eta^{\gamma}_{SC}| < 1.4442$}
\newcommand{\cutNine}{$q_T^{\mu\mu}/m_{\mu\mu\gamma} > 0.3$, $E_T^{\gamma}/m_{\mu\mu\gamma} > 0.3$}
\newcommand{\cutTen}{$122 < m_{\mu\mu\gamma} < 128~\GeV$}
\newcommand{\cutFifteen}{$1.566 < |\eta^{\gamma}_{SC}| < 2.5$  }
\newcommand{\cutSixteen}{$q_T^{\mu\mu}/m_{\mu\mu\gamma} > 0.3$, $E_T^{\gamma}/m_{\mu\mu\gamma} > 0.3$}
\newcommand{\cutSeventeen}{$122~\GeV < m_{\mu\mu\gamma} < 128~\GeV$}
\newcommand{\cutEighteen}{\multicolumn{2}{c|}{$20~\GeV < m_{\mu\mu} < 50~\GeV$; $|\eta^{\gamma}_{SC}| < 1.4442$}}
\newcommand{\cutNineteen}{$q_T^{\mu\mu}/m_{\mu\mu\gamma} > 0.3$, $E_T^{\gamma}/m_{\mu\mu\gamma} > 0.3$}
\newcommand{\cutTwenty}{$122 < m_{\mu\mu\gamma} < 128~\GeV$ }

\begin{tabular}{r|c|r|c|ccc}
Category & Selection creteria & Data & Total signal & ggH & vbfH & VH \\ \hline
                    & \cutThree & 1.2M & 7.86 & 6.90 & 0.57 & 0.39 \\
                    & \cutFour  & 79K  & 5.89 & 5.16 & 0.43 & 0.29 \\
                    & \cutFive  & 3196 & 5.85 & 5.13 & 0.43 & 0.28 \\
                    & \cutSix   & 2662 & 5.59 & 4.93 & 0.40 & 0.26 \\
\hline
\multicolumn{7}{c}{}\\
                      \cutSeven     & 1822 & 4.68 & 4.12 & 0.33 & 0.22 \\
\hline
\hline
                    & \cutEight     &  3.67 & 3.24 & 0.27 & 0.16 \\
(1)  \textit{EB}    & \cutNine      &  3.28 & 2.92 & 0.22 & 0.14 \\
                    & \cutTen       &  2.97 & 2.64 & 0.20 & 0.12 \\
\hline
                    & \cutFifteen   & 793 & 1.00 & 0.88 & 0.07 & 0.06 \\
(2) \textit{EE}     & \cutSixteen   & 347 & 0.81 & 0.71 & 0.05 & 0.04 \\
                    & \cutSeventeen & 57  & 0.58 & 0.51 & 0.04 & 0.03 \\
\hline
\multicolumn{7}{c}{}\\
                      \cutEighteen & 512 & 0.72 & 0.63 & 0.05 & 0.03 \\
\hline\hline
(3)  \textit{mll50} & \cutNineteen & 299 & 0.57 & 0.51 & 0.03 & 0.02 \\
                    & \cutTwenty   & 47 & 0.51 & 0.46 & 0.03 & 0.02 \\

\end{tabular}

\end{center}
\end{table}

\clearpage

\clearpage

\clearpage

\clearpage

\clearpage

\clearpage

\clearpage
\begin{figure}[t]
  \centering
  \includegraphics[width=0.42\textwidth]{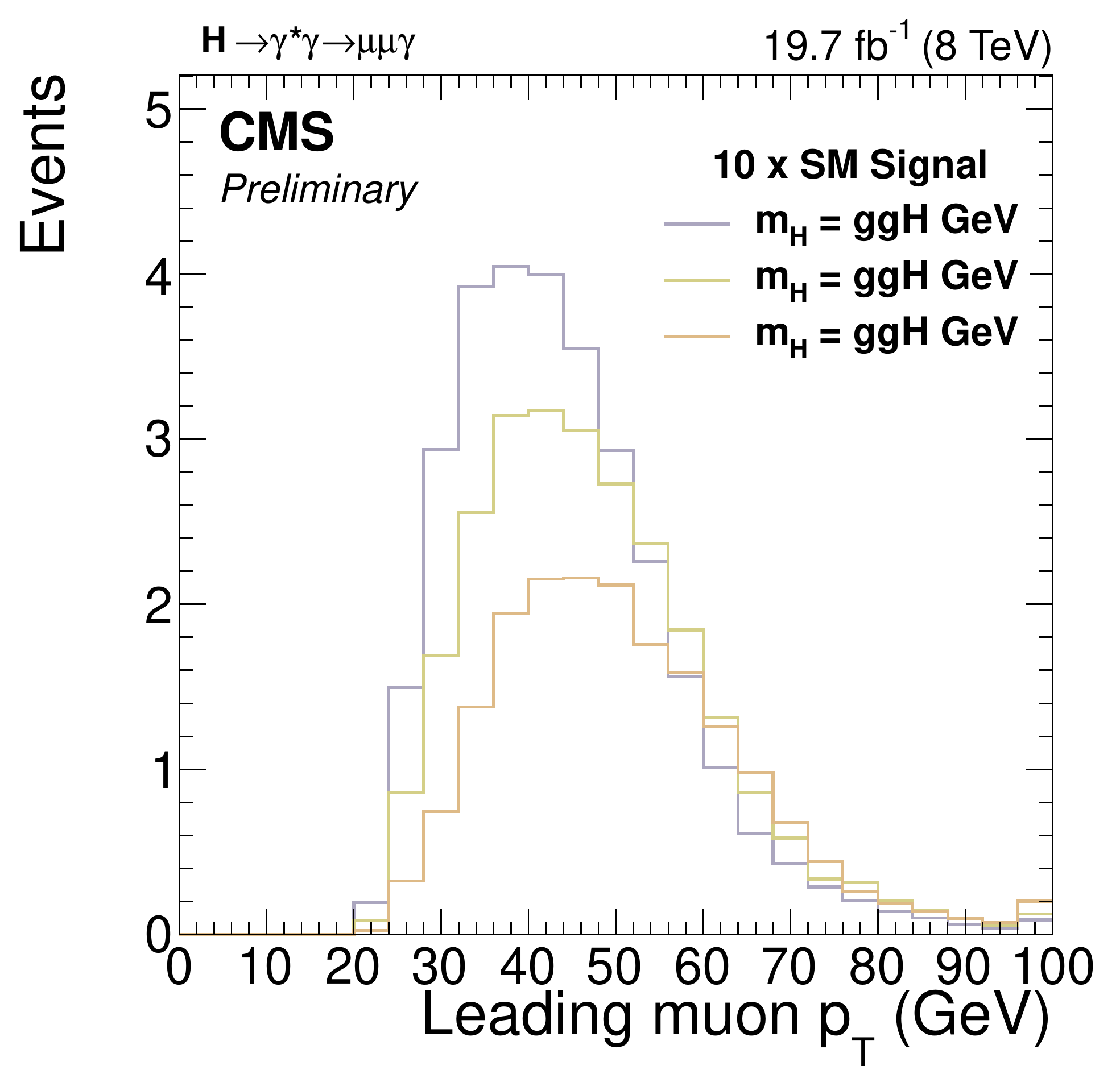}~
  \includegraphics[width=0.42\textwidth]{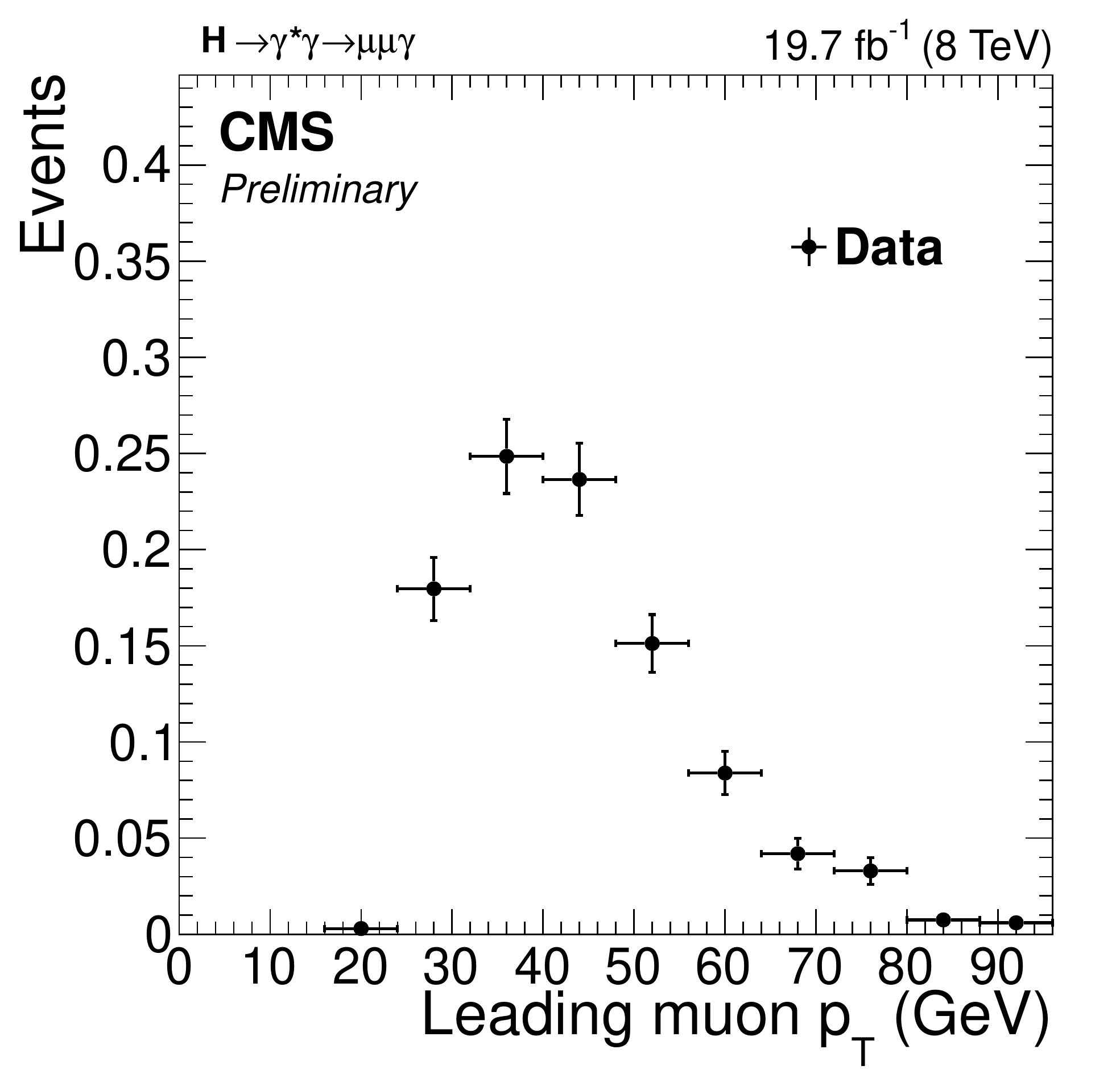}\\
  \includegraphics[width=0.42\textwidth]{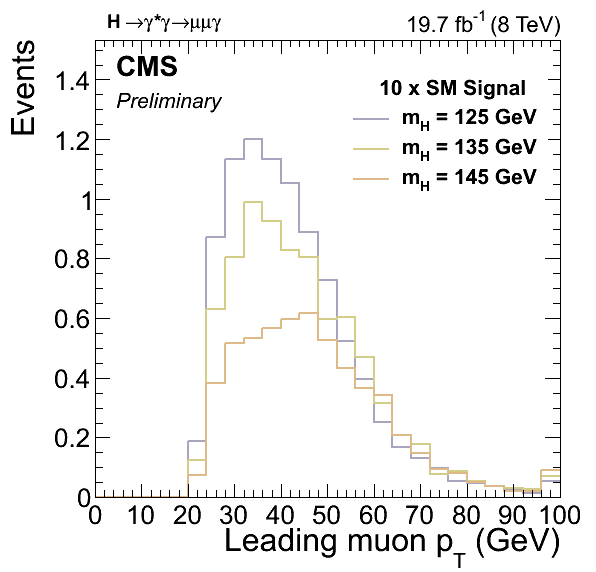}~
  \includegraphics[width=0.42\textwidth]{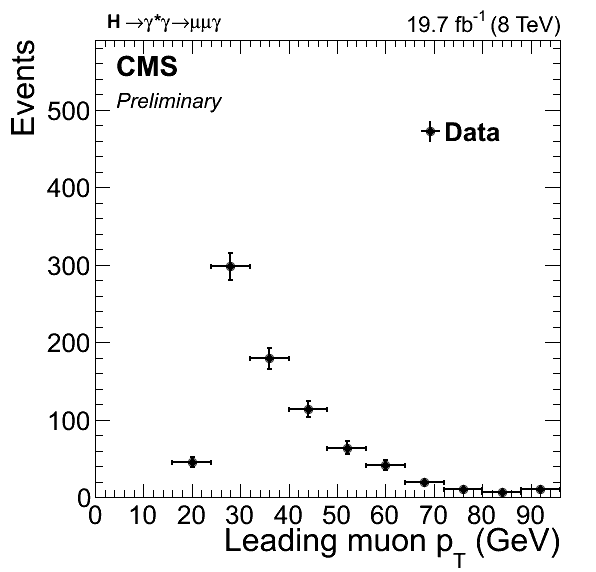}\\
  \includegraphics[width=0.42\textwidth]{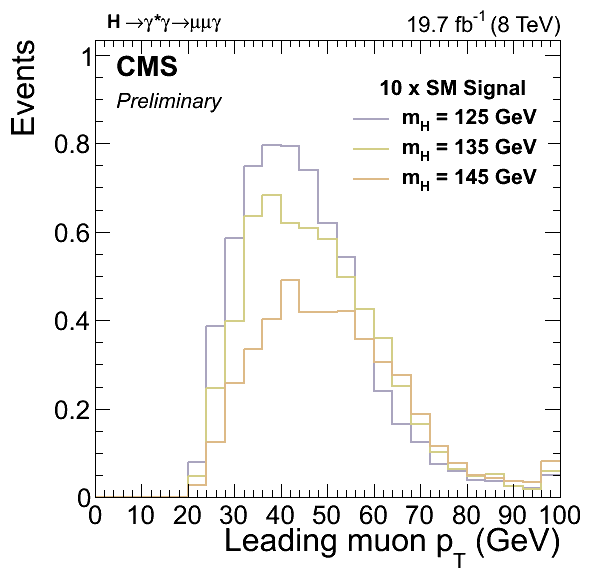}~
  \includegraphics[width=0.42\textwidth]{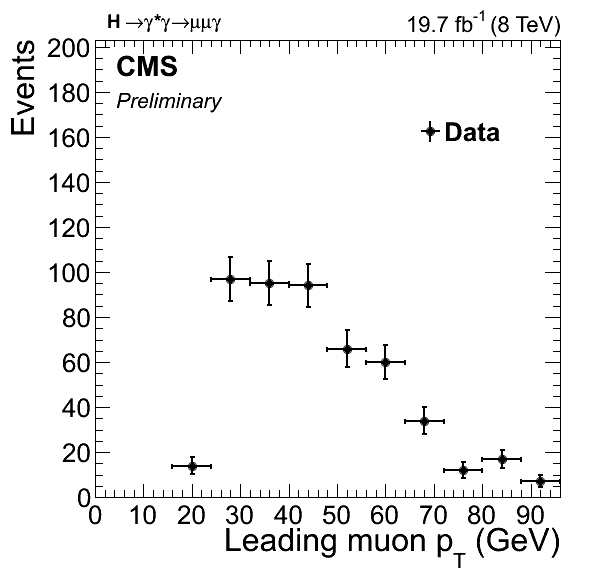}
  \caption[Distributions of the leading muon $\PT$.]{Distributions of the leading muon
    $\PT$, in 3 categories (top to bottom).}
  \label{fig:mu-lpt1}
\end{figure}

\clearpage
\begin{figure}[t]
  \centering
  \includegraphics[width=0.42\textwidth]{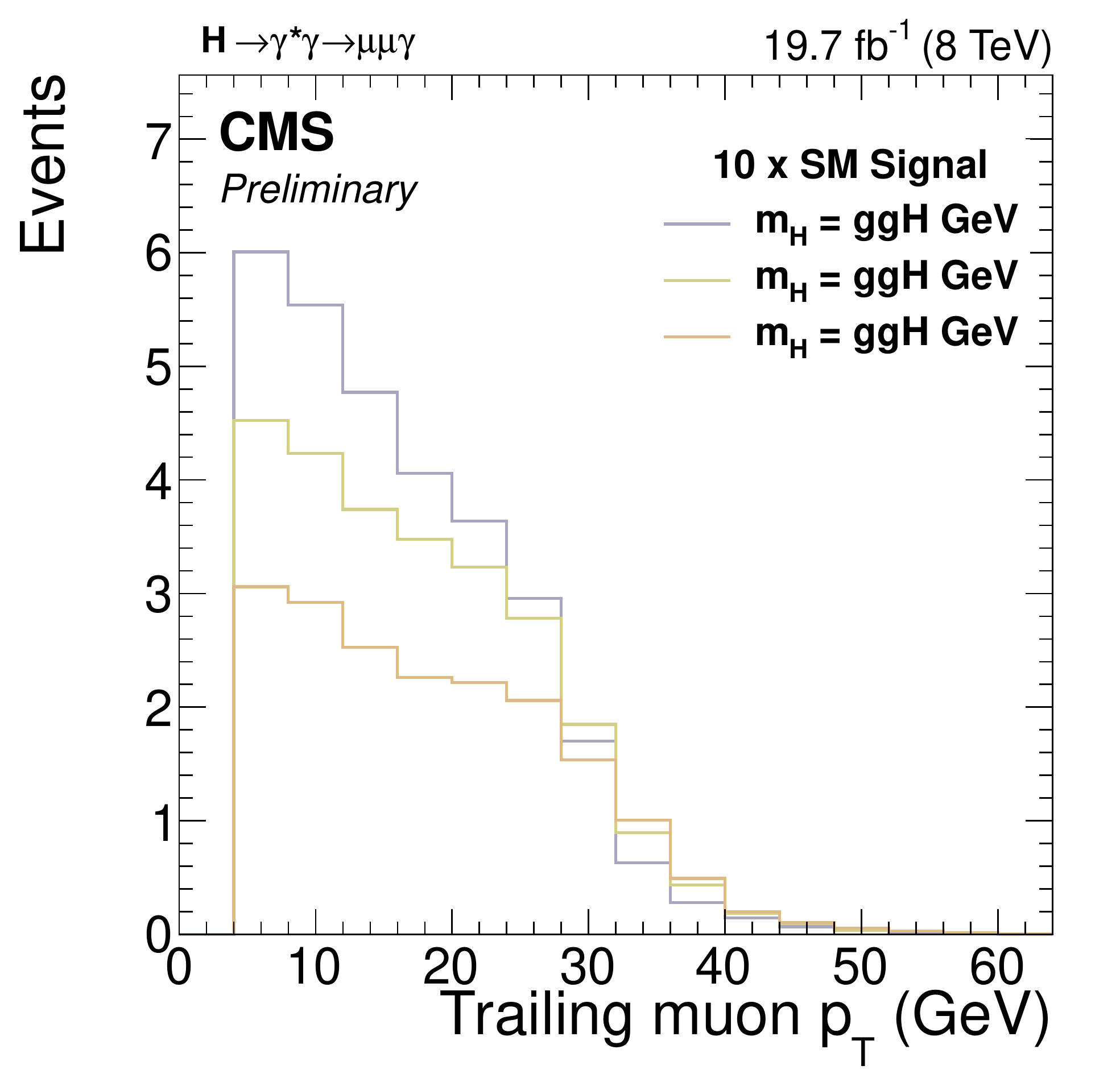}~
  \includegraphics[width=0.42\textwidth]{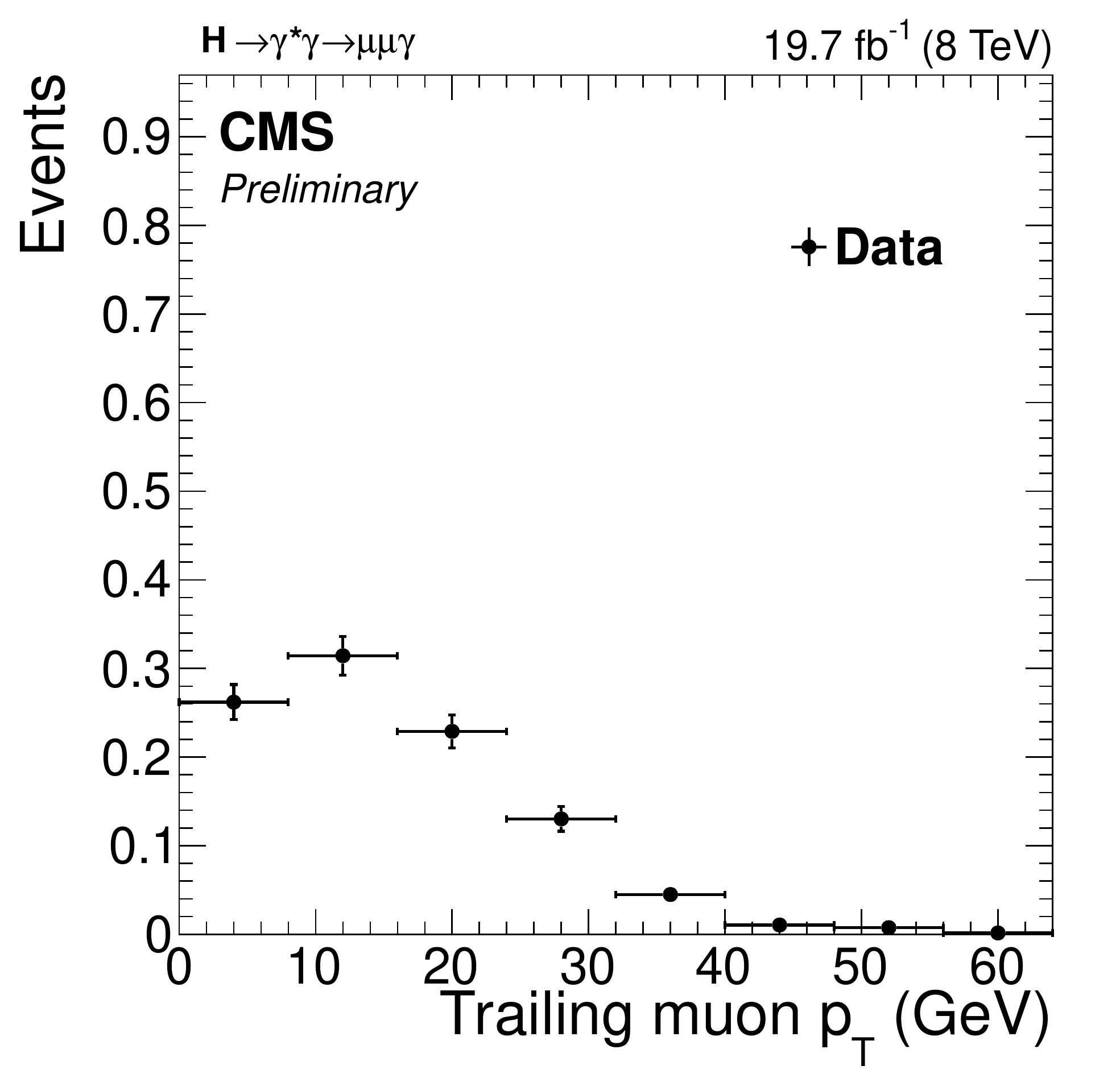}\\
  \includegraphics[width=0.42\textwidth]{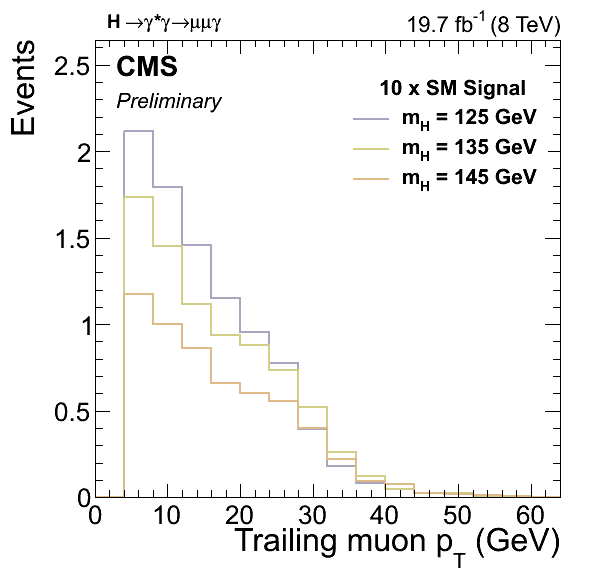}~
  \includegraphics[width=0.42\textwidth]{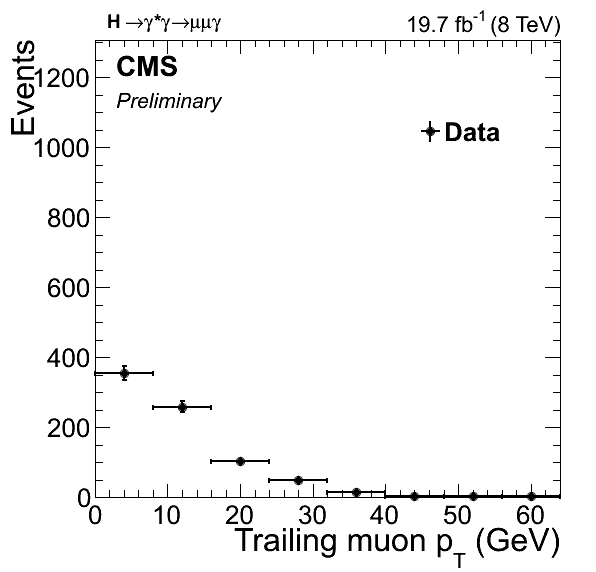}\\
  \includegraphics[width=0.42\textwidth]{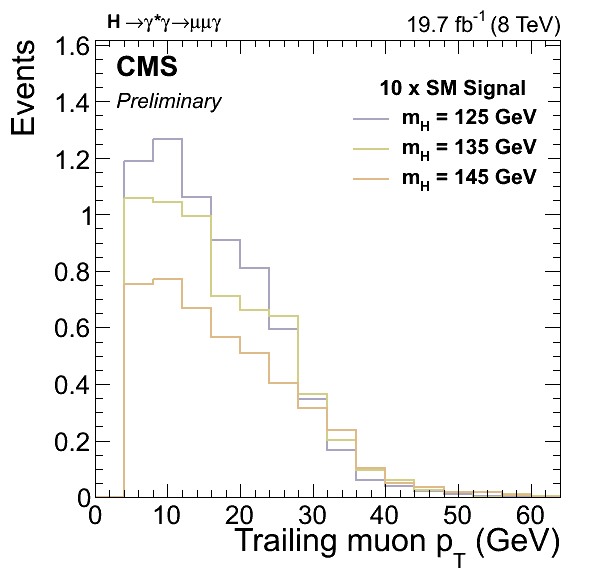}~
  \includegraphics[width=0.42\textwidth]{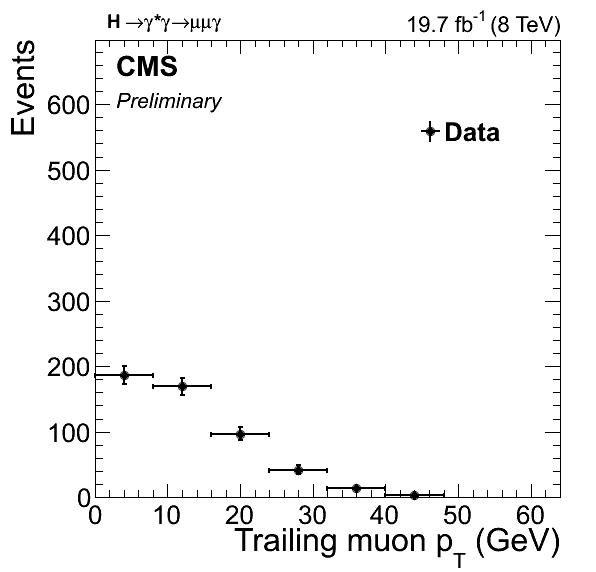}
  \caption[Distributions of the sub-leading muon $\PT$]{Distributions of the sub-leading
    muon $\PT$, in 3 categories (top to bottom).}
  \label{fig:mu-lpt1}
\end{figure}

\clearpage

\clearpage

\clearpage

\clearpage
\subsection{Resolution of \texorpdfstring{$m_{\mu\mu}$, $m_{\mu\mu\gamma}$}{m(mu,mu), m(mu,mu,gamma)} and angular variables}
\label{sec:app-mu-res}
\begin{figure}[b]
  \centering
  \includegraphics[width=0.32\textwidth]{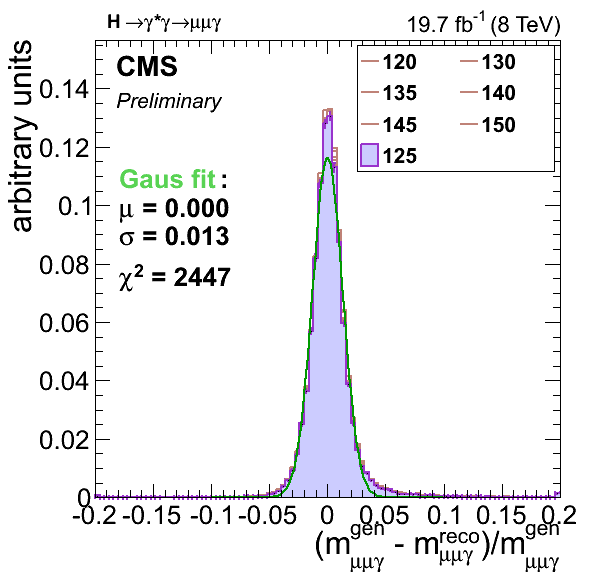}~
  \includegraphics[width=0.32\textwidth]{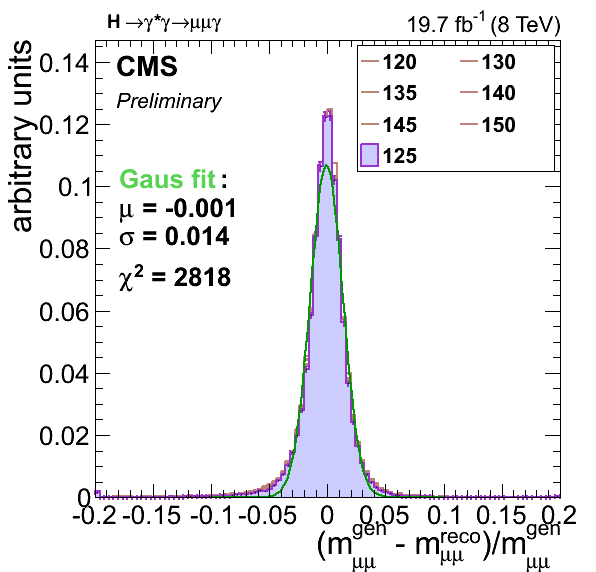}~
  \includegraphics[width=0.32\textwidth]{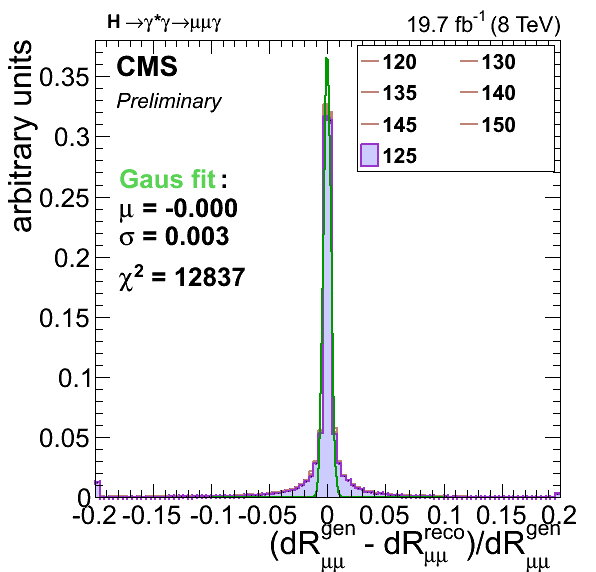}~\\
  \includegraphics[width=0.32\textwidth]{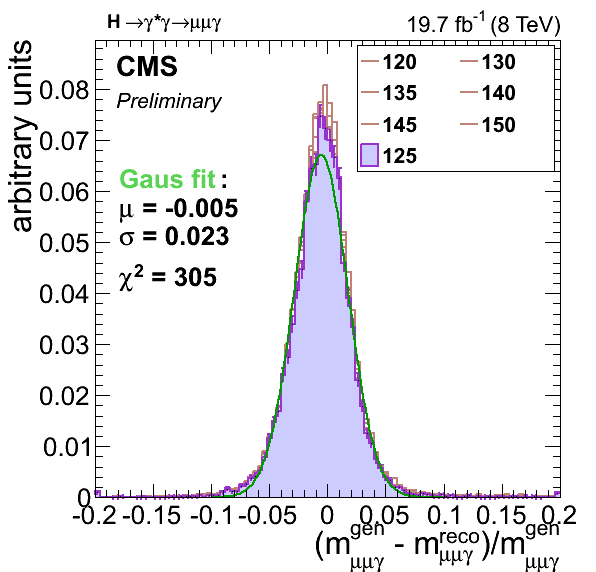}~
  \includegraphics[width=0.32\textwidth]{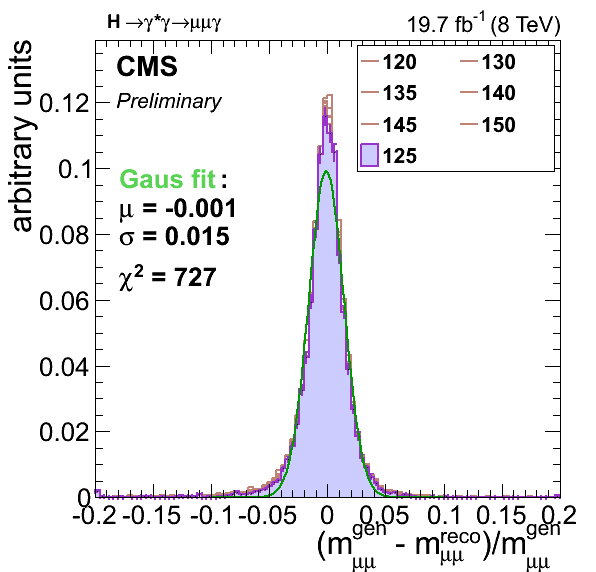}~
  \includegraphics[width=0.32\textwidth]{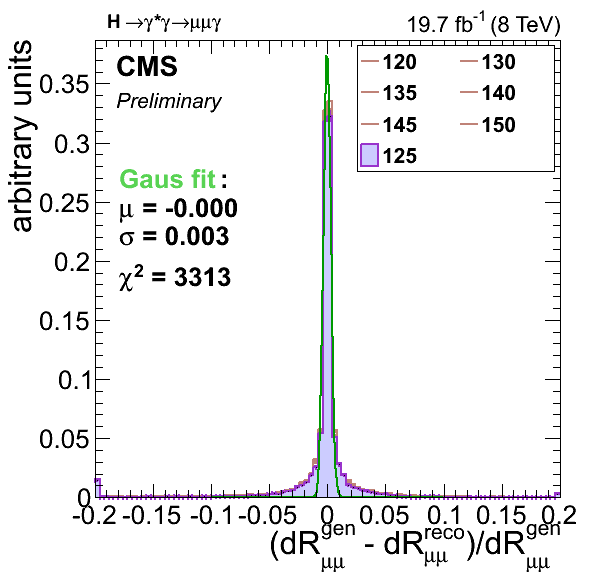}~\\
  \caption[Resolutions of $m_{\mu\mu}$, $m_{\mu\mu\gamma}$ and $\DR(\mu\mu)$.]{Resolutions
    of $m_{\mu\mu}$, $m_{\mu\mu\gamma}$ and $\DR(\mu\mu)$.  Top plots for EB category
    (photon in the Barrel), bottom plots for EE (photon in the Endcap). }
  \label{fig:res-main}
\end{figure}

Resolution of the dimuon invariant mass, three-body mass and $\DR(\mu\mu)$ are determined
from the signal MC samples.  Results using the samples from all of the Higgs boson masses
are presented in Fig.~\ref{fig:res-main} for two event categories: \textit{EB} and
\textit{EE}.  The distributions are fitted with the Gaussian function, which width is
taken as resolution (even though the fit itself may not be very good).  One can see, for
example, a degradation of the $m_{\mu\mu\gamma}$ resolution in the Endcap region.

\begin{figure}[b]
  \centering
  \includegraphics[width=0.32\textwidth]{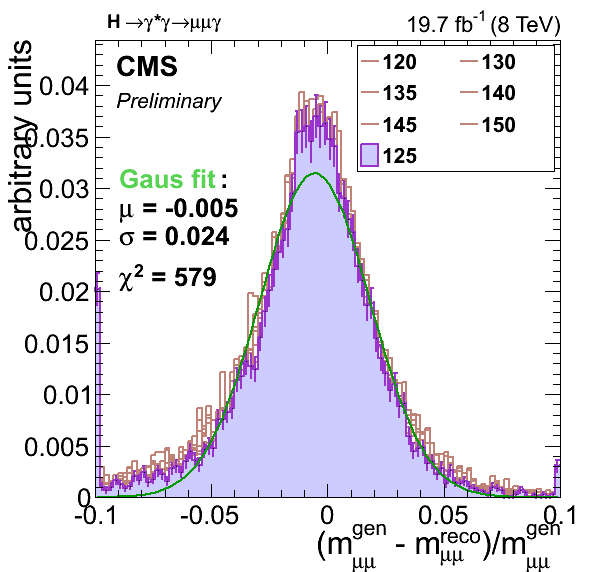}~
  \includegraphics[width=0.32\textwidth]{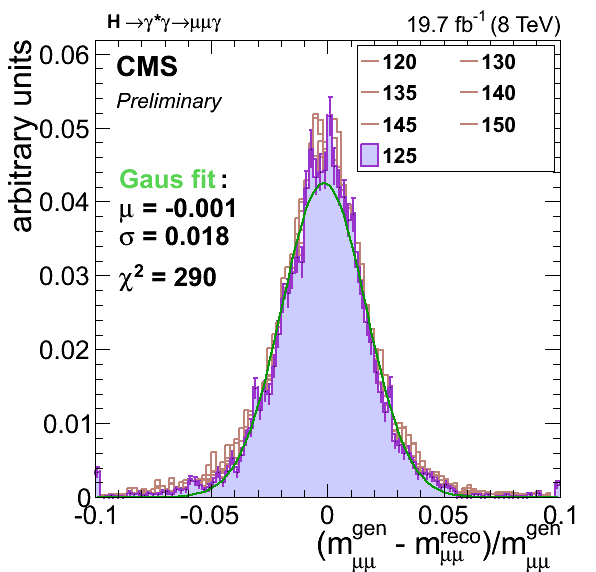}~
  \includegraphics[width=0.32\textwidth]{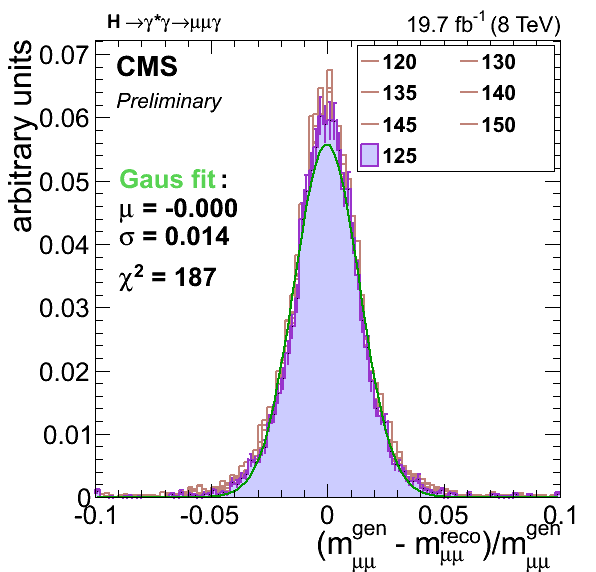}\\
  \includegraphics[width=0.32\textwidth]{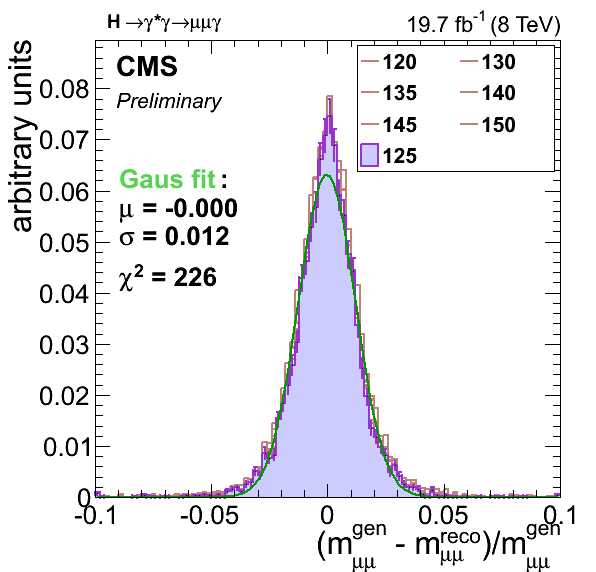}~
  \includegraphics[width=0.32\textwidth]{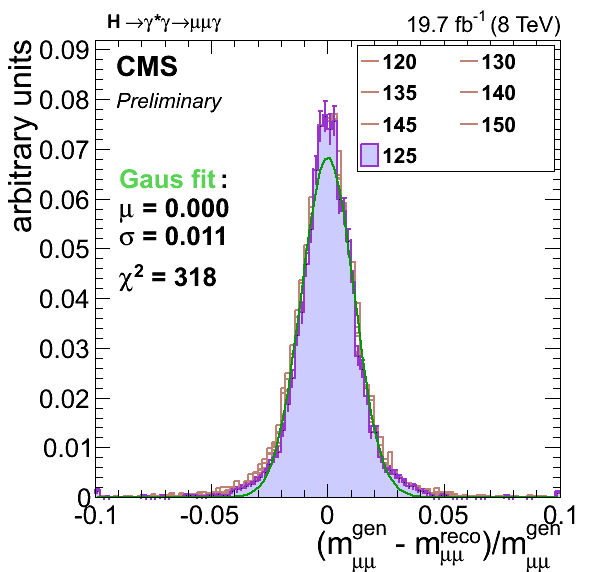}~
  \includegraphics[width=0.32\textwidth]{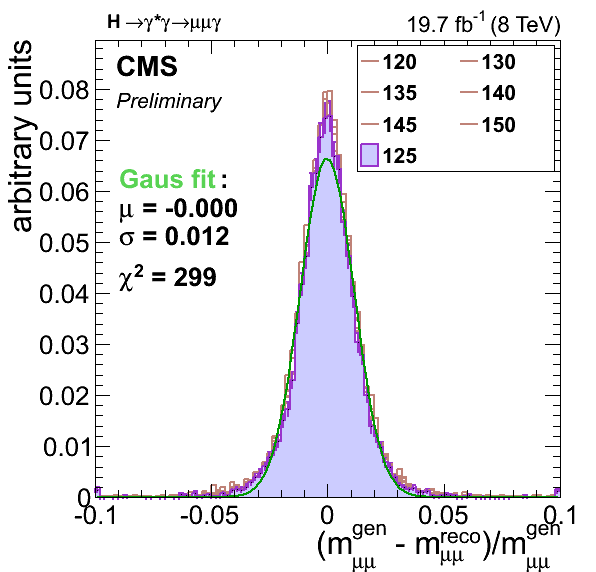}\\
  \caption[Resolution of $m_{\mu\mu}$, in 6 bins of $m_{\mu\mu}$: 0.2 -- 0.5 -- 1.0 -- 2.0
  -- 4.0 -- 9.0 -- 20\,\GeV.]{Resolution of $m_{\mu\mu}$, in 6 bins of $m_{\mu\mu}$: 0.2
    -- 0.5 -- 1.0 -- 2.0 -- 4.0 -- 9.0 -- 20\,\GeV (ordered from left-to-right, then
    top-to-bottom).}
  \label{fig:res-mll}
\end{figure}

Figure~\ref{fig:res-mll} shows the resolution of dilepton invariant mass in different bins
of $m_{\mu\mu}$.  This result is relevant for the limit on differential cross section
presented in Section~\ref{sec:results}.  One can see that there is a decrease of
resolution in low di-lepton mass: 2.4\% in the lowest bin, $m_{\mu\mu}$ =[0.2,0.5]\,\GeV
and 1.2\% in the highest, $m_{\mu\mu}$ =[0.2,0.5]\,\GeV bin.

\clearpage
\section{Electron channel}
Table~\ref{tab:yield-el} shows the event yields after each selection in the electron channel.

\begin{figure}[t]
  \centering
  \includegraphics[width=0.45\textwidth]{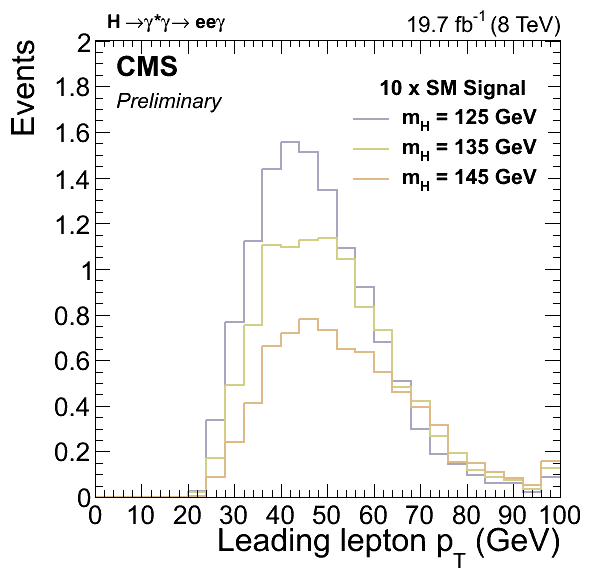}~
  \includegraphics[width=0.45\textwidth]{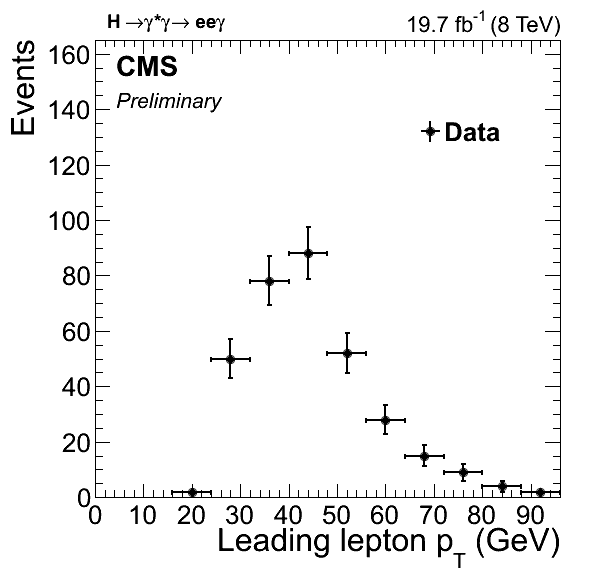}\\
  \includegraphics[width=0.45\textwidth]{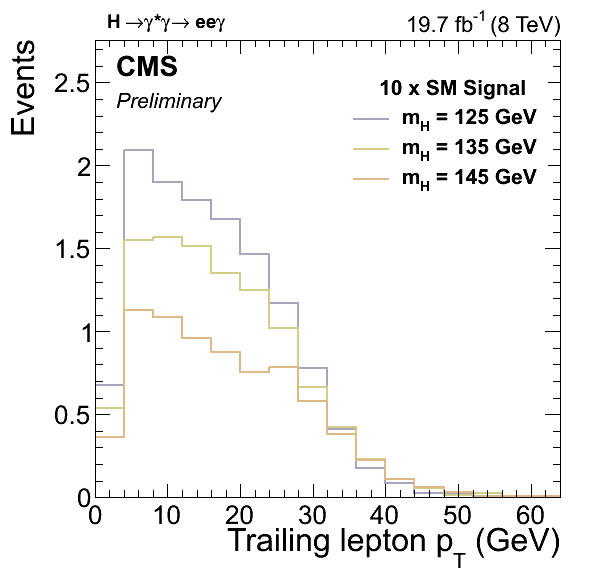}~
  \includegraphics[width=0.45\textwidth]{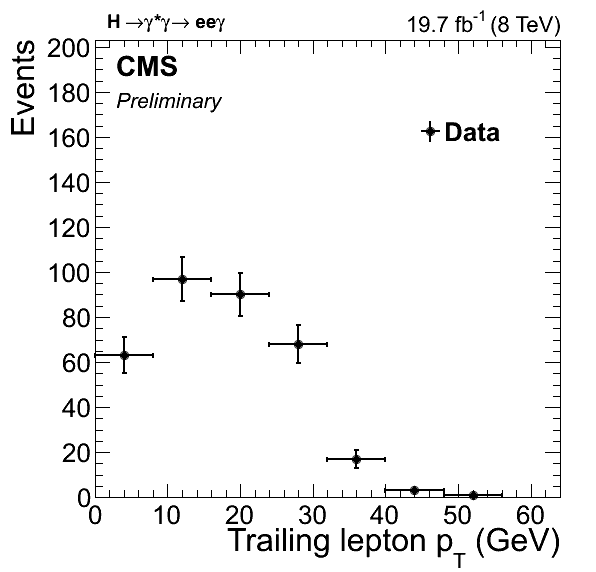}
  \caption[Transverse momenta of the leading and trailing GSF tracks inside the Dalitz
  electron object.]{Transverse momenta of the leading and trailing GSF tracks inside the
    Dalitz electron object (see text of Sec.\ref{sec:reco-el}).}
  \label{fig:el-pt}
\end{figure}

\begin{figure}[t]
  \centering
  \includegraphics[width=0.45\textwidth]{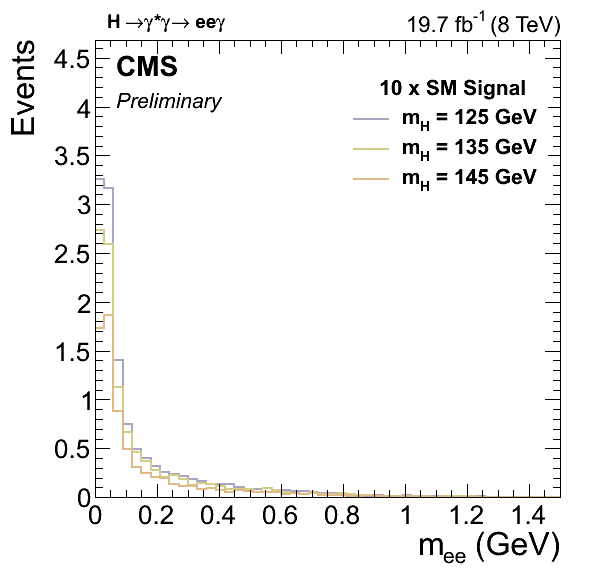}~
  \includegraphics[width=0.45\textwidth]{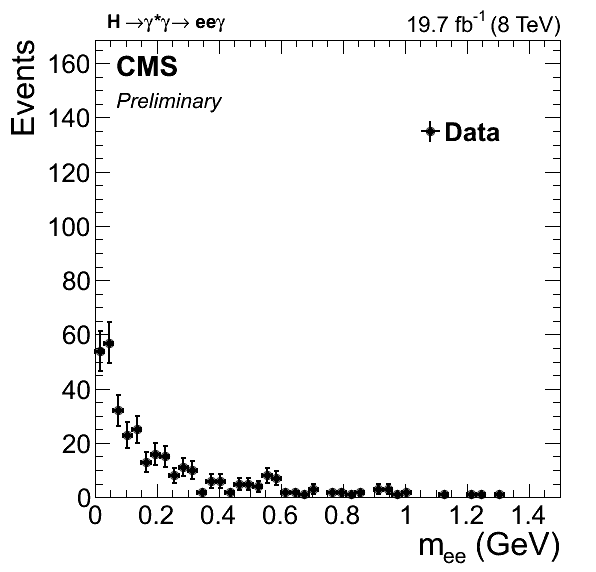}\\
  \includegraphics[width=0.45\textwidth]{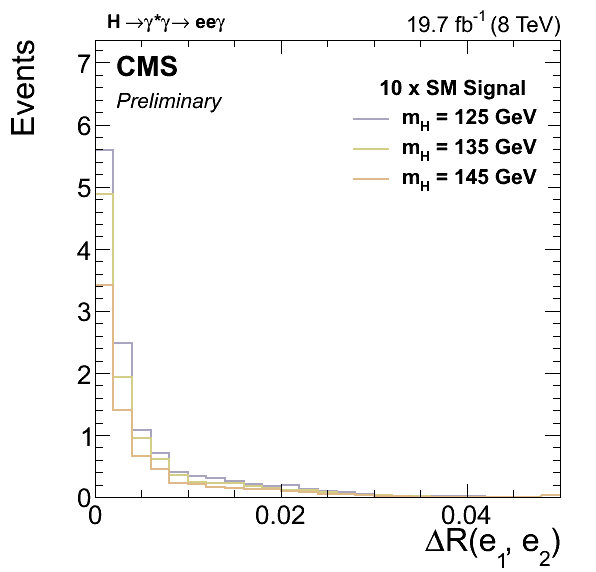}~
  \includegraphics[width=0.45\textwidth]{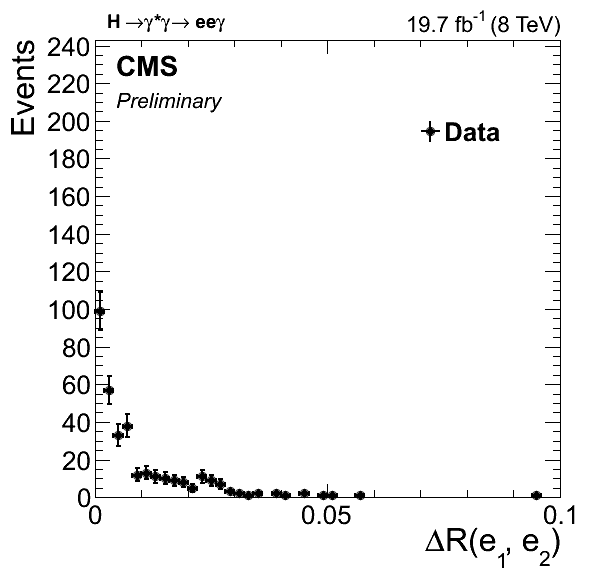}
  \caption{Invariant mass and $\Delta R_{\eta\phi}$ of the two GSF tracks.}
  \label{fig:el-mee}
\end{figure}

\begin{figure}[t]
  \centering
  \includegraphics[width=0.45\textwidth]{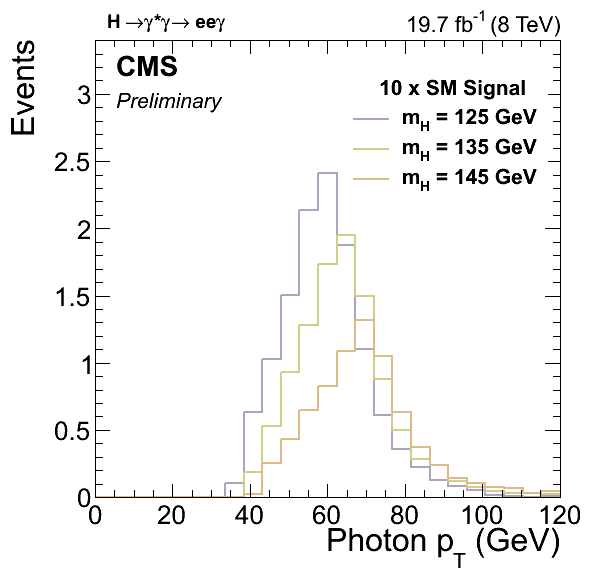}~
  \includegraphics[width=0.45\textwidth]{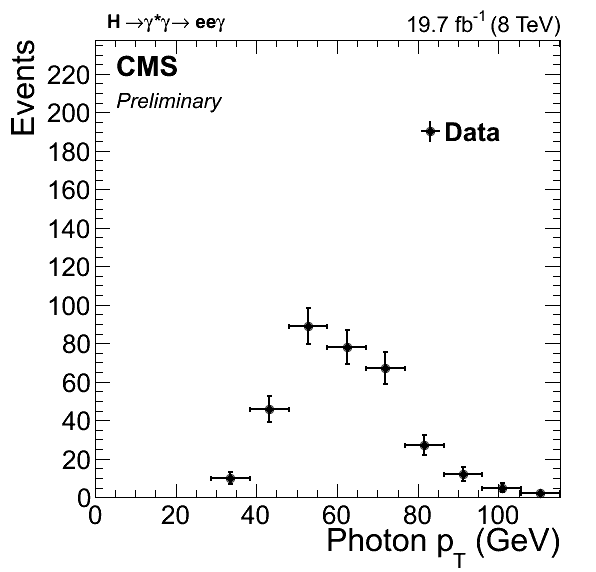}\\
  \includegraphics[width=0.45\textwidth]{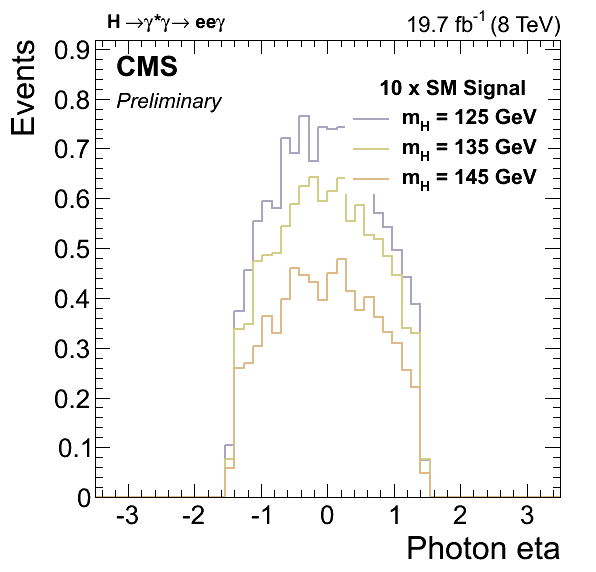}~
  \includegraphics[width=0.45\textwidth]{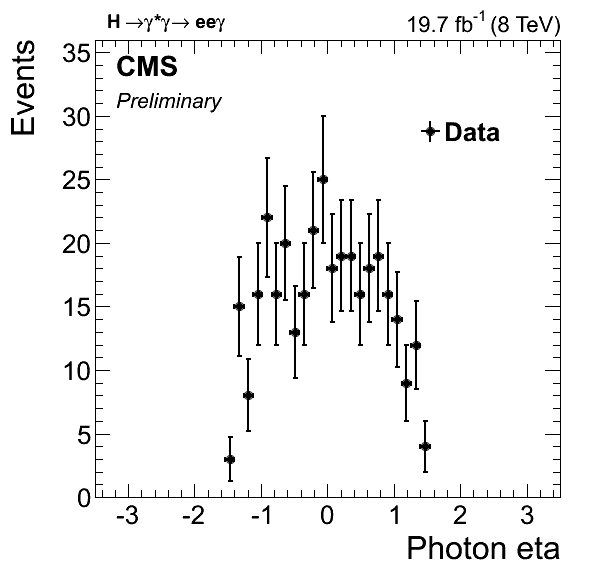}\\
  \includegraphics[width=0.45\textwidth]{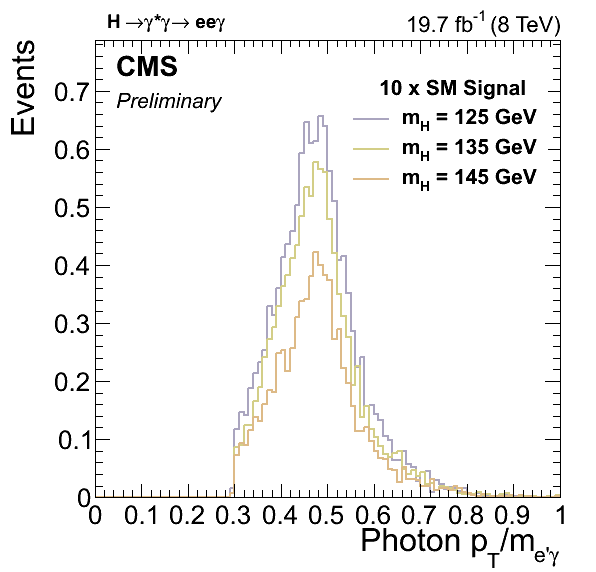}~
  \includegraphics[width=0.45\textwidth]{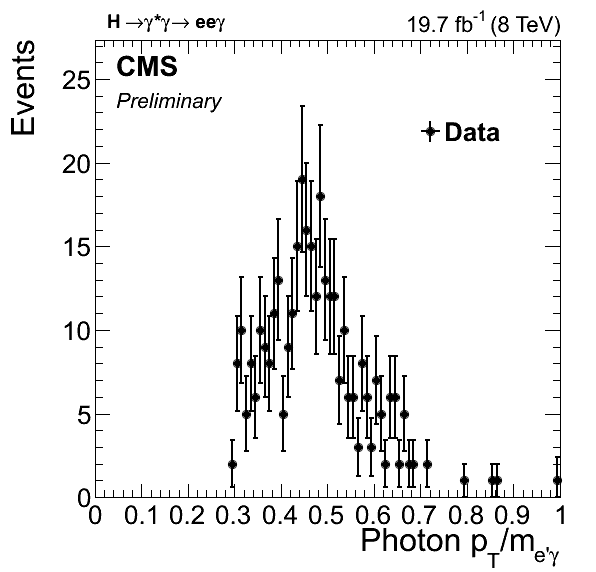}
  \caption{Photon distributions: $\PT, \eta, \PT/m_{e'\gamma}$.}
  \label{fig:el-gamma}
\end{figure}

\begin{figure}[t]
  \centering
  \includegraphics[width=0.45\textwidth]{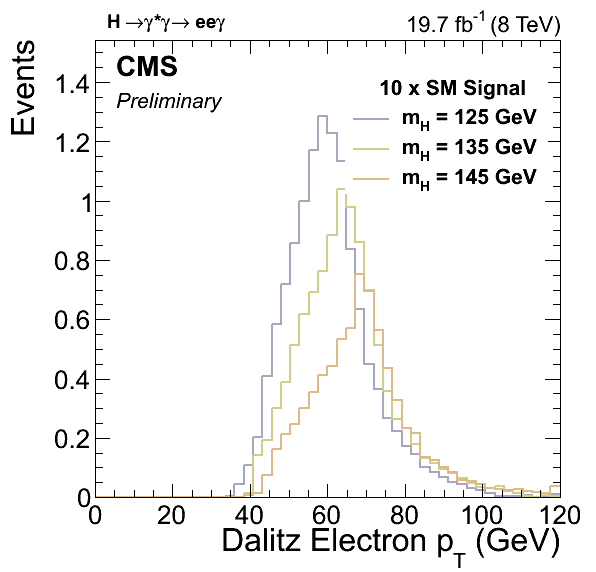}~
  \includegraphics[width=0.45\textwidth]{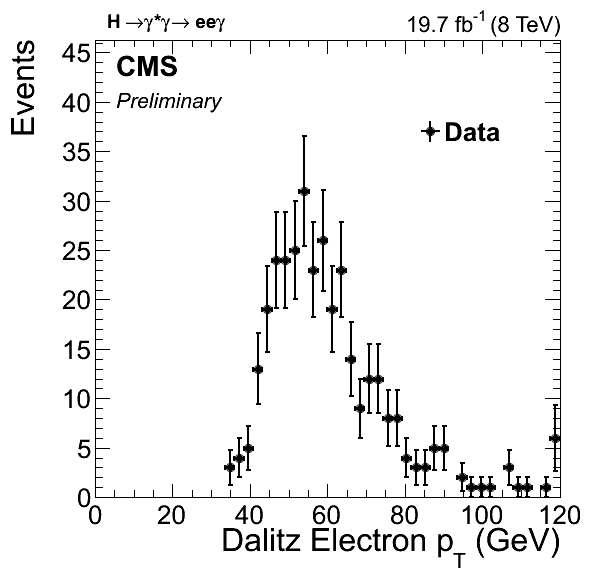}\\
  \includegraphics[width=0.45\textwidth]{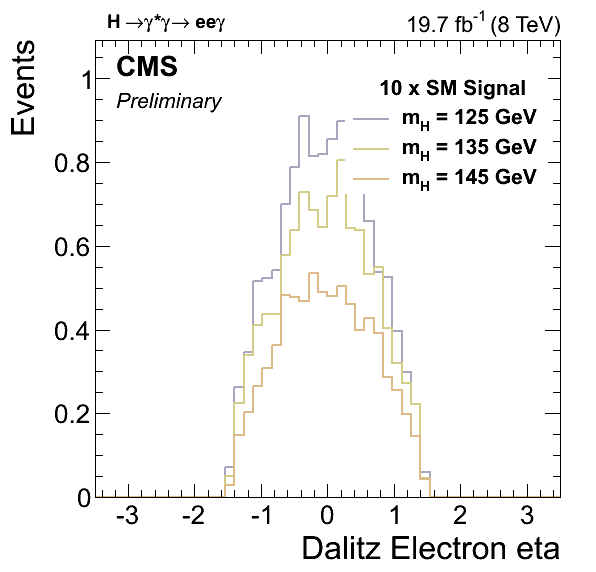}~
  \includegraphics[width=0.45\textwidth]{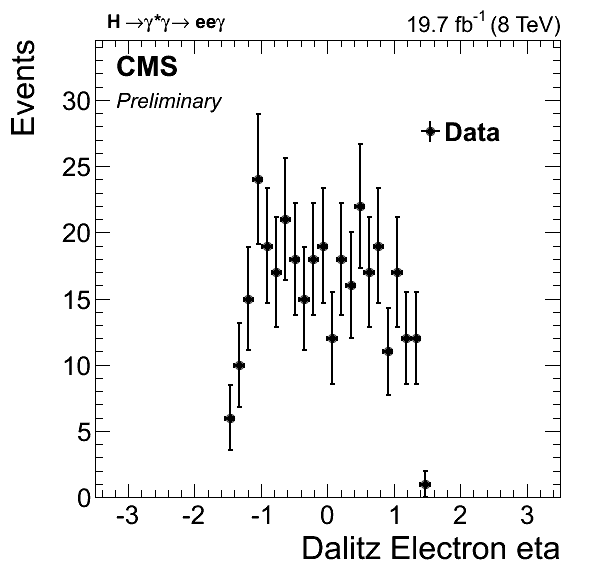}
  \caption{Dalitz electron object  distributions: $\PT, \eta$.}
  \label{fig:el-dale}
\end{figure}

\begin{figure}[t]
  \centering
  \includegraphics[width=0.45\textwidth]{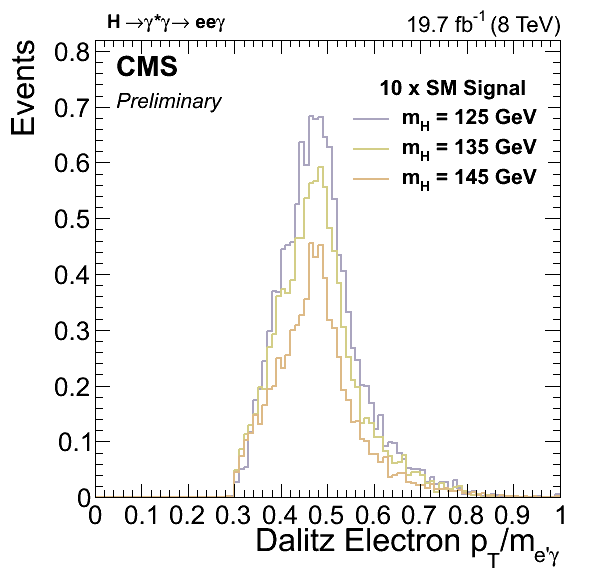}~
  \includegraphics[width=0.45\textwidth]{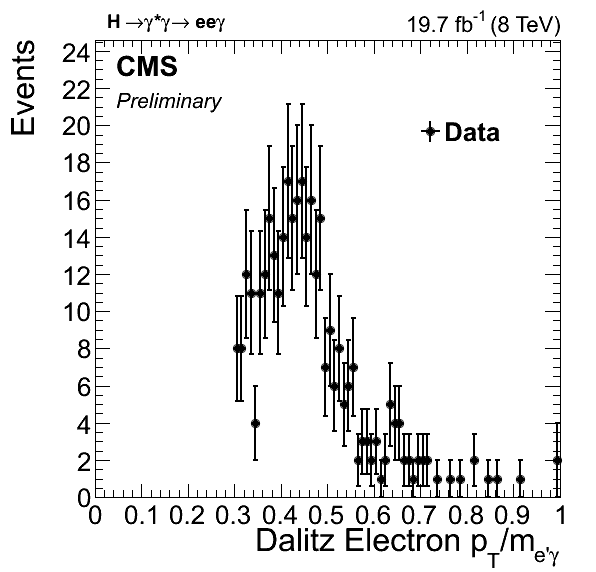}\\
  \includegraphics[width=0.45\textwidth]{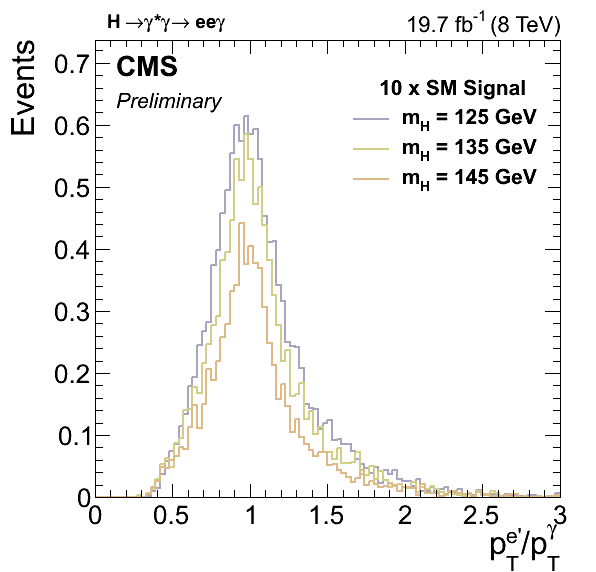}~
  \includegraphics[width=0.45\textwidth]{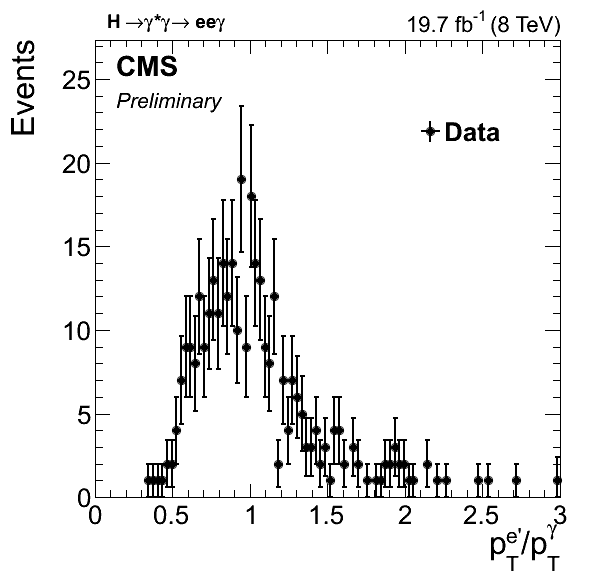}
  \caption{Dalitz electron object  distributions: $\PT/m_{e'\gamma}, \PT^{e'}/\PT^{\gamma}$.}
  \label{fig:el-dale-pt}
\end{figure}

Figures~\ref{fig:el-pt}--\ref{fig:el-dale-pt} show various kinematic
distributions, plotted after the selection denoted by the star~(*) in
Table~\ref{tab:yield-el}.

\begin{figure}[t]
  \centering
  \includegraphics[width=0.40\textwidth]{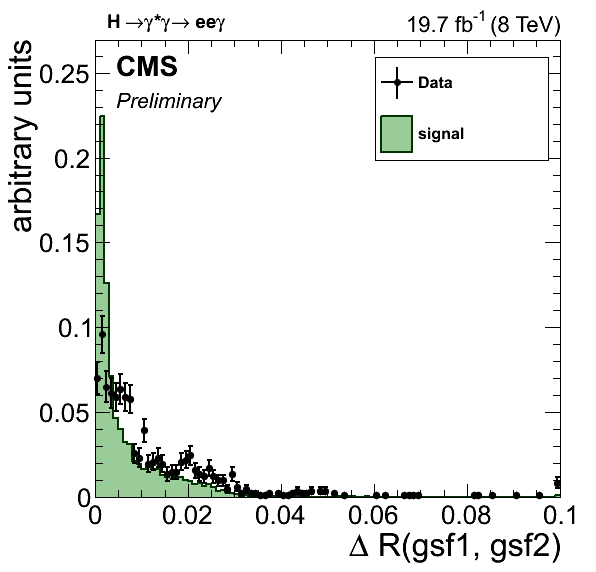}~
  \includegraphics[width=0.40\textwidth]{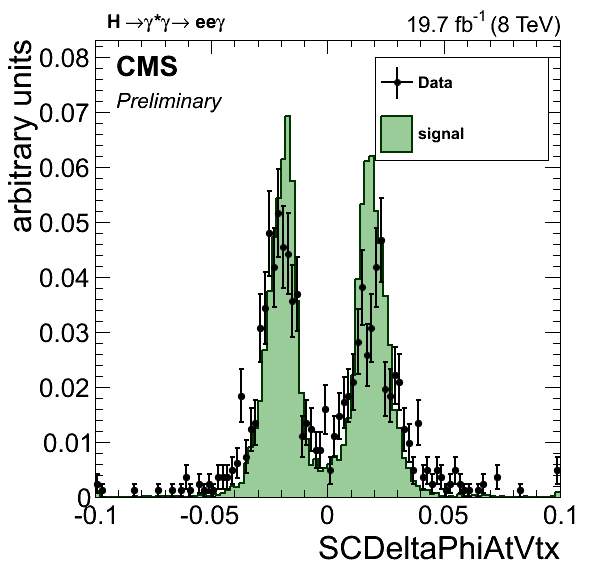}\\
  \includegraphics[width=0.40\textwidth]{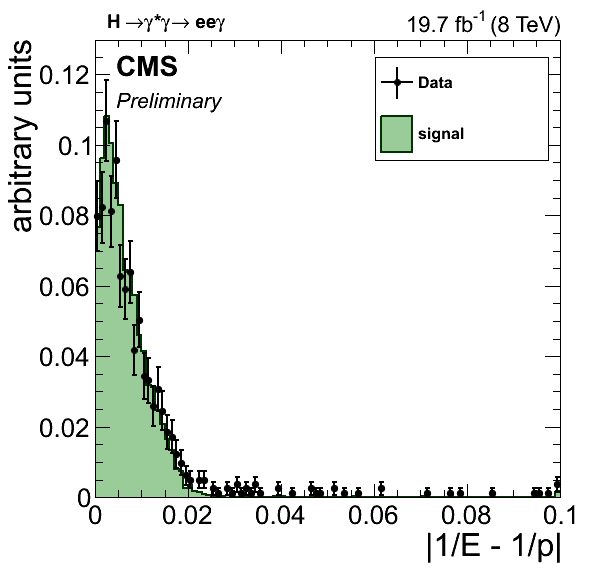}~
  \includegraphics[width=0.40\textwidth]{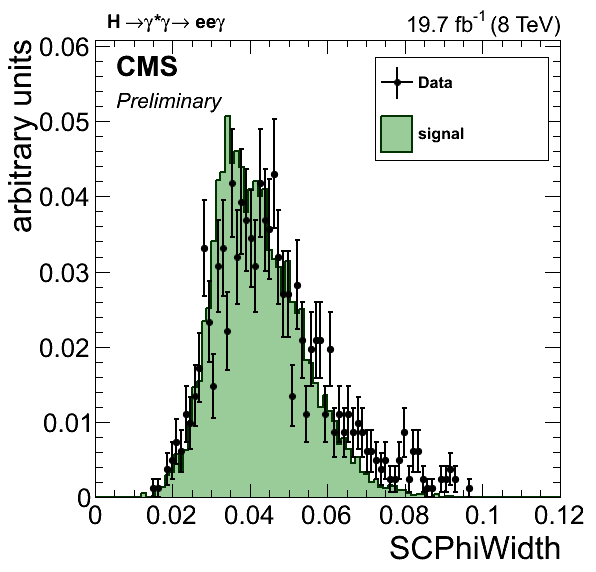}\\
  \includegraphics[width=0.40\textwidth]{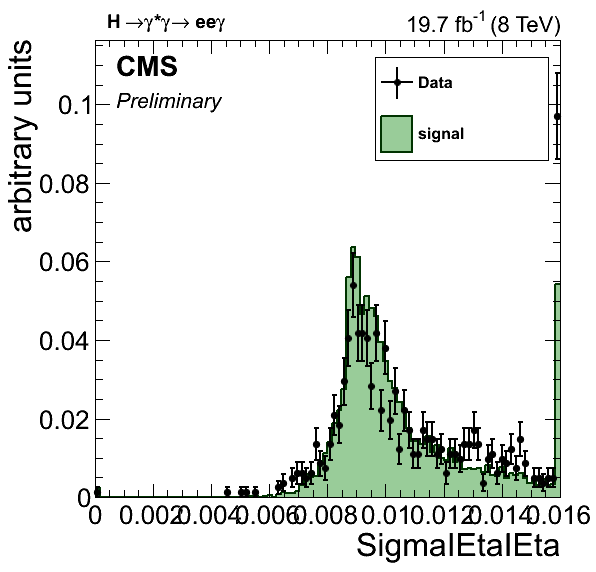}~
  \includegraphics[width=0.40\textwidth]{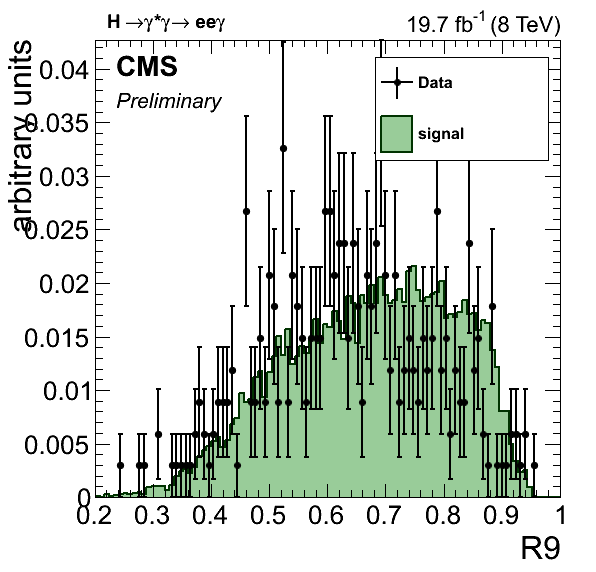}
  \caption[Dalitz electron MVA ID input variables I.]{MVA ID input variables I. $\Delta R$
    between two GSF tracks and $\Delta \phi_{in}$ (top); $|1/E - 1/p|$ and SC $\phi$-width
    (middle); $\sigma_{i\eta i\eta}$ of the second most energetic basic cluster and $R_9$
    (bottom).}
    \label{fig:el-mva1}
\end{figure}

\begin{figure}[t]
  \centering
  \includegraphics[width=0.40\textwidth]{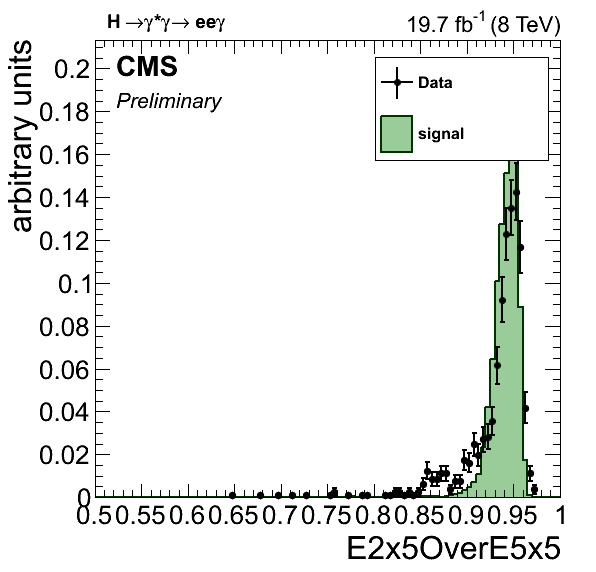}~
  \includegraphics[width=0.40\textwidth]{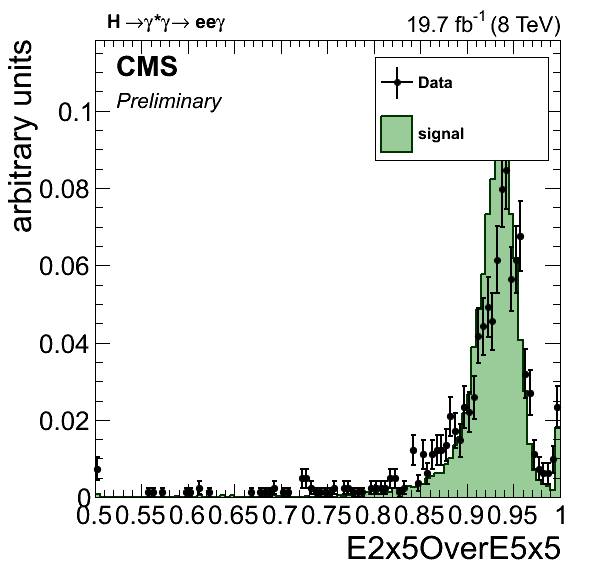}\\
  \includegraphics[width=0.40\textwidth]{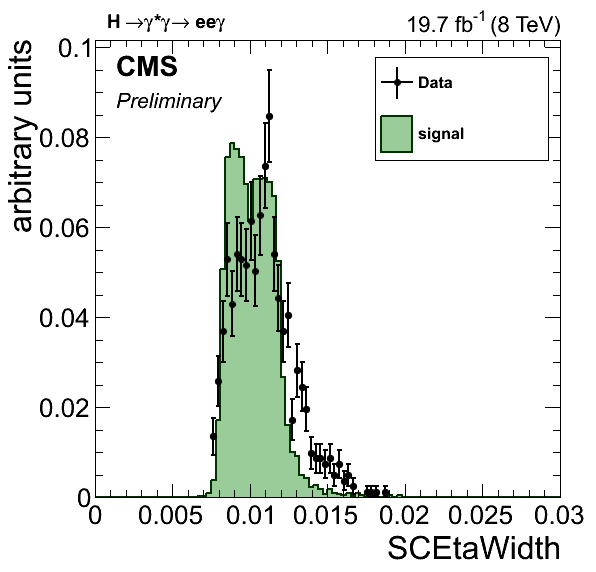}~
  \includegraphics[width=0.40\textwidth]{figs/El/mva/DalitzEle-cut15-EGamma-egm_SCEtaWidth}\\
  \includegraphics[width=0.40\textwidth]{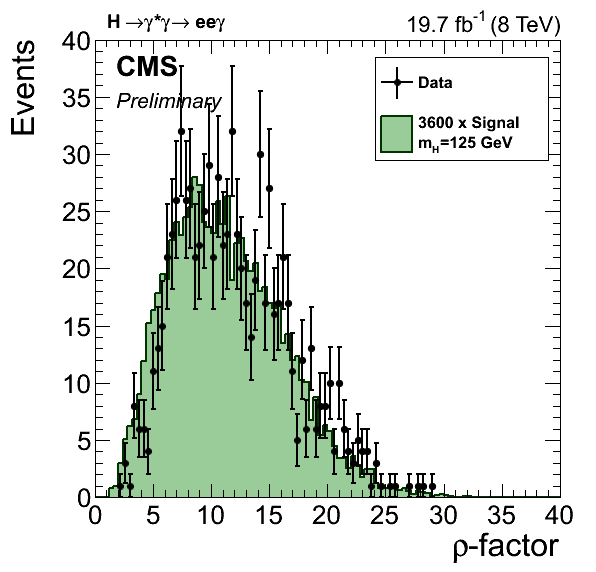}~
  \includegraphics[width=0.40\textwidth]{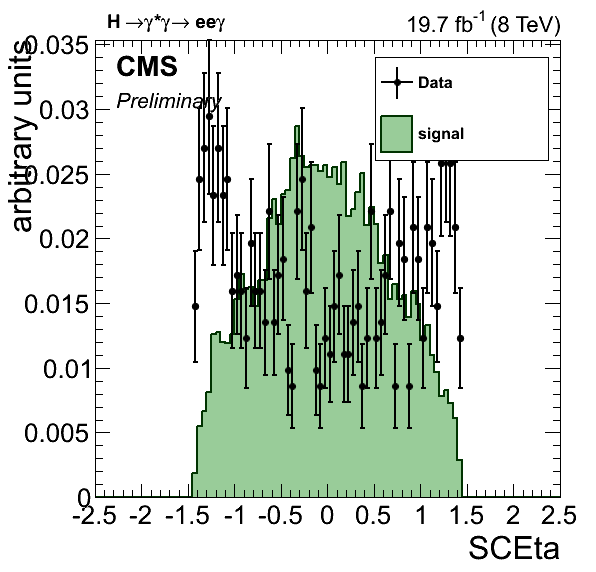}~
  \caption[Dalitz electron MVA ID input variables II.]{MVA ID input variables
    II. $E_{2\times5}/E_{5\times5}$ of the most energetic basic cluster and
    $E_{2\times5}/E_{5\times5}$ of the second most energetic basic cluster (top); H/E and
    SC $\eta$-width (middle); $\rho$ and $\eta^{SC}$ (bottom).}
  \label{fig:el-mva2}
\end{figure}

\begin{figure}[t]
  \centering
  \includegraphics[width=0.40\textwidth]{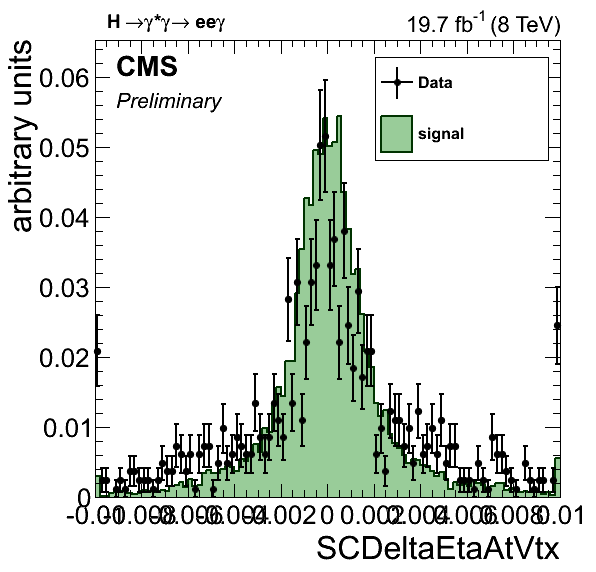}~
  \includegraphics[width=0.40\textwidth]{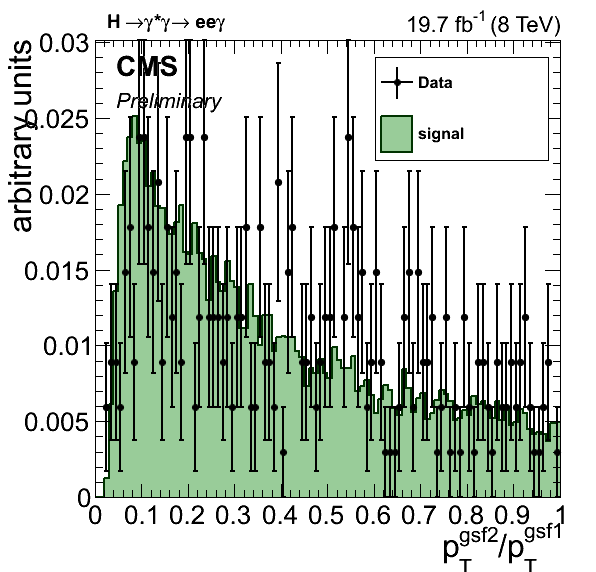}\\
  \includegraphics[width=0.40\textwidth]{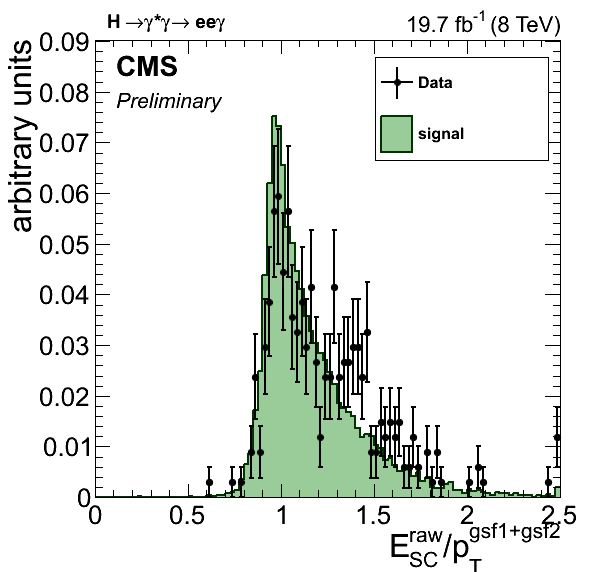}~
  \caption[Dalitz electron MVA ID input variables III.]{MVA input variables III. $\Delta
    \eta_{in}$ and the ratio of $p_T$ between second highest $p_T$ GSF track to highest
    $p_T$ GSF track raw $E^{SC}$ over $p_T$ of two GSF tracks (middle).}
  \label{fig:el-mva3}
\end{figure}

It is also useful to look at the input variables to the MVA ID used for the merged
electron object, after the selection.  Those are shown in Figures~\ref{fig:el-mva1},
\ref{fig:el-mva2} and \ref{fig:el-mva3}.  The distributions from the signal MC of $m_\PH =
125\GeV$ are normalized to the data and shown as green histograms.

\label{sec:app-el}
\begin{table}[h]
  \caption{Event yield after each selection criteria for data and signal with $m_\PH = 125\GeV$ for $L = 19.7\fbinv$ in the electron channel.}
  \label{tab:yield-el}
\begin{center}
\begin{tabular}{ll|rr|cc} 
 & Selection & Data & Sig: total &ggH & vbfH\\
 \hline 
 & Pass Trigger                                & 87M & 23.25 & 21.46 & 1.78 \\
 & DALectron selection (MVA ID cut at 0.12)    & 58K & 3.76 & 3.48 & 0.27\\ 
 & $p_T^{e_1} + p_T^{e_2} > 44\GeV$, $m_{\Pe\Pe} < 1.5\GeV$  & 26K & 3.12 & 2.89 & 0.22\\ 
 & Photon $p_T > 30\GeV; |\eta_{SC}|<1.4442$  & 1566 & 2.06 & 1.91 & 0.14 \\ 
 & $110 < m_{e'\gamma} < 170\GeV$       & 436 & 2.05 & 1.91 & 0.14 \\
* & $ p_T^{e'}/m_{e'\gamma} > 0.3$ and $p_T^{\gamma}/m_{e'\gamma} > 0.3$ & 337 & 1.95 & 1.82 & 0.13 \\ 
\end{tabular}

\end{center}
\end{table}

\clearpage

\clearpage

\clearpage

\clearpage

\clearpage

\clearpage

\clearpage

\clearpage

\clearpage
\section{\texorpdfstring{$\PH\to(\JPsi)\gamma$}{Higgs to JPsi gamma}}
\begin{figure}[ht]
  \centering
  \includegraphics[width=0.44\textwidth]{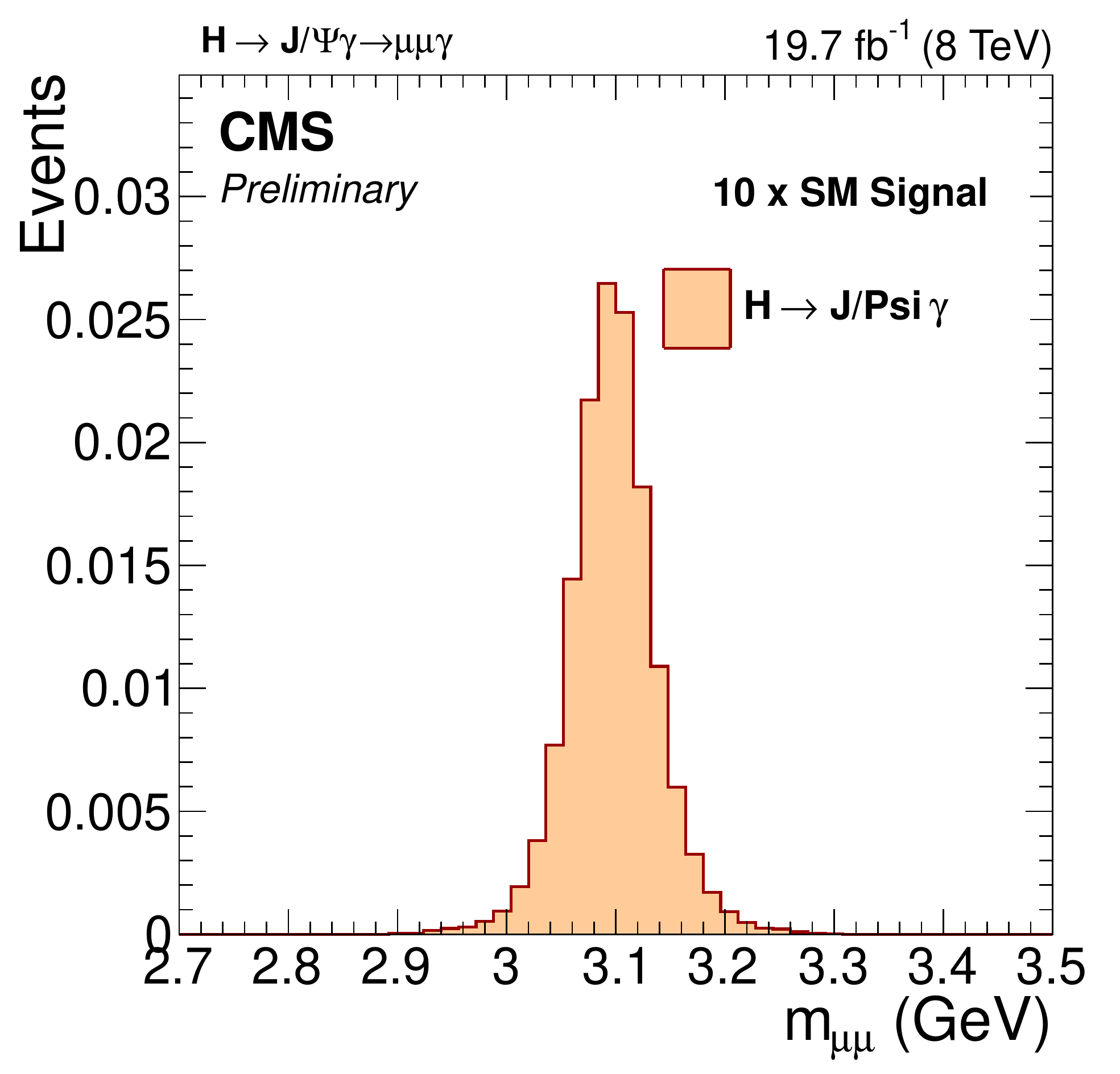}~
  \includegraphics[width=0.44\textwidth]{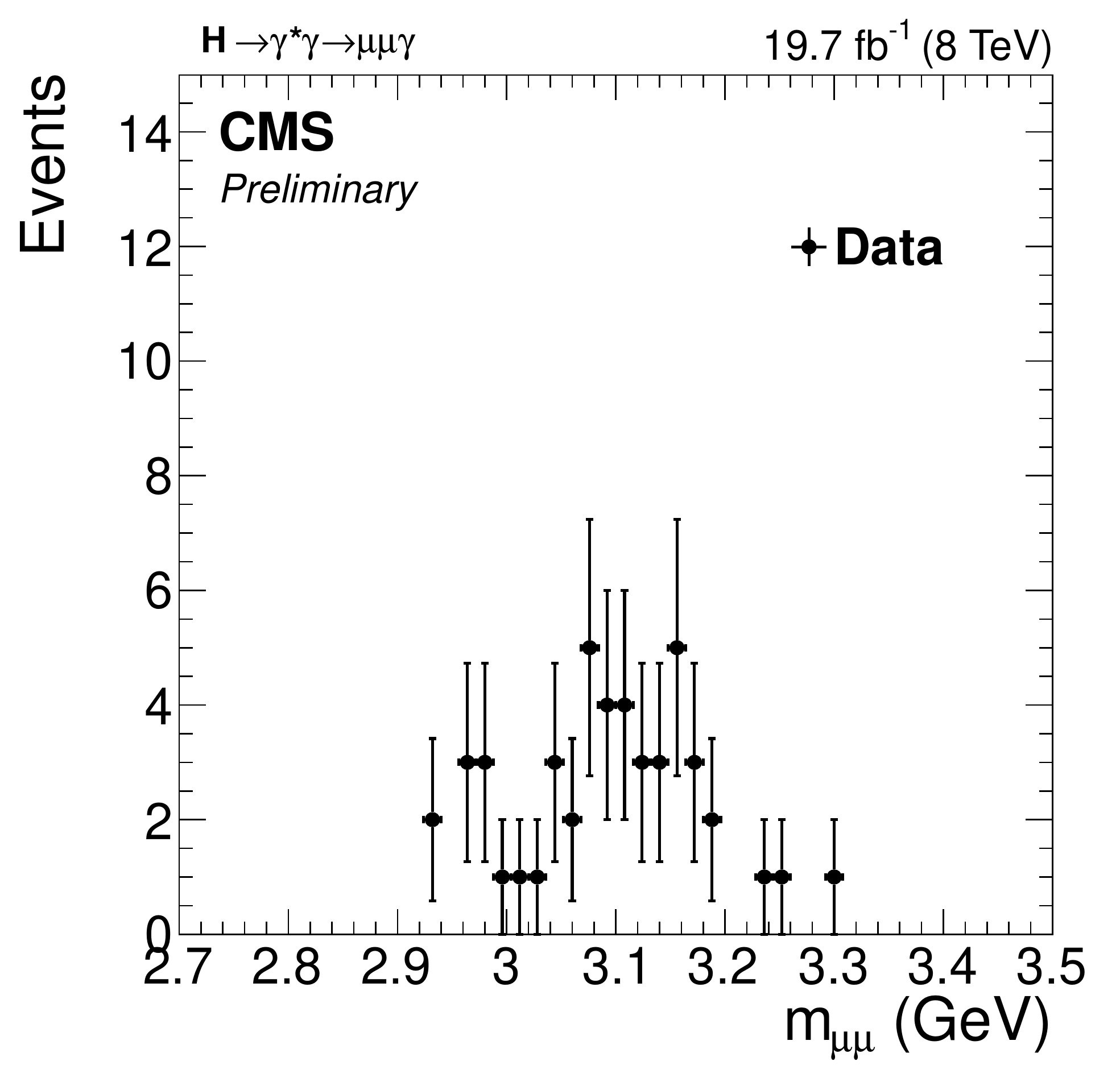}\\
  \includegraphics[width=0.44\textwidth]{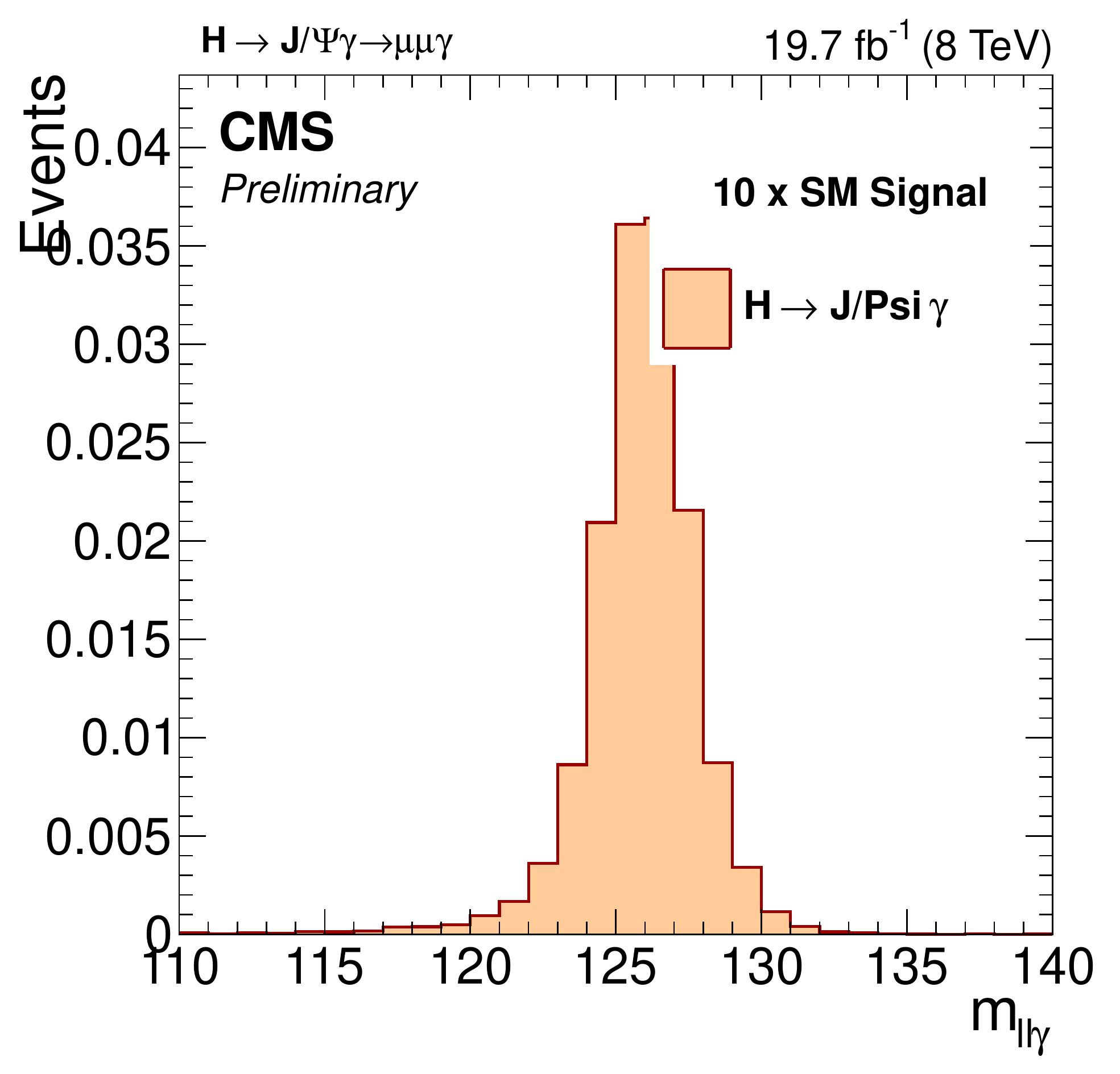}~
  \includegraphics[width=0.44\textwidth]{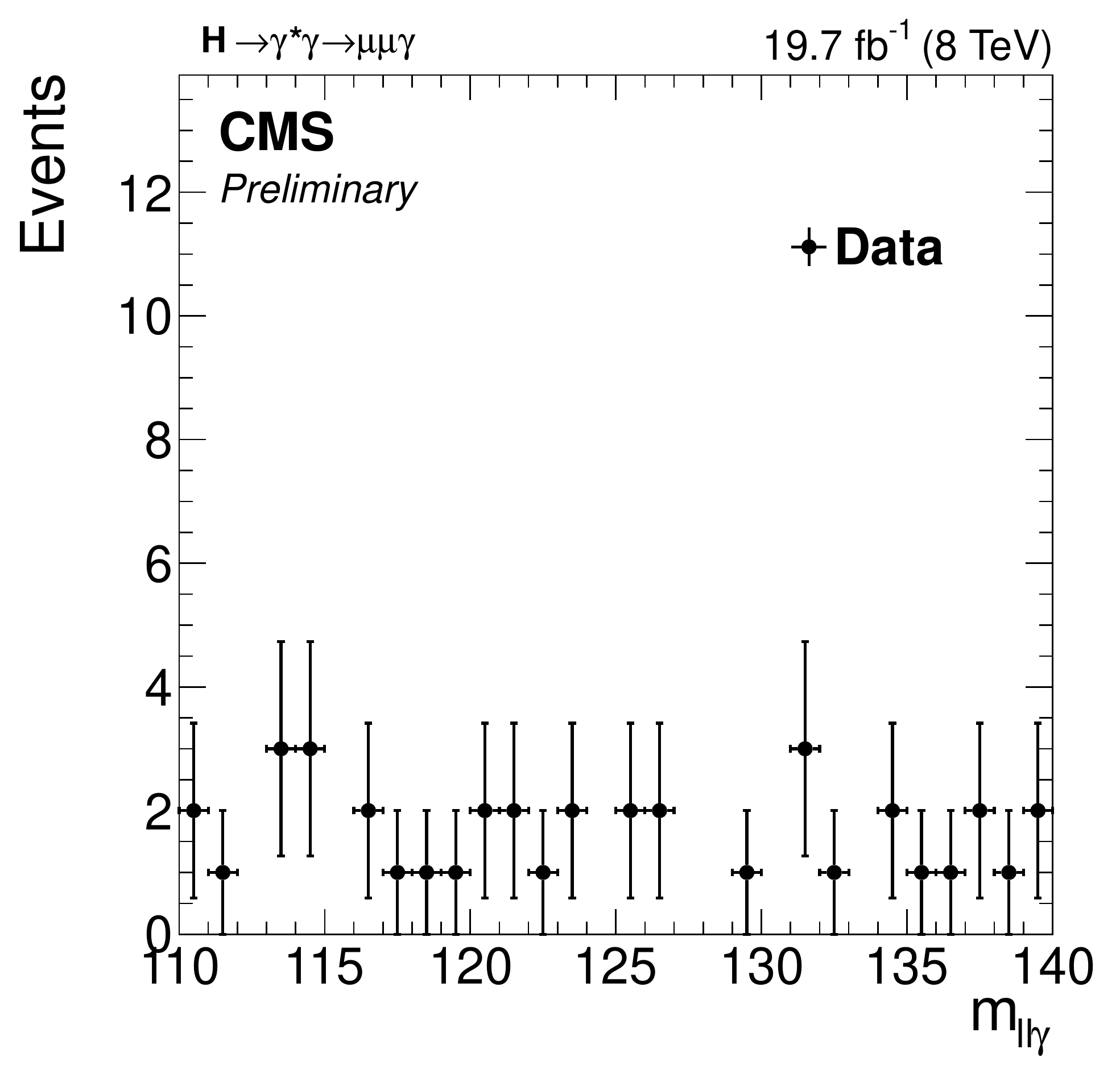}
  \caption[$\JPsi\to\mu\mu$ peak and the three-body mass.]{$\JPsi\to\mu\mu$ peak and the
    three-body mass. The signal distributions (left) are normalized to the total number of
    events in data (right).}
  \label{fig:dist-1}
\end{figure}

\begin{figure}[ht]
  \centering
  \includegraphics[width=0.3\textwidth]{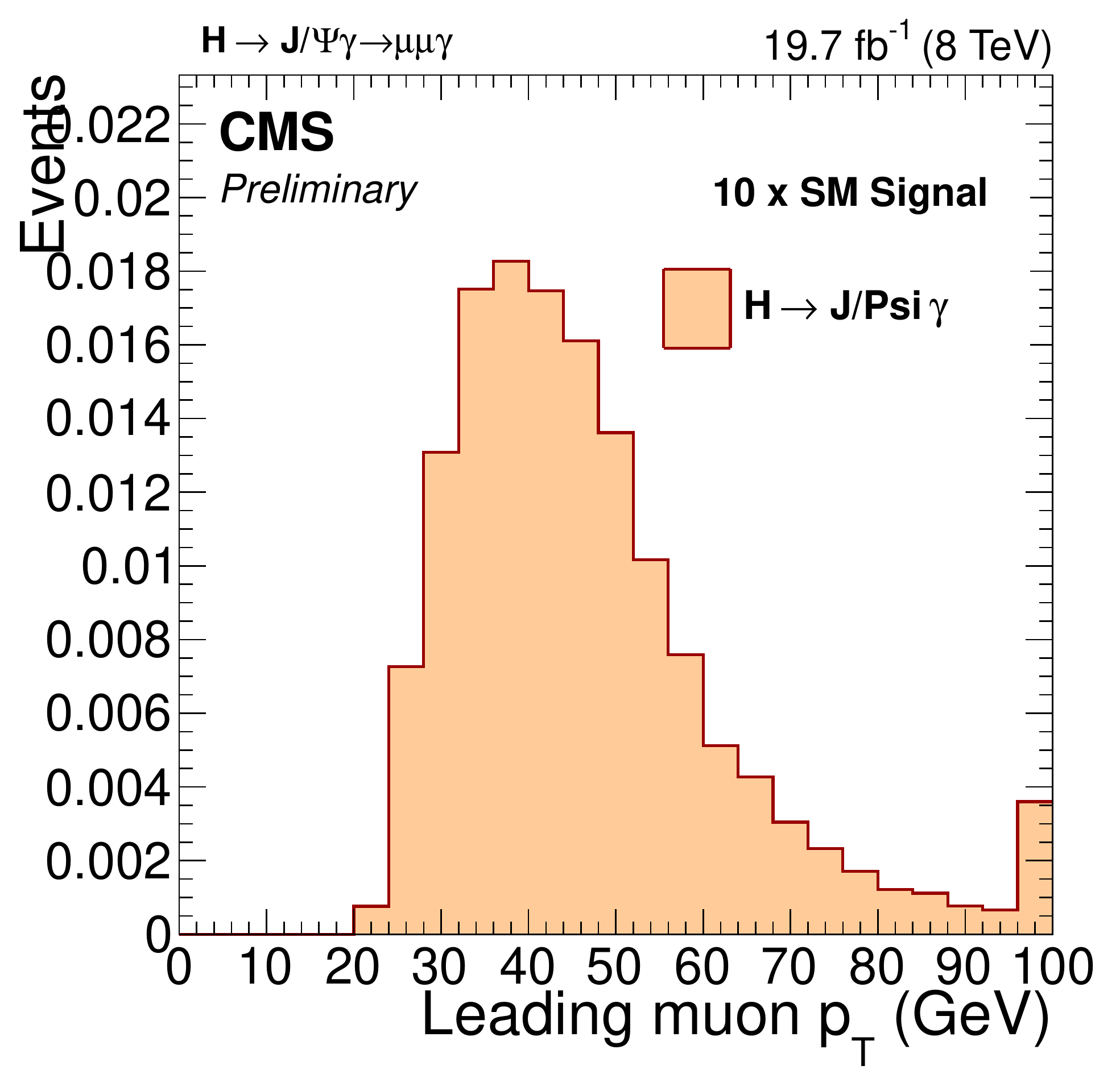}~
  \includegraphics[width=0.3\textwidth]{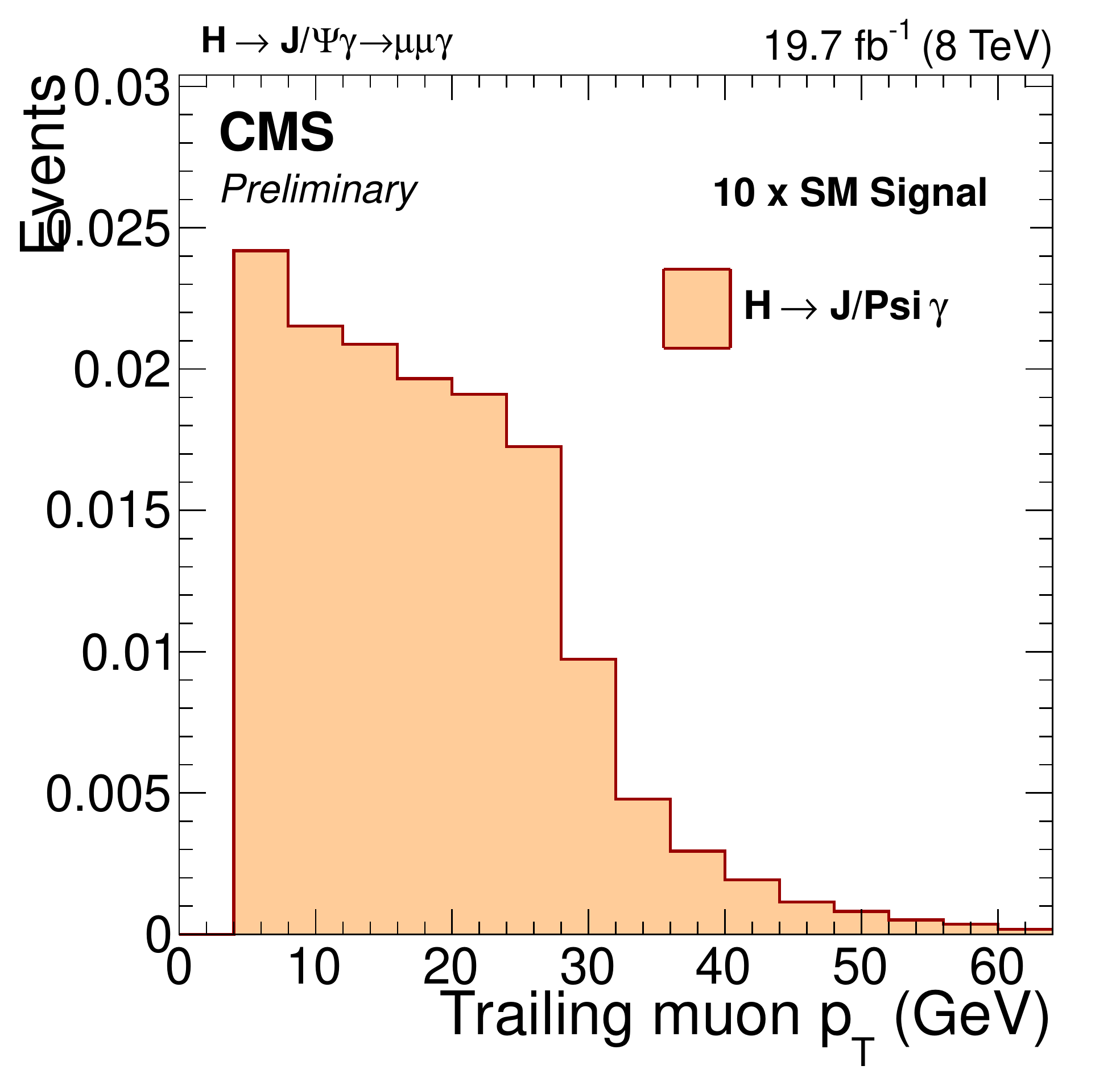}~
  \includegraphics[width=0.3\textwidth]{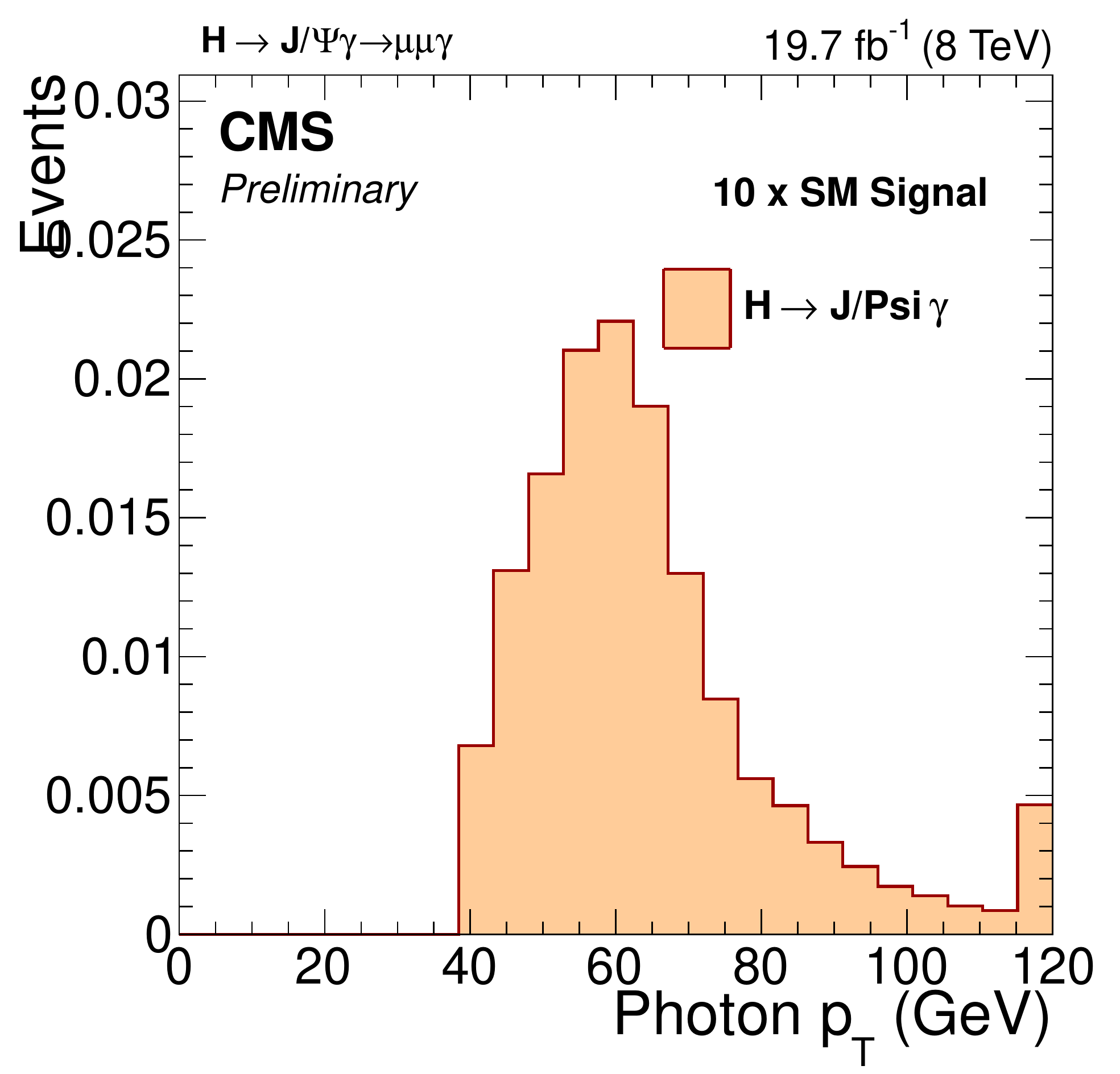}\\
  \includegraphics[width=0.3\textwidth]{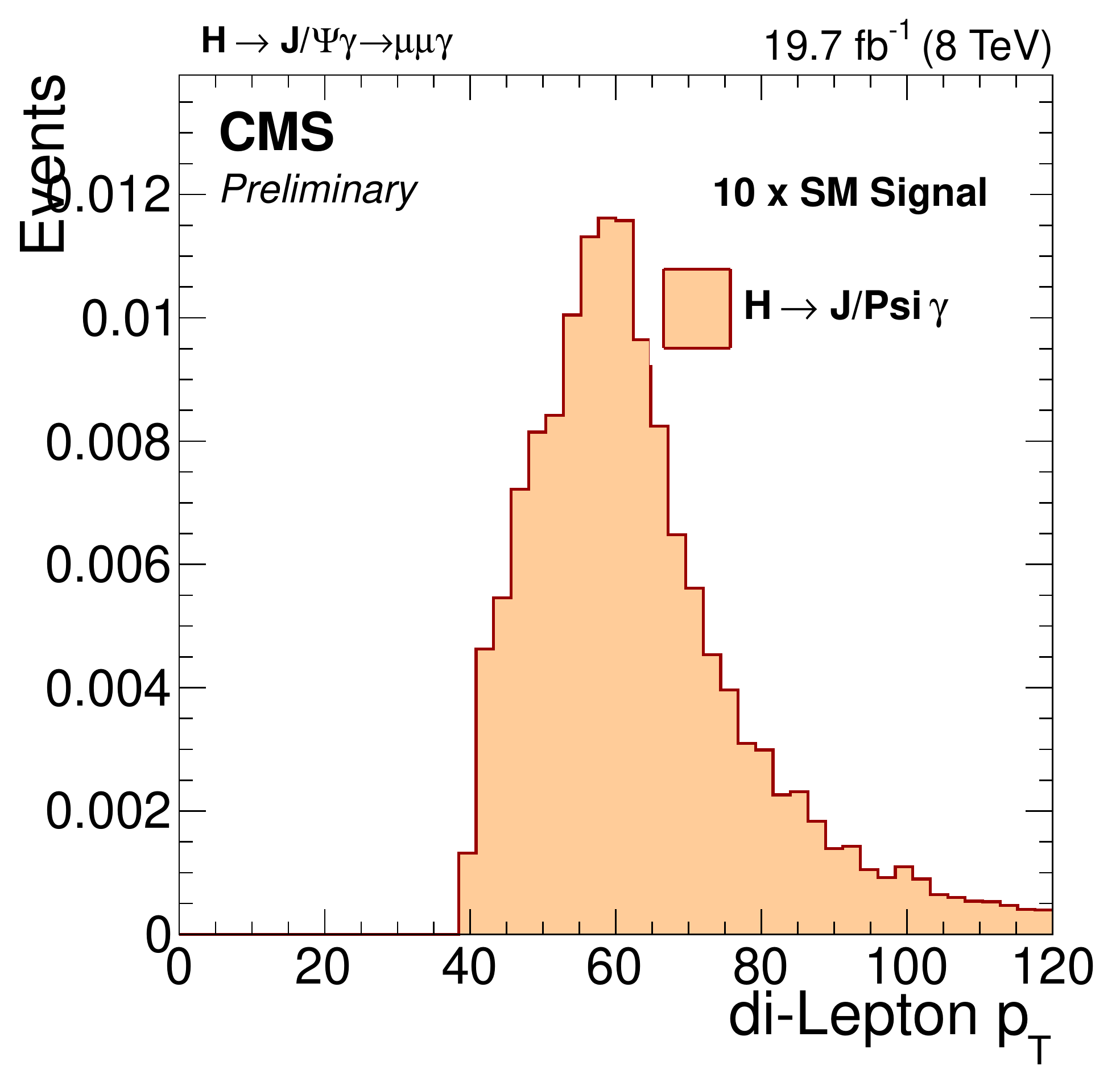}~
  \includegraphics[width=0.3\textwidth]{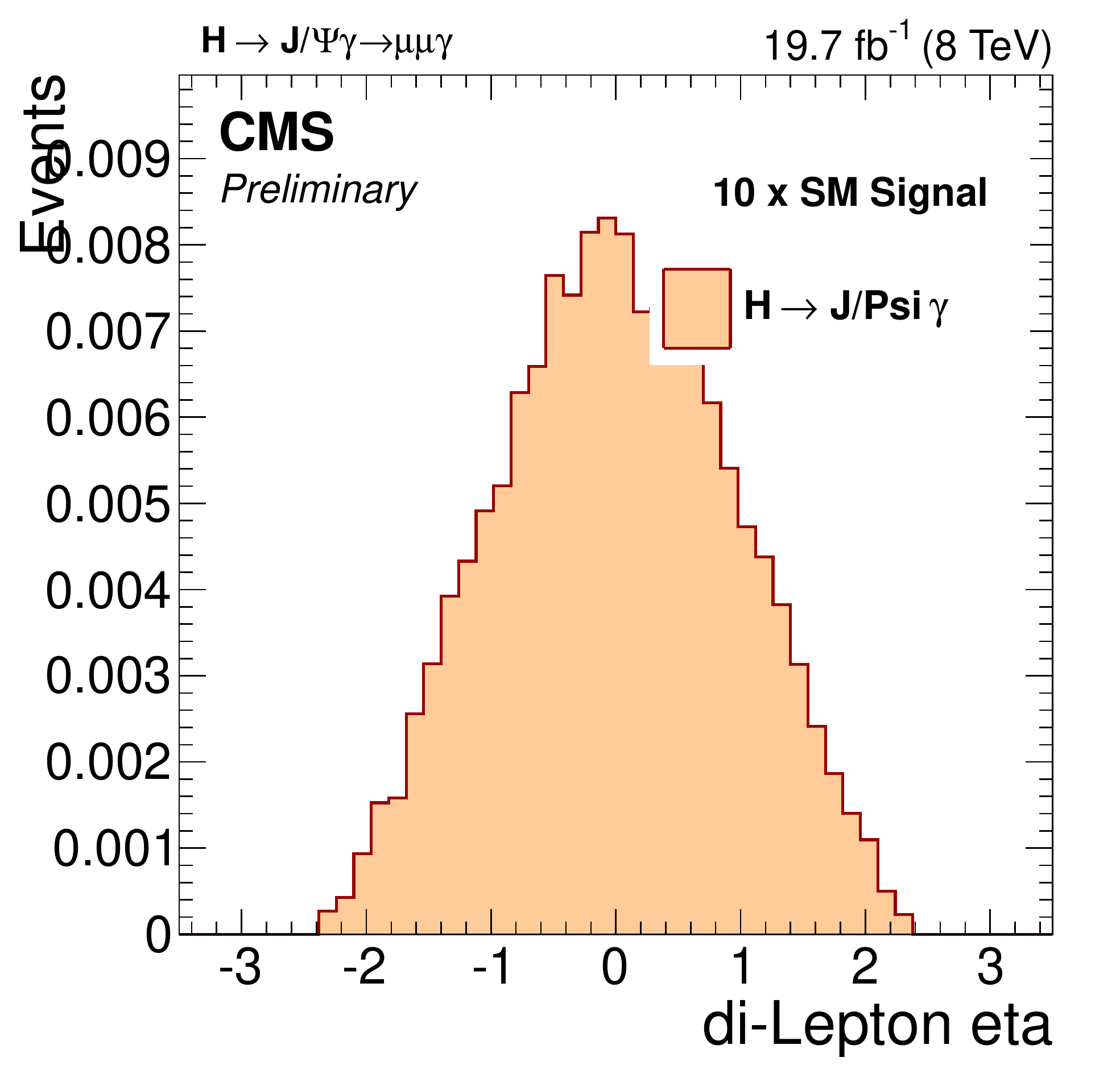}~
  \includegraphics[width=0.3\textwidth]{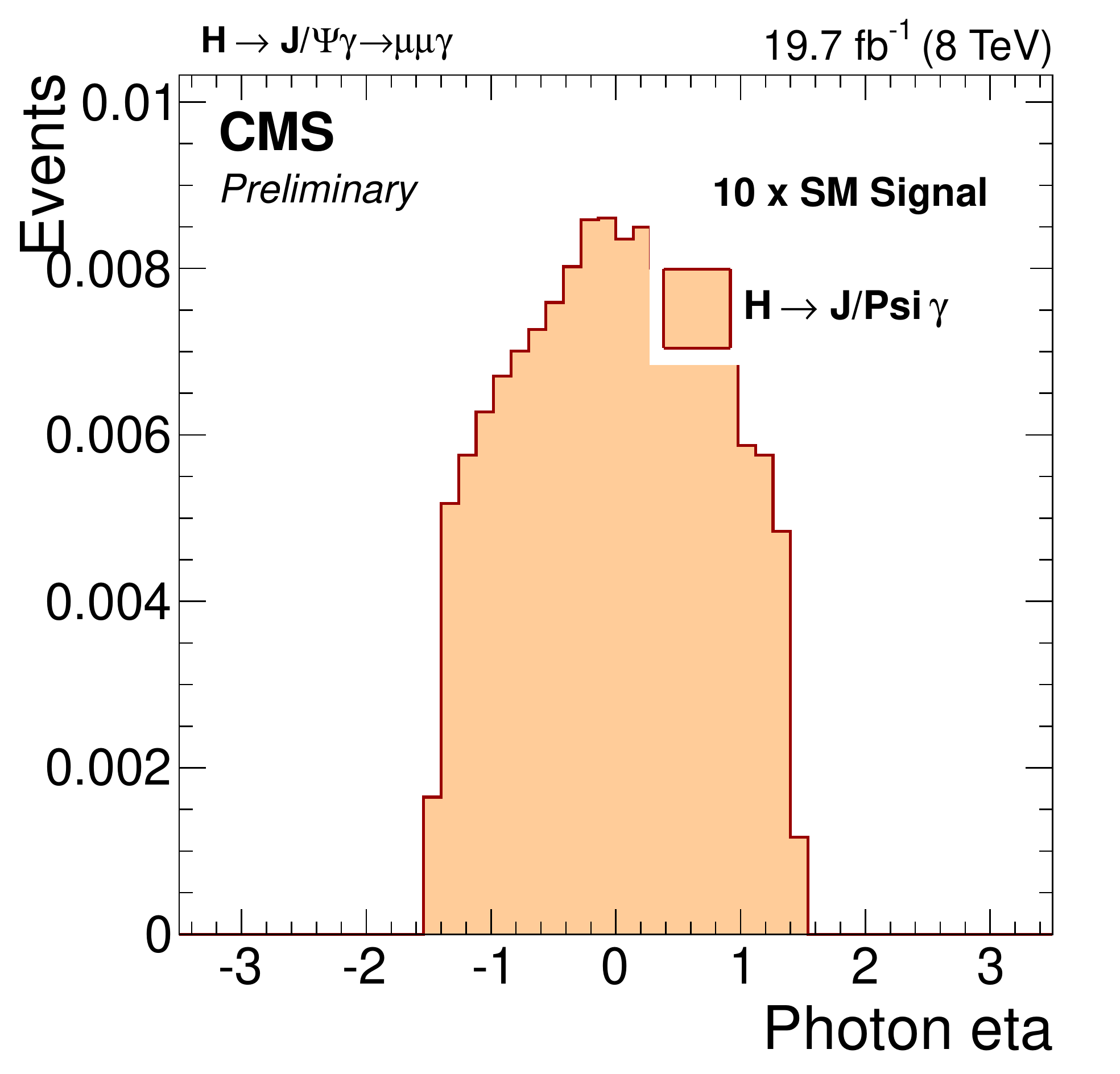}\\
  \includegraphics[width=0.3\textwidth]{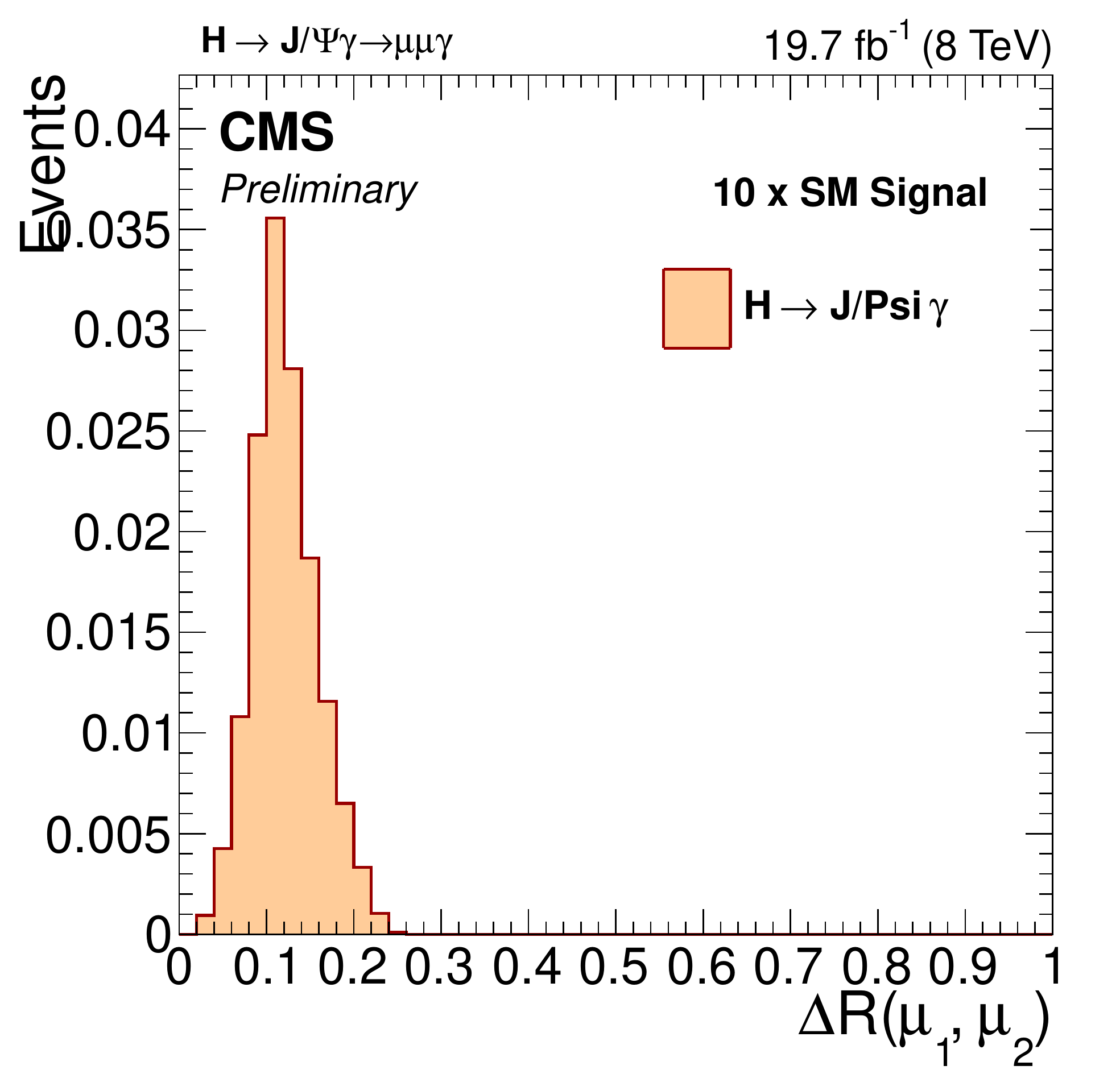}~
  \includegraphics[width=0.3\textwidth]{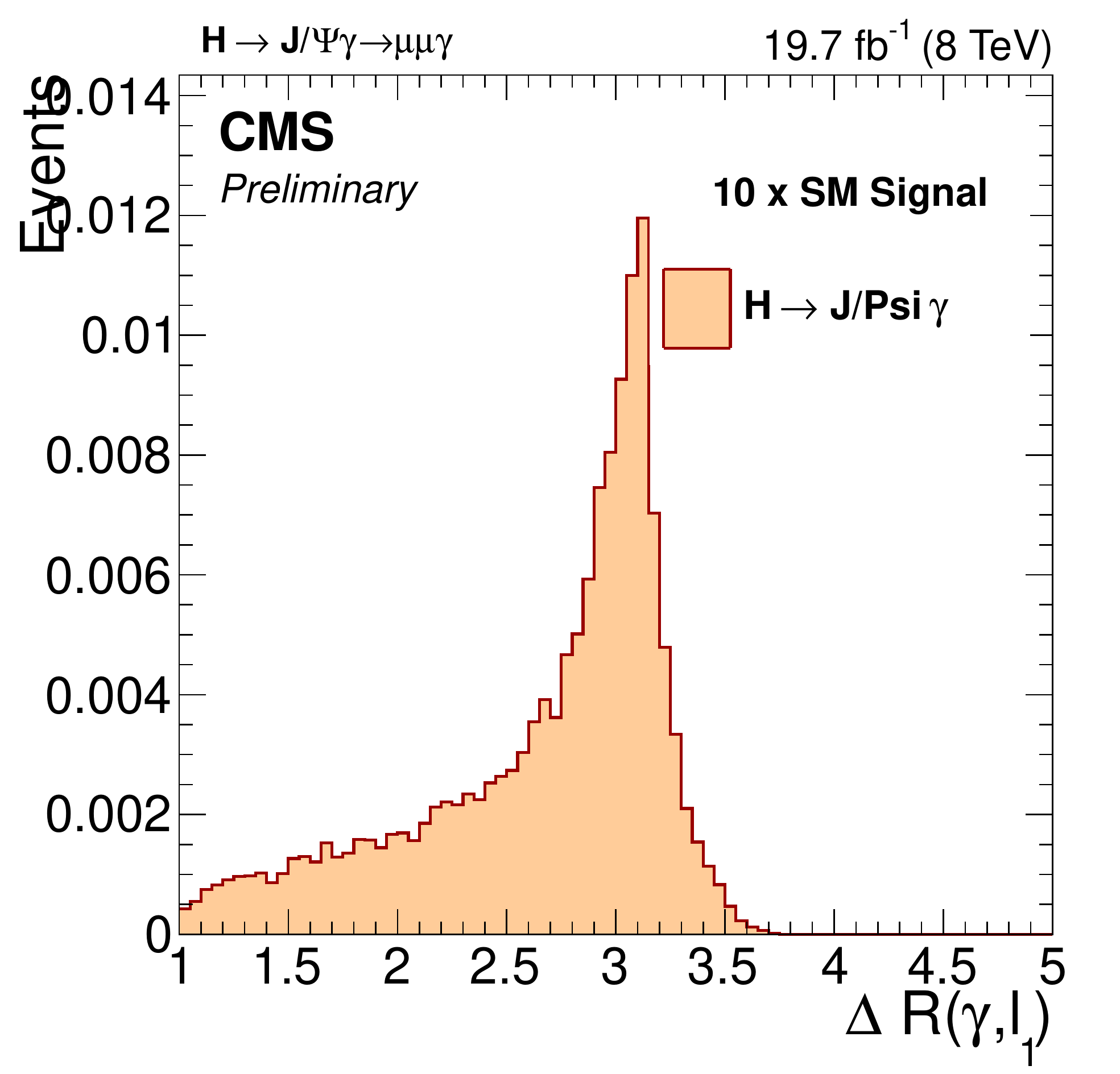}~
  \includegraphics[width=0.3\textwidth]{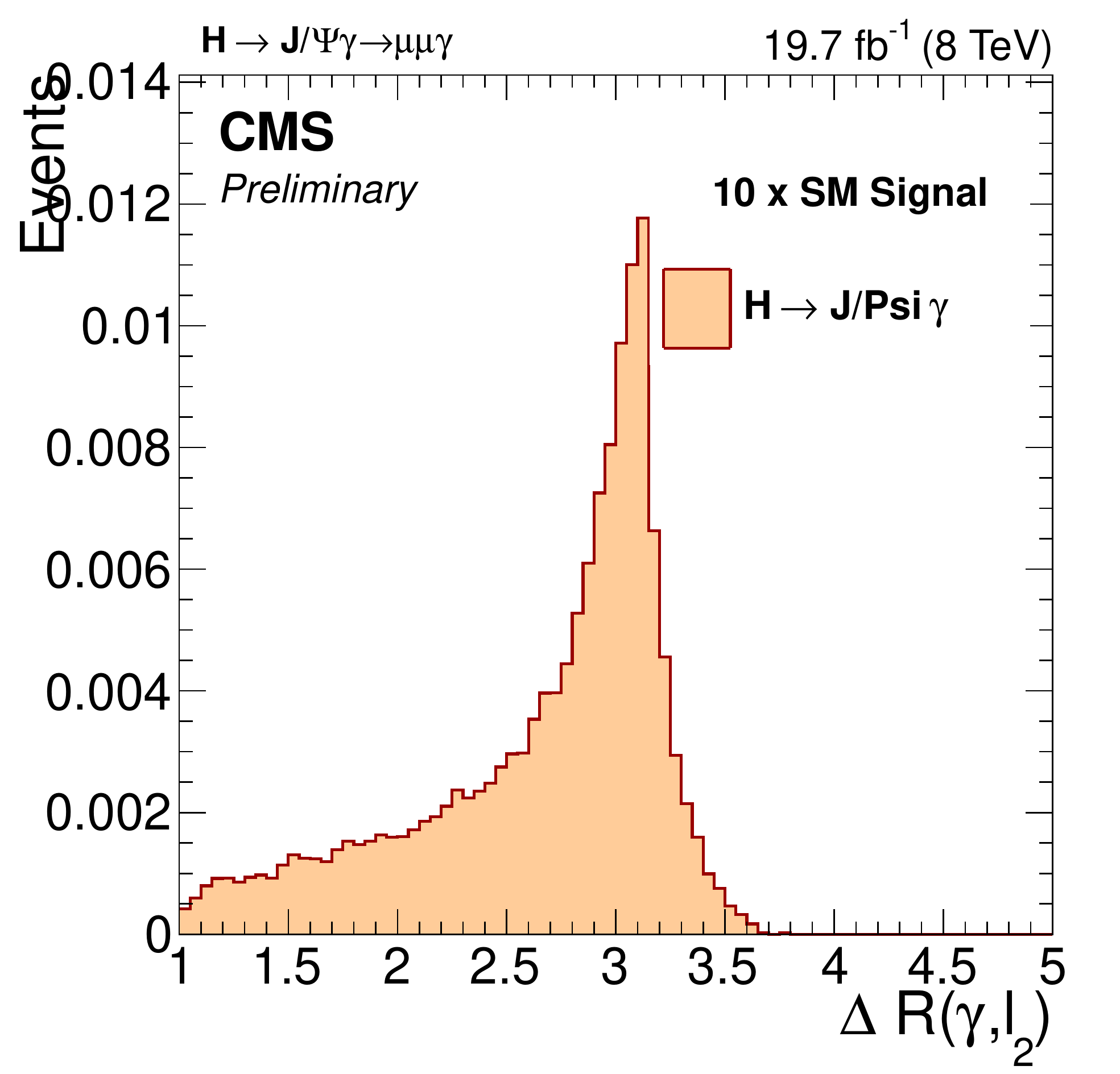}~
  \caption[Distributions of the key variables from the $\PH\to(\JPsi)\gamma$ signal
  process.]{Distributions of the key variables from the $\PH\to(\JPsi)\gamma$ signal
    process.  Transverse momenta of the muons, the dimuon system and the photon;
    pseudorapidity of the photon and the dimuon system; distances $\DR_{\eta\phi}$ between
    the objects.}
  \label{fig:dist-2}
\end{figure}

Table~\ref{tab:yields-jp} shows the yields obtained in the
$\PH\to(\JPsi)\gamma\to\mu\mu\gamma$ search. The signals both from $\JPsi$ and Dalitz
decay channels are shown. Figures~\ref{fig:dist-1} and \ref{fig:dist-2} show the
distributions of the relevant kinematic observables after the full event selection
(denoted by a (*) in Table~\ref{tab:yields-jp}).

\begin{table*}[htb]
  \begin{center}
  \caption{Event yield after each selection criteria for data and $\PH\to(\JPsi)\gamma\to\mu\mu\gamma$ signal with $m_\PH = 125\GeV$ for $L = 19.7\fbinv$.}
  \label{tab:yields-jp}
    \resizebox{\textwidth}{!}{
      \begin{tabular}{c lc|c|c}
        \hline
 &       & Observed events &  \multicolumn{2}{c}{Expected signal events} \\
 &       \multicolumn{1}{c}{Selection requirement}  &  in data   &  \multicolumn{1}{c|}{from $\PH\to\JPsi\gamma$} &\multicolumn{1}{c}{from $\PH\to\gamma^*\gamma$}\\
        \hline
 &       HLT; Photon: $p_T^{\gamma} > 25\,\GeV$, $|\eta^{\gamma}| < 1.444$    & 0.6M  & 0.023 & 5.31 \\
 &       Muon selection:  $p_T^{\mu 1}> 23\GeV$ and $p_T^{\mu 2} > 4\GeV$     & 57K   & 0.018 & 4.01 \\
 &       $m_{\mu\mu}< 20\GeV$, $\DR(\gamma,\mu) > 1$                          & 5714  & 0.018 & 3.37 \\
 &       Loose \JPsi selection: $2.5 < m_{\mu\mu} < 3.7\,\GeV$                & 820   & 0.017 & 0.27 \\
 &       $p_T^{\gamma} > 40\,\GeV$  and  $p_T^{\mu\mu} > 40\,\GeV$            & 221   & 0.015 & 0.23 \\\hline
*&       Tight \JPsi selection: $2.9 < m_{\mu\mu} < 3.3\,\GeV$                & 129   & 0.015 & 0.08 \\
 &       Fit region: $110 < m_{\mu\mu\gamma} < 150\,\GeV$                     & 48    & 0.015 & 0.08 \\
 &       Higgs mass: $122 < m_{\mu\mu\gamma} < 128\,\GeV$                     & 7     & 0.013 & 0.07 \\
        \hline
      \end{tabular}
    }
  \end{center}
\end{table*}

\begin{figure}[ht]
  \centering
  \includegraphics[width=0.3\textwidth]{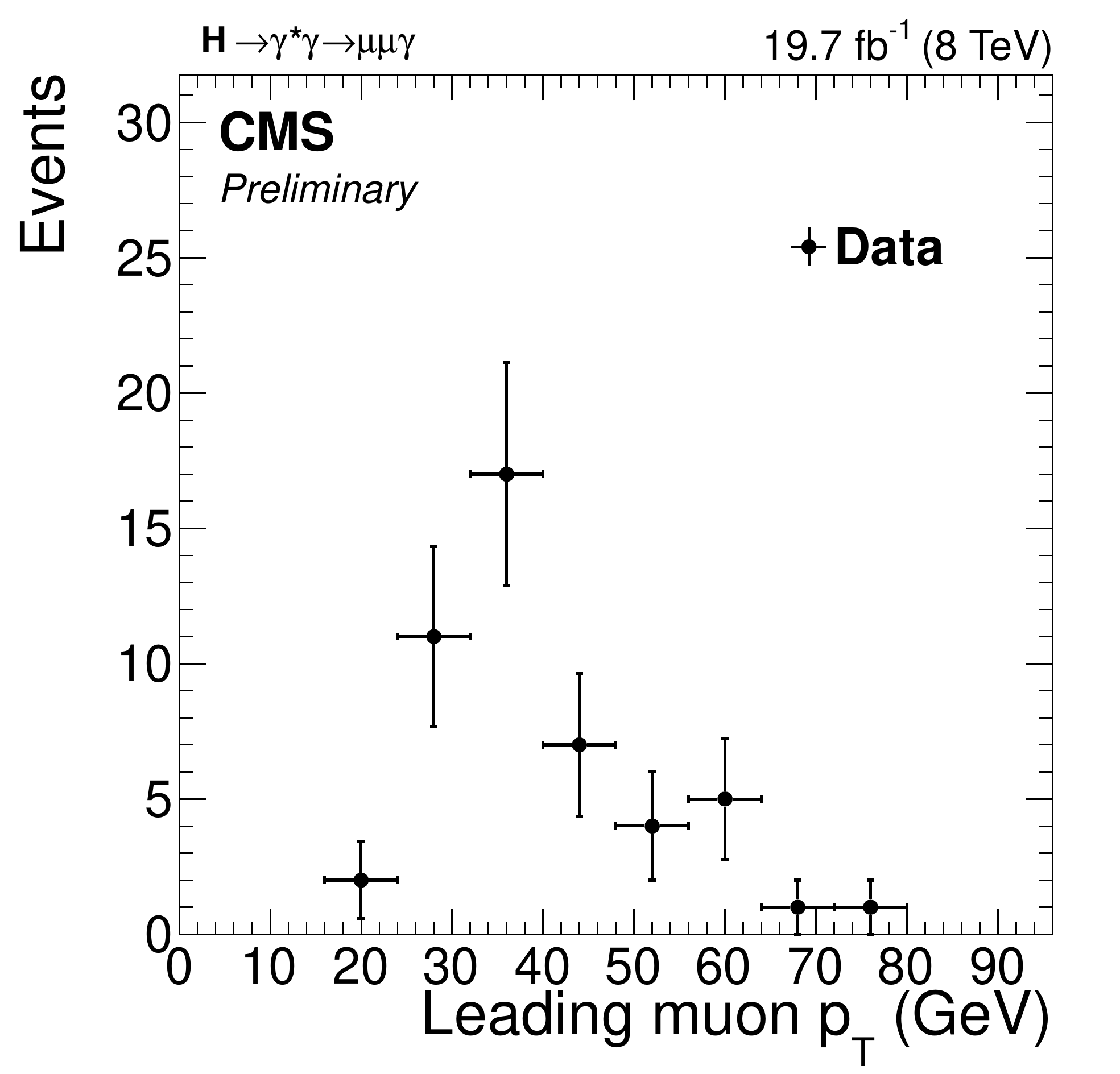}~
  \includegraphics[width=0.3\textwidth]{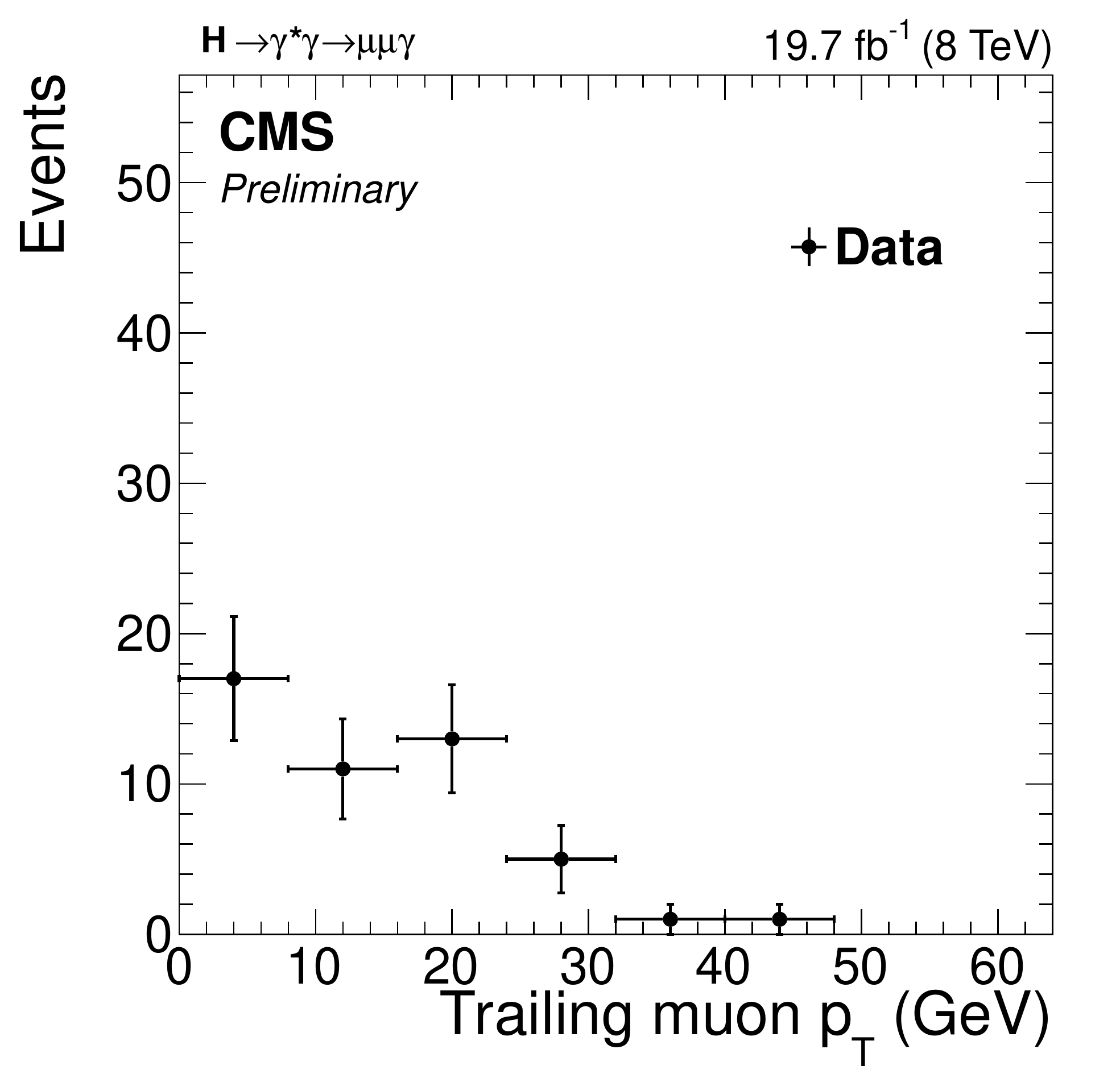}~
  \includegraphics[width=0.3\textwidth]{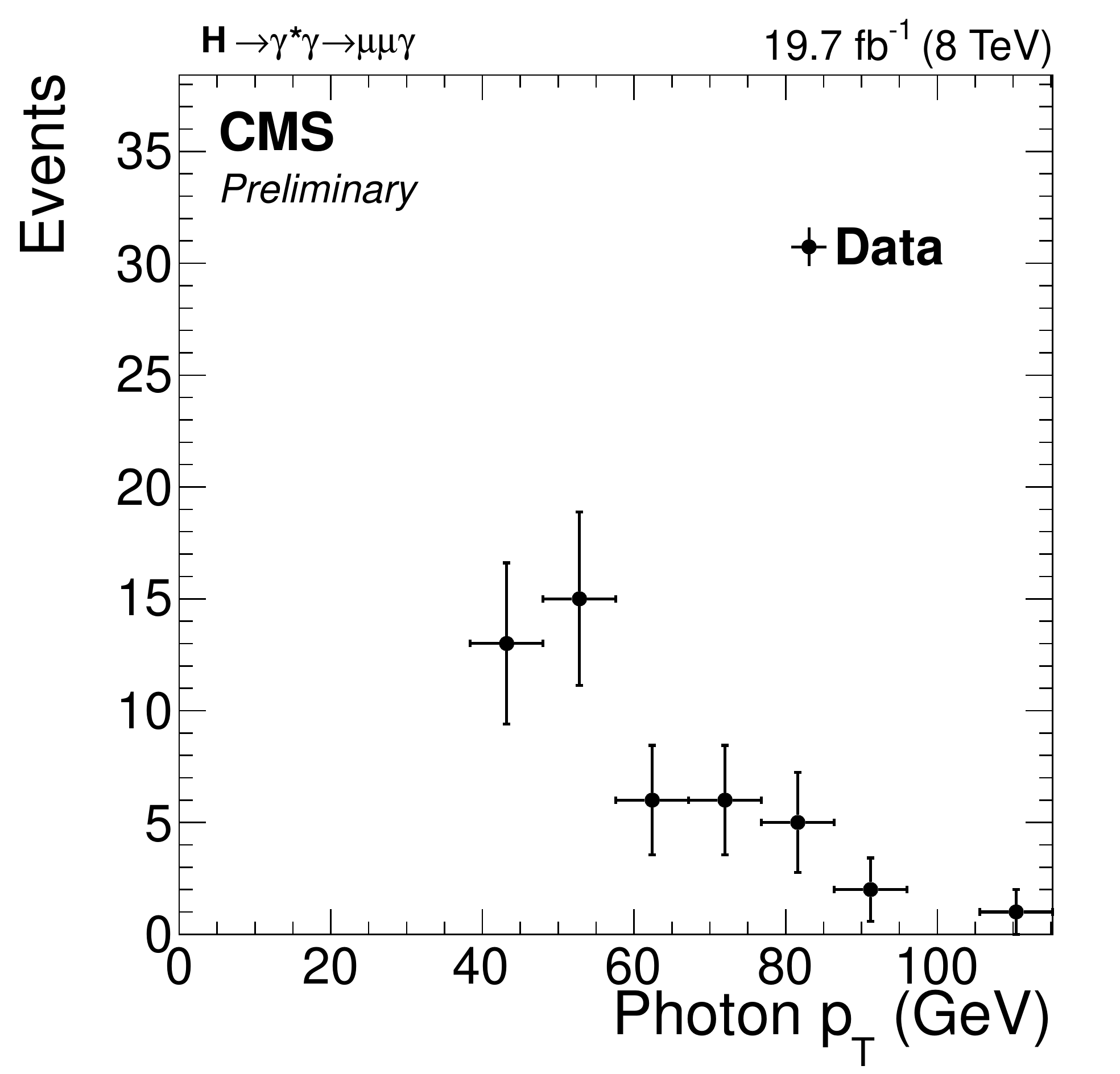}\\
  \includegraphics[width=0.3\textwidth]{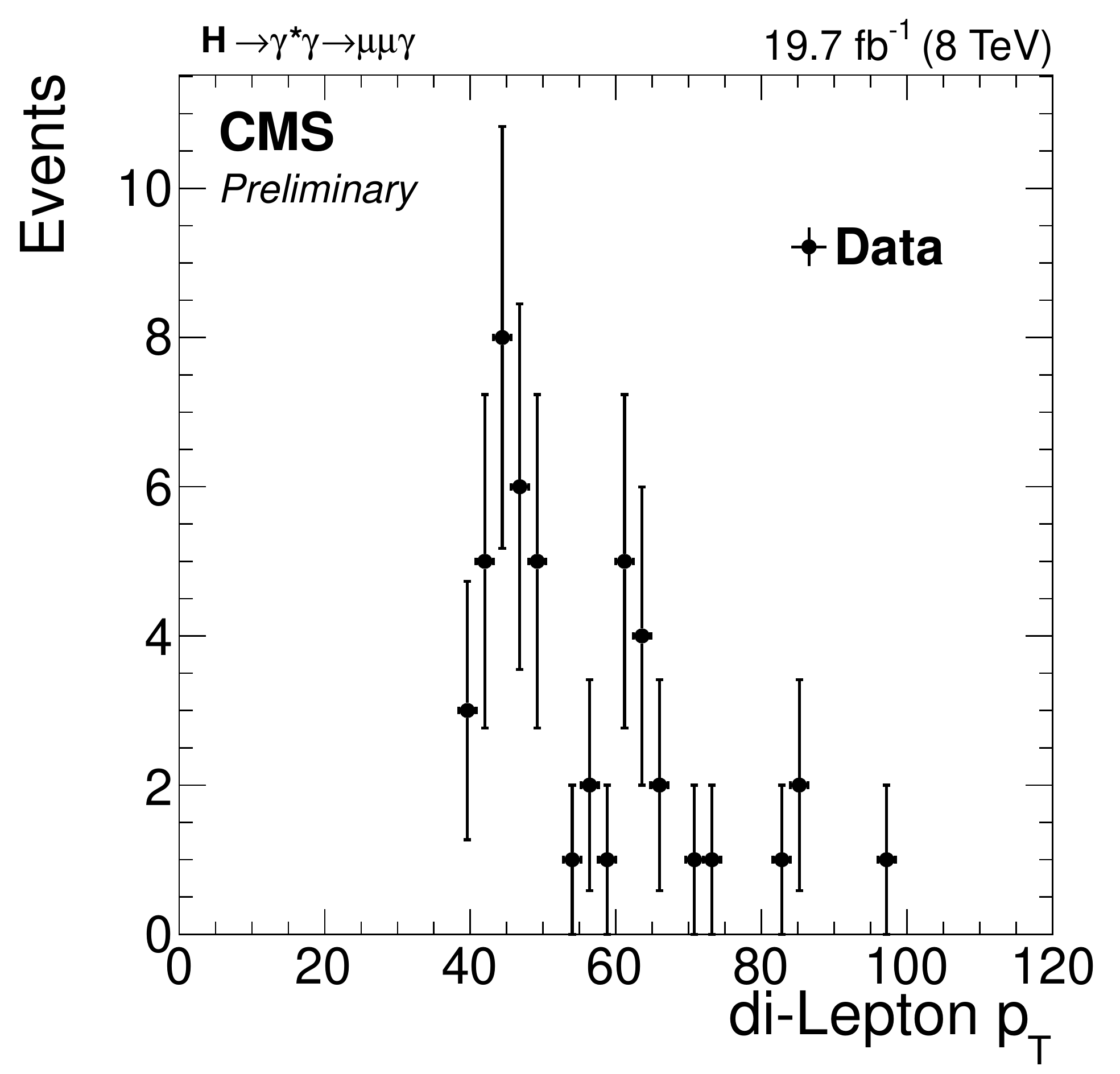}~
  \includegraphics[width=0.3\textwidth]{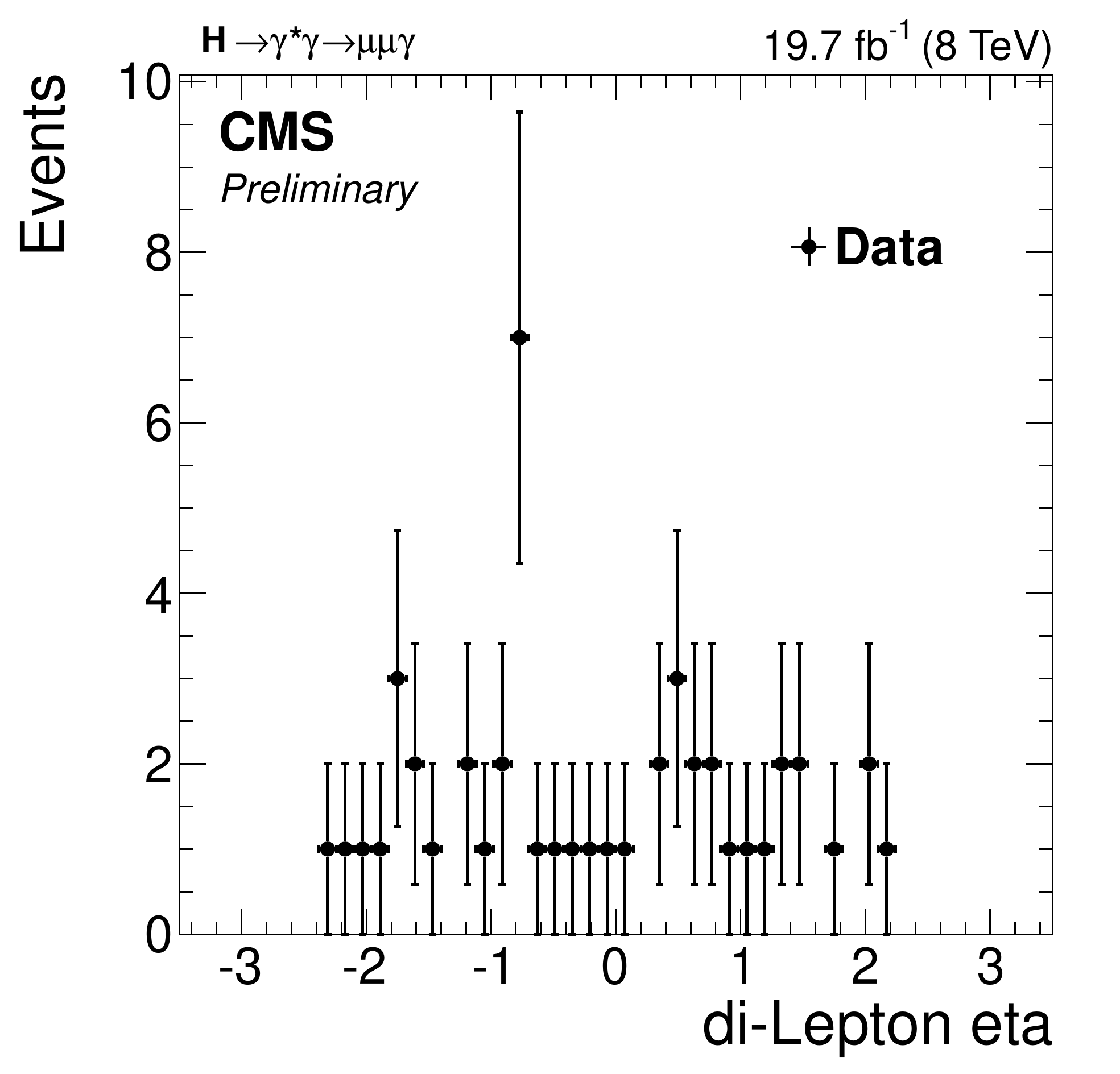}~
  \includegraphics[width=0.3\textwidth]{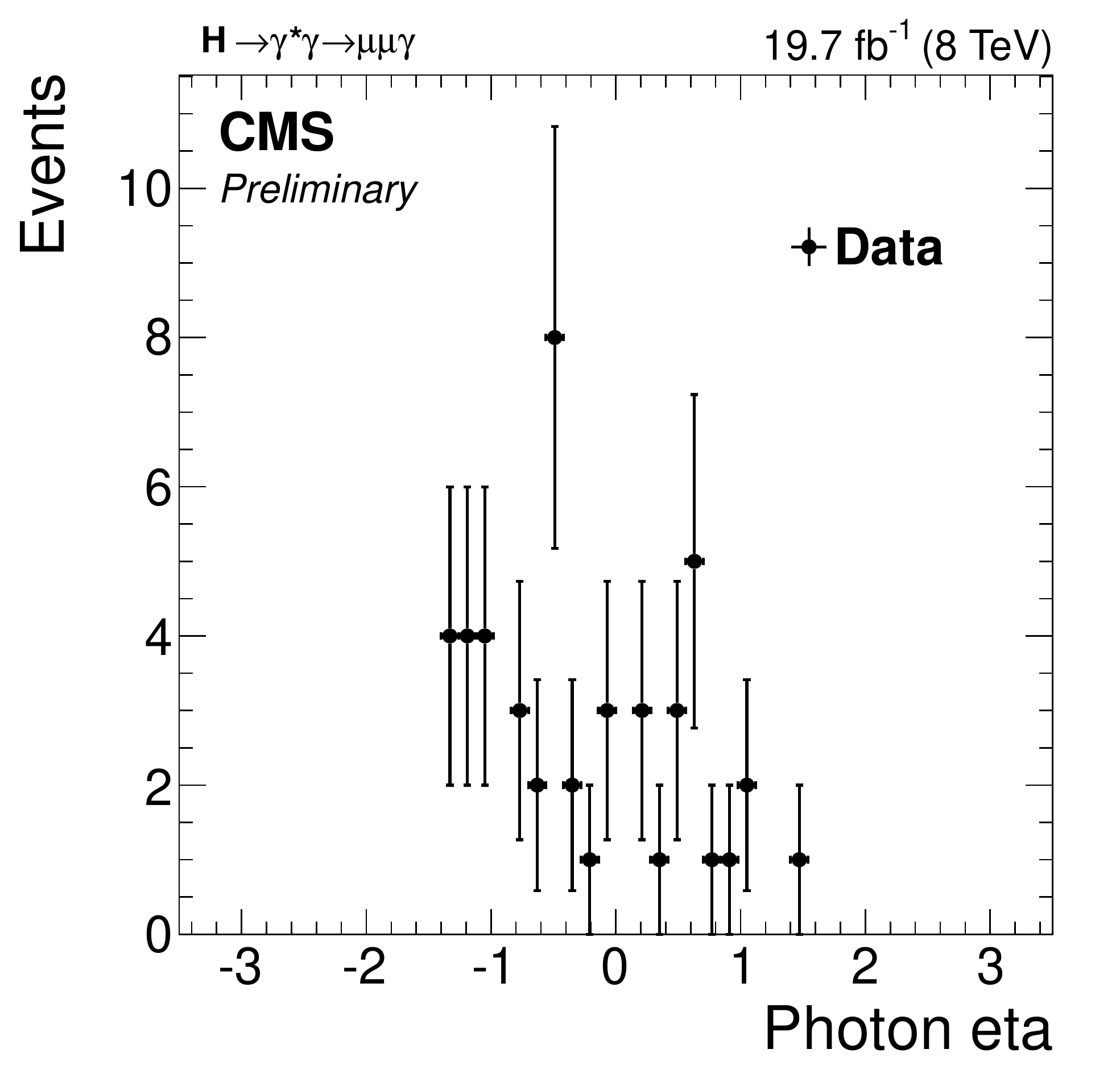}\\
  \includegraphics[width=0.3\textwidth]{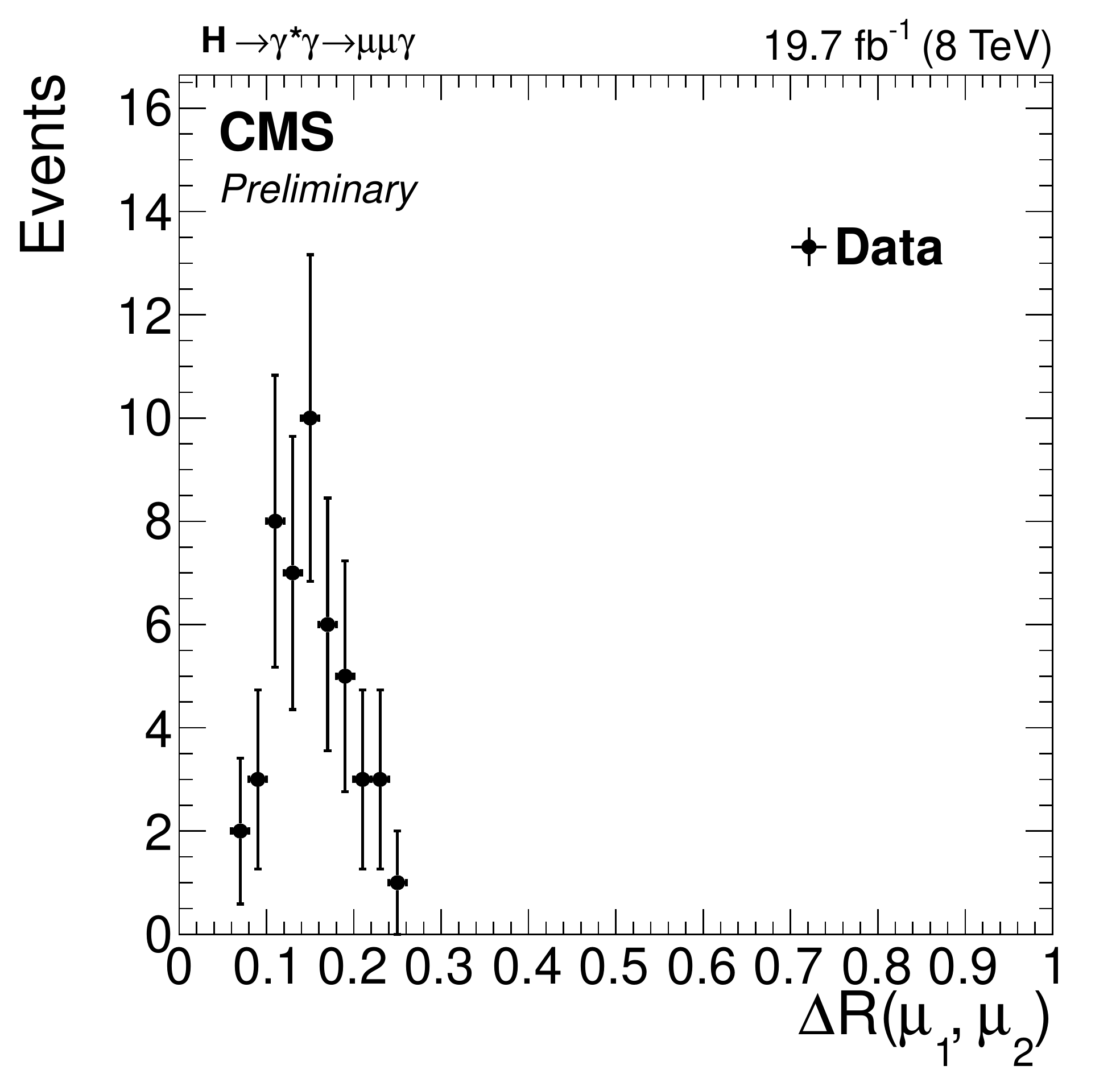}~
  \includegraphics[width=0.3\textwidth]{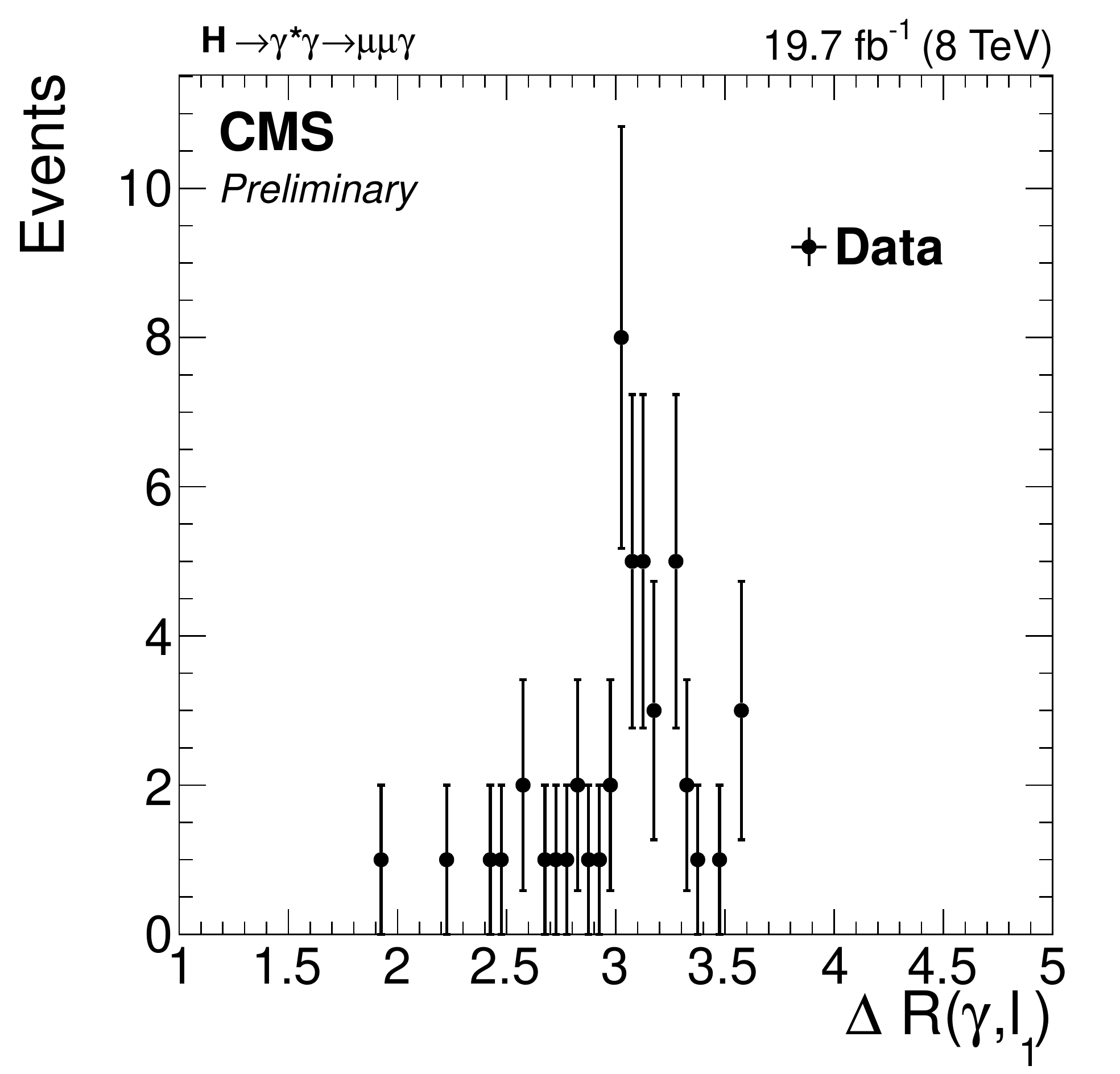}~
  \includegraphics[width=0.3\textwidth]{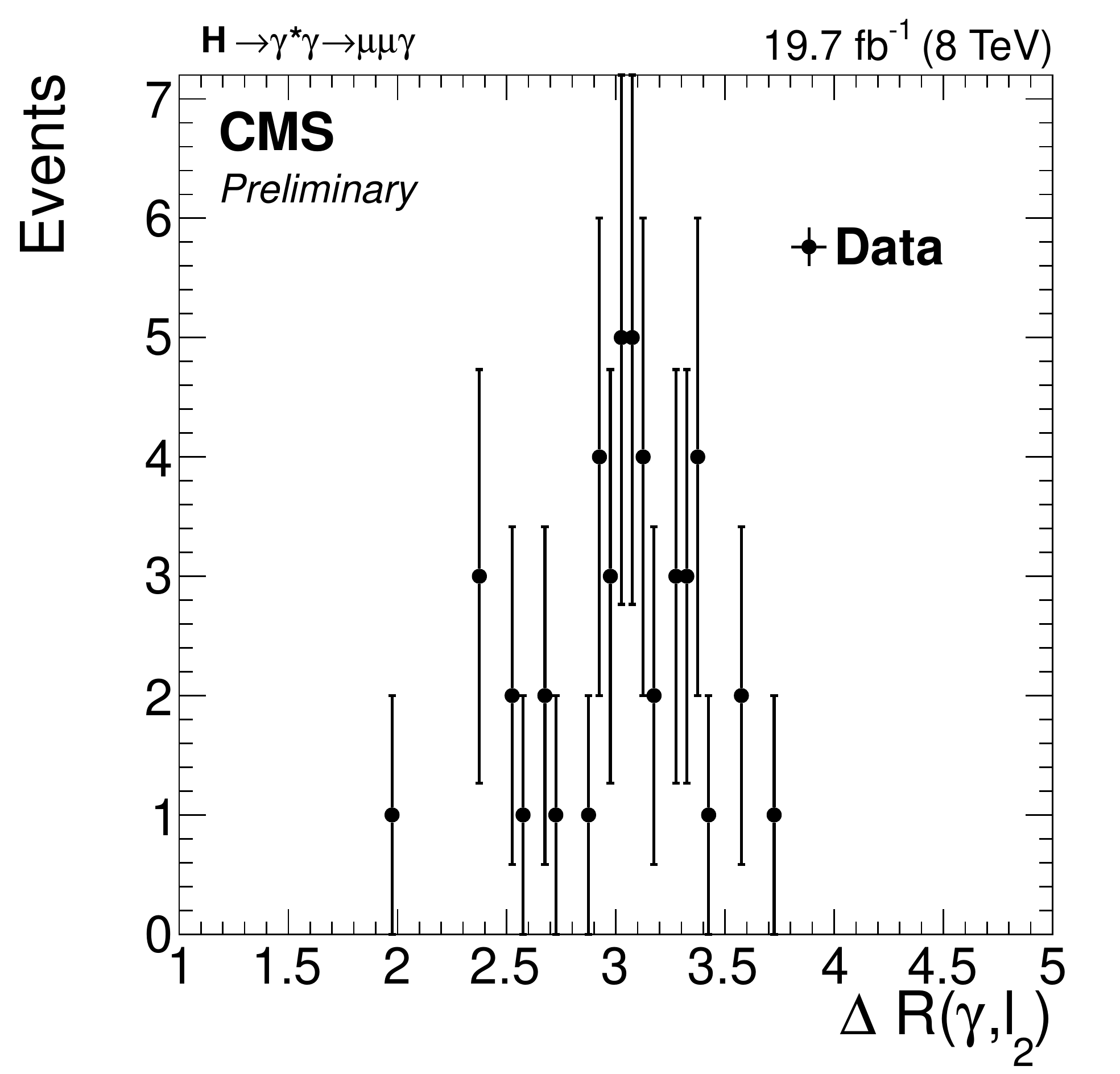}~
  \caption[Distributions of the key variables in data events after the full selection of
  the $\PH\to(\JPsi)\gamma$ search.]{Distributions of the key variables in data events
    after the full selection of the $\PH\to(\JPsi)\gamma$ search.  Transverse momenta of
    the muons, the dimuon system and the photon; pseudorapidity of the photon and the
    dimuon system; distances $\DR_{\eta\phi}$ between the objects.}
  \label{fig:dist-3}
\end{figure}

\clearpage
\chapter{Effect of the Systematics on the Expected Limits}
\label{sec:lim-syst}
Here I compare the expected limits in the Dalitz search analysis with and without the
systematic uncertainties enabled in the limit setting procedure.  I only consider the most
sensitive \textit{EB} category in muon channel.  I find that applying the systematic
uncertainties change the median of the expected limit by less than 4--5\%, and also widens
the error band by a little bit. Overall, the analysis sensitivity is limited by the size
of the data sample, and not by the systematic uncertainties.

\begin{figure}[ht]
  \centering
  \includegraphics[width=0.41\textwidth]{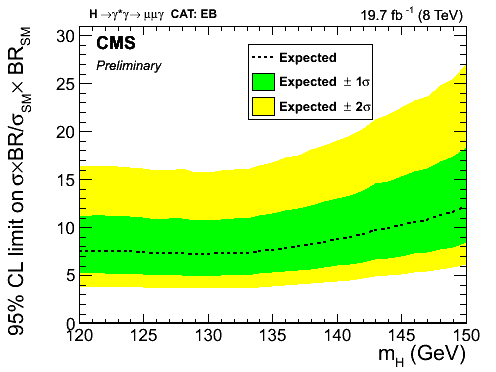}~
  \includegraphics[width=0.41\textwidth]{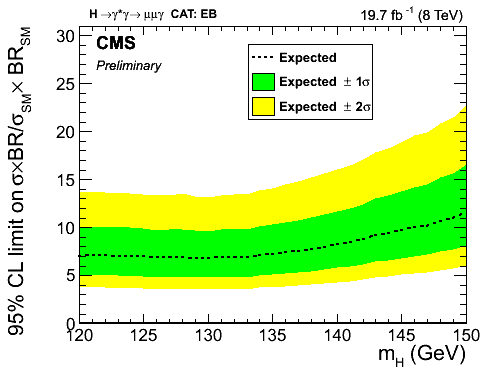}\\
  \includegraphics[width=0.41\textwidth]{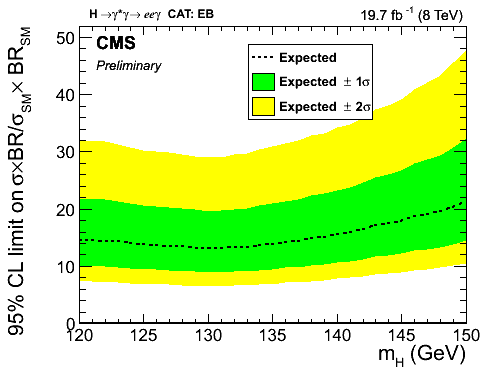}~
  \includegraphics[width=0.41\textwidth]{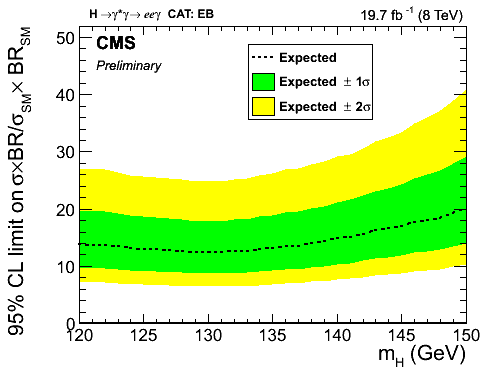}\\
  \includegraphics[width=0.41\textwidth]{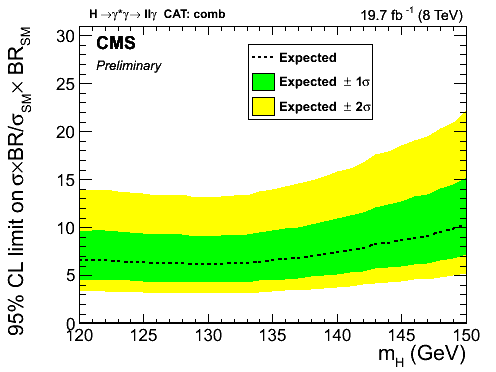}~
  \includegraphics[width=0.41\textwidth]{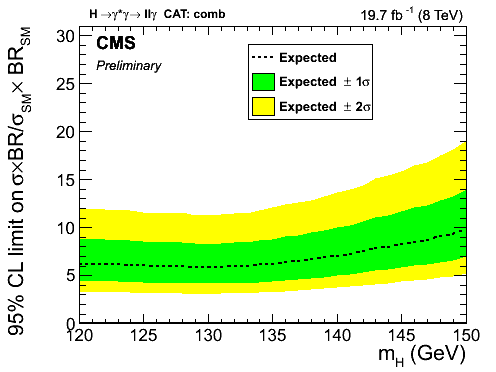}\\
  \caption[Expected limits vs $m_\PH$ with and without systematic uncertainties.]{Expected
    limits vs $m_\PH$, left: default; right: without systematics.  Top: muon channel in EB
    category; middle: electron channel; bottom: combined.}
  \label{fig:sys-lim}
\end{figure}

%

\end{document}